\documentclass[12pt, draftclsnofoot, onecolumn]{IEEEtran}
\usepackage{algorithm}
\usepackage{algpseudocode}
\usepackage{amsfonts}
\usepackage{amsmath}
\usepackage{amssymb}
\usepackage{bm}
\usepackage{breqn}
\usepackage[font=small,labelsep=period]{caption}
\usepackage{cases}
\usepackage{cite}
\usepackage{color}
\usepackage{enumerate}
\usepackage{epsfig}
\usepackage{epstopdf}
\usepackage{exscale}
\usepackage{float}
\usepackage{graphics}
\usepackage{graphicx}
\usepackage{latexsym}
\usepackage{multicol}
\usepackage{multirow}
\usepackage{relsize}
\usepackage{setspace}
\usepackage{stfloats}
\usepackage{subfigure}
\usepackage{url}

\begin{document}

%\captionsetup[figure]{labelformat=simple, labelsep=period}

\title{Performance Analysis on RIS-Aided Wideband Massive MIMO OFDM Systems with Low-Resolution ADCs}
\IEEEoverridecommandlockouts

%%%%%%%%%%%%%%%%%%%%%%%%%%%%%%%%%%%%%%%%%%%%%%%%%%%%%%%%%%%%%%%%%%%%%%%%%%%%%%
\author{Xianzhe~Chen,
        Hong~Ren,~\emph{Member},~\emph{IEEE},
        Cunhua~Pan,~\emph{Senior~Member},~\emph{IEEE},
        Zhangjie~Peng,
        Kangda~Zhi,
        Yong~Liu,
        Xiaojun~Xi,
        Ana~Garcia~Armada,~\emph{Fellow},~\emph{IEEE},
        and~Cheng-Xiang~Wang,~\emph{Fellow},~\emph{IEEE}

\thanks{\emph{(Corresponding author: Cunhua Pan)}}

\thanks{X. Chen, H. Ren and C. Pan are with the National Mobile Communications Research Laboratory, Southeast University, Nanjing 210096, China (e-mail: 230248154@seu.edu.cn; hren@seu.edu.cn; cpan@seu.edu.cn).}

\thanks{Z. Peng is with the College of Information, Mechanical and Electrical Engineering, Shanghai Normal University, Shanghai 200234, China, also with the National Mobile Communications Research Laboratory, Southeast University, Nanjing 210096, China, and also with the Shanghai Engineering Research Center of Intelligent Education and Bigdata, Shanghai Normal University, Shanghai 200234, China (e-mail: pengzhangjie@shnu.edu.cn).}

\thanks{Kangda Zhi is with the School of Electrical Engineering and Computer Science, Technical University of Berlin, 10623 Berlin (e-mail: kdzhi@mail.ustc.edu.cn).}

\thanks{Yong Liu and Xiaojun Xi are with the Wireless Technology Laboratory, Huawei Technologies Co., Ltd., Shanghai 201206, China (e-mail: liu.liuyong@huawei.com; xixiaojun@huawei.com).}

\thanks{Ana~Garcia~Armada  is with the Department of Signal Theory and Communications, University Carlos III of Madrid, 28911 Leganes, Spain (e-mail: agarcia@tsc.uc3m.es).}

\thanks{Cheng-Xiang Wang is with the National Mobile Communications Research Laboratory, School Information of Science and Engineering, Southeast University, Nanjing 210096, China, and also with the Pervasive Communication Research Center, Purple Mountain Laboratories, Nanjing 211111, China (e-mail: chxwang@seu.edu.cn).}
}

\maketitle

\newtheorem{lemma}{Lemma}
\newtheorem{theorem}{Theorem}
\newtheorem{remark}{Remark}
\newtheorem{corollary}{Corollary}
\newtheorem{proposition}{Proposition}

\begin{abstract}
This paper investigates a reconfigurable intelligent surface (RIS)-aided wideband massive multiple-input multiple-output (MIMO) orthogonal frequency division multiplexing (OFDM) system with low-resolution analog-to-digital converters (ADCs).
Frequency-selective Rician fading channels are considered, and the OFDM data transmission process is presented in time domain.
This paper derives the closed-form approximate expression of the uplink achievable rate,
based on which the asymptotic system performance is analyzed when the number of the antennas at the base station and the number of reflecting elements at the RIS grow to infinity.
Besides, the power scaling laws of the considered system are revealed to provide energy-saving insights.
Furthermore, this paper proposes a gradient ascent-based algorithm to design the phase shifts of the RIS for maximizing the minimum user rate.
Finally, numerical results are presented to verify the correctness of analytical conclusions and draw insights.


\end{abstract}

\section{Introduction}

In recent years, the reconfigurable intelligent surface (RIS) has been regarded as a promising technique in the future sixth generation (6G) communication networks \cite{9140329,9237116,9328501,9415664,9475160,9828501,9847080,10054381}.
Composed of multiple programmable reflecting elements, the RIS could be used to extend the coverage and improve the performance of communication systems.
Specifically, the RIS customizes the wireless propagation environment by imposing a flexible phase shift on incident signals independently at each reflecting element,
such that it could strengthen the desired signals or weaken the interference signals.
Additionally, the RIS is free from radio frequency (RF) chains and amplifiers, which makes its power consumption and hardware cost much lower than that of the conventional relay.

Thanks to these attractive merits, the RIS has drawn extensive research interest.
For example,
an RIS-aided multi-user downlink system was investigated in \cite{8982186},
where the beamforming matrix and the phase shift vector were jointly designed for maximizing the weighted sum rate of all users.
In \cite{9180053}, the optimal beamforming was obtained for minimizing the transmit power, based on imperfect cascaded channels.
For RIS-aided multi-user millimeter-wave (mmWave) systems,
a low-overhead channel estimation strategy was proposed in \cite{9732214},
exploiting the correlation and the sparsity of multi-user channels.
%In \cite{9309152}, the authors investigated an RIS-aided multi-user uplink system,
%and focused on the tradeoff between the energy and spectral efficiency.
A closed-form expression was derived for the achievable rate of an RIS-aided multi-pair device-to-device system in \cite{9366346},
and the phase shift vector was optimized for maximizing the sum rate.
In \cite{9734027}, the benefits of active RIS-aided systems were quantified.

On the other hand,
the massive multiple-input multiple-output (MIMO) technique has been widely studied in recent years, due to its advantages of mitigating co-channel interference and reducing transmit power \cite{6736761,6816003,6832435}.
To meet the increasing demands for future communication systems,
recent studies have integrated RIS into massive MIMO systems for further improving the system performance of spectral and energy efficiency.
A closed-form expression for the sum secrecy rate was derived in \cite{9935294},
and a genetic algorithm was proposed to design the phase shift vector for maximizing the rate.
When spatially correlated channels and artificial noise (AN) were considered,
the sum secrecy rate was derived in closed form in \cite{10015836}.
Based on that, a gradient-based algorithm was developed to jointly optimize the power fraction of AN and the phase shift vector for maximizing the rate.
In \cite{9973349}, spatially correlated Rician fading channels and electromagnetic interference were considered,
and a closed-form expression was derived for the uplink achievable rate under imperfect channel state information (CSI).
Then, a gradient ascent-based algorithm was further proposed to solve the minimum user rate maximization problem.

However, conventional massive MIMO systems require a large number of high-resolution analog-to-digital converters/digital-to-analog converters (ADCs/DACs),
leading to excessive power consumption and hardware cost.
To tackle this issue, the low-resolution ADC/DAC scheme was proposed as a promising solution \cite{7979627,9075256,8629287}.
For example, efficient channel estimation methods were proposed for RIS-aided massive MIMO systems in the presence of low-resolution ADCs \cite{9999186,10029363}.
Besides, a closed-form expression was derived for the uplink achievable rate with low-resolution ADCs in \cite{9833357}, under the consideration of the transceiver hardware impairment and the phase error at the RIS.
Then the authors optimized the phase shift vector for maximizing the sum rate.
In \cite{9483943}, the downlink achievable rate was derived in closed form with low-resolution DACs, and the phase shift vector was designed for maximizing the sum rate.

It should be noted that the aforementioned works focused on narrowband communication systems with frequency-flat channels.
However, frequency-selective channels are more common in practical scenarios due to the different delays of the multi-path channels \cite{9786750}.
To combat the frequency-selective fading, the orthogonal frequency division multiplexing (OFDM) is commonly used in wideband communication systems.
Specifically, the asymptotic uplink rate was derived in \cite{7600443} with infinite number of channel taps for wideband massive MIMO OFDM systems with one-bit ADCs,
while in \cite{9356722}, a closed-form expression of the uplink rate was derived with finite number of channel taps for the systems with arbitrary-resolution ADCs.
An RIS-aided wideband single-input single-output (SISO) OFDM system was investigated in \cite{8937491}
where an element-grouping scheme for the RIS was proposed and the reflection coefficients of the RIS were optimized to maximize the rate.
%In \cite{8937491}, the authors investigated an RIS-aided wideband single-input single-output (SISO) OFDM system.
%They proposed an element-grouping scheme at the RIS, and optimized the reflection coefficients of the RIS for maximizing the rate.
Furthermore, the power allocation and the RIS phase shifts were jointly optimized for maximizing the rate in \cite{9039554}.
In \cite{9610122}, a majorization-minimization-based algorithm was proposed to optimize the RIS phase shifts for maximizing the data rate with low computational costs.
In \cite{9389801}, an RIS-aided wideband MIMO OFDM system was studied, and the precoding matrix at the base station (BS) and the phase shifts vector at the RIS were jointly optimized to maximize the downlink rate.

To the best of our knowledge, only a few studies have investigated RIS-aided wideband massive MIMO OFDM systems.
Although the works \cite{9918631} and \cite{10041805} have studied RIS-aided wideband massive MIMO OFDM systems,
they merely focused on the mmWave bands.
There are no studies considering the RIS-aided wideband massive MIMO OFDM systems in sub-6 GHz frequency band,
which is a typical application scenario in 5G and beyond wireless systems.
Furthermore, low-resolution ADCs/DACs were not considered in the existing works on RIS-aided wideband massive MIMO OFDM systems,
and therefore their impacts on the system performance are fully unknown.
Since the quantization at ADCs/DACs is a nonlinear operation \cite{4407763},
the results under frequency-flat channels cannot be extended directly to the wideband systems with frequency-selective channels.
Therefore, it is of significance to investigate an RIS-aided wideband massive MIMO system with low-resolution ADCs/DACs.

Motivated by the above reasons,
in this paper, we study an RIS-aided wideband massive MIMO OFDM system with low-resolution ADCs.
Under frequency-selective Rician fading channels,
we formulate the OFDM data transmission process,
derive the achievable rate in closed form,
analyze the properties of various key system parameters,
and optimize the phase shifts.
The main contributions of this paper are summarized as follows:

\begin{itemize}
  \item   We model an RIS-aided wideband massive MIMO OFDM system in the sub-6 GHz frequency band. Low-resolution ADCs are equipped at the antennas of the BS.
         All channels are assumed to be frequency-selective Rician fading channels.
         Besides, since most existing works only gave a brief introduction for the time-domain transmission of RIS-aided OFDM systems, we formulate the OFDM data transmission process in the time domain in detail. Thus, the equations for the time domain transmission can be introduced more reasonably and naturally.
  \item We derive the closed-form approximate expression for the uplink rate of the considered system.
       Based on that, we analyze the asymptotic performance of the considered system,
       when both the number of the antennas, $N_{\mathrm{b}}$, and the number of reflecting elements, $N_{\mathrm{r}}$, grow to infinity.
      Furthermore, we unveil the power scaling laws for the considered system to draw energy-saving insights.
      It is proved that when $N_{\mathrm{b}}$ and $N_{\mathrm{r}}$ grow to infinity,
      if the RIS is not aligned with any users, the transmit power of each user can be scaled down at most proportionally to $\frac{1}{N_{\mathrm{b}}N_{\mathrm{r}}}$ while the considered system maintains a non-zero converging rate.
      Moreover, when the RIS is aligned with User $n$,
      the transmit power of User $n$ can be further reduced proportionally to $\frac{1}{N_{\mathrm{b}}N^2_{\mathrm{r}}}$,
      while guaranteeing certain system performance.
      The main analytical conclusions are verified by the numerical results.
      %We find that when the number of the antennas $N_{\mathrm{b}}$ and the number of reflecting elements $N_{\mathrm{r}}$ go to infinity, if the RIS is aligned with User $n$, the asymptotic rate of User $n$ is on the order of $\mathcal{O} \left( \log _2\left( N_{\mathrm{r}} \right) \right)$
  \item To improve the system performance of the considered system, we address the phase shift optimization problem with the aim of maximizing the minimum user rate.
        We propose a gradient ascent-based algorithm with accelerated convergence rate.
        The numerical results show that the proposed algorithm can greatly improve the performance of the considered system, which verifies its effectiveness.
\end{itemize}

The remainder of this paper is organized as follows:
Section II
presents the model of RIS-aided wideband massive MIMO OFDM systems with low-resolution ADCs under frequency-selective Rician fading channels,
and introduces the OFDM data transmission process.
Section III
derives the closed-form approximate expressions for the uplink achievable rate, and analyzes the system features.
Section IV
proposes a gradient ascent method-based algorithm to solve the phase shift optimization problem for maximizing the minimum user rate.
Section V
provides the numerical results.
Section VI
gives a brief conclusion.

\emph{Notations:}
In this paper,
scalars, vectors and matrices are respectively denoted by lower case letters, bold lower case letters and bold upper case letters.
The matrix inverse, conjugate-transpose, transpose and conjugate operations are respectively denoted by the superscripts ${\left(  \cdot  \right)^ {-1} }$, ${\left(  \cdot  \right)^H}$, ${\left(  \cdot  \right)^T}$ and ${\left(  \cdot  \right)^{*}}$.
We use ${\rm tr}\left(  \cdot  \right)$, $\left\|  \cdot  \right\|$ and ${\rm E}\left\{  \cdot  \right\}$ to denote trace, Euclidean 2-norm and the expectation operations, respectively.
The notation $\odot$ stands for cyclic convolution, and $\otimes$ denotes the Kronecker Product.
Operation $\bmod$ means taking the remainder after division,
and operation $\lfloor \cdot \rfloor$ means taking the integer part.
${\left[ {\bf{A}} \right]_{i,j}}$ denotes the $\left( {i,j} \right)$th element of matrix ${\bf{A}}$.
The matrix ${{\bf{I}}_N}$ denotes an $N \times N$ identity matrix.
We denote a circularly symmetric complex Gaussian vector ${\bf a}$ with zero mean and covariance ${\bf{\Sigma }}$ by ${\bf a} \thicksim {\cal C}{\cal N}\left( {{\bf{0}},{\bf{\Sigma }}} \right)$.
Furthermore, all the subscripts of vectors and matrices are counted starting from zero for the ease of analysis.
%To simplify the analysis in this paper, subscript $L$ is frequently used, and $L \in \left\{ {A,B} \right\}$ for the subscript ${\left(  \cdot  \right)_L}$ in this paper. Furthermore, we assume $\left\{ {L,\bar L} \right\} = \left\{ {A,B} \right\}$ in one specific equation.

\section{System Model}

\begin{figure}[t]
%\vspace{-0.2cm}
  \centering
  %\vspace{-2cm}
%  \includegraphics[scale=0.5]{picture/System.jpg}\\
  \includegraphics[scale=0.90]{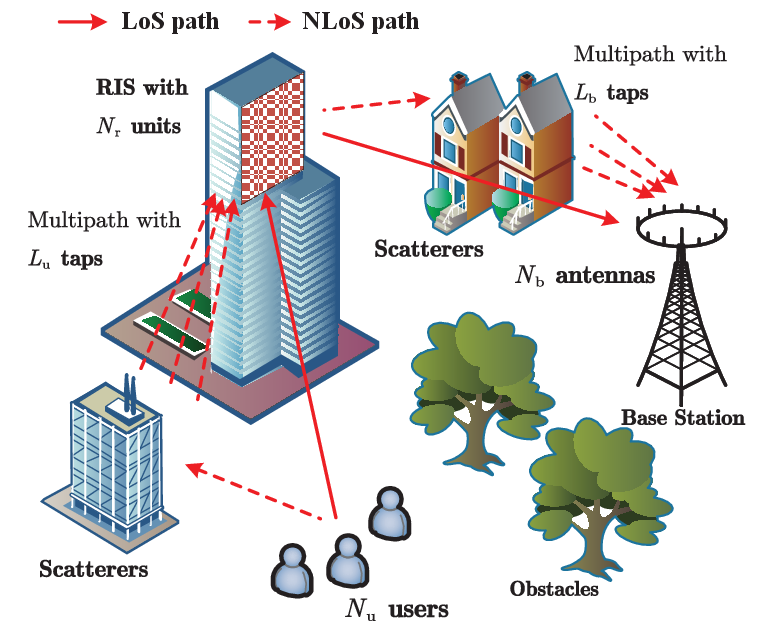}\\  %\vspace{-0.5cm}
  \caption{System Model}\label{p1} \vspace{-0.5cm}
\end{figure}

Fig. \ref{p1} illustrates an RIS-aided multi-user wideband massive MIMO OFDM system with low-resolution ADCs.
Considering the reciprocity of channels, we focus on the uplink communication of the system.
As is shown, each of the $N_{\mathrm{u}}$ single-antenna users sends $N_{\mathrm{c}}$-sub-carrier OFDM signals to an $N_{\mathrm{b}}$-antenna BS with the aid of an $N_{\mathrm{r}}$-unit RIS,
while the direct link from the users to the BS is neglected due to the severe blockage.

\subsection{Channel Model: Channel Impulse Response}

Due to the different delays of the multi-path channels in practical scenarios, frequency-selective Rician fading channels are taken into consideration in this paper, which is a general channel model consisting of both line-of-sight (LoS) and non-line-of-sight (NLoS) parts.

We first present the array response of the uniform square planar array (USPA),
which is deployed at both the RIS and the BS.
For the USPA with a size of $\sqrt{X} \times \sqrt{X}$, the array response of Unit $i$ is modeled as \cite{9743440}
\begin{equation}\label{Array_response}
  \left[ \mathbf{a}_X\left( \phi ^1,\phi ^2 \right) \right] _i=e^{j\frac{2\pi d}{\lambda}\left( x_i\sin \phi ^1\sin \phi ^2+y_i\cos \phi ^2 \right)}\triangleq a_{X,i}\left( \phi ^1,\phi ^2 \right),
\end{equation}
where $x_i=  i  \; \mathrm{mod} \; \sqrt{X}$,
$y_i= \lfloor \frac{i}{\sqrt{X}} \rfloor$, $i=0,1,...,X-1$.
$d$ is the unit spacing, and $\lambda$ is the carrier wavelength.
Meanwhile, we assume that the channels from the users to the RIS and those from the RIS to the BS have  $L_{\mathrm{u}}$ and $L_{\mathrm{b}}$ channel impulse response taps, respectively.
Then, we introduce the channel impulse responses for the User-to-RIS link and the RIS-to-BS link.

For the User-to-RIS link pair $(i,j)$, i.e., the pair comprised of RIS Element $i$ ($i=0,1,...,N_{\mathrm{r}}-1$) and User $j$ ($j=0,1,...,N_{\mathrm{u}}-1$),
the $l$-th, $0\leq l < L_{\mathrm{u}}$, channel tap is \cite{TSE2005,4357323,9699402}
\begin{equation}\label{R-U_tap}
  h_{\mathrm{u},l}^{\left( i,j \right)}=\begin{cases}
	\sqrt{\frac{K_{\mathrm{u},j}}{K_{\mathrm{u},j}+1}} \sigma _{\mathrm{u},0,j} \bar{h}_{\mathrm{u}}^{\left( i,j \right)}+\sqrt{\frac{1}{K_{\mathrm{u},j}+1}}\tilde{h}_{\mathrm{u},0}^{\left( i,j \right)}, \;\;\; l=0\\
	\tilde{h}_{\mathrm{u},l}^{\left( i,j \right)}, \;\;\; 1\leqslant l\leqslant L_{\mathrm{u}}-1\\
\end{cases}.
\end{equation}
It is noted that the first tap, i.e., the tap with $l=0$, corresponds to the superposition of the channels with the shortest delays,
which contains the LoS path and follows the Rician fading,
while the other taps only contain the NLoS paths and follow the Rayleigh fading.
In \eqref{R-U_tap}, the LoS component $\bar{h}_{\mathrm{u}}^{\left( i,j \right)}$ in the first tap is given by
\begin{equation}\label{R-U_tap_LoS}
  \bar{h}_{\mathrm{u}}^{\left( i,j \right)}=a_{N_{\mathrm{r}},i}\left( \phi _{\mathrm{r},j}^{aa},\phi _{\mathrm{r},j}^{ea} \right),
\end{equation}
with $\phi _{\mathrm{r},j}^{aa}$ and $\phi _{\mathrm{r},j}^{ea}$ being the azimuth and elevation angles of arrival (AoAs) from User $j$ to the RIS, respectively.
The NLoS component $\tilde{h}_{\mathrm{u},l}^{\left( i,j \right)}$ follows the distribution of $\mathcal{C} \mathcal{N} \left( 0,\sigma _{\mathrm{u},l,j}^{2} \right)$,
which is subject to the normalized power constraint $\sum_{l=0}^{L_{\mathrm{u}}-1}{\sigma _{\mathrm{u},l,j}^{2}}=1$.
Additionally, $K_{\mathrm{u},j}$ represents the Rician factor.

For the RIS-to-BS link pair $(i,j)$, i.e., the pair comprised of BS Antenna $i$ ($i=0,1,...,N_{\mathrm{b}}-1$) and RIS Element $j$ ($j=0,1,...,N_{\mathrm{r}}-1$),
the $l$-th, $0\leq l < L_{\mathrm{b}}$, channel tap is
\begin{equation}\label{B-R_tap}
  h_{\mathrm{b},l}^{\left( i,j \right)}=\begin{cases}
	\sqrt{\frac{K_{\mathrm{b}}}{K_{\mathrm{b}}+1}}\sigma _{\mathrm{b},0}\bar{h}_{\mathrm{b}}^{\left( i,j \right)}+\sqrt{\frac{1}{K_{\mathrm{b}}+1}}\tilde{h}_{\mathrm{b},0}^{\left( i,j \right)}, \;\;\; l=0\\
	\tilde{h}_{\mathrm{b},l}^{\left( i,j \right)}, \;\;\; 1\leqslant l\leqslant L_{\mathrm{b}}-1\\
\end{cases},
\end{equation}
with the LoS component $\bar{h}_{\mathrm{b}}^{\left( i,j \right)}$ in the first tap given by
\begin{equation}\label{B-R_tap_LoS}
  \bar{h}_{\mathrm{b}}^{\left( i,j \right)}=a_{N_{\mathrm{b}},i}\left( \phi _{\mathrm{b}}^{aa},\phi _{\mathrm{b}}^{ea} \right) a_{N_{\mathrm{r}},j}^{\ast}\left( \phi _{\mathrm{r}}^{ad},\phi _{\mathrm{r}}^{ed} \right),
\end{equation}
where $\phi _{\mathrm{b}}^{aa}$ and $\phi _{\mathrm{b}}^{ea}$ stand for the azimuth and elevation AoAs from the RIS to the BS, respectively.
$\phi _{\mathrm{r}}^{ad}$ and $\phi _{\mathrm{r}}^{ed}$ are the azimuth and elevation angles of departure, respectively.
Besides, in \eqref{B-R_tap}, the NLoS component $\tilde{h}_{\mathrm{b},l}^{\left( i,j \right)}\sim \mathcal{C} \mathcal{N} \left( 0,\sigma _{\mathrm{b},l}^{2} \right)$ is subject to $\sum_{l=0}^{L_{\mathrm{b}}-1}{\sigma _{\mathrm{b},l}^{2}}=1$.
Besides, $K_{\mathrm{b}}$ represents the Rician factor.

Additionally, for the ease of expression, we define the tap vector $\mathbf{h}_{\mathrm{x}}^{\left( i,j \right)}$ as
\begin{equation}\label{h_ij}
  \mathbf{h}_{\mathrm{x}}^{\left( i,j \right)}\triangleq \left[ h_{\mathrm{x},0}^{\left( i,j \right)},h_{\mathrm{x},1}^{\left( i,j \right)},...,h_{\mathrm{x},L_{\mathrm{x}}-1}^{\left( i,j \right)},\underset{N_{\mathrm{c}}-L_{\mathrm{x}}}{\underbrace{0,0,...0}} \right] ^T, \;\;\; \mathrm{x}\in \left\{ \mathrm{u},\mathrm{b} \right\} .
\end{equation}

\subsection{Data Transmission}

In this subsection, we introduce the OFDM data transmission in the time domain.
We denote a time-domain OFDM symbol of length $N_{\mathrm c}$ transmitted from User $n$ as
\begin{equation}\label{TD_su}
  \mathbf{s}_n=\left[ s_n\left[ 0 \right] ,...,s_n\left[ N_{\mathrm{c}}-1 \right] \right] ^T.
\end{equation}
To combat the frequency selectivity of the channels, an $N_{\mathrm{cp}}$-length cyclic prefix (CP) is attached to each OFDM symbol,
forming the following symbol
%Thus, we denote the $N_{\mathrm{c}}$-length OFDM symbol with the $N_{\mathrm{cp}}$-length CP as
\begin{equation}\label{TD_du}
  \mathbf{d}_n = \left[ \overbrace{ {\underbrace{s_n\left[ N_{\mathrm{c}}-N_{\mathrm{cp}} \right] }_{d_n\left[ 0 \right]}  },..., {\underbrace{s_n\left[ N_{\mathrm{c}}-1 \right] }_ {d_n\left[ N_{\mathrm{cp}}-1 \right]}  } }^ {N_{\mathrm{cp}}-{\mathrm{length\;CP}}  } ,  {\underbrace{s_n\left[ 0 \right] }_ {d_n\left[ N_{\mathrm{cp}} \right]}  },...,   {\underbrace{s_n\left[ N_{\mathrm{c}}-1 \right] }_ {d_n\left[ N_{\mathrm{c}}+N_{\mathrm{cp}}-1 \right]} } \right] ^T ,
  m\in \left[ 0,N_{\mathrm{c}}+N_{\mathrm{cp}}-1 \right].
\end{equation}
It is worth noted that
\begin{equation}\label{TD_du_p1}
  d_n\left[ m \right] =s_n\left[ \left( m-N_{\mathrm{cp}} \right) \; \mathrm{mod} \; N_{\mathrm{c}} \right],  m\in \left[ 0,N_{\mathrm{c}}+N_{\mathrm{cp}}-1 \right],
\end{equation}
\begin{equation}\label{TD_du_p2}
  d_n\left[ m \right] =d_n\left[ m+N_{\mathrm{c}} \right] ,  m\in \left[ 0,N_{\mathrm{cp}}-1 \right].
\end{equation}

At time instant $m$ of an OFDM symbol (including CP),
the time-domain OFDM signal from User $n$ to Element $r$ of the RIS is expressed as
\begin{equation}\label{TD_utor_1}
  y_{\mathrm{ris},r}\left[ m \right] =\sum_{l_1=0}^{L_{\mathrm{u}}-1}{\sqrt{\beta _{\mathrm{u},n}}h_{\mathrm{u},l_1}^{\left( r,n \right)}\sqrt{p_n}d_n\left[ m-l_1 \right]}, \;
  m\in \left[ 0,N_{\mathrm{c}}+N_{\mathrm{cp}}+L_{\mathrm{u}}-2 \right],
\end{equation}
where $p_n$ is the transmit power of User $n$.
Scalar $\beta _{\mathrm{u},n}$ stands for large-scale fading coefficient.
The corresponding signal vector of \eqref{TD_utor_1} is denoted as
\begin{equation}\label{TD_utor_2}
  \mathbf{y}_{\mathrm{ris},r}=\left[ \left( \mathbf{y}_{\mathrm{ris},r}^{1} \right) ^T,\left( \mathbf{y}_{\mathrm{ris},r}^{2} \right) ^T,\left( \mathbf{y}_{\mathrm{ris},r}^{3} \right) ^T \right] ^T,
\end{equation}
with
\begin{equation}\label{TD_utor_3}
  \mathbf{y}_{\mathrm{ris},r}^{1}=\left[ y_{\mathrm{ris},r}\left[ 0 \right] ,...,y_{\mathrm{ris},r}\left[ L_{\mathrm{u}}-2 \right] \right] ^T,
\end{equation}
\begin{equation}\label{TD_utor_4}
  \mathbf{y}_{\mathrm{ris},r}^{2}=\left[ y_{\mathrm{ris},r}\left[ L_{\mathrm{u}}-1 \right] ,...,y_{\mathrm{ris},r}\left[ N_{\mathrm{cp}} \right] ,...,y_{\mathrm{ris},r}\left[ N_{\mathrm{c}}+N_{\mathrm{cp}}-1 \right] \right] ^T,
\end{equation}
\begin{equation}\label{TD_utor_5}
  \mathbf{y}_{\mathrm{ris},r}^{3}=\left[ y_{\mathrm{ris},r}\left[ N_{\mathrm{c}}+N_{\mathrm{cp}} \right] ,...,y_{\mathrm{ris},r}\left[ N_{\mathrm{c}}+N_{\mathrm{cp}}+L_{\mathrm{u}}-2 \right] \right] ^T.
\end{equation}
It can be observed that $\mathbf{y}_{\mathrm{ris},r}^{1}$ is related to the previous OFDM symbol, and $\mathbf{y}_{\mathrm{ris},r}^{3}$ depends on the next OFDM symbol.
Considering the interval of $m\in \left[ N_{\mathrm{cp}},N_{\mathrm{c}}+N_{\mathrm{cp}}-1 \right]$, we have
\begin{equation}\label{TD_utor_6}
  y_{\mathrm{ris},r}\left[ m \right] = \! \sum_{l_1=0}^{L_{\mathrm{u}}-1}{\sqrt{\beta _{\mathrm{u},n}}h_{\mathrm{u},l_1}^{\left( r,n \right)}\sqrt{p_n}d_n\left[ m-l_1 \right]}
  \overset{\left( a \right)}{=} \! \sum_{l_1=0}^{L_{\mathrm{u}}-1}{\sqrt{\beta _{\mathrm{u},n}}h_{\mathrm{u},l_1}^{\left( r,n \right)}\sqrt{p_n}s_n\left[ \left( m-N_{\mathrm{cp}}-l_1 \right) \; \mathrm{mod} \; N_{\mathrm{c}} \right]},
\end{equation}
where step $\left( a \right)$ is based on \eqref{TD_du_p1}.
Denote by
\begin{equation}\label{TD_utor_7}
  \bar{\mathbf{y}}_{\mathrm{ris},r}=\left[  {\underbrace{y_{\mathrm{ris},r}\left[ N_{\mathrm{cp}} \right] }_{\bar{y}_{\mathrm{ris},r}\left[ 0 \right]} },..., {\underbrace{y_{\mathrm{ris},r}\left[ N_{\mathrm{c}}+N_{\mathrm{cp}}-1 \right] }_ {\bar{y}_{\mathrm{ris},r}\left[ N_{\mathrm{c}}-1 \right]}  } \right] ^T.
\end{equation}
From \eqref{h_ij} and \eqref{TD_utor_7}, Equation \eqref{TD_utor_6} can be transformed into
\begin{equation}\label{TD_utor_8}
  \bar{\mathbf{y}}_{\mathrm{ris},r}=\sqrt{p_n\beta _{\mathrm{u},n}}\mathbf{h}_{\mathrm{u}}^{\left( r,n \right)}\odot \mathbf{s}_n ,
\end{equation}
where the notation $\odot$ stands for cyclic convolution.
Using the property of the discrete Fourier transform (DFT), we have
\begin{equation}\label{TD_utor_9}
  \mathrm{DFT}\left( \bar{\mathbf{y}}_{\mathrm{ris},r} \right) _i=\sqrt{p_n\beta _{\mathrm{u},n}}\sqrt{N_{\mathrm{c}}}\mathrm{DFT}\left( \mathbf{h}_{\mathrm{u}}^{\left( r,n \right)} \right) _i\cdot \mathrm{DFT}\left( \mathbf{s}_n \right) _i, \; i=0,1,...,N_{\mathrm c}-1.
\end{equation}
Thus we obtain
\begin{equation*}
  \mathbf{F}\bar{\mathbf{y}}_{\mathrm{ris},r}=\sqrt{p_n\beta _{\mathrm{u},n}}\sqrt{N_{\mathrm{c}}}\mathrm{diag}\left( \mathbf{Fh}_{\mathrm{u}}^{\left( r,n \right)} \right) \mathbf{Fs}_n \Longrightarrow
\end{equation*}
\begin{equation}\label{TD_utor_10}
  \bar{\mathbf{y}}_{\mathrm{ris},r}=\sqrt{p_n}\mathbf{F}^H\mathrm{diag}\left( \sqrt{N_{\mathrm{c}}\beta _{\mathrm{u},n}}\mathbf{Fh}_{\mathrm{u}}^{\left( r,n \right)} \right) \mathbf{Fs}_n
  \triangleq \sqrt{p_n}\mathbf{F}^H\mathbf{G}_{\mathrm{u}}^{\left( r,n \right)}\mathbf{Fs}_n
  \triangleq \sqrt{p_n}\mathbf{H}_{\mathrm{u}}^{\left( r,n \right)}\mathbf{s}_n,
\end{equation}
where
matrix $\mathbf{F}$ represents the DFT matrix, with its entries expressed as
\begin{equation}\label{DFT_fij}
  \left[ \mathbf{F} \right] _{i,j}=\frac{1}{\sqrt{N_{\mathrm{c}}}}e^{-j\frac{2\pi}{N_{\mathrm{c}}}ij}\triangleq f_{i,j},  \;\;\;  i,j=0,1,...,N_{\mathrm{c}}-1 .
\end{equation}
Matrix $\mathbf{G}_{\mathrm{u}}^{\left( r,n \right)}$ is the frequency-domain channel given by
\begin{equation}\label{R-U_FD_channel_ij}
  \mathbf{G}_{\mathrm{u}}^{\left( r,n \right)}=\mathrm{diag}\left( \sqrt{N_{\mathrm{c}}\beta _{\mathrm{u},n}}\mathbf{Fh}_{\mathrm{u}}^{\left( r,n \right)} \right).
\end{equation}
Matrix $\mathbf{H}_{\mathrm{u}}^{\left( r,n \right)}\triangleq \mathbf{F}^H\mathbf{G}_{\mathrm{u}}^{\left( r,n \right)}\mathbf{F}$ is the circulant time-domain channel between User $n$ and Element $r$ of the RIS.

Then, after being transmitted from User $n$ and reflected by Element $r$ of the RIS,
for the time-domain OFDM signal received by Antenna $b$ of the BS,
the $N_{\mathrm{cp}}$-length CP is ignored, and the $N_{\mathrm{c}}$-length interval of $m\in \left[ N_{\mathrm{cp}},N_{\mathrm{c}}+N_{\mathrm{cp}}-1 \right]$ is selected as
\begin{equation}\label{TD_utortob_1}
  y_{\mathrm{bs},b}\left[ m \right] =\sum_{l_2=0}^{L_{\mathrm{b}}-1}{\sqrt{\beta _{\mathrm{b}}}h_{\mathrm{b},l_2}^{\left( b,r \right)}e^{j\theta _r}y_{\mathrm{ris},r}\left[ m-l_2 \right]}+z_b\left[ m \right],
\end{equation}
where
$\beta _{\mathrm{b}}$ stands for large-scale fading coefficient.
$\theta _r$ is the phase shift imposed by Element $r$ of the RIS,
and $z_b\left[ m \right]$ is the additive white Gaussian noise (AWGN).
Additionally, the corresponding vector of the received signal is denoted as
\begin{equation}\label{TD_utortob_2}
  \bar{\mathbf{y}}_{\mathrm{bs},b}=\left[ {\underbrace{y_{\mathrm{bs},b}\left[ N_{\mathrm{cp}} \right] }_ {\bar{y}_{\mathrm{bs},b}\left[ 0 \right]} },...,
  {\underbrace{y_{\mathrm{bs},b}\left[ N_{\mathrm{c}}+N_{\mathrm{cp}}-1 \right] }_ {\bar{y}_{\mathrm{bs},b}\left[ N_{\mathrm{c}}-1 \right]}   } \right] ^T.
\end{equation}

Note from \eqref{TD_utortob_1} that
\begin{equation}\label{TD_utortob_3}
  y_{\mathrm{bs},b}\left[ N_{\mathrm{cp}} \right] =\sum_{l_2=0}^{L_{\mathrm{b}}-1}{\sqrt{\beta _{\mathrm{b}}}h_{\mathrm{b},l_2}^{\left( b,r \right)}e^{j\theta _r}y_{\mathrm{ris},r}\left[ N_{\mathrm{cp}}-l_2 \right]}+z_b\left[ N_{\mathrm{cp}} \right] .
\end{equation}
It can be observed that the earliest signal in time on the right hand side of \eqref{TD_utortob_3} is $y_{\mathrm{ris},r} [ N_{\mathrm{cp}}-L_{\mathrm{b}}+1  ]$.
To ensure that $y_{\mathrm{ris},r} [ N_{\mathrm{cp}}-L_{\mathrm{b}}+1  ]$ is independent of the previous OFDM signal,
from \eqref{TD_utor_3}, the CP length $N_{\mathrm{cp}}$ should satisfy
\begin{equation}\label{TD_utortob_4}
  N_{\mathrm{cp}}-L_{\mathrm{b}}+1\geqslant L_{\mathrm{u}}-1\Longrightarrow N_{\mathrm{cp}}\geqslant L_{\mathrm{b}}+L_{\mathrm{u}}-2.
\end{equation}

From \eqref{TD_du_p2} and \eqref{TD_utor_1}, we have \vspace{0.4cm}
\begin{equation*}
\hspace{-7.2cm}
  y_{\mathrm{ris},r}\left[ m+N_{\mathrm{c}} \right] =\sum_{l_1=0}^{L_{\mathrm{u}}-1}{\sqrt{\beta _{\mathrm{u},n}}h_{\mathrm{u},l_1}^{\left( r,n \right)}\sqrt{p_n}d_n\left[ m+N_{\mathrm{c}}-l_1 \right]}
\end{equation*}
\begin{equation}\label{TD_utortob_5}
\hspace{0.6cm}
  =\sum_{l_1=0}^{L_{\mathrm{u}}-1}{\sqrt{\beta _{\mathrm{u},n}}h_{\mathrm{u},l_1}^{\left( r,n \right)}\sqrt{p_n}d_n\left[ m-l_1 \right]}=y_{\mathrm{ris},r}\left[ m \right],
  m\in \left[ L_{\mathrm{u}}-1,N_{\mathrm{cp}}-1 \right].
\end{equation}
From \eqref{TD_utor_7} and \eqref{TD_utortob_5}, we obtain
\begin{equation}\label{TD_utortob_6}
  y_{\mathrm{ris},r}\left[ m \right] =\bar{y}_{\mathrm{ris},r}\left[ \left( m-N_{\mathrm{cp}} \right) \; \mathrm{mod} \; N_{\mathrm{c}} \right] ,  m\in \left[ L_{\mathrm{u}}-1,N_{\mathrm{c}}+N_{\mathrm{cp}}-1 \right].
\end{equation}
Observing from $y_{\mathrm{ris},r}\left[ m-l_2 \right]$ in \eqref{TD_utortob_1},
from \eqref{TD_utortob_4}, we obtain that $m-l_2$ belongs to $[ L_{\mathrm{u}}-1,N_{\mathrm{c}}+N_{\mathrm{cp}}-1 ]$.
Thus we have \vspace{0.3cm}
\begin{equation*}
\hspace{-7.7cm}
  y_{\mathrm{bs},b}\left[ m \right] =\sum_{l_2=0}^{L_{\mathrm{b}}-1}{\sqrt{\beta _{\mathrm{b}}}h_{\mathrm{b},l_2}^{\left( b,r \right)}e^{j\theta _r}y_{\mathrm{ris},r}\left[ m-l_2 \right]}+z_b\left[ m \right]
\end{equation*}
\begin{equation}\label{TD_utortob_7}
\hspace{1.4cm}
  =\sum_{l_2=0}^{L_{\mathrm{b}}-1}{\sqrt{\beta _{\mathrm{b}}}h_{\mathrm{b},l_2}^{\left( b,r \right)}e^{j\theta _r}\bar{y}_{\mathrm{ris},r}\left[ \left( m-N_{\mathrm{cp}}-l_2 \right) \; \mathrm{mod} \; N_{\mathrm{c}} \right]}+z_b\left[ m \right] ,    m\in \left[ N_{\mathrm{cp}},N_{\mathrm{c}}+N_{\mathrm{cp}}-1 \right].
\end{equation}
From \eqref{h_ij} and \eqref{TD_utortob_7}, the received signal within the interval of $m\in \left[ N_{\mathrm{cp}},N_{\mathrm{c}}+N_{\mathrm{cp}}-1 \right]$ can be expressed as
\begin{equation}\label{TD_utortob_8}
  \bar{\mathbf{y}}_{\mathrm{bs},b}=e^{j\theta _r}\sqrt{\beta _{\mathrm{b}}}\mathbf{h}_{\mathrm{b}}^{\left( b,r \right)}\odot \bar{\mathbf{y}}_{\mathrm{ris},r}+\mathbf{z}_b.
\end{equation}
Similarly, using the property of the DFT, we obtain
\begin{equation}\label{TD_utortob_9}
  \mathrm{DFT}\left( \bar{\mathbf{y}}_{\mathrm{bs},b} \right) _i=e^{j\theta _r}\sqrt{\beta _{\mathrm{b}}N_{\mathrm{c}}}\mathrm{DFT}\left( \mathbf{h}_{\mathrm{b}}^{\left( b,r \right)} \right) _i\cdot \mathrm{DFT}\left( \bar{\mathbf{y}}_{\mathrm{ris},r} \right) _i+\mathrm{DFT}\left( \mathbf{z}_b \right) _i, i=0,1,...,N_{\mathrm{c}}.
\end{equation}
Thus, from \eqref{TD_utor_10}, we have
\begin{equation*}
  \bar{\mathbf{y}}_{\mathrm{bs},b}=e^{j\theta _r}\mathbf{F}^H\mathrm{diag}\left( \sqrt{N_{\mathrm{c}}\beta _{\mathrm{b}}}\mathbf{Fh}_{\mathrm{b}}^{\left( b,r \right)} \right) \mathbf{F}\bar{\mathbf{y}}_{\mathrm{ris},r}+\mathbf{z}_b
  \triangleq e^{j\theta _r}\mathbf{F}^H\mathbf{G}_{\mathrm{b}}^{\left( b,r \right)}\mathbf{F}\bar{\mathbf{y}}_{\mathrm{ris},r}+\mathbf{z}_b
\end{equation*}
\begin{equation}\label{TD_utortob_10}
\hspace{-2.4cm}
  \triangleq e^{j\theta _r}\mathbf{H}_{\mathrm{b}}^{\left( b,r \right)}\bar{\mathbf{y}}_{\mathrm{ris},r}+\mathbf{z}_b=\sqrt{p_u}e^{j\theta _r}\mathbf{H}_{\mathrm{b}}^{\left( b,r \right)}\mathbf{H}_{\mathrm{u}}^{\left( r,n \right)}\mathbf{s}_n+\mathbf{z}_b,
\end{equation}
where
matrix $\mathbf{G}_{\mathrm{b}}^{\left( b,r \right)}$ is the frequency-domain channel given by
\begin{equation}\label{B-R_FD_channel_ij}
  \mathbf{G}_{\mathrm{b}}^{\left( b,r \right)}=\mathrm{diag}\left( \sqrt{N_{\mathrm{c}}\beta _{\mathrm{b}}}\mathbf{Fh}_{\mathrm{b}}^{\left( b,r \right)} \right).
\end{equation}
Matrix $\mathbf{H}_{\mathrm{b}}^{\left( b,r \right)}\triangleq \mathbf{F}^H\mathbf{G}_{\mathrm{b}}^{\left( b,r \right)}\mathbf{F}$ is the circulant time-domain channel between Element $r$ of the RIS and Antenna $b$ of the BS.

%It should be aware that to combat the frequency selectivity, an $N_{\mathrm{cp}}$-length ($N_{\mathrm{cp}} \geq L_{\mathrm{u}} + L_{\mathrm{b}} - 2$) cyclic prefix (CP) is attached to each OFDM signal before the transmission and is removed after the reception.
%Due to that, for the time-domain signal $s_u\left[ n  \right]$ in \eqref{TD_utor_1} and \eqref{TD_utortob_1}, we have
%\begin{equation}\label{CP_1}
%  s_u\left[ n \right] =s_u\left[ n+N_{\mathrm{c}} \right] , -N_{\mathrm{cp}}-1<n<0.
%\end{equation}

Therefore, from \eqref{TD_utortob_10}, the time-domain OFDM signals $\mathbf{y}\in \mathbb{C} ^{N_{\mathrm{b}}N_{\mathrm{c}}\times 1}$
received at the antennas of the BS can be expressed as
\begin{equation}\label{TD_y}
  \mathbf{y}=\mathbf{H}_{\mathrm{b}}\mathbf{\Phi H}_{\mathrm{u}}\mathbf{Ps}+\mathbf{z}=\mathbf{H}_{\mathrm{b}}\mathbf{\Phi H}_{\mathrm{u}}\mathbf{P}\left( \mathbf{I}_{N_{\mathrm{u}}}\otimes \mathbf{F}^H \right) \mathbf{x}+\mathbf{z},
\end{equation}
where
$\mathbf{s}=\left[ \mathbf{s}_{0}^{T},\mathbf{s}_{1}^{T},...,\mathbf{s}_{N_{\mathrm{u}}-1}^{T} \right] ^T\in \mathbb{C} ^{N_{\mathrm{u}}N_{\mathrm{c}}\times 1}$
is the time-domain signal vector.
And $\mathbf{x}\in \mathbb{C} ^{N_{\mathrm{u}}N_{\mathrm{c}}\times 1}$ is the frequency-domain signal vector
which, for the ease of derivation, is assumed to have unit norm entries.
Besides, the circulant time-domain channels $\mathbf{H}_{\mathrm{u}}\in \mathbb{C} ^{N_{\mathrm{r}}N_{\mathrm{c}}\times N_{\mathrm{u}}N_{\mathrm{c}}}$
and $\mathbf{H}_{\mathrm{b}}\in \mathbb{C} ^{N_{\mathrm{b}}N_{\mathrm{c}}\times N_{\mathrm{r}}N_{\mathrm{c}}}$ are respectively defined as
\begin{equation}\label{R-U_TD_channel}
\mathbf{H}_{\mathrm{u}}\triangleq \left( \mathbf{I}_{N_{\mathrm{r}}}\otimes \mathbf{F}^H \right) \mathbf{G}_{\mathrm{u}}\left( \mathbf{I}_{N_{\mathrm{u}}}\otimes \mathbf{F} \right) =\left[ \begin{matrix}
	\mathbf{H}_{\mathrm{u}}^{\left( 0,0 \right)}&		\cdots&		\mathbf{H}_{\mathrm{u}}^{\left( 0,N_{\mathrm{u}}-1 \right)}\\
	\vdots&		\ddots&		\vdots\\
	\mathbf{H}_{\mathrm{u}}^{\left( N_{\mathrm{r}}-1,0 \right)}&		\cdots&		\mathbf{H}_{\mathrm{u}}^{\left( N_{\mathrm{r}}-1,N_{\mathrm{u}}-1 \right)}\\
\end{matrix} \right] ,
\end{equation}
\begin{equation}\label{B-R_TD_channel}
  \mathbf{H}_{\mathrm{b}}\triangleq \left( \mathbf{I}_{N_{\mathrm{b}}}\otimes \mathbf{F}^H \right) \mathbf{G}_{\mathrm{b}}\left( \mathbf{I}_{N_{\mathrm{r}}}\otimes \mathbf{F} \right) =\left[ \begin{matrix}
	\mathbf{H}_{\mathrm{b}}^{\left( 0,0 \right)}&		\cdots&		\mathbf{H}_{\mathrm{b}}^{\left( 0,N_{\mathrm{r}}-1 \right)}\\
	\vdots&		\ddots&		\vdots\\
	\mathbf{H}_{\mathrm{b}}^{\left( N_{\mathrm{b}}-1,0 \right)}&		\cdots&		\mathbf{H}_{\mathrm{b}}^{\left( N_{\mathrm{b}}-1,N_{\mathrm{r}}-1 \right)}\\
\end{matrix} \right] ,
\end{equation}
with frequency-domain channels $\mathbf{G}_{\mathrm{u}}$ and $\mathbf{G}_{\mathrm{b}}$ given by
\begin{equation}\label{R-U_FD_channel}
  \mathbf{G}_{\mathrm{u}}=\left[ \begin{matrix}
	\mathbf{G}_{\mathrm{u}}^{\left( 0,0 \right)}&		\cdots&		\mathbf{G}_{\mathrm{u}}^{\left( 0,N_{\mathrm{u}}-1 \right)}\\
	\vdots&		\ddots&		\vdots\\
	\mathbf{G}_{\mathrm{u}}^{\left( N_{\mathrm{r}}-1,0 \right)}&		\cdots&		\mathbf{G}_{\mathrm{u}}^{\left( N_{\mathrm{r}}-1,N_{\mathrm{u}}-1 \right)}\\
\end{matrix} \right] ,
\end{equation}
\begin{equation}\label{B-R_FD_channel}
  \mathbf{G}_{\mathrm{b}}=\left[ \begin{matrix}
	\mathbf{G}_{\mathrm{b}}^{\left( 0,0 \right)}&		\cdots&		\mathbf{G}_{\mathrm{b}}^{\left( 0,N_{\mathrm{r}}-1 \right)}\\
	\vdots&		\ddots&		\vdots\\
	\mathbf{G}_{\mathrm{b}}^{\left( N_{\mathrm{b}}-1,0 \right)}&		\cdots&		\mathbf{G}_{\mathrm{b}}^{\left( N_{\mathrm{b}}-1,N_{\mathrm{r}}-1 \right)}\\
\end{matrix} \right] .
\end{equation}
Diagonal matrix $\mathbf{\Phi }\in \mathbb{C} ^{N_{\mathrm{r}}N_{\mathrm{c}}\times N_{\mathrm{r}}N_{\mathrm{c}}}$,
which represents the phase shifts of the RIS,
and $\mathbf{P}\in \mathbb{C} ^{N_{\mathrm{u}}N_{\mathrm{c}}\times N_{\mathrm{u}}N_{\mathrm{c}}}$,
which stands for transmit power of users,
are respectively given by
\begin{equation}\label{RIS_phase}
  \mathbf{\Phi }=\left[ \begin{matrix}
	\mathbf{\Phi }_0&		&		\\
	&		\ddots&		\\
	&		&		\mathbf{\Phi }_{N_{\mathrm{r}}-1}\\
\end{matrix} \right] , \;\;\; \mathbf{\Phi }_r=e^{j\theta _r}\mathbf{I}_{N_{\mathrm{c}}}, \;\;\; r=0,1,...,N_{\mathrm{r}}-1 ,
\end{equation}
\begin{equation}\label{power_matrix}
  \mathbf{P}=\left[ \begin{matrix}
	\mathbf{P}_0&		&		\\
	&		\ddots&		\\
	&		&		\mathbf{P}_{N_{\mathrm{u}}-1}\\
\end{matrix} \right] , \;\;\;  \mathbf{P}_u=\sqrt{p_u}\mathbf{I}_{N_{\mathrm{u}}}, \;\;\;  u=0,1,...,N_{\mathrm{u}}-1.
\end{equation}
Additionally, $\mathbf{z}\in \mathbb{C} ^{N_{\mathrm{b}}N_{\mathrm{c}}\times 1}$ is the AWGN vector with its entries following the distribution of $\mathcal{C} \mathcal{N} \left( 0,\sigma _{\mathrm{noise}}^{2} \right)$.

\begin{table}[b]
\center
\caption{Values of $\rho$}\label{values_for_rho} \vspace{0.1cm}
\renewcommand\arraystretch{1.6}
\begin{tabular}{|c|c|c|c|c|c|c|}
\hline
 $b$    & 1      & 2      & 3       & 4        & 5       & $>$5   \\ \hline
 $\rho$    & 0.3634 & 0.1175 & 0.03454 & 0.009497 & 0.002499  & $\frac{\pi \sqrt{3}}{2}\cdot 2^{-2b}$ \\ \hline
\end{tabular}
\end{table}

To reduce the power consumption, low-resolution ADCs are equipped at the BS.
In this paper, we adopt the additive quantization noise model (AQNM) to characterize the quantization of OFDM signals \cite{7979627,8629287,9356722}.
Thus, the quantified signal can be expressed as
\begin{equation}\label{TD_yq}
  \mathbf{y}_{\mathrm{q}}=\alpha \mathbf{y}+\mathbf{z}_{\mathrm{q}}=\alpha \mathbf{H}_{\mathrm{b}}\mathbf{\Phi H}_{\mathrm{u}}\mathbf{P}\left( \mathbf{I}_{N_{\mathrm{u}}}\otimes \mathbf{F}^H \right) \mathbf{x}+\alpha \mathbf{z}+\mathbf{z}_{\mathrm{q}} ,
\end{equation}
where $\alpha =1-\rho$, and $\rho$ is the inverse of the signal-to-quantization-noise ratio
with its value being a function of quantization bits, $b$, as given by Table \ref{values_for_rho}.
Vector $\mathbf{z}_{\mathrm{q}}\sim \mathcal{C} \mathcal{N} \left( 0,\mathbf{R}_{\mathbf{z}_{\mathrm{q}}} \right)$ is the additive Gaussian quantization noise uncorrelated to $\mathbf{y}$, which satisfies
\begin{equation}\label{Rzq}
  \mathbf{R}_{\mathbf{z}_{\mathrm{q}}}\approx \alpha \left( 1-\alpha \right) \mathrm{diag}\left( \mathbb{E} \left\{ \mathbf{yy}^H \right\} \right)
  =\alpha \left( 1-\alpha \right) \mathrm{diag}\left( \mathbf{H}_{\mathrm{b}}\mathbf{\Phi H}_{\mathrm{u}}\mathbf{PP}^H\mathbf{H}_{\mathrm{u}}^{H}\mathbf{\Phi }^H\mathbf{H}_{\mathrm{b}}^{H}+\sigma _{\mathrm{noise}}^{2}\mathbf{I}_{N_{\mathrm{b}}N_{\mathrm{c}}} \right).
\end{equation}

Then, the received signal is transformed to the frequency-domain, which is given by
\begin{equation}\label{FD-yqf}
  \mathbf{y}_{\mathrm{qf}}=\left( \mathbf{I}_{N_{\mathrm{b}}}\otimes \mathbf{F} \right) \mathbf{y}_{\mathrm{q}}
  =\alpha \mathbf{G}_{\mathrm{b}}\mathbf{\Phi G}_{\mathrm{u}}\mathbf{Px}+\left( \mathbf{I}_{N_{\mathrm{b}}}\otimes \mathbf{F} \right) \left( \alpha \mathbf{z}+\mathbf{z}_{\mathrm{q}} \right).
\end{equation}
We assume the  maximal-ratio-combining (MRC) processing is adopted at the BS.
Thus, the processed signal can be obtained as
\begin{equation}\label{FD-yqf_mrc}
  \mathbf{r}=\mathbf{W}_{\mathrm{mrc}}\mathbf{y}_{\mathrm{qf}}=\alpha \mathbf{G}_{\mathrm{u}}^{H}\mathbf{\Phi }^H\mathbf{G}_{\mathrm{b}}^{H}\mathbf{G}_{\mathrm{b}}\mathbf{\Phi G}_{\mathrm{u}}\mathbf{Px}+\mathbf{G}_{\mathrm{u}}^{H}\mathbf{\Phi }^H\mathbf{G}_{\mathrm{b}}^{H}\left( \mathbf{I}_{N_{\mathrm{b}}}\otimes \mathbf{F} \right) \left( \alpha \mathbf{z}+\mathbf{z}_{\mathrm{q}} \right),
\end{equation}
where $\mathbf{W}_{\mathrm{mrc}}=\mathbf{G}_{\mathrm{u}}^{H}\mathbf{\Phi }^H\mathbf{G}_{\mathrm{b}}^{H}$ is the MRC beamforming matrix.

\section{Uplink Rate Analysis}

In this section, we derive the closed-form approximate expression of the uplink achievable rate for the RIS-aided multi-user wideband massive MIMO OFDM system with low-resolution ADCs.
Based on that, we will analyze the asymptotic performance and reveal the power scaling laws of the considered system to provide useful insights for the system design.

Without loss of generality, we focus on the received signal from User $n$ ($n=0,1,...,N_{\mathrm{u}}-1$) on Sub-carrier $t$ ($t=0,1,...,N_{\mathrm{c}}-1$),
which, according to \eqref{FD-yqf_mrc}, is given by
\begin{equation*}
\hspace{-2.4cm}
  \mathbf{r}_{nt}=\alpha \mathbf{g}_{\mathrm{u},nt}^{H}\mathbf{\Phi }^H\mathbf{G}_{\mathrm{b}}^{H}\mathbf{G}_{\mathrm{b}}\mathbf{\Phi G}_{\mathrm{u}}\mathbf{Px}+\mathbf{g}_{\mathrm{u},nt}^{H}\mathbf{\Phi }^H\mathbf{G}_{\mathrm{b}}^{H}\left( \mathbf{I}_{N_{\mathrm{b}}}\otimes \mathbf{F} \right) \left( \alpha \mathbf{z}+\mathbf{z}_{\mathrm{q}} \right)
\end{equation*}
\begin{equation*}
  =\alpha \sum_{j=0}^{N_{\mathrm{u}}-1}{\sqrt{p_j}\mathbf{g}_{\mathrm{u},nt}^{H}\mathbf{\Phi }^H\mathbf{G}_{\mathrm{b}}^{H}\mathbf{G}_{\mathrm{b}}\mathbf{\Phi g}_{\mathrm{u},jt}x_{jt}}+\mathbf{g}_{\mathrm{u},nt}^{H}\mathbf{\Phi }^H\mathbf{G}_{\mathrm{b}}^{H}\left( \mathbf{I}_{N_{\mathrm{b}}}\otimes \mathbf{F} \right) \left( \alpha \mathbf{z}+\mathbf{z}_{\mathrm{q}} \right)
\end{equation*}
\begin{equation*}
\hspace{-0.3cm}
  = \underbrace{ \alpha \sqrt{p_n}\mathbf{g}_{\mathrm{u},nt}^{H}\mathbf{\Phi }^H\mathbf{G}_{\mathrm{b}}^{H}\mathbf{G}_{\mathrm{b}}\mathbf{\Phi g}_{\mathrm{u},nt}x_{nt} }_{ \mathrm{term}_{\mathrm{r}}^{1} }
  + \underbrace{ \alpha \sum_{j\ne n}^{N_{\mathrm{u}}-1}{\sqrt{p_j}\mathbf{g}_{\mathrm{u},nt}^{H}\mathbf{\Phi }^H\mathbf{G}_{\mathrm{b}}^{H}\mathbf{G}_{\mathrm{b}}\mathbf{\Phi g}_{\mathrm{u},jt}x_{jt}} }_{ \mathrm{term}_{\mathrm{r}}^{2} }
\end{equation*}
\begin{equation}\label{FD-yqf_mrc_nt}
\hspace{-3.0cm}
  + \underbrace{ \alpha \mathbf{g}_{\mathrm{u},nt}^{H}\mathbf{\Phi }^H\mathbf{G}_{\mathrm{b}}^{H}\left( \mathbf{I}_{N_{\mathrm{b}}}\otimes \mathbf{F} \right) \mathbf{z} }_{ \mathrm{term}_{\mathrm{r}}^{3} }
  + \underbrace{ \mathbf{g}_{\mathrm{u},nt}^{H}\mathbf{\Phi }^H\mathbf{G}_{\mathrm{b}}^{H}\left( \mathbf{I}_{N_{\mathrm{b}}}\otimes \mathbf{F} \right) \mathbf{z}_{\mathrm{q}} }_{ \mathrm{term}_{\mathrm{r}}^{4} } ,
\end{equation}
where the subscript $\left( \cdot \right) _{nt}$ denotes the index of $nt\triangleq nN_{\mathrm{c}}+t$,
and $\mathbf{g}_{\mathrm{u},nt}^{H}$ stands for the $\left( nN_{\mathrm{c}}+t \right)$-th column vector of $\mathbf{G}_{\mathrm{u}}$.
The first term of $\mathbf{r}_{nt}$, i.e., $\mathrm{term}_{\mathrm{r}}^{1}$, represents the desired signal.
$\mathrm{term}_{\mathrm{r}}^{2}$ stands for the interference from other users on the same Sub-carrier $t$.
$\mathrm{term}_{\mathrm{r}}^{3}$ is from the noise at the receiving antenna.
As for the last term $\mathrm{term}_{\mathrm{r}}^{4}$,
it is generated from the quantization noise brought by the low-resolution ADCs,
which is related to the power of signals from the other users and sub-carriers.

%\textcolor[rgb]{1.00,0.00,0.00}{(Although in \eqref{FD-yqf_mrc_nt} we focus on the signals from one user and on one sub-carrier, the interference is from the other users and sub-carriers.
%Thus, we think the results based on that have generality.
%Besides, It should be noted that due to the added CP to each OFDM symbol, the inter-channel interference (ICI) is theoretically eliminated.
%However, because of the non-linear quantization at the low-resolution ADCs,
%the interference from the quantization noise is related to the power of signals from the other users and sub-carriers.}

%\textcolor[rgb]{1.00,0.00,0.00}{In the following part, we derive the achievable rate of User $n$ on all sub-carriers instead of only on one sub-carrier.)}

Based on \eqref{FD-yqf_mrc_nt}, the uplink achievable rate of User $n$ is expressed as \cite{8937491,9039554,6457363}
\begin{equation}\label{R_n_sim}
  R_n=\frac{1}{N_{\mathrm{cp}}+N_{\mathrm{c}}}\sum_{t=0}^{N_{\mathrm{c}}-1}{R_{n,t}},
\end{equation}
where
\begin{equation}\label{R_nt_sim}
\hspace{-6.7cm}
  R_{n,t}=\mathbb{E} \left\{ \log _2\left( 1+\frac{\mathcal{S}}{\mathcal{I}} \right) \right\},
\end{equation}
\begin{equation}\label{R_nt_S_sim}
\hspace{-5.3cm}
  \mathcal{S} =\alpha ^2p_n\left| \mathbf{g}_{\mathrm{u},nt}^{H}\mathbf{\Phi }^H\mathbf{G}_{\mathrm{b}}^{H}\mathbf{G}_{\mathrm{b}}\mathbf{\Phi g}_{\mathrm{u},nt} \right|^2,
\end{equation}
\begin{equation*}
  \mathcal{I} =\alpha ^2\sum_{u\ne n}^{N_{\mathrm{u}}-1}{p_u\left| \mathbf{g}_{\mathrm{u},nt}^{H}\mathbf{\Phi }^H\mathbf{G}_{\mathrm{b}}^{H}\mathbf{G}_{\mathrm{b}}\mathbf{\Phi g}_{\mathrm{u},ut} \right|^2}+\sigma _{\mathrm{noise}}^{2}\alpha ^2\left\| \mathbf{g}_{\mathrm{u},nt}^{H}\mathbf{\Phi }^H\mathbf{G}_{\mathrm{b}}^{H} \right\| ^2
\end{equation*}
\begin{equation}\label{R_nt_I_sim}
  +\mathbf{g}_{\mathrm{u},nt}^{H}\mathbf{\Phi }^H\mathbf{G}_{\mathrm{b}}^{H}\left( \mathbf{I}_{N_{\mathrm{b}}}\otimes \mathbf{F} \right) \mathbf{R}_{\mathbf{z}_{\mathrm{q}}}\left( \mathbf{I}_{N_{\mathrm{b}}}\otimes \mathbf{F}^H \right) \mathbf{G}_{\mathrm{b}}\mathbf{\Phi g}_{\mathrm{u},nt}.
\end{equation}
To enable an insightful analysis,
we aim to obtain a closed-form expression of the uplink achievable rate $R_{n}$.
However, it is observed that the expectation operator in \eqref{R_nt_sim} is intractable to handle.
To address that, by using Lemma~1 in \cite{6816003},
we transform $R_{n,t}$ in \eqref{R_nt_sim} into a tractable approximation as\footnote{\cite{6816003} has proved that the approximation is valid when $\mathcal{S}$ and $\mathcal{I}$ are both the sum of nonnegative random variables. Besides, it becomes more accurate with a larger number of BS antennas.}
\begin{equation}\label{R_nt_hat_1}
  R_{n,t}=\mathbb{E} \left\{ \log _2\left( 1+\frac{\mathcal{S}}{\mathcal{I}} \right) \right\} \approx \tilde{R}_{n,t}=\log _2\left( 1+\frac{\mathbb{E} \left\{ \mathcal{S} \right\}}{\mathbb{E} \left\{ \mathcal{I} \right\}} \right).
\end{equation}
Thus, the uplink achievable rate $R_{n}$ of User $n$ can be approximated as
\begin{equation}\label{R_n_hat_1}
  R_n\approx \tilde{R}_n=\frac{1}{N_{\mathrm{cp}}+N_{\mathrm{c}}}\sum_{t=0}^{N_{\mathrm{c}}-1}{\tilde{R}_{n,t}} .
\end{equation}
Then,
we obtain a closed-form approximate expression for $R_{n}$ in the following theorem.

\begin{theorem} \label{theorem_R_nt_mrc_appro}
With MRC processing, the approximate expression for the uplink achievable rate of User $n$ in an RIS-aided multi-user wideband massive MIMO OFDM system with low-resolution ADCs under Rician fading channels can be expressed as
\begin{equation*}
\hspace{-8.2cm}
  R_n\approx \tilde{R}_n=\frac{1}{N_{\mathrm{cp}}+N_{\mathrm{c}}}\sum_{t=0}^{N_{\mathrm{c}}-1}{\tilde{R}_{n,t}}
\end{equation*}
\begin{equation}\label{R_nt_appro}
  = \frac{N_{\mathrm{c}}}{N_{\mathrm{cp}}+N_{\mathrm{c}}}  \log _2\left( 1+\frac{\alpha ^2p_n\varpi _n}{\alpha ^2\sum_{u\ne n}^{N_{\mathrm{u}}-1}{p_u\eta _{n,u}}+\alpha _n\left( 1-\alpha \right) \xi _n+\sigma _{\mathrm{noise}}^{2}\alpha \epsilon _n} \right) ,
\end{equation}
where
$\varpi _n$, $\eta _{n,u}$, $\xi _n$, and $\epsilon _n$ are respectively defined in
\eqref{omega_bar_n}, \eqref{eta_n_u}, \eqref{xi_n}, and \eqref{epsilon_n} in Appendix~\ref{app_A}.
\end{theorem}

\begin{IEEEproof}
See Appendix \ref{app_A}.
\end{IEEEproof}

\begin{remark}\label{Rayleigh_case}
%\addtocounter{equation}{1}
When the Rician factors in \eqref{R_nt_appro} become zero, i.e., $K_{\mathrm{u},n},K_{\mathrm{b}}=0$,
indicating the rich-scattering environment without LoS path,
the Rician channels degenerate to the Rayleigh fading case.
In this context, the uplink rate is expressed as
\begin{equation}\label{R_nt_appro_Rayleigh}
  \tilde{R}_{n}^{\mathrm{Rayleigh}}=  \frac{N_{\mathrm{c}}}{N_{\mathrm{cp}}+N_{\mathrm{c}}}  \log _2\left( 1+\frac{\alpha ^2p_n\beta _{\mathrm{u},n}\left( N_{\mathrm{b}}+1 \right) \left( N_{\mathrm{r}}+1 \right)}{\alpha ^2\sum_{u\ne n}^{N_{\mathrm{u}}-1}{p_u\beta _{\mathrm{u},u}\left( N_{\mathrm{b}}+N_{\mathrm{r}} \right)}+  \alpha \left( 1-\alpha \right) c^{\mathrm{Rayleigh}}+\frac{\sigma _{\mathrm{noise}}^{2}\alpha}{\beta _{\mathrm{b}}}} \right)  ,
\end{equation}
where $c^{\mathrm{Rayleigh}}$ is defined as
\begin{equation*}
  c^{\mathrm{Rayleigh}}=N_{\mathrm{r}}\sum_{u=0}^{N_{\mathrm{u}}-1}{p_u\beta _{\mathrm{u},u}}+\sum_{u=0}^{N_{\mathrm{u}}-1}{p_u\beta _{\mathrm{u},u}}\sum_{k=0}^{L_{\mathrm{b}}-1}{\sigma _{\mathrm{b},k}^{4}}+p_n\beta _{\mathrm{u},n}\sum_{k=0}^{L_{\mathrm{u}}-1}{\sigma _{\mathrm{u},k,n}^{4}}+p_n\beta _{\mathrm{u},n}N_{\mathrm{r}}\sum_{k_1=0}^{L_{\mathrm{b}}-1}{\sigma _{\mathrm{b},k_1}^{4}}\sum_{k_2=0}^{L_{\mathrm{u}}-1}{\sigma _{\mathrm{u},k_2,n}^{4}}
\end{equation*}
\begin{equation}\label{c_Rayleigh}
  +2 p_n \beta _{\mathrm{u},n}N_{\mathrm{r}}\sum_{k_1=0}^{L_{\mathrm{b}}-1}{\sum_{k_2=k_1+1}^{L_{3}^{\min}-1}{\sum_{k_3=k_2-k_1}^{L_{\mathrm{u}}-1}{\sigma _{\mathrm{b},k_1}^{2}\sigma _{\mathrm{b},k_2}^{2}\sigma _{\mathrm{u},k_3,n}^{2}\sigma _{\mathrm{u},k_1-k_2+k_3,n}^{2}}}},
\end{equation}
with $L_{3}^{\min}=\min \left\{ L_{\mathrm{b}},L_{\mathrm{u}}+k_1 \right\}$.
\end{remark}

%Remark~\ref{Rayleigh_case}
It can be observed from Theorem~\ref{theorem_R_nt_mrc_appro} that
the uplink achievable rate of RIS-aided wideband massive MIMO OFDM systems with low-resolution ADCs is related to the power of taps of the frequency-selective channels.
Specifically, the quantization noise term $\xi _n$ in \eqref{R_nt_appro} is closely related to the power of channel taps.
By contrast, the expressions of the power of channel taps in $\varpi _n$, $\eta _{n,u}$, and $\epsilon _n$ are eliminated
by the power constraints $\sum_{l=0}^{L_{\mathrm{b}}-1}{\sigma _{\mathrm{b},l}^{2}}=1$ and
$\sum_{l=0}^{L_{\mathrm{u}}-1}{\sigma _{\mathrm{u},l,j}^{2}}=1$, $ j= 0,1,..., N_{\mathrm{u}}-1$,
due to the perfect CSI assumption in this paper.
It can be expected that $\varpi _n$, $\eta _{n,u}$, and $\epsilon _n$ will be related to the power of channel taps when imperfect CSI is considered.

%\textcolor[rgb]{1.00,0.00,0.00}{(Different from the derivations in systems with frequency-flat channels, the taps of the frequency-selective channels need to be considered in this paper, causing the derivation much more challenging.)}

Furthermore, Theorem \ref{theorem_R_nt_mrc_appro} indicates that
the uplink rate depends largely on the number of the antennas $N_{\mathrm{b}}$ and the number of reflecting elements $N_{\mathrm{r}}$.
According to \eqref{PHI_n} and \eqref{Phi_n_r}, we have $\left| \varPhi _{N_{\mathrm{r}}}\left( n \right) \right|\leqslant N_{\mathrm{r}}$.
The equality holds when $\varPhi \left( n,r \right) =1$, $\forall r$,
which represents the case that the RIS is aligned with User $n$ to enhance its channel quality.
Otherwise,
when the RIS is not aligned with User $n$,
we assume $\left| \varPhi _{N_{\mathrm{r}}}\left( n \right) \right| $ is bounded since $\left| \varPhi _{N_{\mathrm{r}}}\left( n \right) \right| < N_{\mathrm{r}}$.
Thus, as $N_{\mathrm{b}}, N_{\mathrm{r}} \rightarrow \infty$, we propose the following corollary:

\begin{corollary}\label{N_b_N_r}
%\addtocounter{equation}{1}
With MRC processing, as $N_{\mathrm{b}}, N_{\mathrm{r}} \rightarrow \infty$,
the asymptotic performance of the uplink achievable rate of User $n$ in the RIS-aided multi-user wideband massive MIMO OFDM system with low-resolution ADCs under Rician fading channels is given by
\begin{equation}\label{N_b_N_r_converge}
\tilde{R}_{n}\rightarrow \begin{cases}
	\mathrm{the}\; \mathrm{RIS}\; \mathrm{is}\; \mathrm{not}\; \mathrm{aligned}\; \mathrm{with}\; \mathrm{any}\; \mathrm{users}:\\
	\frac{N_{\mathrm{c}}}{N_{\mathrm{cp}}+N_{\mathrm{c}}}\log _2\left( 1+\frac{\sigma _{\mathrm{b},0}^{4}\varsigma _{\mathrm{u},n}^{2}K_{\mathrm{b}}^{2}+\left( \sigma _{\mathrm{b},0}^{2}\varsigma _{\mathrm{u},n}K_{\mathrm{b}}+\sigma _{\mathrm{u},0,n}^{2}\varsigma _{\mathrm{b}}K_{\mathrm{u},n}+\varsigma _{\mathrm{u},n}\varsigma _{\mathrm{b}} \right) ^2}{\sum_{u\ne n}^{N_{\mathrm{u}}-1}{\frac{p_u\beta _{\mathrm{u},u}\left( K_{\mathrm{u},n}+1 \right)}{p_n\beta _{\mathrm{u},n}\left( K_{\mathrm{u},u}+1 \right)}\sigma _{\mathrm{b},0}^{4}\varsigma _{\mathrm{u},n}\varsigma _{\mathrm{u},u}K_{\mathrm{b}}^{2}}} \right) ,\\
	\mathrm{the}\; \mathrm{RIS}\; \mathrm{is}\; \mathrm{aligned}\; \mathrm{with}\; \mathrm{User}\; n:\\
	\frac{N_{\mathrm{c}}}{N_{\mathrm{cp}}+N_{\mathrm{c}}}\log _2\left( 1+\frac{\sigma _{\mathrm{u},0,n}^{2}K_{\mathrm{u},n}N_{\mathrm{r}}}{\sum_{u\ne n}^{N_{\mathrm{u}}-1}{\frac{p_u\beta _{\mathrm{u},u}\left( K_{\mathrm{u},n}+1 \right)}{p_n\beta _{\mathrm{u},n}\left( K_{\mathrm{u},u}+1 \right)}\varsigma _{\mathrm{u},u}}} \right) ,\\
\end{cases}.
\end{equation}
\end{corollary}

It can be observed from Corollary \ref{N_b_N_r} that the asymptotic uplink rate for $\tilde{R}_{n}$ remains limited when the RIS is not aligned with any users.
If the RIS is aligned with User $n$, the asymptotic rate is on the order of $\mathcal{O} \left( \log _2\left( N_{\mathrm{r}} \right) \right)$,
showing the potential gain brought by the RIS.
By contrast, in both cases, the asymptotic rates are independent of $N_{\mathrm{b}}$,
which means when $N_{\mathrm{b}} \rightarrow \infty$, $\tilde{R}_{n}$ will converge to a limited positive value.

It is known that one important advantage of massive MIMO technique is that the transmit power of users can be effectively reduced proportionally to the number of antennas.
We thus investigate the power scaling laws of the considered system to provide more insights on energy savings.

\begin{corollary}\label{Scaling_law_Nb}
%\addtocounter{equation}{1}
With MRC processing,
as $N_{\mathrm{b}} \rightarrow \infty$,
the transmit power of each user can be scaled down at most to $p_j=\frac{E_j}{N_{\mathrm{b}}}$ with fixed $E_j$,
$j=0,1,...,N_{\mathrm{u}}-1$.
In this case, the uplink achievable rate of User $n$ in the RIS-aided multi-user wideband massive MIMO OFDM system with low-resolution ADCs under Rician fading channels converges to
\begin{equation}\label{Scaling_law_Nb_rate}
  \tilde{R}_{n}\rightarrow  \frac{N_{\mathrm{c}}}{N_{\mathrm{cp}}+N_{\mathrm{c}}}
 \log _2\left( 1+\frac{\varGamma _{n}^{1}}{\sum_{u\ne n}^{N_{\mathrm{u}}-1}{\frac{E_u\beta _{\mathrm{u},u}\left( K_{\mathrm{u},n}+1 \right)}{E_n\beta _{\mathrm{u},n}\left( K_{\mathrm{u},u}+1 \right)}\varGamma _{n,u}^{2}}+\frac{\sigma _{\mathrm{noise}}^{2}\left( K_{\mathrm{u},n}+1 \right) \left( K_{\mathrm{b}}+1 \right)}{\alpha \beta _{\mathrm{u},n}\beta _{\mathrm{b}}E_n}\varGamma _{n}^{3}} \right) ,
\end{equation}
with \vspace{0.2cm}
\begin{equation*}
\hspace{-1.2cm}
  \varGamma _{n}^{1}
  = \sigma _{\mathrm{u},0,n}^{4}\sigma _{\mathrm{b},0}^{4}K_{\mathrm{u},n}^{2}K_{\mathrm{b}}^{2}\left| \varPhi _{N_{\mathrm{r}}}\left( n \right) \right|^4
  + \left( 2\sigma _{\mathrm{u},0,n}^{2}\varsigma _{\mathrm{u},n}\varsigma _{\mathrm{b}}^{2}K_{\mathrm{u},n}+2\sigma _{\mathrm{b},0}^{2}\varsigma _{\mathrm{u},n}^{2}\varsigma _{\mathrm{b}}K_{\mathrm{b}}+\varsigma _{\mathrm{u},n}^{2}\varsigma _{\mathrm{b}}^{2} \right) N_{\mathrm{r}}
\end{equation*}
\begin{equation*}
  +2\sigma _{\mathrm{u},0,n}^{2}\sigma _{\mathrm{b},0}^{2}K_{\mathrm{u},n}K_{\mathrm{b}}\left| \varPhi _{N_{\mathrm{r}}}\left( n \right) \right|^2\left( \left( 2\sigma _{\mathrm{b},0}^{2}\varsigma _{\mathrm{u},n}K_{\mathrm{b}}+\sigma _{\mathrm{u},0,n}^{2}\varsigma _{\mathrm{b}}K_{\mathrm{u},n}+\varsigma _{\mathrm{u},n}\varsigma _{\mathrm{b}} \right) N_{\mathrm{r}}+2\varsigma _{\mathrm{u},n}\varsigma _{\mathrm{b}} \right)
\end{equation*}
\begin{equation}\label{Scaling_law_Nb_rate_p1}
\hspace{-3.9cm}
  +\left( \sigma _{\mathrm{b},0}^{4}\varsigma _{\mathrm{u},n}^{2}K_{\mathrm{b}}^{2}+\left( \sigma _{\mathrm{b},0}^{2}\varsigma _{\mathrm{u},n}K_{\mathrm{b}}+\sigma _{\mathrm{u},0,n}^{2}\varsigma _{\mathrm{b}}K_{\mathrm{u},n}+\varsigma _{\mathrm{u},n}\varsigma _{\mathrm{b}} \right) ^2 \right) N_{\mathrm{r}}^{2} ,
\end{equation}
\begin{equation*}
\hspace{-2.9cm}
  \varGamma _{n,u}^{2}
  = \sigma _{\mathrm{u},0,n}^{2}\sigma _{\mathrm{u},0,u}^{2}\sigma _{\mathrm{b},0}^{4}K_{\mathrm{u},n}K_{\mathrm{u},u}K_{\mathrm{b}}^{2}\left| \varPhi _{N_{\mathrm{r}}}\left( n \right) \right|^2\left| \varPhi _{N_{\mathrm{r}}}\left( u \right) \right|^2
  + \varsigma _{\mathrm{u},n}\varsigma _{\mathrm{u},u}\sigma _{\mathrm{b},0}^{4}K_{\mathrm{b}}^{2}N_{\mathrm{r}}^{2}
\end{equation*}
\begin{equation*}
\hspace{-4.6cm}
  +\left( \sigma _{\mathrm{b},0}^{2}\varsigma _{\mathrm{u},u}K_{\mathrm{b}}N_{\mathrm{r}}+2\varsigma _{\mathrm{u},u}\varsigma _{\mathrm{b}} \right) \sigma _{\mathrm{u},0,n}^{2}\sigma _{\mathrm{b},0}^{2}K_{\mathrm{u},n}K_{\mathrm{b}}\left| \varPhi _{N_{\mathrm{r}}}\left( n \right) \right|^2
\end{equation*}
\begin{equation*}
\hspace{-4.6cm}
  +\left( \sigma _{\mathrm{b},0}^{2}\varsigma _{\mathrm{u},n}K_{\mathrm{b}}N_{\mathrm{r}}+2\varsigma _{\mathrm{u},n}\varsigma _{\mathrm{b}} \right) \sigma _{\mathrm{u},0,u}^{2}\sigma _{\mathrm{b},0}^{2}K_{\mathrm{u},u}K_{\mathrm{b}}\left| \varPhi _{N_{\mathrm{r}}}\left( u \right) \right|^2
\end{equation*}
\begin{equation*}
\hspace{-2.1cm}
  +\left( \sigma _{\mathrm{u},0,u}^{2}\varsigma _{\mathrm{u},n}\varsigma _{\mathrm{b}}^{2}K_{\mathrm{u},u}+\sigma _{\mathrm{u},0,n}^{2}\varsigma _{\mathrm{u},u}\varsigma _{\mathrm{b}}^{2}K_{\mathrm{u},n}+\varsigma _{\mathrm{u},n}\varsigma _{\mathrm{u},u}\left( 2\sigma _{\mathrm{b},0}^{2}\varsigma _{\mathrm{b}}K_{\mathrm{b}}+\varsigma _{\mathrm{b}}^{2} \right) \right) N_{\mathrm{r}}
\end{equation*}
\begin{equation*}
\hspace{-6.9cm}
  +\sigma _{\mathrm{u},0,n}^{2}\sigma _{\mathrm{u},0,u}^{2}\varsigma _{\mathrm{b}}^{2}K_{\mathrm{u},n}K_{\mathrm{u},u}\left| \left( \bar{\mathbf{h}}_{\mathrm{u}}^{\left( \cdot ,n \right)} \right) ^H\bar{\mathbf{h}}_{\mathrm{u}}^{\left( \cdot ,u \right)} \right|^2
\end{equation*}
\begin{equation}\label{Scaling_law_Nb_rate_p2}
\hspace{-1.6cm}
  +2\sigma _{\mathrm{u},0,n}^{2}\sigma _{\mathrm{u},0,u}^{2}\sigma _{\mathrm{b},0}^{2}\varsigma _{\mathrm{b}}K_{\mathrm{u},n}K_{\mathrm{u},u}K_{\mathrm{b}}\mathrm{Re}\left( \left( \varPhi _{N_{\mathrm{r}}}\left( n \right) \right) ^{\ast}\varPhi _{N_{\mathrm{r}}}\left( u \right) \left( \bar{\mathbf{h}}_{\mathrm{u}}^{\left( \cdot ,u \right)} \right) ^H\bar{\mathbf{h}}_{\mathrm{u}}^{\left( \cdot ,n \right)} \right) ,
\end{equation}
\begin{equation}\label{Scaling_law_Nb_rate_p3}
\hspace{-2.2cm}
  \varGamma _{n}^{3}
  = \sigma _{\mathrm{u},0,n}^{2}K_{\mathrm{u},n}\sigma _{\mathrm{b},0}^{2}K_{\mathrm{b}}\left| \varPhi _{N_{\mathrm{r}}}\left( n \right) \right|^2+\left( \sigma _{\mathrm{u},0,n}^{2}K_{\mathrm{u},n}\varsigma _{\mathrm{b}}+\sigma _{\mathrm{b},0}^{2}K_{\mathrm{b}}\varsigma _{\mathrm{u},n}+\varsigma _{\mathrm{u},n}\varsigma _{\mathrm{b}} \right) N_{\mathrm{r}} .
\end{equation}

\end{corollary}

Corollary \ref{Scaling_law_Nb} shows that when $N_{\mathrm{b}} \rightarrow \infty$, the transmit power of the users can be scaled down proportionally to $\frac{1}{N_{\mathrm{b}}}$ and the system maintains a constant uplink achievable rate as in \eqref{Scaling_law_Nb_rate}.
In the following, we will prove that benefited from the gain of the RIS,
the power scaling law in Corollary \ref{Scaling_law_Nb} can be extended,
where the transmit power of the users can be further reduced proportionally to the number of reflecting elements $N_{\mathrm{r}}$,
while guaranteeing certain system performance.

\begin{corollary}\label{Scaling_law_Nb_Nr}
%\addtocounter{equation}{1}
With MRC processing,
when the RIS is not aligned with any users,
as $N_{\mathrm{b}}, N_{\mathrm{r}} \rightarrow \infty$,
the transmit power of all users can be scaled down at most to $p_j=\frac{E_j}{N_{\mathrm{b}}N_{\mathrm{r}}}$ with fixed $E_j$,
$j=0,1,...,N_{\mathrm{u}}-1$.
In this case, the uplink achievable rate of User $n$ in the RIS-aided multi-user wideband massive MIMO OFDM system with low-resolution ADCs under Rician fading channels converges to
\begin{equation}\label{Scaling_law_Nb_Nr_rate}
  \tilde{R}_{n}\rightarrow \frac{N_{\mathrm{c}}}{N_{\mathrm{cp}}+N_{\mathrm{c}}}
 \log _2\left( 1+\frac{\bar{\varGamma}_{n}^{1}}{\sum_{u\ne n}^{N_{\mathrm{u}}-1}{\frac{E_u\beta _{\mathrm{u},u}\left( K_{\mathrm{u},n}+1 \right)}{E_n\beta _{\mathrm{u},n}\left( K_{\mathrm{u},u}+1 \right)}\bar{\varGamma}_{n,u}^{2}}+\frac{\sigma _{\mathrm{noise}}^{2}\left( K_{\mathrm{u},n}+1 \right) \left( K_{\mathrm{b}}+1 \right)}{\alpha \beta _{\mathrm{u},n}\beta _{\mathrm{b}}E_n}\bar{\varGamma}_{n}^{3}} \right)  ,
\end{equation}
where
\begin{equation}\label{Scaling_law_Nb_Nr_rate_p1}
  \bar{\varGamma}_{n}^{1}=\sigma _{\mathrm{b},0}^{4}\varsigma _{\mathrm{u},n}^{2}K_{\mathrm{b}}^{2}+\left( \sigma _{\mathrm{b},0}^{2}\varsigma _{\mathrm{u},n}K_{\mathrm{b}}+\sigma _{\mathrm{u},0,n}^{2}\varsigma _{\mathrm{b}}K_{\mathrm{u},n}+\varsigma _{\mathrm{u},n}\varsigma _{\mathrm{b}} \right) ^2
\end{equation}
\begin{equation}\label{Scaling_law_Nb_Nr_rate_p2}
  \bar{\varGamma}_{n,u}^{2}=\varsigma _{\mathrm{u},n}\varsigma _{\mathrm{u},u}\sigma _{\mathrm{b},0}^{4}K_{\mathrm{b}}^{2}
\end{equation}
\begin{equation}\label{Scaling_law_Nb_Nr_rate_p3}
  \bar{\varGamma}_{n}^{3}=\sigma _{\mathrm{u},0,n}^{2}K_{\mathrm{u},n}\varsigma _{\mathrm{b}}+\sigma _{\mathrm{b},0}^{2}K_{\mathrm{b}}\varsigma _{\mathrm{u},n}+\varsigma _{\mathrm{u},n}\varsigma _{\mathrm{b}}.
\end{equation}

\end{corollary}

The power scaling law in Corollary \ref{Scaling_law_Nb_Nr} indicates that when the RIS is not aligned with any users,
as $N_{\mathrm{b}}, N_{\mathrm{r}} \rightarrow \infty$, the transmit power of each user can be reduced proportionally to $\frac{1}{N_{\mathrm{b}}N_{\mathrm{r}}}$ while the considered system keeps a constant converging rate.
Moreover, when the RIS is aligned with User $n$,
the transmit power $p_n$ of User $n$ can be further reduced proportionally to $\frac{1}{N_{\mathrm{b}}N^2_{\mathrm{r}}}$,
while guaranteeing certain system performance,
which is shown in the following corollary.

\begin{corollary}\label{Scaling_law_aligned}
%\addtocounter{equation}{1}
With MRC processing,
when the RIS is aligned with User $n$,
as $N_{\mathrm{b}}, N_{\mathrm{r}} \rightarrow \infty$,
the transmit power $p_n$ of User $n$ can be scaled down at most to $p_n=\frac{E_n}{N_{\mathrm{b}}N_{\mathrm{r}}^{2}}$ with fixed $E_n$,
while the transmit power of the other users can be scaled down at most to $p_u=\frac{E_u}{N_{\mathrm{b}}N_{\mathrm{r}}}$ with fixed $E_u$,
$u\neq n$.
In this case, the uplink achievable rate of User $n$ in the RIS-aided multi-user wideband massive MIMO OFDM system with low-resolution ADCs under Rician fading channels converges to
\begin{equation}\label{Scaling_law_aligned_p1}
  \tilde{R}_{n}\rightarrow  \frac{N_{\mathrm{c}}}{N_{\mathrm{cp}}+N_{\mathrm{c}}}
  \log _2\left( 1+\frac{\sigma _{\mathrm{u},0,n}^{2}\sigma _{\mathrm{b},0}^{2}K_{\mathrm{u},n}K_{\mathrm{b}}}{\sum_{u\ne n}^{N_{\mathrm{u}}-1}{\frac{E_u\beta _{\mathrm{u},u}\left( K_{\mathrm{u},n}+1 \right)}{E_n\beta _{\mathrm{u},n}\left( K_{\mathrm{u},u}+1 \right)}\sigma _{\mathrm{b},0}^{2}\varsigma _{\mathrm{u},u}K_{\mathrm{b}}}+\frac{\sigma _{\mathrm{noise}}^{2}\left( K_{\mathrm{u},n}+1 \right) \left( K_{\mathrm{b}}+1 \right)}{\alpha \beta _{\mathrm{u},n}\beta _{\mathrm{b}}E_n}} \right) .
\end{equation}

\end{corollary}

\section{RIS Phase Shift Design}

Since the RIS tailors the communication propagation environment by adjusting the phase of the signal through its reflecting elements,
it is crucial to design the RIS phase shifts for optimizing the achievable rate performance in the considered systems.

Considering fairness requirements, we aim to design the RIS phase shifts to maximize the minimum user rate.
We define the set of the users as ${\mathcal{N}} = \left\{ 0, 1, ..., N_{{\mathrm{u}}-1}  \right\}$,
and the angle vector of the RIS phase shifts as
$\bm{\theta}   \triangleq \left[ \theta _0,\theta _1,...,\theta _{N_{\mathrm{r}}-1} \right] ^T\in \mathbb{C} ^{N_{\mathrm{r}}\times 1}$.
Then, the phase shift optimization problem can be formulated as
\begin{equation*}
  \max_{\bm{\theta }} \,\,  \min_{n\in \mathcal{N}}  \tilde{R}_{n}\left( \bm{\theta } \right)
\end{equation*}
\begin{equation}\label{opt_P1}
  \mathrm{s}.\mathrm{t}. \;\;\;  \theta _i\in \left[ 0,2\pi \right) ,\forall i.
\end{equation}
%It is noted from \eqref{R_nt_appro} that $\tilde{R}_{nt}$ is independent of $t$.
%Thus, the objective function of Problem \eqref{opt_P1} can be transformed to
%\begin{equation}\label{min_R_nt}
%  \min_{n\in \mathcal{N}} \sum_{t=0}^{N_{\mathrm{c}}-1}{\tilde{R}_{nt}\left( \bm{\theta } \right)}
%  \Longrightarrow
%  \min_{n\in \mathcal{N}} \tilde{R}_{nt}\left( \bm{\theta } \right).
%\end{equation}
For the ease of expression, we define
\begin{equation}\label{SINR_nt}
  \mathrm{SINR}_{n}\triangleq \frac{\alpha ^2p_n\varpi _n}{\alpha ^2 \sum_{u\ne n}^{N_{\mathrm{u}}-1}{ p_u \eta _{n,u}}+ \alpha \left( 1-\alpha \right) \xi _n+\sigma _{\mathrm{noise}}^{2}\alpha \epsilon _n} .
\end{equation}
Then, from \eqref{R_nt_appro}, $\tilde{R}_{n}$ can be expressed as
\begin{equation}\label{R_nt_appro_SINR}
  \tilde{R}_{n}= \frac{N_{\mathrm{c}}}{N_{\mathrm{cp}}+N_{\mathrm{c}}}
  \log _2\left( 1+\mathrm{SINR}_{n} \right).
\end{equation}
Therefore, Problem \eqref{opt_P1} can be recast as
\begin{equation*}
  \max_{\bm{\theta }} \,\,  \min_{n\in \mathcal{N}} \mathrm{SINR}_{n}\left( \bm{\theta } \right)
\end{equation*}
\begin{equation}\label{opt_P2}
  \mathrm{s}.\mathrm{t}. \;\;\;  \theta _i\in \left[ 0,2\pi \right) ,\forall i.
\end{equation}

The non-differentiable objective function of Problem \eqref{opt_P2} can be approximated as \cite{1992An,9973349}
\begin{equation}\label{obj_F}
  \min_{n\in \mathcal{N}} \mathrm{SINR}_{n}\left( \bm{\theta } \right)
  \approx -\frac{1}{\mu}\ln \left\{ \sum_{n=0}^{N_{\mathrm{u}}-1}{\exp \left\{ -\mu \mathrm{SINR}_{n}\left( \bm{\theta } \right) \right\}} \right\} \triangleq F\left( \bm{\theta } \right),
\end{equation}
where the constant parameter $\mu$ controls the approximation accuracy with the error smaller than $\frac{ \ln {  N_{\mathrm{u}}   }  }{\mu}$.
Based on that, Problem \eqref{opt_P2} can be transformed to
\begin{equation*}
  \max_{\bm{\theta }} \,\,  F\left( \bm{\theta } \right)
\end{equation*}
\begin{equation}\label{opt_P3}
  \mathrm{s}.\mathrm{t}. \;\;\;  \theta _i\in \left[ 0,2\pi \right) ,\forall i.
\end{equation}

\begin{algorithm}[t]   %算法开始
\caption{Gradient Ascent Algorithm for RIS Phase Shift Design} %算法的题目
\label{optAlgo} %算法的标签
\begin{algorithmic}[1] %此处的[1]控制一下算法中的每句前面都有标号
  \State {\bf Initialization:}
  Randomly generate $\bm{\theta }_1$.
  Set $a_1=1$, $\mathbf{x}_1=\bm{\theta }_1$, $\alpha =0.3$, $\beta =0.8$.
  Set iteration number $i=1$.

 \While{1}

    \State Calculate the gradient of $F\left( \bm{\theta } \right)$ at $\bm{\theta }=\bm{\theta }_i$ as $\nabla F\left( \bm{\theta }_i \right) \triangleq \left. \frac{\partial}{\partial \bm{\theta }}F\left( \bm{\theta } \right) \right|_{\bm{\theta }=\bm{\theta }_i}$;

    \State Set step size $k=1$;

    \While{$F\left( \bm{\theta }_i+k\nabla F\left( \bm{\theta }_i \right) \right) <F\left( \bm{\theta }_i \right) +\alpha k\left\| \nabla F\left( \bm{\theta }_i \right) \right\| ^2$}
    \State $k=\beta k$;
    \EndWhile

    \State $\mathbf{x}_{i+1}=\bm{\theta }_i+k\nabla F\left( \bm{\theta }_i \right)$;

    \State $a_{i+1}=\frac{1+\sqrt{4a_{i}^{2}+1}}{2}$;

    \State $\bm{\theta }_{i+1}=\mathbf{x}_{i+1}+\frac{\left( a_i-1 \right) \left( \mathbf{x}_{i+1}-\mathbf{x}_i \right)}{a_{i+1}}$;

    \If{$F\left( \bm{\theta }_{i+1} \right) -F\left( \bm{\theta }_i \right) <10^{-4}$}
    \State $\bm{\theta }^{\mathrm{opt}}=\bm{\theta }_{i+1}$;
    \State Break;
    \EndIf

    \State $i=i+1$;

\EndWhile

\State Output the optimal phase shift design $\bm{\theta }^{\mathrm{opt}}$.

\end{algorithmic}
\end{algorithm}

Herein, we propose a gradient ascent-based algorithm to solve Problem \eqref{opt_P3}, which is given by Algorithm \ref{optAlgo}.
Steps 4-7 in Algorithm \ref{optAlgo} is to obtain a suitable step size based on the backtracking line search \cite{Convexopt2004}.
Steps 9-10 is to increase the convergence speed of the algorithm based on Nesterov's accelerated gradient method \cite{1983Nesterov}.
Based on \eqref{obj_F}, the gradient of the objective function $F\left( \bm{\theta } \right)$ can be calculated as follows:
\begin{equation}\label{obj_F_gradient}
  \frac{\partial F\left( \bm{\theta } \right)}{\partial \bm{\theta }}=\frac{\sum_{n=0}^{N_{\mathrm{u}}-1}{\exp \left\{ -\mu \mathrm{SINR}_{n}\left( \bm{\theta } \right) \right\}}\frac{\partial \mathrm{SINR}_{n}\left( \bm{\theta } \right)}{\partial \bm{\theta }}}{\sum_{n=0}^{N_{\mathrm{u}}-1}{\exp \left\{ -\mu \mathrm{SINR}_{n}\left( \bm{\theta } \right) \right\}}},
\end{equation}
where, using \eqref{SINR_nt}, $\frac{\partial \mathrm{SINR}_{n}\left( \bm{\theta } \right)}{\partial \bm{\theta }}$ is further expressed as
\begin{equation*}
  \frac{\partial \mathrm{SINR}_{n}\left(  \bm{\theta } \right)}{\partial  \bm{\theta } }=\frac{\alpha ^2 p_n \frac{\partial \varpi _n\left(  \bm{\theta } \right)}{\partial  \bm{\theta } }}{\alpha ^2\sum_{u\ne n}^{N_{\mathrm{u}}-1}{  p_u \eta _{n,u}\left(  \bm{\theta } \right)}+  \alpha \left( 1-\alpha \right) \xi _n\left(  \bm{\theta } \right) +\sigma _{\mathrm{noise}}^{2}\alpha \epsilon _n\left(  \bm{\theta } \right)}
\end{equation*}
\begin{equation}\label{SINR_nt_gradient}
  -\frac{\alpha ^2 p_n \varpi _n\left(  \bm{\theta } \right) \left( \alpha ^2 \sum_{u\ne n}^{N_{\mathrm{u}}-1}{  p_u \frac{\partial \eta _{n,u}\left(  \bm{\theta } \right)}{\partial  \bm{\theta } }}+  \alpha \left( 1-\alpha \right) \frac{\partial \xi _n\left(  \bm{\theta } \right)}{\partial  \bm{\theta } }+\sigma _{\mathrm{noise}}^{2}\alpha \frac{\partial \epsilon _n\left(  \bm{\theta } \right)}{\partial  \bm{\theta } } \right)}{\left( \alpha ^2 \sum_{u\ne n}^{N_{\mathrm{u}}-1}{ p_u \eta _{n,u}\left(  \bm{\theta } \right)}+  \alpha \left( 1-\alpha \right) \xi _n\left(  \bm{\theta } \right) +\sigma _{\mathrm{noise}}^{2}\alpha \epsilon _n\left(  \bm{\theta } \right) \right) ^2}
  ,
\end{equation}
and $\frac{\partial}{\partial \theta _i}\varpi _n\left( \bm{\theta } \right)$, $\frac{\partial}{\partial \theta _i}\eta _{n,u}\left( \bm{\theta } \right)$,
$\frac{\partial}{\partial \theta _i}\xi _n\left( \bm{\theta } \right)$, and $\frac{\partial}{\partial \theta _i}\epsilon _n\left( \bm{\theta } \right)$
are respectively given by
\begin{equation*}
  \frac{\partial}{\partial \theta _i}\varpi _n\left( \bm{\theta } \right) =
  \frac{-4\beta _{\mathrm{u},n}^{2}\beta _{\mathrm{b}}^{2}\sigma _{\mathrm{u},0,n}^{2}\sigma _{\mathrm{b},0}^{2}K_{\mathrm{u},n}K_{\mathrm{b}}N_{\mathrm{b}}}{\left( K_{\mathrm{u},n}+1 \right) ^2\left( K_{\mathrm{b}}+1 \right) ^2}c_{\mathrm{gra}}^{1}\left( n \right)
  \times
  \bigg(
  \sigma _{\mathrm{u},0,n}^{2}\sigma _{\mathrm{b},0}^{2}K_{\mathrm{u},n}K_{\mathrm{b}}N_{\mathrm{b}}\left| \varPhi _{N_{\mathrm{r}}}\left( n \right) \right|^2
\end{equation*}
\begin{equation}\label{omega_bar_n_derivative_i}
  +2\sigma _{\mathrm{b},0}^{2}\varsigma _{\mathrm{u},n}K_{\mathrm{b}}N_{\mathrm{b}}N_{\mathrm{r}}
  +\left( \sigma _{\mathrm{u},0,n}^{2}\varsigma _{\mathrm{b}}K_{\mathrm{u},n}+\varsigma _{\mathrm{u},n}\varsigma _{\mathrm{b}} \right) \left( N_{\mathrm{b}}+1 \right) N_{\mathrm{r}}
  +2\varsigma _{\mathrm{u},n}\varsigma _{\mathrm{b}}\left( N_{\mathrm{b}}+1 \right)
  \bigg),
\end{equation}
\begin{equation*}
\hspace{-6.2cm}
  \frac{\partial}{\partial \theta _i}\eta _{n,u}\left( \bm{\theta } \right) =
  \frac{-2\beta _{\mathrm{u},n}\beta _{\mathrm{u},u}\beta _{\mathrm{b}}^{2}\sigma _{\mathrm{b},0}^{2}K_{\mathrm{b}}N_{\mathrm{b}}}{\left( K_{\mathrm{u},n}+1 \right) \left( K_{\mathrm{u},u}+1 \right) \left( K_{\mathrm{b}}+1 \right) ^2}
  \times
\end{equation*}
\begin{equation*}
\hspace{-1.5cm}
  \Bigg(
  \sigma _{\mathrm{u},0,n}^{2}\sigma _{\mathrm{u},0,u}^{2}\sigma _{\mathrm{b},0}^{2}K_{\mathrm{u},n}K_{\mathrm{u},u}K_{\mathrm{b}}N_{\mathrm{b}}\left( \left| \varPhi _{N_{\mathrm{r}}}\left( u \right) \right|^2c_{\mathrm{gra}}^{1}\left( n \right) +\left| \varPhi _{N_{\mathrm{r}}}\left( n \right) \right|^2c_{\mathrm{gra}}^{1}\left( u \right) \right)
\end{equation*}
\begin{equation*}
  +\left( \left( \sigma _{\mathrm{b},0}^{2}\varsigma _{\mathrm{u},u}K_{\mathrm{b}}N_{\mathrm{b}}+\sigma _{\mathrm{u},0,u}^{2}\varsigma _{\mathrm{b}}K_{\mathrm{u},u}+\varsigma _{\mathrm{u},u}\varsigma _{\mathrm{b}} \right) N_{\mathrm{r}}+2\varsigma _{\mathrm{u},u}\varsigma _{\mathrm{b}}N_{\mathrm{b}} \right) \sigma _{\mathrm{u},0,n}^{2}K_{\mathrm{u},n}c_{\mathrm{gra}}^{1}\left( n \right)
\end{equation*}
\begin{equation*}
  +\left( \left( \sigma _{\mathrm{b},0}^{2}\varsigma _{\mathrm{u},n}K_{\mathrm{b}}N_{\mathrm{b}}+\sigma _{\mathrm{u},0,n}^{2}\varsigma _{\mathrm{b}}K_{\mathrm{u},n}+\varsigma _{\mathrm{u},n}\varsigma _{\mathrm{b}} \right) N_{\mathrm{r}}+2\varsigma _{\mathrm{u},n}\varsigma _{\mathrm{b}}N_{\mathrm{b}} \right) \sigma _{\mathrm{u},0,u}^{2}K_{\mathrm{u},u}c_{\mathrm{gra}}^{1}\left( u \right)
\end{equation*}
\begin{equation}\label{eta_n_u_derivative_i}
\hspace{-2.8cm}
  +\sigma _{\mathrm{u},0,n}^{2}\sigma _{\mathrm{u},0,u}^{2}\varsigma _{\mathrm{b}}K_{\mathrm{u},n}K_{\mathrm{u},u}N_{\mathrm{b}}\mathrm{Im}\left( c_{\mathrm{gra}}^{2}\left( n,u \right) \left( \bar{\mathbf{h}}_{\mathrm{u}}^{\left( \cdot ,u \right)} \right) ^H\bar{\mathbf{h}}_{\mathrm{u}}^{\left( \cdot ,n \right)} \right)
  \Bigg),
\end{equation}
\begin{equation*}
  \frac{\partial}{\partial \theta _i}\xi _n\left( \bm{\theta } \right) =
  \frac{-2\beta _{\mathrm{u},n}\beta _{\mathrm{b}}^{2}\sigma _{\mathrm{b},0}^{2}K_{\mathrm{b}}N_{\mathrm{b}}}{\left( K_{\mathrm{u},n}+1 \right) \left( K_{\mathrm{b}}+1 \right) ^2}
  \times
  \Bigg(
  p_n \frac{2\beta _{\mathrm{u},n}\sigma _{\mathrm{u},0,n}^{4}K_{\mathrm{u},n}}{K_{\mathrm{u},n}+1}\left( \sigma _{\mathrm{b},0}^{2}K_{\mathrm{b}}N_{\mathrm{r}}+\sigma _{\mathrm{b},0}^{2}N_{\mathrm{r}}+2\varsigma _{\mathrm{b}} \right) c_{\mathrm{gra}}^{1}\left( n \right)
\end{equation*}
\begin{equation*}
\hspace{-1.5cm}
  +\sigma _{\mathrm{u},0,n}^{2}\sigma _{\mathrm{b},0}^{2}K_{\mathrm{b}}K_{\mathrm{u},n}\sum_{u=0}^{N_{\mathrm{u}}-1}{  p_u  \frac{\beta _{\mathrm{u},u}K_{\mathrm{u},u}}{K_{\mathrm{u},u}+1}\sigma _{\mathrm{u},0,u}^{2}\left( \left| \varPhi _{N_{\mathrm{r}}}\left( u \right) \right|^2c_{\mathrm{gra}}^{1}\left( n \right) +\left| \varPhi _{N_{\mathrm{r}}}\left( n \right) \right|^2c_{\mathrm{gra}}^{1}\left( u \right) \right)}
\end{equation*}
\begin{equation*}
  +\left( \sigma _{\mathrm{u},0,n}^{2}\varsigma _{\mathrm{b}}K_{\mathrm{u},n}N_{\mathrm{r}}+\sigma _{\mathrm{b},0}^{2}\varsigma _{\mathrm{u},n}K_{\mathrm{b}}N_{\mathrm{r}}+\varsigma _{\mathrm{u},n}\varsigma _{\mathrm{b}}N_{\mathrm{r}}+2\sigma _{\mathrm{b},0}^{2}\varsigma _{\mathrm{u},n} \right) \sum_{u=0}^{N_{\mathrm{u}}-1}{  p_u  \frac{\beta _{\mathrm{u},u}K_{\mathrm{u},u}}{K_{\mathrm{u},u}+1}\sigma _{\mathrm{u},0,u}^{2}c_{\mathrm{gra}}^{1}\left( u \right)}
\end{equation*}
\begin{equation*}
\hspace{-3.1cm}
  +\sigma _{\mathrm{u},0,n}^{2}\sigma _{\mathrm{b},0}^{2}K_{\mathrm{u},n}\mathrm{Im}\left( \sum_{u=0}^{N_{\mathrm{u}}-1}{   p_u   \sigma _{\mathrm{u},0,u}^{2}\frac{\beta _{\mathrm{u},u}K_{\mathrm{u},u}}{K_{\mathrm{u},u}+1}c_{\mathrm{gra}}^{2}\left( n,u \right) \left( \bar{\mathbf{h}}_{\mathrm{u}}^{\left( \cdot ,u \right)} \right) ^H\bar{\mathbf{h}}_{\mathrm{u}}^{\left( \cdot ,n \right)}} \right)
\end{equation*}
\begin{equation*}
\hspace{-7.0cm}
  +\sigma _{\mathrm{u},0,n}^{2}\varsigma _{\mathrm{b}}K_{\mathrm{u},n}N_{\mathrm{r}}c_{\mathrm{gra}}^{1}\left( n \right) \sum_{u=0}^{N_{\mathrm{u}}-1}{  p_u  \frac{\beta _{\mathrm{u},u}K_{\mathrm{u},u}}{K_{\mathrm{u},u}+1}\sigma _{\mathrm{u},0,u}^{2}}
\end{equation*}
\begin{equation}\label{xi_n_derivative_i}
\hspace{-3.4cm}
  +\sigma _{\mathrm{u},0,n}^{2}K_{\mathrm{u},n}\left( \varsigma _{\mathrm{b}}N_{\mathrm{r}}+2\sigma _{\mathrm{b},0}^{2}+\sigma _{\mathrm{b},0}^{2}K_{\mathrm{b}}N_{\mathrm{r}} \right) c_{\mathrm{gra}}^{1}\left( n \right) \sum_{u=0}^{N_{\mathrm{u}}-1}{  p_u  \frac{\beta _{\mathrm{u},u}\varsigma _{\mathrm{u},u}}{K_{\mathrm{u},u}+1}}
  \Bigg) ,
\end{equation}
\begin{equation}\label{epsilon_n_derivative_i}
  \frac{\partial}{\partial \theta _i}\epsilon _n\left( \bm{\theta } \right)
  =\frac{-2\beta _{\mathrm{u},n}\beta _{\mathrm{b}}\sigma _{\mathrm{u},0,n}^{2}\sigma _{\mathrm{b},0}^{2}K_{\mathrm{u},n}K_{\mathrm{b}}N_{\mathrm{b}}}{\left( K_{\mathrm{u},n}+1 \right) \left( K_{\mathrm{b}}+1 \right)}c_{\mathrm{gra}}^{1}\left( n \right) ,
\end{equation}
with
\begin{equation}\label{c_gra_1}
  c_{\mathrm{gra}}^{1}\left( n \right) =\mathrm{Im}\left( \varPhi \left( n,i \right) \sum_{r\ne i}^{N_{\mathrm{r}}-1}{\left( \varPhi \left( n,r \right) \right) ^{\ast}} \right),
\end{equation}
\begin{equation}\label{c_gra_2}
  c_{\mathrm{gra}}^{2}\left( n,u \right) =\left( \varPhi _{N_{\mathrm{r}}}\left( n \right) \right) ^{\ast}\varPhi \left( u,i \right) -\left( \varPhi \left( n,i \right) \right) ^{\ast}\varPhi _{N_{\mathrm{r}}}\left( u \right).
\end{equation}

Moreover,
since we need to calculate the objective function and its gradient in each iteration of Algorithm \ref{optAlgo},
the complexity of each iteration of Algorithm \ref{optAlgo} is on the order of $\mathcal{O} \left(   N_{\mathrm{r}}N^2_{\mathrm{u}}   \right)$ when assuming the steps of the backtracking line search is limited.

\section{Numerical Results}
In this section, numerical results are presented to verify the main results of this paper.
Referring to \cite{9973349} and \cite{9356722}, in our simulation, unless otherwise stated,
the distance between the BS and the RIS is set to be 200 m,
and $N_{\mathrm{u}}=4$ users are distributed at a circle centered at the RIS with the radius of 30 m.
The numbers of the channel taps of the RIS-to-BS link and of the User-to-RIS links are $L_{\mathrm{u}}=5$ with an attenuation step of 2.5 dB and $L_{\mathrm{b}}=4$ with an attenuation step of 5 dB, respectively.
The number of sub-carriers is $N_{\mathrm{c}}=32$.
All the AoAs and AoDs are generated from a uniform distribution in $(0, 2\pi)$,
and the array spacing is set as $ d = \lambda / 2 $.
The large-scale coefficient follows  $\beta =10^{-3}a^{-2.8}$,
where $a$ stands for the distance.
The Rician factors are set to $K_{\mathrm{b}}=10$ dB for the RIS-to-BS link,
and $K_{\mathrm{u},i}=3$ dB, $\forall i$, for the User-to-RIS links.
Besides, the transmit power on each sub-carrier is set to $p=30$ dBm,
and the noise power is $\sigma _{\mathrm{noise}}^{2}=-104$ dBm.
Furthermore, all the simulation results calculated from the Monte-Carlo method are obtained by averaging over 2000 random channel realizations.

%\begin{figure}
%\vspace{-0.6cm}
%    \centering
%    % Requires \usepackage{graphicx}
%    \includegraphics[scale=0.6]{picture/rician/SEComparison.eps}\\  \vspace{-0.1cm}
%    \caption{Sum SEs versus $M$}\label{p_SEwithM}
%\vspace{-0.5cm}
%\end{figure}

\begin{figure}
%\vspace{-0.6cm}
    \centering
    \includegraphics[scale=0.8]{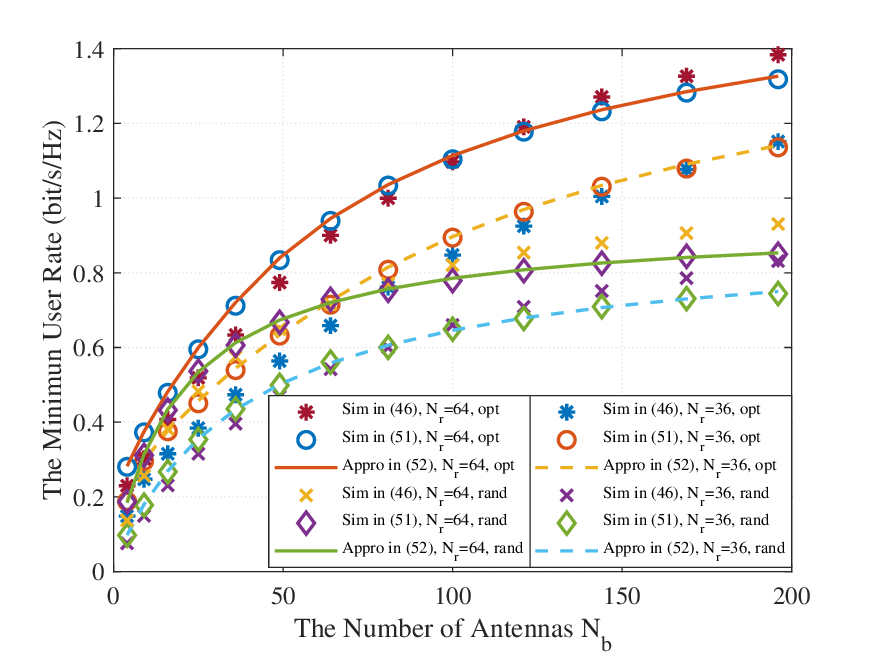}\\  \vspace{-0.1cm}
    \caption{The Minimum User Rate versus $N_{\mathrm{b}}$}\label{p_rate_Nb}
%\vspace{-0.5cm}
\end{figure}

Fig. \ref{p_rate_Nb}
investigates the impact of the number of the antennas $N_{\mathrm{b}}$ at the BS on the minimum uplink user rate.
The legend ``opt" means that the phase shifts of the RIS are optimized by the proposed Algorithm \ref{optAlgo},
while the legend ``rand" means that the phase shifts of the RIS are randomly generated.
As previously discussed, it is challenging to obtain a closed-form expression of \eqref{R_n_sim}.
Thus, we resort to the approximation \eqref{R_n_hat_1},
and derive the closed-form expression \eqref{R_nt_appro} in Theorem \ref{theorem_R_nt_mrc_appro}.
It is readily seen that the curves of results from \eqref{R_n_sim} and from \eqref{R_n_hat_1} are close,
and the curves of results from \eqref{R_n_hat_1} and from \eqref{R_nt_appro} are perfectly matched,
which verifies the correctness of Theorem \ref{theorem_R_nt_mrc_appro}.
Furthermore, we find that the uplink rates first grow fast as $N_{\mathrm{b}}$ increases,
and then tend to be saturated with large $N_{\mathrm{b}}$,
which reveals that when $N_{\mathrm{b}}$ is large enough,
the performance gain brought by enlarging the scale of antenna array is marginal.
It can be seen that in that case, the achievable rate can be greatly improved by increasing the number of RIS elements, $N_{\mathrm{r}}$.
Besides, Fig. \ref{p_rate_Nb} indicates that the proposed Algorithm \ref{optAlgo} effectively improves the system performance compared with the un-optimized cases.
Moreover, we find that the rate with the optimized RIS phase shifts do not saturate as fast as that with the random RIS phase shifts, which means the optimized RIS phase shifts could also improve the performance gain brought by number of the BS antennas.

\begin{figure}
%\vspace{-0.6cm}
    \centering
    \includegraphics[scale=0.8]{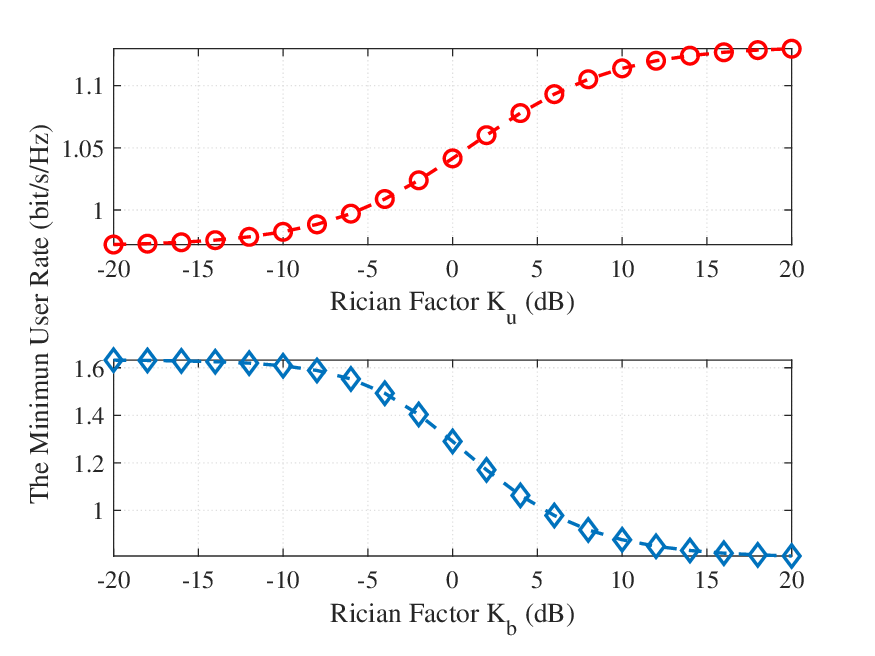}\\  \vspace{-0.1cm}
    \caption{The Minimum User Rate versus The Rician Factors}\label{p_rate_Ku_Kb}
%\vspace{-0.5cm}
\end{figure}

Fig. \ref{p_rate_Ku_Kb}
shows how the minimum user rate varies with
the Rician factors $K_{\mathrm{u}}$ of the User-to-RIS link and $K_{\mathrm{b}}$ of the RIS-to-BS link.
We fix the number of the antennas to $N_{\mathrm{b}} = 100$
and the number of the RIS reflecting elements to $N_{\mathrm{r}} = 64$.
It is readily observed from Fig. \ref{p_rate_Ku_Kb} (a) that
the minimum uplink user rate increases with $K_{\mathrm{u}}$,
and decreases with the $K_{\mathrm{b}}$.
This is because the LoS matrix $\bar{\mathbf{G}}_{\mathrm{u}}$ (defined in \eqref{FD_channels}) of the User-to-RIS link is full-rank,
while the LoS matrix $\bar{\mathbf{G}}_{\mathrm{b}}$ of the RIS-to-BS link is rank-deficient which cannot effectively support multi-user transmission.
As the Rician factor increase, the LoS components become dominant, which results in the observations of Fig. \ref{p_rate_Ku_Kb}.
%It is readily observed from Fig. \ref{p_rate_Ku_Kb} (a) that
%the minimum uplink user rate grows
%from 0.82 bit/s/Hz to 0.96 bit/s/Hz, when the Rician factor $K_{\mathrm{u}}$ of the User-to-RIS link increases from -15 dB to 15 dB.
%On the contrary, Fig. \ref{p_rate_Ku_Kb} (b) indicates that
%the minimum uplink user rate drops from 1.4 bit/s/Hz to 0.7 bit/s/Hz, when the Rician factor $K_{\mathrm{b}}$ of the RIS-to-BS link increases from -15 dB to 15 dB.
It can be concluded that
to improve the performance of the considered system,
the RIS should be equipped at the place where the User-to-RIS link has a high-qulity LoS path,
and the RIS-to-BS link has adequate scatterers.

\begin{figure}
%\vspace{-0.6cm}
    \centering
    \includegraphics[scale=0.8]{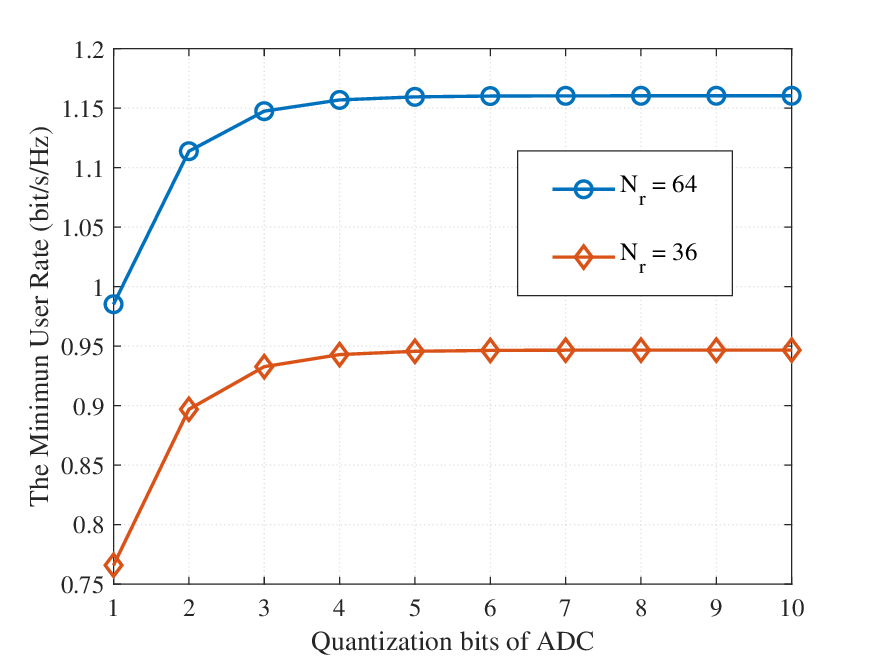}\\  \vspace{-0.1cm}
    \caption{The Minimum User Rate versus The Quantization Bit of ADC}\label{p_rate_quantization}
%\vspace{-0.5cm}
\end{figure}

Fig. \ref{p_rate_quantization}
depicts the relationship between the minimum uplink user rate and the number of quantization bits $b$ of the ADC.
In both cases of $N_{\mathrm{r}} = 64$ and $N_{\mathrm{r}} = 36$,
the minimum uplink user rate first grows fast with $b$, and then becomes saturated after $b = 4$.
This means that the performance loss caused by low-resolution ADCs is marginal with 4-bit quantization precision.
Besides, the power consumption and hardware cost usually grow rapidly with $b$.
Thus, for the considered system,
setting $b = 4$ would be a suitable choice to balance the performance and the cost.

\begin{figure}
%\vspace{-0.6cm}
    \centering
    \includegraphics[scale=0.8]{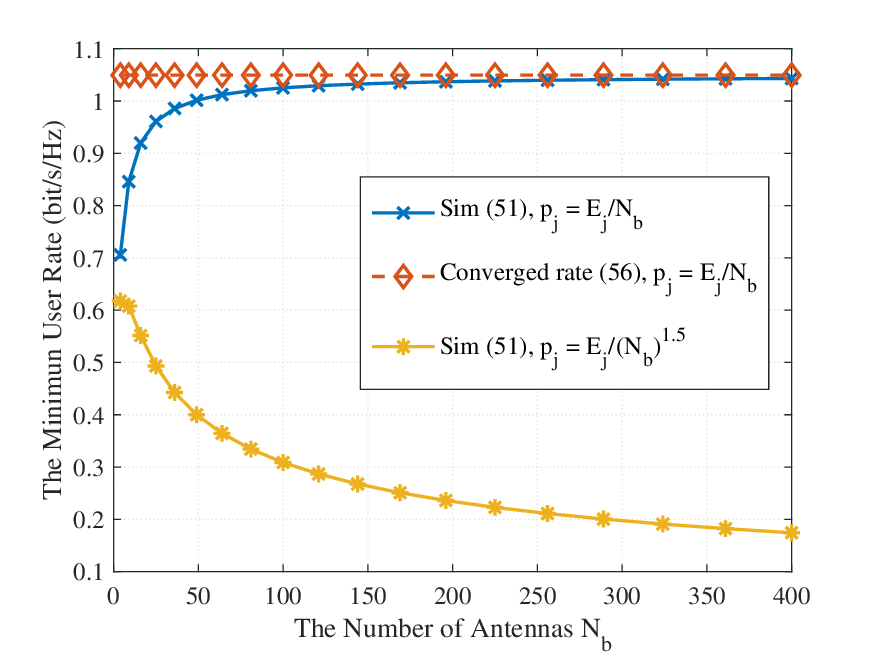}\\  \vspace{-0.1cm}
    \caption{The Power Scaling Law Related to $N_{\mathrm{b}}$}\label{p_rate_scaling}
%\vspace{-0.5cm}
\end{figure}

Fig. \ref{p_rate_scaling}
verifies the power scaling law proposed in Corollary \ref{Scaling_law_Nb}.
We set the number of the RIS reflecting elements to $N_{\mathrm{r}} = 64$,
and $E_j=E=50$ dBm, $j=0,1,...,N_{\mathrm{u}}-1$.
It is observed that
when $N_{\mathrm{b}}$ becomes large and $p_j$ is scaled down to $p_j = \frac{E_j}{N_{\mathrm{b}}}$,
the rate obtained by \eqref{R_nt_hat_1} converges to a non-zero value,
which is consistent with the analytical result obtained in \eqref{Scaling_law_Nb_rate}.
Furthermore, we illustrate the case when $p_j$ is scaled down heavier as $p_j = \frac{E_j}{{\left( N_{\mathrm{b}} \right)}^{1.5} }$.
It can be seen that the minimum user rate tends to be zero as $N_{\mathrm{b}} \rightarrow \infty$.
Thus, the simulation verifies the correctness of the conclusion in Corollary \ref{Scaling_law_Nb} that
the power of the users can be scaled down proportionally at most to $\frac{1}{N_{\mathrm{b}}}$,
while guaranteeing required system performance.

\begin{figure}
%\vspace{-0.6cm}
    \centering
    \includegraphics[scale=0.8]{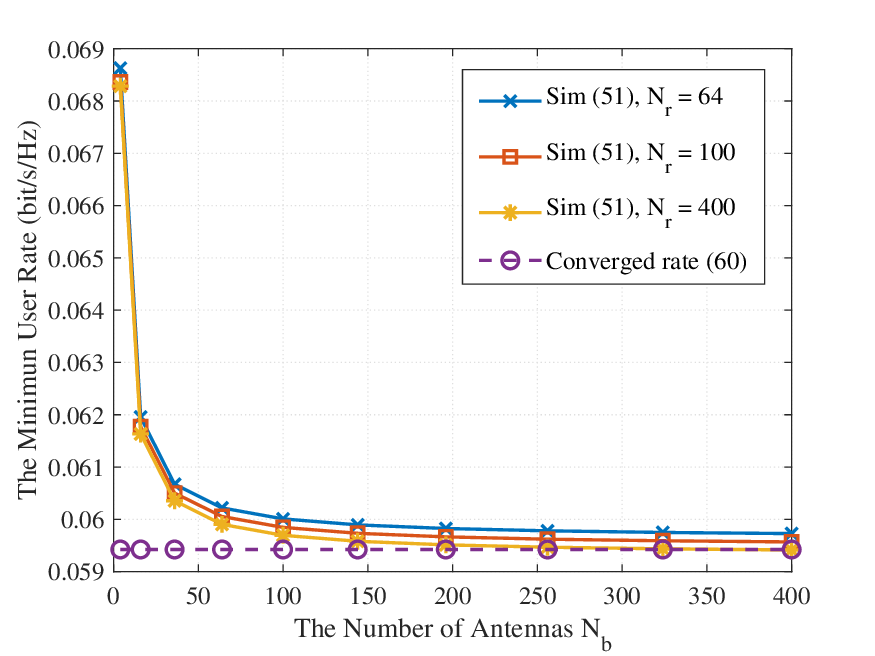}\\  \vspace{-0.1cm}
    \caption{The Power Scaling Law Related to $N_{\mathrm{b}}$ and $N_{\mathrm{r}}$}\label{p_rate_scaling2}
%\vspace{-0.5cm}
\end{figure}

Fig. \ref{p_rate_scaling2}
investigates the power scaling law in Corollary \ref{Scaling_law_Nb_Nr}.
The power of the users are scaled down to $p_j = \frac{E_j}{N_{\mathrm{b}} N_{\mathrm{r}} }$,
$E_j=E=50$ dBm, $j=0,1,...,N_{\mathrm{u}}-1$.
Three cases are considered:
1) $N_{\mathrm{r}} = 64$,
2) $N_{\mathrm{r}} = 100$,
and 3) $N_{\mathrm{r}} = 400$.
As $N_{\mathrm{b}} $ becomes large,
the minimum user rates in the three cases all converge to non-zero values,
which further verify the correctness of the conclusion in Corollary \ref{Scaling_law_Nb}.
From Case 1 to Case 3, it is observed that as $N_{\mathrm{r}}$ grows large,
the non-zero values in the three cases become closer to the asymptotic rate obtained by \eqref{Scaling_law_Nb_Nr_rate},
which well supports the conclusion in Corollary \ref{Scaling_law_Nb_Nr}.

\section{Conclusion}
This paper studied an RIS-aided wideband massive MIMO OFDM system with low-resolution ADCs.
The AQNM was adopted for low-resolution ADCs,
and all the channels were assumed to follow frequency-selective Rician fading.
This paper presented the time-domain OFDM data transmission process.
Then, with MRC processing, the closed-form approximate expression for the uplink achievable user rate was derived in Theorem \ref{theorem_R_nt_mrc_appro}.
Based on that, this paper analyzed the asymptotic performance in Corollary \ref{N_b_N_r} when the number of the antennas $N_{\mathrm{b}}$ and the number of reflecting elements $N_{\mathrm{r}}$ went to infinity.
It showed that the asymptotic rate of User $n$ is on the order of $\mathcal{O} \left( \log _2\left( N_{\mathrm{r}} \right) \right)$
when the RIS is aligned with User $n$, which means that we can improve the system performance for a specific user by adjusting the RIS phase shifts and increasing $N_{\mathrm{r}}$.
Furthermore, the power scaling laws were investigated in Corollaries \ref{Scaling_law_Nb}-\ref{Scaling_law_aligned} for the power saving designs of the considered system.
They indicated that when $N_{\mathrm{b}}$ and $N_{\mathrm{r}}$ go to infinity, if the RIS is not aligned with any users, the transmit power of each user can be scaled down at most proportionally to $\frac{1}{N_{\mathrm{b}}N_{\mathrm{r}}}$ while the considered system keeps a constant asymptotic rate.
On the other hand, when the RIS is aligned with User $n$,
the transmit power of User $n$ can be further reduced proportionally to $\frac{1}{N_{\mathrm{b}}N^2_{\mathrm{r}}}$,
while guaranteeing certain system performance.
Furthermore, this paper proposed a gradient ascent-based algorithm, i.e., Algorithm \ref{optAlgo}, to design the optimal RIS phase shifts with the maximized minimum user rate.
The numerical results were presented to validate the main conclusions of this paper.
Besides, we found that to improve the system performance, the RIS should be equipped at the place where the User-to-RIS link has a high-qulity LoS path, and the RIS-to-BS link has adequate scatterers.
Additionally, to trade-off the performance and the cost of the system, we found that a setting of ADC quantization bits $b=4$ would be a suitable choice.

\begin{appendices}

%\section{Lemmas}\label{app_lemmma} \vspace{-0.2cm}
%First, we review some key preliminary results given in the following lemmas.
%
%\textcolor[rgb]{0.00,0.07,1.00}{\emph{Lemma 1:}
%  Suppose $\mathbf{W}=\mathbf{H}^H\mathbf{H}$ is a non-central Wishart matrix with the distribution of $\mathcal{W} _N\left( M,\mathbf{P},\mathbf{\Sigma } \right)$, where $\mathbf{\Sigma }$ is the covariance matrix of the row vectors of $\mathbf{H}$, and $\mathbf{P}$ is the mean matrix of $\mathbf{H}$.
%  Then, we can approximate $\mathbf{W}=\mathbf{H}^H\mathbf{H}$ by a central Wishart distribution $\mathcal{W} _N\left( M,\mathbf{0},\mathbf{\bar{\Sigma}} \right)$ with $\mathbf{\bar{\Sigma}}=\mathbf{\Sigma }+\frac{1}{M}\mathbf{P}^H\mathbf{P}$ \cite{5474635,6210404}.}
%
%
%\emph{Lemma 2:} \cite[\emph{Lemma 1}]{6816003}
%  If  $X = \sum\nolimits_{i = 1}^{{t_1}} {{X_i}} $ and $Y = \sum\nolimits_{j = 1}^{{t_2}} {{Y_j}} $  are both sums of nonnegative random variables ${X_i}$  and  ${Y_j}$, then we get the following approximation
%  \vspace{-0.3cm}\begin{equation}\label{9-1}
%    {\rm E}\left\{ {{{\log }_2}\left( {1 + \frac{X}{Y}} \right)} \right\} \approx {\log _2}\left( {1 + \frac{{{\rm E}\left\{ X \right\}}}{{{\rm E}\left\{ Y \right\}}}} \right).\vspace{-0.3cm}
%  \end{equation}
%  Note that it is not necessary for the random variables  $X$ and $Y$  to be independent.
%  In addition, the approximation becomes more accurate as ${t_1}$  and  ${t_2}$ increase.

\section{}\label{app_A}

We commence by focusing on the expressions of the frequency-domain channels $\mathbf{G}_{\mathrm{u}}$ and $\mathbf{G}_{\mathrm{b}}$.
According to \eqref{R-U_FD_channel_ij} and \eqref{B-R_FD_channel_ij}, we have
\begin{equation*}
  \mathbf{G}_{\mathrm{x}}^{\left( i,j \right)}=\mathrm{diag}\left( \sqrt{N_{\mathrm{c}}\beta _{\mathrm{x},j}}\mathbf{Fh}_{\mathrm{x}}^{\left( i,j \right)} \right) =\sqrt{N_{\mathrm{c}}\beta _{\mathrm{x},j}}
  \left[ \begin{matrix}
	\sum_{k=0}^{N_{\mathrm{c}}-1}{f_{0,k}h_{\mathrm{x},k}^{\left( i,j \right)}}&		0&		0\\
	0&		\ddots&		0\\
	0&		0&		\sum_{k=0}^{N_{\mathrm{c}}-1}{f_{\left( N_{\mathrm{c}}-1 \right) ,k}h_{\mathrm{x},k}^{\left( i,j \right)}}\\
\end{matrix} \right]
\end{equation*}
\begin{equation}\label{expand_FD_channel_ij}
\hspace{-1cm}
  =\sqrt{N_{\mathrm{c}}\beta _{\mathrm{x},j}}
  \left[ \begin{matrix}
	\sum_{k=0}^{L_{\mathrm{x}}-1}{f_{0,k}h_{\mathrm{x},k}^{\left( i,j \right)}}&		0&		0\\
	0&		\ddots&		0\\
	0&		0&		\sum_{k=0}^{L_{\mathrm{x}}-1}{f_{\left( N_{\mathrm{c}}-1 \right) ,k}h_{\mathrm{x},k}^{\left( i,j \right)}}\\
\end{matrix} \right] , \;\;\;    \mathrm{x}\in \left\{ \mathrm{u},\mathrm{b} \right\}.
\end{equation}
Therefore, the $t$-th element of the diagonal matrix $\mathbf{G}_{\mathrm{u}}^{\left( i,j \right)}$ can be further expressed as
\begin{equation*}
\hspace{-11cm}
  \left[ \mathbf{G}_{\mathrm{u}}^{\left( i,j \right)} \right] _{t,t}=\sqrt{N_{\mathrm{c}}\beta _{\mathrm{u},j}}\sum_{k=0}^{L_{\mathrm{u}}-1}{f_{t,k}h_{\mathrm{u},k}^{\left( i,j \right)}}
\end{equation*}
\begin{equation*}
  =\sqrt{N_{\mathrm{c}}\beta _{\mathrm{u},j}}\left( \sqrt{\frac{K_{\mathrm{u},j}}{K_{\mathrm{u},j}+1}}f_{t,0}\sigma _{\mathrm{u},0,j}\bar{h}_{\mathrm{u}}^{\left( i,j \right)}+\sqrt{\frac{1}{K_{\mathrm{u},j}+1}}f_{t,0}\tilde{h}_{\mathrm{u},0}^{\left( i,j \right)}+\sum_{k=1}^{L_{\mathrm{u}}-1}{f_{t,k}\tilde{h}_{\mathrm{u},k}^{\left( i,j \right)}} \right)
\end{equation*}
\begin{equation}\label{R-U_FD_channel_ij_tt}
\hspace{1.4cm}
  = \underbrace{\sqrt{\frac{\beta _{\mathrm{u},j}K_{\mathrm{u},j}}{K_{\mathrm{u},j}+1}}\sigma _{\mathrm{u},0,j}a_{N_{\mathrm{r}},i}\left( \phi _{\mathrm{r},j}^{aa},\phi _{\mathrm{r},j}^{ea} \right)}_{\bar{g}_{\mathrm{u}}^{\left( i,j \right)}}
  + \underbrace{\sqrt{\frac{\beta _{\mathrm{u},j}}{K_{\mathrm{u},j}+1}}\tilde{h}_{\mathrm{u},0}^{\left( i,j \right)}+\sqrt{N_{\mathrm{c}}\beta _{\mathrm{u},j}}\sum_{k=1}^{L_{\mathrm{u}}-1}{f_{t,k}\tilde{h}_{\mathrm{u},k}^{\left( i,j \right)}}} _{\tilde{g}_{\mathrm{u},t,t}^{\left( i,j \right)}}
  \triangleq g_{\mathrm{u},t,t}^{\left( i,j \right)} ,
\end{equation}
where $\bar{g}_{\mathrm{u}}^{\left( i,j \right)}$ (independent of $t$) stands for the LoS component, and $\tilde{g}_{\mathrm{u},t,t}^{\left( i,j \right)}$ represents the NLoS component.
Since $\tilde{h}_{\mathrm{u},l}^{\left( i,j \right)}$ follows the distribution of $\mathcal{C} \mathcal{N} \left( 0,\sigma _{\mathrm{u},l,j}^{2} \right)$ with $\sum_{l=0}^{L_{\mathrm{u}}-1}{\sigma _{\mathrm{u},l,j}^{2}}=1$,
from \eqref{DFT_fij},
we have
\begin{equation*}
  \tilde{g}_{\mathrm{u},t,t}^{\left( i,j \right)}=\sqrt{\frac{\beta _{\mathrm{u},j}}{K_{\mathrm{u},j}+1}}\tilde{h}_{\mathrm{u},0}^{\left( i,j \right)}+\sqrt{N_{\mathrm{c}}\beta _{\mathrm{u},j}}\sum_{k=1}^{L_{\mathrm{u}}-1}{f_{t,k}\tilde{h}_{\mathrm{u},k}^{\left( i,j \right)}}\sim \mathcal{C} \mathcal{N} \left( 0,\frac{\beta _{\mathrm{u},j}\sigma _{\mathrm{u},0,j}^{2}}{K_{\mathrm{u},j}+1}+N_{\mathrm{c}}\beta _{\mathrm{u},j}\sum_{k=1}^{L_{\mathrm{u}}-1}{\left| f_{t,k} \right|^2\sigma _{\mathrm{u},k,j}^{2}} \right)
\end{equation*}
\begin{equation*}
\hspace{-0.9cm}
  \overset{\left( a \right)}{=} \mathcal{C} \mathcal{N} \left( 0,\frac{\beta _{\mathrm{u},j}\sigma _{\mathrm{u},0,j}^{2}}{K_{\mathrm{u},j}+1}+\beta _{\mathrm{u},j}\sum_{k=1}^{L_{\mathrm{u}}-1}{\sigma _{\mathrm{u},k,j}^{2}} \right) =\mathcal{C} \mathcal{N} \left( 0,\frac{\beta _{\mathrm{u},j}\sigma _{\mathrm{u},0,j}^{2}}{K_{\mathrm{u},j}+1}+\beta _{\mathrm{u},j}\left( 1-\sigma _{\mathrm{u},0,j}^{2} \right) \right)
\end{equation*}
\begin{equation}\label{R-U_FD_channel_ij_tt_NLoS}
\hspace{-5.6cm}
  = \mathcal{C} \mathcal{N} \left( 0,\frac{\beta _{\mathrm{u},j}\varsigma _{\mathrm{u},j}}{K_{\mathrm{u},j}+1} \right), \;\;\;
  \varsigma _{\mathrm{u},j}\triangleq \left( 1-\sigma _{\mathrm{u},0,j}^{2} \right) K_{\mathrm{u},j}+1,
\end{equation}
where step $(a)$ is based on $\left| f_{t,k} \right|^2=\frac{1}{N_{\mathrm{c}}}$.

Similarly, the $t$-th element of the diagonal matrix $\mathbf{G}_{\mathrm{b}}^{\left( i,j \right)}$ is given by
\begin{equation*}
\hspace{-10cm}
  \left[ \mathbf{G}_{\mathrm{b}}^{\left( i,j \right)} \right] _{t,t}=\sqrt{N_{\mathrm{c}}\beta _{\mathrm{b}}}\sum_{k=0}^{L_{\mathrm{b}}-1}{f_{t,k}h_{\mathrm{b},k}^{\left( i,j \right)}}
\end{equation*}
\begin{equation*}
\hspace{0.4cm}
  =\sqrt{N_{\mathrm{c}}\beta _{\mathrm{b}}}\left( \sqrt{\frac{K_{\mathrm{b}}}{K_{\mathrm{b}}+1}}f_{t,0}\sigma _{\mathrm{b},0}\bar{h}_{\mathrm{b}}^{\left( i,j \right)}+\sqrt{\frac{1}{K_{\mathrm{b}}+1}}f_{t,0}\tilde{h}_{\mathrm{b},0}^{\left( i,j \right)}+\sum_{k=1}^{L_{\mathrm{b}}-1}{f_{t,k}\tilde{h}_{\mathrm{b},k}^{\left( i,j \right)}} \right)
\end{equation*}
\begin{equation*}
\hspace{-4.2cm}
  =  \underbrace{\sqrt{\frac{\beta _{\mathrm{b}}K_{\mathrm{b}}}{K_{\mathrm{b}}+1}}\sigma _{\mathrm{b},0}a_{N_{\mathrm{b}},i}\left( \phi _{\mathrm{b}}^{aa},\phi _{\mathrm{b}}^{ea} \right) a_{N_{\mathrm{r}},j}^{\ast}\left( \phi _{\mathrm{r}}^{ad},\phi _{\mathrm{r}}^{ed} \right) } _  {\bar{g}_{\mathrm{b}}^{\left( i,j \right)}}
\end{equation*}
\begin{equation}\label{B-R_FD_channel_ij_tt}
\hspace{-3.7cm}
  +  \underbrace{\sqrt{\frac{\beta _{\mathrm{b}}}{K_{\mathrm{b}}+1}}\tilde{h}_{\mathrm{b},0}^{\left( i,j \right)}+\sqrt{N_{\mathrm{c}}\beta _{\mathrm{b}}}\sum_{k=1}^{L_{\mathrm{b}}-1}{f_{t,k}\tilde{h}_{\mathrm{b},k}^{\left( i,j \right)}}} _{\tilde{g}_{\mathrm{b},t,t}^{\left( i,j \right)}}
  \triangleq g_{\mathrm{b},t,t}^{\left( i,j \right)} ,
\end{equation}
where $\bar{g}_{\mathrm{b}}^{\left( i,j \right)}$ is the LoS component,
and $\tilde{g}_{\mathrm{b},t,t}^{\left( i,j \right)}$ is the NLoS component.
Because of \eqref{DFT_fij} and $\tilde{h}_{\mathrm{b},l}^{\left( i,j \right)}\sim \mathcal{C} \mathcal{N} \left( 0,\sigma _{\mathrm{b},l}^{2} \right) $ subject to $\sum_{l=0}^{L_{\mathrm{u}}-1}{\sigma _{\mathrm{b},l}^{2}}=1 $,
the NLoS component $\tilde{g}_{\mathrm{b},t,t}^{\left( i,j \right)}$ satisfies
\begin{equation}\label{B-R_FD_channel_ij_tt_NLoS}
  \tilde{g}_{\mathrm{b},t,t}^{\left( i,j \right)}\sim \mathcal{C} \mathcal{N} \left( 0,\frac{\beta _{\mathrm{b}}\varsigma _{\mathrm{b}}}{K_{\mathrm{b}}+1} \right) , \;\;\;
   \varsigma _{\mathrm{b}}=\left( 1-\sigma _{\mathrm{b},0}^{2} \right) K_{\mathrm{b}}+1.
\end{equation}
It can be observed from \eqref{R-U_FD_channel_ij_tt} and \eqref{B-R_FD_channel_ij_tt} that
$\tilde{g}_{\mathrm{u},t,t}^{\left( i,j \right)}$ (or $\tilde{g}_{\mathrm{b},t,t}^{\left( i,j \right)}$)
is independent for each pair $(i,j)$, but is correlated for each sub-carrier $t$.

Based on \eqref{R-U_FD_channel_ij_tt} and \eqref{B-R_FD_channel_ij_tt},
the frequency-domain channels can be divided into two parts as follows:
\begin{equation}\label{FD_channels}
  \mathbf{G}_{\mathrm{x}}=\bar{\mathbf{G}}_{\mathrm{x}}+\tilde{\mathbf{G}}_{\mathrm{x}}, \;\;\;
  \mathrm{x}\in \left\{ \mathrm{u},\mathrm{b} \right\},
\end{equation}
where $\bar{\mathbf{G}}_{\mathrm{x}}$ represents the LoS components, and $\tilde{\mathbf{G}}_{\mathrm{x}}$ stands for the NLoS components.

Then, we start to introduce the derivation for Theorem \ref{theorem_R_nt_mrc_appro}.
To obtain a closed-form expression for the approximation of $\tilde{R}_{n,t}$ in \eqref{R_nt_hat_1}, we need to calculate the expectations $\mathbb{E} \left\{ \mathcal{S} \right\}$ and $\mathbb{E} \left\{ \mathcal{I} \right\}$.
From \eqref{R_nt_S_sim} and \eqref{R_nt_I_sim}, these two expectations are expressed as
\begin{equation}\label{R_nt_S_E}
  \mathbb{E} \left\{ \mathcal{S} \right\} =\alpha ^2p_n \mathbb{E} \left\{ \left| \mathbf{g}_{\mathrm{u},nt}^{H}\mathbf{\Phi }^H\mathbf{G}_{\mathrm{b}}^{H}\mathbf{G}_{\mathrm{b}}\mathbf{\Phi g}_{\mathrm{u},nt} \right|^2 \right\},
\end{equation}
\begin{equation*}
  \mathbb{E} \left\{ \mathcal{I} \right\} =\alpha ^2\sum_{u\ne n}^{N_{\mathrm{u}}-1}{ p_u \mathbb{E} \left\{ \left| \mathbf{g}_{\mathrm{u},nt}^{H}\mathbf{\Phi }^H\mathbf{G}_{\mathrm{b}}^{H}\mathbf{G}_{\mathrm{b}}\mathbf{\Phi g}_{\mathrm{u},ut} \right|^2 \right\}} + \sigma _{\mathrm{noise}}^{2} \alpha ^2\mathbb{E} \left\{ \left\| \mathbf{g}_{\mathrm{u},nt}^{H}\mathbf{\Phi }^H\mathbf{G}_{\mathrm{b}}^{H} \right\| ^2 \right\}
\end{equation*}
\begin{equation}\label{R_nt_I_E}
  +\mathbb{E} \left\{ \mathbf{g}_{\mathrm{u},nt}^{H}\mathbf{\Phi }^H\mathbf{G}_{\mathrm{b}}^{H}\left( \mathbf{I}_{N_{\mathrm{b}}}\otimes \mathbf{F} \right) \mathbf{R}_{\mathbf{z}_{\mathrm{q}}}\left( \mathbf{I}_{N_{\mathrm{b}}}\otimes \mathbf{F}^H \right) \mathbf{G}_{\mathrm{b}}\mathbf{\Phi g}_{\mathrm{u},nt} \right\}.
\end{equation}
Then, we will calculate the expectations in \eqref{R_nt_S_E} and \eqref{R_nt_I_E} one by one.

\subsection{Derivation of $\mathbb{E} \left\{ \left| \mathbf{g}_{\mathrm{u},nt}^{H}\mathbf{\Phi }^H\mathbf{G}_{\mathrm{b}}^{H}\mathbf{G}_{\mathrm{b}}\mathbf{\Phi g}_{\mathrm{u},nt} \right|^2 \right\}$}\label{ES_1}

The term $\mathbf{g}_{\mathrm{u},nt}^{H}\mathbf{\Phi }^H\mathbf{G}_{\mathrm{b}}^{H}\mathbf{G}_{\mathrm{b}}\mathbf{\Phi g}_{\mathrm{u},nt}$ in this expectation can be expanded as
\begin{equation*}
\hspace{-1.5cm}
  \mathbf{g}_{\mathrm{u},nt}^{H}\mathbf{\Phi }^H\mathbf{G}_{\mathrm{b}}^{H}\mathbf{G}_{\mathrm{b}}\mathbf{\Phi g}_{\mathrm{u},nt}=\left( \bar{\mathbf{g}}_{\mathrm{u},nt}^{H}+\tilde{\mathbf{g}}_{\mathrm{u},nt}^{H} \right) \mathbf{\Phi }^H\left( \bar{\mathbf{G}}_{\mathrm{b}}^{H}+\tilde{\mathbf{G}}_{\mathrm{b}}^{H} \right) \left( \bar{\mathbf{G}}_{\mathrm{b}}+\tilde{\mathbf{G}}_{\mathrm{b}} \right) \mathbf{\Phi }\left( \bar{\mathbf{g}}_{\mathrm{u},nt}+\tilde{\mathbf{g}}_{\mathrm{u},nt} \right)
\end{equation*}
\begin{equation}\label{ES_1_term}
\hspace{1cm}
  = \underbrace{\bar{\mathbf{g}}_{\mathrm{u},nt}^{H}\mathbf{A}\bar{\mathbf{g}}_{\mathrm{u},nt}}_{w_{nn}^{1}}
  + \underbrace{\bar{\mathbf{g}}_{\mathrm{u},nt}^{H}\mathbf{A}\tilde{\mathbf{g}}_{\mathrm{u},nt}} _{w_{nn}^{2}}
  + \underbrace{\tilde{\mathbf{g}}_{\mathrm{u},nt}^{H}\mathbf{A}\bar{\mathbf{g}}_{\mathrm{u},nt}} _{w_{nn}^{3}}
  + \underbrace{\tilde{\mathbf{g}}_{\mathrm{u},nt}^{H}\mathbf{A}\tilde{\mathbf{g}}_{\mathrm{u},nt}} _{w_{nn}^{4}} ,
\end{equation}
where $\mathbf{A}\triangleq \mathbf{\Phi }^H\left( \bar{\mathbf{G}}_{\mathrm{b}}^{H}\bar{\mathbf{G}}_{\mathrm{b}}+\bar{\mathbf{G}}_{\mathrm{b}}^{H}\tilde{\mathbf{G}}_{\mathrm{b}}+\tilde{\mathbf{G}}_{\mathrm{b}}^{H}\bar{\mathbf{G}}_{\mathrm{b}}+\tilde{\mathbf{G}}_{\mathrm{b}}^{H}\tilde{\mathbf{G}}_{\mathrm{b}} \right) \mathbf{\Phi }$.
Thus, the expectation can be expressed as
\begin{equation*}
\hspace{-3cm}
  \mathbb{E} \left\{ \left| \mathbf{g}_{\mathrm{u},nt}^{H}\mathbf{\Phi }^H\mathbf{G}_{\mathrm{b}}^{H}\mathbf{G}_{\mathrm{b}}\mathbf{\Phi g}_{\mathrm{u},nt} \right|^2 \right\} =\mathbb{E} \left\{ \left| \sum_{i=1}^4{w_{nn}^{i}} \right|^2 \right\}
  = \sum_{i=1}^4{\mathbb{E} \left\{ \left| w_{nn}^{i} \right|^2 \right\}}
\end{equation*}
\begin{equation}\label{ES_1_ex}
  + 2\mathrm{Re}\left( \sum_{i=1}^4{\sum_{j=i+1}^4{\mathbb{E} \left\{ w_{nn}^{i}\left( w_{nn}^{j} \right) ^{\ast} \right\}}} \right)
  \overset{\left( a \right)}{=}\sum_{i=1}^4{\mathbb{E} \left\{ \left| w_{nn}^{i} \right|^2 \right\}}+2\mathrm{Re}\left( \mathbb{E} \left\{ w_{nn}^{1}\left( w_{nn}^{4} \right) ^{\ast} \right\} \right).
\end{equation}
The step $\left( a \right)$ is obtained by removing the zero-value terms.
Then, we focus on the calculation of the expectations in \eqref{ES_1_ex}, i.e., $\mathbb{E} \left\{ \left| w_{nn}^{i} \right|^2 \right\}$, $i=1,2,3,4$, and $ \mathbb{E} \left\{ w_{nn}^{1}\left( w_{nn}^{4} \right) ^{\ast} \right\} $.

We present the detailed steps for the calculation of $\mathbb{E} \left\{ \left| w_{nn}^{1} \right|^2 \right\}$ as an example.
The term $w_{nn}^{1}$ can be expressed as
\begin{equation*}
\hspace{-1.9cm}
  w_{nn}^{1}=\bar{\mathbf{g}}_{\mathrm{u},nt}^{H}\mathbf{A}\bar{\mathbf{g}}_{\mathrm{u},nt}
  =  \underbrace{\bar{\mathbf{g}}_{\mathrm{u},nt}^{H}\mathbf{\Phi }^H\bar{\mathbf{G}}_{\mathrm{b}}^{H}\bar{\mathbf{G}}_{\mathrm{b}}\mathbf{\Phi }\bar{\mathbf{g}}_{\mathrm{u},nt}} _{w_{nn}^{1,1}}
  +  \underbrace{\bar{\mathbf{g}}_{\mathrm{u},nt}^{H}\mathbf{\Phi }^H\bar{\mathbf{G}}_{\mathrm{b}}^{H}\tilde{\mathbf{G}}_{\mathrm{b}}\mathbf{\Phi }\bar{\mathbf{g}}_{\mathrm{u},nt}} _{w_{nn}^{1,2}}
\end{equation*}
\begin{equation}\label{w_1_nn_expand}
\hspace{1.4cm}
  + \underbrace{\bar{\mathbf{g}}_{\mathrm{u},nt}^{H}\mathbf{\Phi }^H\tilde{\mathbf{G}}_{\mathrm{b}}^{H}\bar{\mathbf{G}}_{\mathrm{b}}\mathbf{\Phi }\bar{\mathbf{g}}_{\mathrm{u},nt}} _{w_{nn}^{1,3}}
  + \underbrace{\bar{\mathbf{g}}_{\mathrm{u},nt}^{H}\mathbf{\Phi }^H\tilde{\mathbf{G}}_{\mathrm{b}}^{H}\tilde{\mathbf{G}}_{\mathrm{b}}\mathbf{\Phi }\bar{\mathbf{g}}_{\mathrm{u},nt}} _{w_{nn}^{1,4}},
\end{equation}
where $w_{nn}^{1,i}$, $i=1,2,3,4$, can be expanded respectively as
\begin{equation*}
\hspace{-2cm}
  w_{nn}^{1,1}=\bar{\mathbf{g}}_{\mathrm{u},nt}^{H}\mathbf{\Phi }^H\bar{\mathbf{G}}_{\mathrm{b}}^{H}\bar{\mathbf{G}}_{\mathrm{b}}\mathbf{\Phi }\bar{\mathbf{g}}_{\mathrm{u},nt}
  =\sum_{b=0}^{N_{\mathrm{b}}-1}{\sum_{r_1=0}^{N_{\mathrm{r}}-1}{\left( e^{j\theta _{r_1}}\bar{g}_{\mathrm{u}}^{\left( r_1,n \right)}\bar{g}_{\mathrm{b}}^{\left( b,r_1 \right)} \right) ^{\ast}}\sum_{r_2=0}^{N_{\mathrm{r}}-1}{e^{j\theta _{r_2}}\bar{g}_{\mathrm{u}}^{\left( r_2,n \right)}\bar{g}_{\mathrm{b}}^{\left( b,r_2 \right)}}}
\end{equation*}
\begin{equation*}
\hspace{-2.9cm}
  \overset{\left( a \right)}{=}\frac{\beta _{\mathrm{u},n}K_{\mathrm{u},n}}{K_{\mathrm{u},n}+1}\frac{\beta _{\mathrm{b}}K_{\mathrm{b}}}{K_{\mathrm{b}}+1}\sum_{b=0}^{N_{\mathrm{b}}-1}{\left| \sum_{r=0}^{N_{\mathrm{r}}-1}{e^{j\theta _r}a_{N_{\mathrm{r}},r}^{\ast}\left( \phi _{\mathrm{r}}^{ad},\phi _{\mathrm{r}}^{ed} \right) a_{N_{\mathrm{r}},r}\left( \phi _{\mathrm{r},n}^{aa},\phi _{\mathrm{r},n}^{ea} \right)} \right|^2}
\end{equation*}
\begin{equation}\label{w_1_1_nn}
\hspace{-7cm}
  =\frac{\beta _{\mathrm{u},n}\beta _{\mathrm{b}}K_{\mathrm{u},n}K_{\mathrm{b}}N_{\mathrm{b}}}{\left( K_{\mathrm{u},n}+1 \right) \left( K_{\mathrm{b}}+1 \right)}\sigma _{\mathrm{u},0,n}^{2}\sigma _{\mathrm{b},0}^{2}\left| \varPhi _{N_{\mathrm{r}}}\left( n \right) \right|^2,
\end{equation}
\begin{equation*}
\hspace{-3.9cm}
  w_{nn}^{1,2}=\bar{\mathbf{g}}_{\mathrm{u},nt}^{H}\mathbf{\Phi }^H\bar{\mathbf{G}}_{\mathrm{b}}^{H}\tilde{\mathbf{G}}_{\mathrm{b}}\mathbf{\Phi }\bar{\mathbf{g}}_{\mathrm{u},nt}
  = \frac{\beta _{\mathrm{u},n}K_{\mathrm{u},n}}{K_{\mathrm{u},n}+1}\sqrt{\frac{\beta _{\mathrm{b}}K_{\mathrm{b}}}{K_{\mathrm{b}}+1}}\sigma _{\mathrm{u},0,n}^{2}\sigma _{\mathrm{b},0}\left( \varPhi _{N_{\mathrm{r}}}\left( n \right) \right) ^{\ast}
\end{equation*}
\begin{equation}\label{w_1_2_nn}
\hspace{-5.4cm}
  \times
  \sum_{b=0}^{N_{\mathrm{b}}-1}{a_{N_{\mathrm{b}},b}^{\ast}\left( \phi _{\mathrm{b}}^{aa},\phi _{\mathrm{b}}^{ea} \right) \sum_{r=0}^{N_{\mathrm{r}}-1}{e^{j\varphi _r}a_{N_{\mathrm{r}},r}\left( \phi _{\mathrm{r},n}^{aa},\phi _{\mathrm{r},n}^{ea} \right) \tilde{g}_{\mathrm{b},t,t}^{\left( b,r \right)}}},
\end{equation}
\begin{equation*}
\hspace{-4.3cm}
  w_{nn}^{1,3}=\bar{\mathbf{g}}_{\mathrm{u},nt}^{H}\mathbf{\Phi }^H\tilde{\mathbf{G}}_{\mathrm{b}}^{H}\bar{\mathbf{G}}_{\mathrm{b}}\mathbf{\Phi }\bar{\mathbf{g}}_{\mathrm{u},nt}\,\,
  = \frac{\beta _{\mathrm{u},n}K_{\mathrm{u},n}}{K_{\mathrm{u},n}+1}\sqrt{\frac{\beta _{\mathrm{b}}K_{\mathrm{b}}}{K_{\mathrm{b}}+1}}\sigma _{\mathrm{u},0,n}^{2}\sigma _{\mathrm{b},0}\varPhi _{N_{\mathrm{r}}}\left( n \right)
\end{equation*}
\begin{equation}\label{w_1_3_nn}
\hspace{-4.7cm}
  \times
  \sum_{b=0}^{N_{\mathrm{b}}-1}{a_{N_{\mathrm{b}},b}\left( \phi _{\mathrm{b}}^{aa},\phi _{\mathrm{b}}^{ea} \right) \sum_{r=0}^{N_{\mathrm{r}}-1}{\left( e^{j\varphi _r}a_{N_{\mathrm{r}},r}\left( \phi _{\mathrm{r},n}^{aa},\phi _{\mathrm{r},n}^{ea} \right) \tilde{g}_{\mathrm{b},t,t}^{\left( b,r \right)} \right) ^{\ast}}},
\end{equation}
\begin{equation*}
\hspace{-11cm}
  w_{nn}^{1,4}=\bar{\mathbf{g}}_{\mathrm{u},nt}^{H}\mathbf{\Phi }^H\tilde{\mathbf{G}}_{\mathrm{b}}^{H}\tilde{\mathbf{G}}_{\mathrm{b}}\mathbf{\Phi }\bar{\mathbf{g}}_{\mathrm{u},nt}\,\,
\end{equation*}
\begin{equation}\label{w_1_4_nn}
  =\frac{\beta _{\mathrm{u},n}K_{\mathrm{u},n}}{K_{\mathrm{u},n}+1}\sigma _{\mathrm{u},0,n}^{2}\sum_{b=0}^{N_{\mathrm{b}}-1}{\sum_{r_1=0}^{N_{\mathrm{r}}-1}{\sum_{r_2=0}^{N_{\mathrm{r}}-1}{\left( e^{j\varphi _{r_1}}a_{N_{\mathrm{r}},r_1}\left( \phi _{\mathrm{r},n}^{aa},\phi _{\mathrm{r},n}^{ea} \right) \tilde{g}_{\mathrm{b},t,t}^{\left( b,r_1 \right)} \right) ^{\ast}e^{j\varphi _{r_2}}a_{N_{\mathrm{r}},r_2}\left( \phi _{\mathrm{r},n}^{aa},\phi _{\mathrm{r},n}^{ea} \right) \tilde{g}_{\mathrm{b},t,t}^{\left( b,r_2 \right)}}}} ,
\end{equation}
where step $\left( a \right)$ in \eqref{w_1_1_nn} is obtained by substituting \eqref{R-U_FD_channel_ij_tt} and \eqref{B-R_FD_channel_ij_tt} into it.
Besides, the following definitions are used
\begin{equation}\label{PHI_n}
  \varPhi _{N_{\mathrm{r}}}\left( n \right) \triangleq \sum_{r=0}^{N_{\mathrm{r}}-1}{\varPhi \left( n,r \right)},
\end{equation}
\begin{equation}\label{Phi_n_r}
  \varPhi \left( n,r \right) \triangleq e^{j\theta _r}a_{N_{\mathrm{r}},r}^{\ast}\left( \phi _{\mathrm{r}}^{ad},\phi _{\mathrm{r}}^{ed} \right) a_{N_{\mathrm{r}},r}\left( \phi _{\mathrm{r},n}^{aa},\phi _{\mathrm{r},n}^{ea} \right) .
\end{equation}
Equation \eqref{w_1_2_nn}-\eqref{w_1_4_nn} are expanded similarly as the expansion of \eqref{w_1_1_nn}.

Based on \eqref{w_1_nn_expand}, the expectation $\mathrm{E}\left\{ \left| w_{nn}^{1} \right|^2 \right\}$ can be expanded as
\begin{equation*}
\hspace{-8cm}
  \mathrm{E}\left\{ \left| w_{nn}^{1} \right|^2 \right\} =\mathrm{E}\left\{ \left| w_{nn}^{1,1}+w_{nn}^{1,2}+w_{nn}^{1,3}+w_{nn}^{1,4} \right|^2 \right\}
\end{equation*}
\begin{equation}\label{E_w_1_nn_expand}
  \overset{\left( a \right)}{=}\mathrm{E}\left\{ \left| w_{nn}^{1,1} \right|^2 \right\} +\mathrm{E}\left\{ \left| w_{nn}^{1,2} \right|^2 \right\} +\mathrm{E}\left\{ \left| w_{nn}^{1,3} \right|^2 \right\} +\mathrm{E}\left\{ \left| w_{nn}^{1,4} \right|^2 \right\} +2\mathrm{E}\left\{ \mathrm{Re}\left( w_{nn}^{1,1}\times \left( w_{nn}^{1,4} \right) ^* \right) \right\} ,
\end{equation}
where step $\left( a \right)$ is obtained by removing the zero-value terms.
Thus, the calculation of the expectation $\mathrm{E}\left\{ \left| w_{nn}^{1} \right|^2 \right\}$ can be divided into five parts.
Herein, we present the calculation of $\mathrm{E}\left\{ \left| w_{nn}^{1,2} \right|^2 \right\}$ as an example:
\begin{equation*}
\hspace{-3cm}
  \mathrm{E}\left\{ \left| w_{nn}^{1,2} \right|^2 \right\}
  =c_{nn}^{1,2}\mathrm{E}\left\{ \left| \sum_{b=0}^{N_{\mathrm{b}}-1}{\sum_{r=0}^{N_{\mathrm{r}}-1}{a_{N_{\mathrm{b}},b}^{\ast}\left( \phi _{\mathrm{b}}^{aa},\phi _{\mathrm{b}}^{ea} \right) e^{j\theta _r}a_{N_{\mathrm{r}},r}\left( \phi _{\mathrm{r},n}^{aa},\phi _{\mathrm{r},n}^{ea} \right) \tilde{g}_{\mathrm{b},t,t}^{\left( b,r \right)}}} \right|^2 \right\}
\end{equation*}
\begin{equation*}
\hspace{-5.1cm}
  =c_{nn}^{1,2}\mathrm{E}\left\{ \sum_{b=0}^{N_{\mathrm{b}}-1}{\left| \sum_{r=0}^{N_{\mathrm{r}}-1}{a_{N_{\mathrm{b}},b}^{\ast}\left( \phi _{\mathrm{b}}^{aa},\phi _{\mathrm{b}}^{ea} \right) e^{j\theta _r}a_{N_{\mathrm{r}},r}\left( \phi _{\mathrm{r},n}^{aa},\phi _{\mathrm{r},n}^{ea} \right) \tilde{g}_{\mathrm{b},t,t}^{\left( b,r \right)}} \right|^2} \right\}
\end{equation*}
\begin{equation*}
\hspace{-1.5cm}
  +c_{nn}^{1,2}\mathrm{E}\left\{ \sum_{b_1=0}^{N_{\mathrm{b}}-1}{\sum_{b_2\ne b_1}^{N_{\mathrm{b}}-1}{\left( \begin{array}{c}
	\left( \sum_{r_1=0}^{N_{\mathrm{r}}-1}{a_{N_{\mathrm{b}},b_1}^{\ast}\left( \phi _{\mathrm{b}}^{aa},\phi _{\mathrm{b}}^{ea} \right) e^{j\theta _{r_1}}a_{N_{\mathrm{r}},r_1}\left( \phi _{\mathrm{r},n}^{aa},\phi _{\mathrm{r},n}^{ea} \right) \tilde{g}_{\mathrm{b},t,t}^{\left( b_1,r_1 \right)}} \right)\\
	\times \left( \sum_{r_2=0}^{N_{\mathrm{r}}-1}{a_{N_{\mathrm{b}},b_2}^{\ast}\left( \phi _{\mathrm{b}}^{aa},\phi _{\mathrm{b}}^{ea} \right) e^{j\theta _{r_2}}a_{N_{\mathrm{r}},r_2}\left( \phi _{\mathrm{r},n}^{aa},\phi _{\mathrm{r},n}^{ea} \right) \tilde{g}_{\mathrm{b},t,t}^{\left( b_2,r_2 \right)}} \right) ^{\ast}\\
\end{array} \right)}} \right\}
\end{equation*}
\begin{equation*}
  \overset{\left( a \right)}{=}c_{nn}^{1,2}\mathrm{E}\left\{ \sum_{b=0}^{N_{\mathrm{b}}-1}{\sum_{r=0}^{N_{\mathrm{r}}-1}{\left| a_{N_{\mathrm{b}},b}^{\ast}\left( \phi _{\mathrm{b}}^{aa},\phi _{\mathrm{b}}^{ea} \right) e^{j\theta _r}a_{N_{\mathrm{r}},r}\left( \phi _{\mathrm{r},n}^{aa},\phi _{\mathrm{r},n}^{ea} \right) \tilde{g}_{\mathrm{b},t,t}^{\left( b,r \right)} \right|^2}} \right\}
  =c_{nn}^{1,2}\mathrm{E}\left\{ \sum_{b=0}^{N_{\mathrm{b}}-1}{\sum_{r=0}^{N_{\mathrm{r}}-1}{\left| \tilde{g}_{\mathrm{b},t,t}^{\left( b,r \right)} \right|^2}} \right\}
\end{equation*}
\begin{equation}\label{E_w_1_2_nn_value}
\hspace{-7cm}
  \overset{\left( b \right)}{=} \frac{K_{\mathrm{u},n}^{2}K_{\mathrm{b}}\beta _{\mathrm{u},n}^{2}\beta _{\mathrm{b}}^{2}}{\left( K_{\mathrm{u},n}+1 \right) ^2\left( K_{\mathrm{b}}+1 \right) ^2}\varsigma _{\mathrm{b}}\sigma _{\mathrm{u},0,n}^{4}\sigma _{\mathrm{b},0}^{2}N_{\mathrm{b}}N_{\mathrm{r}}\left| \varPhi _{N_{\mathrm{r}}}\left( n \right) \right|^2 ,
\end{equation}
where $c_{nn}^{1,2}\triangleq \left( \frac{\beta _{\mathrm{u},n}K_{\mathrm{u},n}}{K_{\mathrm{u},n}+1} \right) ^2\frac{\beta _{\mathrm{b}}K_{\mathrm{b}}}{K_{\mathrm{b}}+1}\sigma _{\mathrm{u},0,n}^{4}\sigma _{\mathrm{b},0}^{2}\left| \varPhi _{N_{\mathrm{r}}}\left( n \right) \right|^2$.
Step $\left( a \right)$ in \eqref{E_w_1_2_nn_value} is obtained by removing the zero-value terms,
and step $\left( b \right)$ is because of \eqref{B-R_FD_channel_ij_tt_NLoS}.
Similarly, the remaining parts are obtained as
\begin{equation}\label{E_w_1_1_nn_value}
\hspace{-4cm}
  \mathrm{E}\left\{ \left| w_{nn}^{1,1} \right|^2 \right\}
  =\frac{K_{\mathrm{u},n}^{2}K_{\mathrm{b}}^{2}\beta _{\mathrm{u},n}^{2}\beta _{\mathrm{b}}^{2}}{\left( K_{\mathrm{u},n}+1 \right) ^2\left( K_{\mathrm{b}}+1 \right) ^2}N_{\mathrm{b}}^{2}\sigma _{\mathrm{u},0,n}^{4}\sigma _{\mathrm{b},0}^{4}\left| \varPhi _{N_{\mathrm{r}}}\left( n \right) \right|^4,
\end{equation}
\begin{equation}\label{E_w_1_3_nn_value}
\hspace{-3.2cm}
  \mathrm{E}\left\{ \left| w_{nn}^{1,3} \right|^2 \right\}
  =\frac{K_{\mathrm{u},n}^{2}K_{\mathrm{b}}\beta _{\mathrm{u},n}^{2}\beta _{\mathrm{b}}^{2}}{\left( K_{\mathrm{u},n}+1 \right) ^2\left( K_{\mathrm{b}}+1 \right) ^2}\varsigma _{\mathrm{b}}\sigma _{\mathrm{u},0,n}^{4}\sigma _{\mathrm{b},0}^{2}N_{\mathrm{b}}N_{\mathrm{r}}\left| \varPhi _{N_{\mathrm{r}}}\left( n \right) \right|^2,
\end{equation}
\begin{equation}\label{E_w_1_4_nn_value}
\hspace{-3.9cm}
  \mathrm{E}\left\{ \left| w_{nn}^{1,4} \right|^2 \right\}
  =\frac{K_{\mathrm{u},n}^{2}\beta _{\mathrm{u},n}^{2}\beta _{\mathrm{b}}^{2}}{\left( K_{\mathrm{u},n}+1 \right) ^2\left( K_{\mathrm{b}}+1 \right) ^2}N_{\mathrm{b}}\left( N_{\mathrm{b}}+1 \right) N_{\mathrm{r}}^{2}\sigma _{\mathrm{u},0,n}^{4}\varsigma _{\mathrm{b}}^{2},
\end{equation}
\begin{equation}\label{E_w_1_1_and_1_4_nn_value}
  \mathrm{E}\left\{ \mathrm{Re}\left( w_{nn}^{1,1}\times \left( w_{nn}^{1,4} \right) ^* \right) \right\}
  =\frac{K_{\mathrm{u},n}^{2}K_{\mathrm{b}}\beta _{\mathrm{u},n}^{2}\beta _{\mathrm{b}}^{2}}{\left( K_{\mathrm{u},n}+1 \right) ^2\left( K_{\mathrm{b}}+1 \right) ^2}\varsigma _{\mathrm{b}}\sigma _{\mathrm{u},0,n}^{4}\sigma _{\mathrm{b},0}^{2}N_{\mathrm{b}}^{2}N_{\mathrm{r}}\left| \varPhi _{N_{\mathrm{r}}}\left( n \right) \right|^2.
\end{equation}
Substituting \eqref{E_w_1_2_nn_value}-\eqref{E_w_1_1_and_1_4_nn_value} into \eqref{E_w_1_nn_expand}, we have
\begin{equation*}
  \mathbb{E} \left\{ \left| w_{nn}^{1} \right|^2 \right\}
  =\frac{\beta _{\mathrm{u},n}^{2}\beta _{\mathrm{b}}^{2}\sigma _{\mathrm{u},0,n}^{4}K_{\mathrm{u},n}^{2}N_{\mathrm{b}}}{\left( K_{\mathrm{u},n}+1 \right) ^2\left( K_{\mathrm{b}}+1 \right) ^2}
  \bigg( \sigma _{\mathrm{b},0}^{4}K_{\mathrm{b}}^{2}N_{\mathrm{b}}\left| \varPhi _{N_{\mathrm{r}}}\left( n \right) \right|^4+2\sigma _{\mathrm{b},0}^{2}\varsigma _{\mathrm{b}}K_{\mathrm{b}}N_{\mathrm{r}}\left| \varPhi _{N_{\mathrm{r}}}\left( n \right) \right|^2
\end{equation*}
\begin{equation}\label{E_w_1_nn_value}
\hspace{-2cm}
  +2\sigma _{\mathrm{b},0}^{2}\varsigma _{\mathrm{b}}K_{\mathrm{b}}N_{\mathrm{b}}N_{\mathrm{r}}\left| \varPhi _{N_{\mathrm{r}}}\left( n \right) \right|^2+\varsigma _{\mathrm{b}}^{2}\left( N_{\mathrm{b}}+1 \right) N_{\mathrm{r}}^{2}
  \bigg) .
\end{equation}

Similar to the calculation of $\mathbb{E} \left\{ \left| w_{nn}^{1} \right|^2 \right\}$, the other expectations in \eqref{ES_1_ex} can be obtained as
\begin{equation*}
\hspace{-0.8cm}
  \mathbb{E} \left\{ \left| w_{nn}^{2} \right|^2 \right\}
  = \frac{\beta _{\mathrm{u},n}^{2}\beta _{\mathrm{b}}^{2}\sigma _{\mathrm{u},0,n}^{2}\varsigma _{\mathrm{u},n}K_{\mathrm{u},n}N_{\mathrm{b}}}{\left( K_{\mathrm{u},n}+1 \right) ^2\left( K_{\mathrm{b}}+1 \right) ^2}
  \bigg( \left( \sigma _{\mathrm{b},0}^{2}K_{\mathrm{b}}N_{\mathrm{b}}N_{\mathrm{r}}+\varsigma _{\mathrm{b}}N_{\mathrm{r}}+2\varsigma _{\mathrm{b}}N_{\mathrm{b}} \right) \sigma _{\mathrm{b},0}^{2}K_{\mathrm{b}}\left| \varPhi _{N_{\mathrm{r}}}\left( n \right) \right|^2
\end{equation*}
\begin{equation}\label{E_w_2_nn_value}
\hspace{-5.8cm}
  +\left( \sigma _{\mathrm{b},0}^{2}K_{\mathrm{b}}\varsigma _{\mathrm{b}}+\varsigma _{\mathrm{b}}^{2} \right) N_{\mathrm{r}}^{2}+\varsigma _{\mathrm{b}}^{2}N_{\mathrm{b}}N_{\mathrm{r}}
  \bigg) ,
\end{equation}
\begin{equation}\label{E_w_3_nn_value}
\hspace{-11.4cm}
  \mathbb{E} \left\{ \left| w_{nn}^{3} \right|^2 \right\} =\mathbb{E} \left\{ \left| w_{nn}^{2} \right|^2 \right\} ,
\end{equation}
\begin{equation}\label{E_w_4_nn_value}
  \mathbb{E} \left\{ \left| w_{nn}^{4} \right|^2 \right\}
  =\frac{\beta _{\mathrm{u},n}^{2}\beta _{\mathrm{b}}^{2}\varsigma _{\mathrm{u},n}^{2}N_{\mathrm{b}}N_{\mathrm{r}}}{\left( K_{\mathrm{u},n}+1 \right) ^2\left( K_{\mathrm{b}}+1 \right) ^2}\left( 2\sigma _{\mathrm{b},0}^{4}K_{\mathrm{b}}^{2}N_{\mathrm{b}}N_{\mathrm{r}}+\left( 2\sigma _{\mathrm{b},0}^{2}K_{\mathrm{b}}\varsigma _{\mathrm{b}}+\varsigma _{\mathrm{b}}^{2} \right) \left( N_{\mathrm{b}}+1 \right) \left( N_{\mathrm{r}}+1 \right) \right),
\end{equation}
\begin{equation*}
\hspace{-0.5cm}
  \mathbb{E} \left\{ w_{nn}^{1}\left( w_{nn}^{4} \right) ^{\ast} \right\}
  = \frac{\beta _{\mathrm{u},n}^{2}\beta _{\mathrm{b}}^{2}\sigma _{\mathrm{u},0,n}^{2}\varsigma _{\mathrm{u},n}K_{\mathrm{u},n}N_{\mathrm{b}}}{\left( K_{\mathrm{u},n}+1 \right) ^2\left( K_{\mathrm{b}}+1 \right) ^2}
  \bigg(  \left( \left( \sigma _{\mathrm{b},0}^{2}K_{\mathrm{b}}+\varsigma _{\mathrm{b}} \right) N_{\mathrm{b}}N_{\mathrm{r}}+2\varsigma _{\mathrm{b}} \right) \sigma _{\mathrm{b},0}^{2}K_{\mathrm{b}}\left| \varPhi _{N_{\mathrm{r}}}\left( n \right) \right|^2
\end{equation*}
\begin{equation}\label{E_w_1_and 4_nn_value}
\hspace{-4.3cm}
  +\left( \sigma _{\mathrm{b},0}^{2}K_{\mathrm{b}}\varsigma _{\mathrm{b}}+\varsigma _{\mathrm{b}}^{2} \right) N_{\mathrm{b}}N_{\mathrm{r}}^{2}+\varsigma _{\mathrm{b}}^{2}N_{\mathrm{r}}
  \bigg) .
\end{equation}
Substituting \eqref{E_w_1_nn_value}-\eqref{E_w_1_and 4_nn_value} into \eqref{ES_1_ex}, the expectation $\mathbb{E} \left\{ \left| \mathbf{g}_{\mathrm{u},nt}^{H}\mathbf{\Phi }^H\mathbf{G}_{\mathrm{b}}^{H}\mathbf{G}_{\mathrm{b}}\mathbf{\Phi g}_{\mathrm{u},nt} \right|^2 \right\}$
can be obtained as
\begin{equation*}
\hspace{-0.9cm}
  \mathbb{E} \left\{ \left| \mathbf{g}_{\mathrm{u},nt}^{H}\mathbf{\Phi }^H\mathbf{G}_{\mathrm{b}}^{H}\mathbf{G}_{\mathrm{b}}\mathbf{\Phi g}_{\mathrm{u},nt} \right|^2 \right\}
  = \frac{\beta _{\mathrm{u},n}^{2}\beta _{\mathrm{b}}^{2}N_{\mathrm{b}}}{\left( K_{\mathrm{u},n}+1 \right) ^2\left( K_{\mathrm{b}}+1 \right) ^2}
  \times \bigg(
  \sigma _{\mathrm{u},0,n}^{4}\sigma _{\mathrm{b},0}^{4}K_{\mathrm{u},n}^{2}K_{\mathrm{b}}^{2}N_{\mathrm{b}}\left| \varPhi _{N_{\mathrm{r}}}\left( n \right) \right|^4
\end{equation*}
\begin{equation*}
\hspace{-1.5cm}
  + 2\sigma _{\mathrm{u},0,n}^{2}\sigma _{\mathrm{b},0}^{2}K_{\mathrm{u},n}K_{\mathrm{b}}\left| \varPhi _{N_{\mathrm{r}}}\left( n \right) \right|^2
  \Big(
  2\sigma _{\mathrm{b},0}^{2}\varsigma _{\mathrm{u},n}K_{\mathrm{b}}N_{\mathrm{b}}N_{\mathrm{r}}
  +\left( \sigma _{\mathrm{u},0,n}^{2}\varsigma _{\mathrm{b}}K_{\mathrm{u},n}+\varsigma _{\mathrm{u},n}\varsigma _{\mathrm{b}} \right) \left( N_{\mathrm{b}}+1 \right) N_{\mathrm{r}}
\end{equation*}
\begin{equation*}
  +2\varsigma _{\mathrm{u},n}\varsigma _{\mathrm{b}}\left( N_{\mathrm{b}}+1 \right) \Big)
  +\left( \sigma _{\mathrm{u},0,n}^{4}\varsigma _{\mathrm{b}}^{2}K_{\mathrm{u},n}^{2}+2\sigma _{\mathrm{u},0,n}^{2}\sigma _{\mathrm{b},0}^{2}\varsigma _{\mathrm{u},n}\varsigma _{\mathrm{b}}K_{\mathrm{u},n}K_{\mathrm{b}} \right) \left( N_{\mathrm{b}}+1 \right) N_{\mathrm{r}}^{2}
  +2\sigma _{\mathrm{b},0}^{4}\varsigma _{\mathrm{u},n}^{2}K_{\mathrm{b}}^{2}N_{\mathrm{b}}N_{\mathrm{r}}^{2}
\end{equation*}
\begin{equation}\label{omega_bar_n}
\hspace{-3.2cm}
  +\left( 2\sigma _{\mathrm{u},0,n}^{2}\varsigma _{\mathrm{u},n}\varsigma _{\mathrm{b}}^{2}K_{\mathrm{u},n}+2\sigma _{\mathrm{b},0}^{2}\varsigma _{\mathrm{u},n}^{2}\varsigma _{\mathrm{b}}K_{\mathrm{b}}+\varsigma _{\mathrm{u},n}^{2}\varsigma _{\mathrm{b}}^{2} \right) \left( N_{\mathrm{b}}+1 \right) N_{\mathrm{r}}\left( N_{\mathrm{r}}+1 \right)
 \bigg)
  \triangleq \varpi _n.
\end{equation}

\subsection{Derivation of $\mathbb{E} \left\{ \left| \mathbf{g}_{\mathrm{u},nt}^{H}\mathbf{\Phi }^H\mathbf{G}_{\mathrm{b}}^{H}\mathbf{G}_{\mathrm{b}}\mathbf{\Phi g}_{\mathrm{u},ut} \right|^2 \right\}$ and $\mathbb{E} \left\{ \left\| \mathbf{g}_{\mathrm{u},nt}^{H}\mathbf{\Phi }^H\mathbf{G}_{\mathrm{b}}^{H} \right\| ^2 \right\}$}\label{EI_1_and_2}

It is noted that
\begin{equation}\label{EI_2_expand}
  \left\| \mathbf{g}_{\mathrm{u},nt}^{H}\mathbf{\Phi }^H\mathbf{G}_{\mathrm{b}}^{H} \right\| ^2
  =w_{nn}^{1}+w_{nn}^{2}+w_{nn}^{3}+w_{nn}^{4}.
\end{equation}
Thus, based on the derivation in Appendix \ref{ES_1}, it is easy to obtain
\begin{equation*}
\hspace{-7cm}
  \mathbb{E} \left\{ \left\| \mathbf{g}_{\mathrm{u},nt}^{H}\mathbf{\Phi }^H\mathbf{G}_{\mathrm{b}}^{H} \right\| ^2 \right\}
  = \frac{\beta _{\mathrm{u},n}\beta _{\mathrm{b}}N_{\mathrm{b}}}{\left( K_{\mathrm{u},n}+1 \right) \left( K_{\mathrm{b}}+1 \right)}
\end{equation*}
\begin{equation}\label{epsilon_n}
  \times
  \left( \sigma _{\mathrm{u},0,n}^{2}K_{\mathrm{u},n}\sigma _{\mathrm{b},0}^{2}K_{\mathrm{b}}\left| \varPhi _{N_{\mathrm{r}}}\left( n \right) \right|^2+\left( \sigma _{\mathrm{u},0,n}^{2}K_{\mathrm{u},n}\varsigma _{\mathrm{b}}+\sigma _{\mathrm{b},0}^{2}K_{\mathrm{b}}\varsigma _{\mathrm{u},n}+\varsigma _{\mathrm{u},n}\varsigma _{\mathrm{b}} \right) N_{\mathrm{r}} \right)
  \triangleq \epsilon _n.
\end{equation}

As for $\mathbb{E} \left\{ \left| \mathbf{g}_{\mathrm{u},nt}^{H}\mathbf{\Phi }^H\mathbf{G}_{\mathrm{b}}^{H}\mathbf{G}_{\mathrm{b}}\mathbf{\Phi g}_{\mathrm{u},ut} \right|^2 \right\}$,
the term $\mathbf{g}_{\mathrm{u},nt}^{H}\mathbf{\Phi }^H\mathbf{G}_{\mathrm{b}}^{H}\mathbf{G}_{\mathrm{b}}\mathbf{\Phi g}_{\mathrm{u},ut}$ can be expanded similar to \eqref{ES_1_term} as follows
\begin{equation*}
\hspace{-1.4cm}
  \mathbf{g}_{\mathrm{u},nt}^{H}\mathbf{\Phi }^H\mathbf{G}_{\mathrm{b}}^{H}\mathbf{G}_{\mathrm{b}}\mathbf{\Phi g}_{\mathrm{u},ut}=\left( \bar{\mathbf{g}}_{\mathrm{u},nt}^{H}+\tilde{\mathbf{g}}_{\mathrm{u},nt}^{H} \right) \mathbf{\Phi }^H\left( \bar{\mathbf{G}}_{\mathrm{b}}^{H}+\tilde{\mathbf{G}}_{\mathrm{b}}^{H} \right) \left( \bar{\mathbf{G}}_{\mathrm{b}}+\tilde{\mathbf{G}}_{\mathrm{b}} \right) \mathbf{\Phi }\left( \bar{\mathbf{g}}_{\mathrm{u},ut}+\tilde{\mathbf{g}}_{\mathrm{u},ut} \right)
\end{equation*}
\begin{equation}\label{EI_1_term}
\hspace{1.1cm}
  = \underbrace{\bar{\mathbf{g}}_{\mathrm{u},nt}^{H}\mathbf{A}\bar{\mathbf{g}}_{\mathrm{u},ut}} _ {w_{nu}^{1}}
  + \underbrace{\bar{\mathbf{g}}_{\mathrm{u},nt}^{H}\mathbf{A}\tilde{\mathbf{g}}_{\mathrm{u},ut}} _{w_{nu}^{2}}
  + \underbrace{\tilde{\mathbf{g}}_{\mathrm{u},nt}^{H}\mathbf{A}\bar{\mathbf{g}}_{\mathrm{u},ut}} _{w_{nu}^{3}}
  + \underbrace{\tilde{\mathbf{g}}_{\mathrm{u},nt}^{H}\mathbf{A}\tilde{\mathbf{g}}_{\mathrm{u},ut}} _{w_{nu}^{4}} .
\end{equation}
Therefore, the expectation is re-expressed as
\begin{equation*}
\hspace{-5.3cm}
  \mathbb{E} \left\{ \left| \mathbf{g}_{\mathrm{u},nt}^{H}\mathbf{\Phi }^H\mathbf{G}_{\mathrm{b}}^{H}\mathbf{G}_{\mathrm{b}}\mathbf{\Phi g}_{\mathrm{u},ut} \right|^2 \right\} =\mathbb{E} \left\{ \left| \sum_{i=1}^4{w_{nu}^{i}} \right|^2 \right\}
\end{equation*}
\begin{equation}\label{EI_1_ex}
  =\sum_{i=1}^4{\mathbb{E} \left\{ \left| w_{nu}^{i} \right|^2 \right\}}
  +2\mathrm{Re}\left( \sum_{i=1}^4{\sum_{j=i+1}^4{\mathbb{E} \left\{ w_{nu}^{i}\left( w_{nu}^{j} \right) ^{\ast} \right\}}} \right)
  \overset{\left( a \right)}{=}\sum_{i=1}^4{\mathbb{E} \left\{ \left| w_{nu}^{i} \right|^2 \right\}},
\end{equation}
where step $\left( a \right)$ is obtained by removing the zero-value terms.
Then, using the similar method in Appendix \ref{ES_1}, we can obtain $\mathbb{E} \left\{ \left| w_{nu}^{i} \right|^2 \right\}$, $i=1,2,3,4$, respectively as follows:
\begin{equation*}
\hspace{-2.5cm}
  \mathbb{E} \left\{ \left| w_{nu}^{1} \right|^2 \right\}
  = \frac{\beta _{\mathrm{u},n}\beta _{\mathrm{u},u}\beta _{\mathrm{b}}^{2}\sigma _{\mathrm{u},0,n}^{2}\sigma _{\mathrm{u},0,u}^{2}K_{\mathrm{u},n}K_{\mathrm{u},u}N_{\mathrm{b}}}{\left( K_{\mathrm{u},n}+1 \right) \left( K_{\mathrm{u},u}+1 \right) \left( K_{\mathrm{b}}+1 \right) ^2}
  \bigg(
  \sigma _{\mathrm{b},0}^{4}K_{\mathrm{b}}^{2}N_{\mathrm{b}}\left| \varPhi _{N_{\mathrm{r}}}\left( n \right) \right|^2\left| \varPhi _{N_{\mathrm{r}}}\left( u \right) \right|^2
\end{equation*}
\begin{equation*}
\hspace{0.6cm}
  +\sigma _{\mathrm{b},0}^{2}\varsigma _{\mathrm{b}}K_{\mathrm{b}}N_{\mathrm{r}}\left( \left| \varPhi _{N_{\mathrm{r}}}\left( n \right) \right|^2+\left| \varPhi _{N_{\mathrm{r}}}\left( u \right) \right|^2 \right)
  +\varsigma _{\mathrm{b}}^{2}\left( N_{\mathrm{r}}^{2}+N_{\mathrm{b}}\left| \left( \bar{\mathbf{h}}_{\mathrm{u}}^{\left( \cdot ,n \right)} \right) ^H\bar{\mathbf{h}}_{\mathrm{u}}^{\left( \cdot ,u \right)} \right|^2 \right)
\end{equation*}
\begin{equation}\label{E_w_1_nu_value}
\hspace{-2.3cm}
  +2\sigma _{\mathrm{b},0}^{2}\varsigma _{\mathrm{b}}K_{\mathrm{b}}N_{\mathrm{b}}\mathrm{Re}\left( \left( \varPhi _{N_{\mathrm{r}}}\left( n \right) \right) ^{\ast}\varPhi _{N_{\mathrm{r}}}\left( u \right) \left( \bar{\mathbf{h}}_{\mathrm{u}}^{\left( \cdot ,u \right)} \right) ^H\bar{\mathbf{h}}_{\mathrm{u}}^{\left( \cdot ,n \right)} \right)
  \bigg),
\end{equation}
\begin{equation*}
\hspace{-2.3cm}
  \mathbb{E} \left\{ \left| w_{nu}^{2} \right|^2 \right\}
  = \frac{\beta _{\mathrm{u},n}\beta _{\mathrm{u},u}\beta _{\mathrm{b}}^{2}\sigma _{\mathrm{u},0,n}^{2}\varsigma _{\mathrm{u},u}K_{\mathrm{u},n}N_{\mathrm{b}}}{\left( K_{\mathrm{u},n}+1 \right) \left( K_{\mathrm{u},u}+1 \right) \left( K_{\mathrm{b}}+1 \right) ^2}
  \bigg(
  \left( \sigma _{\mathrm{b},0}^{2}K_{\mathrm{b}}N_{\mathrm{b}}N_{\mathrm{r}}+\varsigma _{\mathrm{b}}N_{\mathrm{r}}+2\varsigma _{\mathrm{b}}N_{\mathrm{b}} \right)
\end{equation*}
\begin{equation}\label{E_w_2_nu_value}
\hspace{-2.6cm}
  \times
  \sigma _{\mathrm{b},0}^{2}K_{\mathrm{b}}\left| \varPhi _{N_{\mathrm{r}}}\left( n \right) \right|^2
  +\left( \sigma _{\mathrm{b},0}^{2}\varsigma _{\mathrm{b}}K_{\mathrm{b}}+\varsigma _{\mathrm{b}}^{2} \right) N_{\mathrm{r}}^{2}+\varsigma _{\mathrm{b}}^{2}N_{\mathrm{b}}N_{\mathrm{r}}
  \bigg),
\end{equation}
\begin{equation*}
\hspace{-2.3cm}
  \mathbb{E} \left\{ \left| w_{nu}^{3} \right|^2 \right\}
  = \frac{\beta _{\mathrm{u},n}\beta _{\mathrm{u},u}\beta _{\mathrm{b}}^{2}\sigma _{\mathrm{u},0,u}^{2}\varsigma _{\mathrm{u},n}K_{\mathrm{u},u}N_{\mathrm{b}}}{\left( K_{\mathrm{u},n}+1 \right) \left( K_{\mathrm{u},u}+1 \right) \left( K_{\mathrm{b}}+1 \right) ^2}
  \bigg(
  \left( \sigma _{\mathrm{b},0}^{2}K_{\mathrm{b}}N_{\mathrm{b}}N_{\mathrm{r}}+\varsigma _{\mathrm{b}}N_{\mathrm{r}}+2\varsigma _{\mathrm{b}}N_{\mathrm{b}} \right)
\end{equation*}
\begin{equation}\label{E_w_3_nu_value}
\hspace{-2.6cm}
  \times
  \sigma _{\mathrm{b},0}^{2}K_{\mathrm{b}}\left| \varPhi _{N_{\mathrm{r}}}\left( u \right) \right|^2
  +\left( \sigma _{\mathrm{b},0}^{2}\varsigma _{\mathrm{b}}K_{\mathrm{b}}+\varsigma _{\mathrm{b}}^{2} \right) N_{\mathrm{r}}^{2}+\varsigma _{\mathrm{b}}^{2}N_{\mathrm{b}}N_{\mathrm{r}}
  \bigg),
\end{equation}
\begin{equation}\label{E_w_4_nu_value}
  \mathbb{E} \left\{ \left| w_{nu}^{4} \right|^2 \right\}
  =\frac{\beta _{\mathrm{u},n}\beta _{\mathrm{u},u}\beta _{\mathrm{b}}^{2}\varsigma _{\mathrm{u},n}\varsigma _{\mathrm{u},u}N_{\mathrm{b}}N_{\mathrm{r}}}{\left( K_{\mathrm{u},n}+1 \right) \left( K_{\mathrm{u},u}+1 \right) \left( K_{\mathrm{b}}+1 \right) ^2}\left( \sigma _{\mathrm{b},0}^{4}K_{\mathrm{b}}^{2}N_{\mathrm{b}}N_{\mathrm{r}}+\left( 2\sigma _{\mathrm{b},0}^{2}\varsigma _{\mathrm{b}}K_{\mathrm{b}}+\varsigma _{\mathrm{b}}^{2} \right) \left( N_{\mathrm{r}}+N_{\mathrm{b}} \right) \right).
\end{equation}
Substituting \eqref{E_w_1_nu_value}-\eqref{E_w_4_nu_value} into \eqref{EI_1_ex}, we arrive at
\begin{equation*}
\hspace{-4.8cm}
  \mathbb{E} \left\{ \left| \mathbf{g}_{\mathrm{u},nt}^{H}\mathbf{\Phi }^H\mathbf{G}_{\mathrm{b}}^{H}\mathbf{G}_{\mathrm{b}}\mathbf{\Phi g}_{\mathrm{u},ut} \right|^2 \right\}
  =\frac{\beta _{\mathrm{u},n}\beta _{\mathrm{u},u}\beta _{\mathrm{b}}^{2}N_{\mathrm{b}}}{\left( K_{\mathrm{u},n}+1 \right) \left( K_{\mathrm{u},u}+1 \right) \left( K_{\mathrm{b}}+1 \right) ^2}
\end{equation*}
\begin{equation*}
\hspace{-7cm}
  \times
  \bigg(
  \sigma _{\mathrm{u},0,n}^{2}\sigma _{\mathrm{u},0,u}^{2}\sigma _{\mathrm{b},0}^{4}K_{\mathrm{u},n}K_{\mathrm{u},u}K_{\mathrm{b}}^{2}N_{\mathrm{b}}\left| \varPhi _{N_{\mathrm{r}}}\left( n \right) \right|^2\left| \varPhi _{N_{\mathrm{r}}}\left( u \right) \right|^2
\end{equation*}
\begin{equation*}
\hspace{-1.3cm}
  +\left( \left( \sigma _{\mathrm{b},0}^{2}\varsigma _{\mathrm{u},u}K_{\mathrm{b}}N_{\mathrm{b}}+\sigma _{\mathrm{u},0,u}^{2}\varsigma _{\mathrm{b}}K_{\mathrm{u},u}+\varsigma _{\mathrm{u},u}\varsigma _{\mathrm{b}} \right) N_{\mathrm{r}}+2\varsigma _{\mathrm{u},u}\varsigma _{\mathrm{b}}N_{\mathrm{b}} \right) \sigma _{\mathrm{u},0,n}^{2}\sigma _{\mathrm{b},0}^{2}K_{\mathrm{u},n}K_{\mathrm{b}}\left| \varPhi _{N_{\mathrm{r}}}\left( n \right) \right|^2
\end{equation*}
\begin{equation*}
\hspace{-1.3cm}
  +\left( \left( \sigma _{\mathrm{b},0}^{2}\varsigma _{\mathrm{u},n}K_{\mathrm{b}}N_{\mathrm{b}}+\sigma _{\mathrm{u},0,n}^{2}\varsigma _{\mathrm{b}}K_{\mathrm{u},n}+\varsigma _{\mathrm{u},n}\varsigma _{\mathrm{b}} \right) N_{\mathrm{r}}+2\varsigma _{\mathrm{u},n}\varsigma _{\mathrm{b}}N_{\mathrm{b}} \right) \sigma _{\mathrm{u},0,u}^{2}\sigma _{\mathrm{b},0}^{2}K_{\mathrm{u},u}K_{\mathrm{b}}\left| \varPhi _{N_{\mathrm{r}}}\left( u \right) \right|^2
\end{equation*}
\begin{equation*}
  +\left( \varsigma _{\mathrm{u},n}\varsigma _{\mathrm{u},u}\sigma _{\mathrm{b},0}^{4}K_{\mathrm{b}}^{2}N_{\mathrm{b}}+\sigma _{\mathrm{u},0,n}^{2}\sigma _{\mathrm{u},0,u}^{2}\varsigma _{\mathrm{b}}^{2}K_{\mathrm{u},n}K_{\mathrm{u},u}+\left( \sigma _{\mathrm{u},0,n}^{2}\varsigma _{\mathrm{u},u}K_{\mathrm{u},n}+\sigma _{\mathrm{u},0,u}^{2}\varsigma _{\mathrm{u},n}K_{\mathrm{u},u} \right) \sigma _{\mathrm{b},0}^{2}\varsigma _{\mathrm{b}}K_{\mathrm{b}} \right) N_{\mathrm{r}}^{2}
\end{equation*}
\begin{equation*}
\hspace{-2.4cm}
  +\left( \sigma _{\mathrm{u},0,u}^{2}\varsigma _{\mathrm{u},n}\varsigma _{\mathrm{b}}^{2}K_{\mathrm{u},u}+\sigma _{\mathrm{u},0,n}^{2}\varsigma _{\mathrm{u},u}\varsigma _{\mathrm{b}}^{2}K_{\mathrm{u},n}+\varsigma _{\mathrm{u},n}\varsigma _{\mathrm{u},u}\left( 2\sigma _{\mathrm{b},0}^{2}\varsigma _{\mathrm{b}}K_{\mathrm{b}}+\varsigma _{\mathrm{b}}^{2} \right) \right) N_{\mathrm{r}}\left( N_{\mathrm{b}}+N_{\mathrm{r}} \right)
\end{equation*}
\begin{equation*}
\hspace{-8.5cm}
  +\sigma _{\mathrm{u},0,n}^{2}\sigma _{\mathrm{u},0,u}^{2}\varsigma _{\mathrm{b}}^{2}K_{\mathrm{u},n}K_{\mathrm{u},u}N_{\mathrm{b}}\left| \left( \bar{\mathbf{h}}_{\mathrm{u}}^{\left( \cdot ,n \right)} \right) ^H\bar{\mathbf{h}}_{\mathrm{u}}^{\left( \cdot ,u \right)} \right|^2
\end{equation*}
\begin{equation}\label{eta_n_u}
\hspace{-0.0000001cm}
  +2\sigma _{\mathrm{u},0,n}^{2}\sigma _{\mathrm{u},0,u}^{2}\sigma _{\mathrm{b},0}^{2}\varsigma _{\mathrm{b}}K_{\mathrm{u},n}K_{\mathrm{u},u}K_{\mathrm{b}}N_{\mathrm{b}}\mathrm{Re}\left( \left( \varPhi _{N_{\mathrm{r}}}\left( n \right) \right) ^{\ast}\varPhi _{N_{\mathrm{r}}}\left( u \right) \left( \bar{\mathbf{h}}_{\mathrm{u}}^{\left( \cdot ,u \right)} \right) ^H\bar{\mathbf{h}}_{\mathrm{u}}^{\left( \cdot ,n \right)} \right)
 \bigg)
  \triangleq  \eta _{n,u} ,
\end{equation}
where $\bar{\mathbf{h}}_{\mathrm{u}}^{\left( \cdot ,n \right)}$ is defined as
\begin{equation}\label{h_u_LoS_vector}
  \bar{\mathbf{h}}_{\mathrm{u}}^{\left( \cdot ,n \right)}=\left[ \bar{h}_{\mathrm{u}}^{\left( 0,n \right)},...,\bar{h}_{\mathrm{u}}^{\left( N_{\mathrm{r}}-1,n \right)} \right] ^T,
\end{equation}
and $\bar{h}_{\mathrm{u}}^{\left( i,j \right)}$ is given in \eqref{R-U_tap_LoS}.

\subsection{Derivation of $\mathbb{E} \left\{ \mathbf{g}_{\mathrm{u},nt}^{H}\mathbf{\Phi }^H\mathbf{G}_{\mathrm{b}}^{H}\left( \mathbf{I}_{N_{\mathrm{b}}}\otimes \mathbf{F} \right) \mathbf{R}_{\mathbf{z}_{\mathrm{q}}}\left( \mathbf{I}_{N_{\mathrm{b}}}\otimes \mathbf{F}^H \right) \mathbf{G}_{\mathrm{b}}\mathbf{\Phi g}_{\mathrm{u},nt} \right\}$}\label{EI_3}

Substituting \eqref{TD_y} into \eqref{Rzq}, we have
\begin{equation*}
\hspace{-5.2cm}
  \mathbf{R}_{\mathbf{z}_{\mathrm{q}}}\approx
  \alpha \left( 1-\alpha \right) \mathrm{diag}\left( \mathbf{H}_{\mathrm{b}}\mathbf{\Phi H}_{\mathrm{u}}\mathbf{PP}^H\mathbf{H}_{\mathrm{u}}^{H}\mathbf{\Phi }^H\mathbf{H}_{\mathrm{b}}^{H}+\sigma _{\mathrm{noise}}^{2}\mathbf{I}_{N_{\mathrm{b}}N_{\mathrm{c}}} \right)
\end{equation*}
\begin{equation}\label{Rzq_expand}
  \overset{\left( a \right)}{=}\alpha \left( 1-\alpha \right) \mathrm{diag}\left( \left( \mathbf{I}_{N_{\mathrm{b}}}\otimes \mathbf{F}^H \right) \mathbf{G}_{\mathrm{b}}\mathbf{\Phi G}_{\mathrm{u}}\mathbf{PP}^H\mathbf{G}_{\mathrm{u}}^{H}\mathbf{\Phi }^H\mathbf{G}_{\mathrm{b}}^{H}\left( \mathbf{I}_{N_{\mathrm{b}}}\otimes \mathbf{F} \right) \right) +\sigma _{\mathrm{noise}}^{2}\alpha \left( 1-\alpha \right) \mathbf{I}_{N_{\mathrm{b}}N_{\mathrm{c}}},
\end{equation}
where step $\left( a \right)$ is based on \eqref{R-U_TD_channel} and \eqref{B-R_TD_channel}.
It should be noted that $\mathbf{R}_{\mathbf{z}_{\mathrm{q}}}$ is a diagonal matrix.
Therefore, we focus on the diagonal matrix
\begin{equation}\label{gamma_p1}
  \mathrm{diag}\left( \left( \mathbf{I}_{N_{\mathrm{b}}}\otimes \mathbf{F}^H \right) \mathbf{G}_{\mathrm{b}}\mathbf{\Phi G}_{\mathrm{u}}\mathbf{PP}^H\mathbf{G}_{\mathrm{u}}^{H}\mathbf{\Phi }^H\mathbf{G}_{\mathrm{b}}^{H}\left( \mathbf{I}_{N_{\mathrm{b}}}\otimes \mathbf{F} \right) \right) \triangleq \mathbf{\Upsilon }.
\end{equation}
The $(bN_{\mathrm{c}}+t)$-th diagonal element of $\mathbf{\Upsilon }$ can be expressed as
\begin{equation*}
\hspace{-5.6cm}
  \left[ \mathbf{\Upsilon } \right] _{bt,bt}=\left[ \left( \mathbf{I}_{N_{\mathrm{b}}}\otimes \mathbf{F}^H \right) \mathbf{G}_{\mathrm{b}}\mathbf{\Phi G}_{\mathrm{u}}\mathbf{PP}^H\mathbf{G}_{\mathrm{u}}^{H}\mathbf{\Phi }^H\mathbf{G}_{\mathrm{b}}^{H}\left( \mathbf{I}_{N_{\mathrm{b}}}\otimes \mathbf{F} \right) \right] _{bt,bt}
\end{equation*}
\begin{equation*}
  =\left[ \mathbf{F}^H\left[ e^{j\varphi _0}\mathbf{G}_{\mathrm{b}}^{\left( b,0 \right)}, e^{j\varphi _1}\mathbf{G}_{\mathrm{b}}^{\left( b,1 \right)}, ..., e^{j\varphi _{N_{\mathrm{r}}-1}}\mathbf{G}_{\mathrm{b}}^{\left( b,N_{\mathrm{r}}-1 \right)} \right] \mathbf{G}_{\mathrm{u}}\mathbf{PP}^H\mathbf{G}_{\mathrm{u}}^{H}\left[ \begin{array}{c}
	e^{-j\varphi _0}\left( \mathbf{G}_{\mathrm{b}}^{\left( b,0 \right)} \right) ^H\\
	e^{-j\varphi _1}\left( \mathbf{G}_{\mathrm{b}}^{\left( b,1 \right)} \right) ^H\\
	\vdots\\
	e^{-j\varphi _{N_{\mathrm{r}}-1}}\left( \mathbf{G}_{\mathrm{b}}^{\left( b,N_{\mathrm{r}}-1 \right)} \right) ^H\\
\end{array} \right] \mathbf{F} \right] _{t,t}
\end{equation*}
\begin{equation*}
\hspace{-4.2cm}
  =\sum_{s=0}^{N_{\mathrm{c}}-1}{\left| f_{s,t} \right|^2\sum_{u=0}^{N_{\mathrm{u}}-1}{p_u\left( \sum_{r_1=0}^{N_{\mathrm{r}}-1}{e^{j\varphi _{r_1}}g_{\mathrm{b},s,s}^{\left( b,r_1 \right)}g_{\mathrm{u},s,s}^{\left( r_1,u \right)}} \right) \left( \sum_{r_2=0}^{N_{\mathrm{r}}-1}{e^{j\varphi _{r_2}}g_{\mathrm{b},s,s}^{\left( b,r_2 \right)}g_{\mathrm{u},s,s}^{\left( r_2,u \right)}} \right) ^{\ast}}}
\end{equation*}
\begin{equation}\label{gamma_b}
\hspace{-3cm}
  =\frac{1}{N_{\mathrm{c}}}\sum_{s=0}^{N_{\mathrm{c}}-1}{\sum_{u=0}^{N_{\mathrm{u}}-1}{p_u\left( \sum_{r_1=0}^{N_{\mathrm{r}}-1}{e^{j\varphi _{r_1}}g_{\mathrm{b},s,s}^{\left( b,r_1 \right)}g_{\mathrm{u},s,s}^{\left( r_1,u \right)}} \right) \left( \sum_{r_2=0}^{N_{\mathrm{r}}-1}{e^{j\varphi _{r_2}}g_{\mathrm{b},s,s}^{\left( b,r_2 \right)}g_{\mathrm{u},s,s}^{\left( r_2,u \right)}} \right) ^{\ast}}}
  \triangleq \gamma \left( b \right).
\end{equation}
It is observed from \eqref{gamma_b} that $\gamma \left( b \right)$ is dependent with antenna $b$, but is independent of sub-carrier~$t$.
Therefore, we have
\begin{equation}\label{gamma_matrix}
  \mathbf{\Upsilon }
  =\left[ \begin{matrix}
	\gamma \left( 0 \right) \mathbf{I}_{N_{\mathrm{c}}}&		&		&		0\\
	&		\gamma \left( 1 \right) \mathbf{I}_{N_{\mathrm{c}}}&		&		\\
	&		&		\ddots&		\\
	0&		&		&		\gamma \left( N_{\mathrm{b}}-1 \right) \mathbf{I}_{N_{\mathrm{c}}}\\
\end{matrix} \right] .
\end{equation}

From \eqref{Rzq_expand}, \eqref{gamma_p1} and \eqref{epsilon_n}, the expectation can be re-expressed as
\begin{equation*}
\hspace{-1.4cm}
  \mathbb{E} \left\{ \mathbf{g}_{\mathrm{u},nt}^{H}\mathbf{\Phi }^H\mathbf{G}_{\mathrm{b}}^{H}\left( \mathbf{I}_{N_{\mathrm{b}}}\otimes \mathbf{F} \right) \mathbf{R}_{\mathbf{z}_{\mathrm{q}}}\left( \mathbf{I}_{N_{\mathrm{b}}}\otimes \mathbf{F}^H \right) \mathbf{G}_{\mathrm{b}}\mathbf{\Phi g}_{\mathrm{u},nt} \right\}
\end{equation*}
\begin{equation}\label{EI_3_ex}
  =\alpha \left( 1-\alpha \right) \mathbb{E} \left\{ \mathbf{g}_{\mathrm{u},nt}^{H}\mathbf{\Phi }^H\mathbf{G}_{\mathrm{b}}^{H}\mathbf{\Upsilon G}_{\mathrm{b}}\mathbf{\Phi g}_{\mathrm{u},nt} \right\}
  +  \sigma _{\mathrm{noise}}^{2} \alpha \left( 1-\alpha \right) \epsilon _n.
\end{equation}
Then, we focus on the calculation of $\mathbb{E} \left\{ \mathbf{g}_{\mathrm{u},nt}^{H}\mathbf{\Phi }^H\mathbf{G}_{\mathrm{b}}^{H}\mathbf{\Upsilon G}_{\mathrm{b}}\mathbf{\Phi g}_{\mathrm{u},nt} \right\}$.
In \eqref{gamma_p1}, the expression of $\mathbf{\Upsilon }$ contains four $\mathbf{G}$ matrices.
Since each $\mathbf{G}$ matrix can be divided into two parts, i.e. LoS part and NLoS part, we divide $\mathbf{\Upsilon }$ as
\begin{equation}\label{gamma_matrix_expand}
  \mathbf{\Upsilon }=\sum_{i_1=1}^2{\sum_{i_2=1}^2{\sum_{i_3=1}^2{\sum_{i_4=1}^2{\mathbf{\Upsilon }_{i_1i_2i_3i_4}}}}},
\end{equation}
where $i_j=1$ corresponds to the LoS part of the $j$-th $\mathbf{G}$ matrix,
and $i_j=2$ corresponds to the NLoS part of the $j$-th $\mathbf{G}$ matrix, $j=1,2,3,4$.
For example, we have
\begin{equation}\label{gamma_matrix_expand_example}
  \mathbf{\Upsilon }_{1112}=\mathrm{diag}\left( \left( \mathbf{I}_{N_{\mathrm{b}}}\otimes \mathbf{F}^H \right) \bar{\mathbf{G}}_{\mathrm{b}}\mathbf{\Phi }\bar{\mathbf{G}}_{\mathrm{u}}\mathbf{PP}^H\bar{\mathbf{G}}_{\mathrm{u}}^{H}\mathbf{\Phi }^H\tilde{\mathbf{G}}_{\mathrm{b}}^{H}\left( \mathbf{I}_{N_{\mathrm{b}}}\otimes \mathbf{F} \right) \right) .
\end{equation}
Based on \eqref{gamma_matrix_expand}, the expectation $\mathbb{E} \left\{ \mathbf{g}_{\mathrm{u},nt}^{H}\mathbf{\Phi }^H\mathbf{G}_{\mathrm{b}}^{H}\mathbf{\Upsilon G}_{\mathrm{b}}\mathbf{\Phi g}_{\mathrm{u},nt} \right\}$ can be re-expressed as
\begin{equation}\label{EI_3_gamma_expand}
  \mathbb{E} \left\{ \mathbf{g}_{\mathrm{u},nt}^{H}\mathbf{\Phi }^H\mathbf{G}_{\mathrm{b}}^{H}\mathbf{\Upsilon G}_{\mathrm{b}}\mathbf{\Phi g}_{\mathrm{u},nt} \right\}
  =\sum_{i_1=1}^2{\sum_{i_2=1}^2{\sum_{i_3=1}^2{\sum_{i_4=1}^2{\mathbb{E} \left\{ \mathbf{g}_{\mathrm{u},nt}^{H}\mathbf{\Phi }^H\mathbf{G}_{\mathrm{b}}^{H}\mathbf{\Upsilon }_{i_1i_2i_3i_4}\mathbf{G}_{\mathrm{b}}\mathbf{\Phi g}_{\mathrm{u},nt} \right\}}}}}.
\end{equation}

Before we start to calculate the decomposed expectations in \eqref{EI_3_gamma_expand} one by one,
we present the following two useful lemmas.

\begin{lemma}\label{lemma_sum}

Due to the periodicity of the function $e^{j\frac{2\pi}{N_{\mathrm{c}}}sk}$ respect to $s$, we have
\begin{equation}\label{lemma_sum_eq1}
    \sum_{s=0}^{N_{\mathrm{c}}-1}{e^{j\frac{2\pi}{N_{\mathrm{c}}}sk}}=\begin{cases}
	0, \;\;\;\;\; k\ne 0\\
	N_{\mathrm{c}}, \;\;\; k=0\\
\end{cases}.
\end{equation}
Furthermore, noting that $N_{\mathrm{c}}\gg L_{\mathrm{u}},L_{\mathrm{b}}$,
we obtain
\begin{equation}\label{lemma_sum_eq2}
\hspace{-5.4cm}
  \sum_{k_1=1}^{L_{\mathrm{u}}-1}{\sum_{k_2=1}^{L_{\mathrm{b}}-1}{\sum_{s=0}^{N_{\mathrm{c}}-1}{e^{-j\frac{2\pi}{N_{\mathrm{c}}}\left( k_1-k_2 \right) \left( s-t \right)}}\sigma _{\mathrm{u},k_1,n}^{2}\sigma _{\mathrm{b},k_2}^{2}}}=N_{\mathrm{c}}\sum_{k=1}^{L_{1}^{\min}-1}{\sigma _{\mathrm{u},k,n}^{2}\sigma _{\mathrm{b},k}^{2}} ,
\end{equation}
\begin{equation}\label{lemma_sum_eq3}
 \sum_{k_1=1}^{L_{\mathrm{u}}-1}{\sum_{k_2\ne k_1}^{L_{\mathrm{u}}-1}{\sum_{k_3=1}^{L_{\mathrm{b}}-1}{\sum_{s=0}^{N_{\mathrm{c}}-1}{e^{-j\frac{2\pi}{N_{\mathrm{c}}}\left( k_2-k_1-k_3 \right) \left( s-t \right)}}\sigma _{\mathrm{u},k_1,n}^{2}\sigma _{\mathrm{u},k_2,n}^{2}\sigma _{\mathrm{b},k_3}^{2}}}}
  =
  N_{\mathrm{c}}\sum_{k_1=1}^{L_{\mathrm{u}}-1}{\sum_{k_2=k_1+1}^{L_{2}^{\min}-1}{\sigma _{\mathrm{u},k_1,n}^{2}\sigma _{\mathrm{u},k_2,n}^{2}\sigma _{\mathrm{b},k_2-k_1}^{2}}},
\end{equation}
\begin{equation}\label{lemma_sum_eq4}
\hspace{-0.6cm}
  \sum_{k_1=1}^{L_{\mathrm{b}}-1}{\sum_{k_2\ne k_1}^{L_{\mathrm{b}}-1}{\sum_{k_3=1}^{L_{\mathrm{u}}-1}{\sum_{s=0}^{N_{\mathrm{c}}-1}{e^{-j\frac{2\pi}{N_{\mathrm{c}}}\left( k_1-k_2-k_3 \right) \left( s-t \right)}}\sigma _{\mathrm{b},k_1}^{2}\sigma _{\mathrm{b},k_2}^{2}\sigma _{\mathrm{u},k_3,n}^{2}}}}
  =
  N_{\mathrm{c}}\sum_{k_1=1}^{L_{\mathrm{b}}-1}{\sum_{k_2=k_1+1}^{L_{3}^{\min}-1}{\sigma _{\mathrm{b},k_1}^{2}\sigma _{\mathrm{b},k_2}^{2}\sigma _{\mathrm{u},k_2-k_1,n}^{2}}},
\end{equation}
\begin{equation*}
\hspace{-4.7cm}
  \sum_{k_1=1}^{L_{\mathrm{b}}-1}{\sum_{k_2\ne k_1}^{L_{\mathrm{b}}-1}{\sum_{k_3=1}^{L_{\mathrm{u}}-1}{\sum_{k_4\ne k_3}^{L_{\mathrm{u}}-1}{\sum_{s=0}^{N_{\mathrm{c}}-1}{e^{-j\frac{2\pi}{N_{\mathrm{c}}}\left( k_1-k_2+k_3-k_4 \right) \left( s-t \right)}\sigma _{\mathrm{b},k_1}^{2}\sigma _{\mathrm{b},k_2}^{2}\sigma _{\mathrm{u},k_3,n}^{2}\sigma _{\mathrm{u},k_4,n}^{2}}}}}}
\end{equation*}
\begin{equation}\label{lemma_sum_eq5}
\hspace{-6.3cm}
  = 2N_{\mathrm{c}}\sum_{k_1=1}^{L_{\mathrm{b}}-1}{\sum_{k_2=k_1+1}^{L_{4}^{\min}-1}{\sum_{k_3=k_2-k_1+1}^{L_{\mathrm{u}}-1}{\sigma _{\mathrm{b},k_1}^{2}\sigma _{\mathrm{b},k_2}^{2}\sigma _{\mathrm{u},k_3,n}^{2}\sigma _{\mathrm{u},k_1-k_2+k_3,n}^{2}}}},
\end{equation}
where $L_{1}^{\min}=\min \left\{ L_{\mathrm{b}},L_{\mathrm{u}} \right\}$,
$L_{2}^{\min}=\min \left\{ L_{\mathrm{u}},L_{\mathrm{b}}+k_1 \right\}$,
$L_{3}^{\min}=\min \left\{ L_{\mathrm{b}},L_{\mathrm{u}}+k_1 \right\}$
and $L_{4}^{\min}=\min \left\{ L_{\mathrm{b}},L_{\mathrm{u}}+k_1-1 \right\}$.

\end{lemma}

\begin{lemma}\label{lemma_expectation}

Given the random variables $\tilde{g}_{\mathrm{b},t,t}^{\left( i,j \right)}$ in \eqref{R-U_FD_channel_ij_tt_NLoS} and $\tilde{g}_{\mathrm{b},t,t}^{\left( i,j \right)}$ in \eqref{B-R_FD_channel_ij_tt_NLoS}, the related expectations can be calculated as
\begin{equation}\label{lemma_expectation_eq1}
\hspace{-6.1cm}
  \mathbb{E} \left\{ \left( \tilde{g}_{\mathrm{b},s,s}^{\left( b,r \right)} \right) ^{\ast}\tilde{g}_{\mathrm{b},t,t}^{\left( b,r \right)} \right\} =\frac{\beta _{\mathrm{b}}\sigma _{\mathrm{b},0}^{2}}{K_{\mathrm{b}}+1}+\beta _{\mathrm{b}}\sum_{k=1}^{L_{\mathrm{b}}-1}{e^{-j\frac{2\pi}{N_{\mathrm{c}}}k\left( t-s \right)}\sigma _{\mathrm{b},k}^{2}},
\end{equation}
\begin{equation}\label{lemma_expectation_eq2}
\hspace{-5.3cm}
  \mathbb{E} \left\{ \left( \tilde{g}_{\mathrm{u},s,s}^{\left( r,n \right)} \right) ^{\ast}\tilde{g}_{\mathrm{u},t,t}^{\left( r,n \right)} \right\} =\frac{\beta _{\mathrm{u},n}\sigma _{\mathrm{u},0,n}^{2}}{K_{\mathrm{u},n}+1}+\beta _{\mathrm{u},n}\sum_{k=1}^{L_{\mathrm{u}}-1}{e^{-j\frac{2\pi}{N_{\mathrm{c}}}k\left( t-s \right)}\sigma _{\mathrm{u},k,n}^{2}},
\end{equation}
\begin{equation*}
  \mathbb{E} \left\{ \left( \tilde{g}_{\mathrm{u},s,s}^{\left( r,n \right)} \right) ^{\ast}\tilde{g}_{\mathrm{u},t,t}^{\left( r,n \right)}\left( \tilde{g}_{\mathrm{b},s,s}^{\left( b,r \right)} \right) ^{\ast}\tilde{g}_{\mathrm{b},t,t}^{\left( b,r \right)} \right\}
  = \frac{\beta _{\mathrm{u},n}\beta _{\mathrm{b}}\sigma _{\mathrm{u},0,n}^{2}\sigma _{\mathrm{b},0}^{2}}{\left( K_{\mathrm{u},n}+1 \right) \left( K_{\mathrm{b}}+1 \right)}+\frac{\beta _{\mathrm{u},n}\beta _{\mathrm{b}}\sigma _{\mathrm{u},0,n}^{2}}{K_{\mathrm{u},n}+1}\sum_{k=1}^{L_{\mathrm{b}}-1}{e^{-j\frac{2\pi}{N_{\mathrm{c}}}k\left( t-s \right)}\sigma _{\mathrm{b},k}^{2}}
\end{equation*}
\begin{equation}\label{lemma_expectation_eq3}
  +\frac{\beta _{\mathrm{u},n}\beta _{\mathrm{b}}\sigma _{\mathrm{b},0}^{2}}{K_{\mathrm{b}}+1}\sum_{k=1}^{L_{\mathrm{u}}-1}{e^{-j\frac{2\pi}{N_{\mathrm{c}}}k\left( t-s \right)}\sigma _{\mathrm{u},k,n}^{2}}+\beta _{\mathrm{u},n}\beta _{\mathrm{b}}\sum_{k_1=1}^{L_{\mathrm{u}}-1}{\sum_{k_2=1}^{L_{\mathrm{b}}-1}{e^{-j\frac{2\pi}{N_{\mathrm{c}}}\left( k_1+k_2 \right) \left( t-s \right)}\sigma _{\mathrm{u},k_1,n}^{2}\sigma _{\mathrm{b},k_2}^{2}}},
\end{equation}
\begin{equation*}
\hspace{-1.1cm}
  \mathbb{E} \left\{ \left( \tilde{g}_{\mathrm{b},s,s}^{\left( b,r \right)} \right) ^{\ast}\tilde{g}_{\mathrm{b},s,s}^{\left( b,r \right)}\left( \tilde{g}_{\mathrm{b},t,t}^{\left( b,r \right)} \right) ^{\ast}\tilde{g}_{\mathrm{b},t,t}^{\left( b,r \right)} \right\}
  = \beta _{\mathrm{b}}^{2} \bigg(
  \tau _{\mathrm{b}}+\sum_{k=1}^{L_{\mathrm{b}}-1}{\sigma _{\mathrm{b},k}^{4}}+\frac{\sigma _{\mathrm{b},0}^{2}}{K_{\mathrm{b}}+1}\sum_{k=1}^{L_{\mathrm{b}}-1}{e^{-j\frac{2\pi}{N_{\mathrm{c}}}k\left( s-t \right)}\sigma _{\mathrm{b},k}^{2}}
\end{equation*}
\begin{equation}\label{lemma_expectation_eq4}
  +\frac{\sigma _{\mathrm{b},0}^{2}}{K_{\mathrm{b}}+1}\sum_{k=1}^{L_{\mathrm{b}}-1}{e^{j\frac{2\pi}{N_{\mathrm{c}}}k\left( s-t \right)}\sigma _{\mathrm{b},k}^{2}}+\sum_{k_1=1}^{L_{\mathrm{b}}-1}{\sum_{k_2\ne k_1}^{L_{\mathrm{b}}-1}{e^{-j\frac{2\pi}{N_{\mathrm{c}}}\left( k_1-k_2 \right) \left( s-t \right)}\sigma _{\mathrm{b},k_1}^{2}\sigma _{\mathrm{b},k_2}^{2}}}
  \bigg) ,
\end{equation}
\begin{equation*}
  \mathbb{E} \left\{ \left( \tilde{g}_{\mathrm{u},t,t}^{\left( r,n \right)} \right) ^{\ast}\tilde{g}_{\mathrm{u},t,t}^{\left( r,n \right)}\left( \tilde{g}_{\mathrm{u},s,s}^{\left( r,n \right)} \right) ^{\ast}\tilde{g}_{\mathrm{u},s,s}^{\left( r,n \right)} \right\}
  = \beta _{\mathrm{u},n}^{2} \bigg(
  \tau _{\mathrm{u},n}+\sum_{k=1}^{L_{\mathrm{u}}-1}{\sigma _{\mathrm{u},k,n}^{4}}+\frac{\sigma _{\mathrm{u},0,n}^{2}}{K_{\mathrm{u},n}+1}\sum_{k=1}^{L_{\mathrm{u}}-1}{e^{j\frac{2\pi}{N_{\mathrm{c}}}k\left( s-t \right)}\sigma _{\mathrm{u},k,n}^{2}}
\end{equation*}
\begin{equation}\label{lemma_expectation_eq5}
  +\frac{\sigma _{\mathrm{u},0,n}^{2}}{K_{\mathrm{u},n}+1}\sum_{k=1}^{L_{\mathrm{u}}-1}{e^{-j\frac{2\pi}{N_{\mathrm{c}}}k\left( s-t \right)}\sigma _{\mathrm{u},k,n}^{2}}+\sum_{k_1=1}^{L_{\mathrm{u}}-1}{\sum_{k_2\ne k_1}^{L_{\mathrm{u}}-1}{e^{-j\frac{2\pi}{N_{\mathrm{c}}}\left( k_1-k_2 \right) \left( s-t \right)}\sigma _{\mathrm{u},k_1,n}^{2}\sigma _{\mathrm{u},k_2,n}^{2}}}
  \bigg) ,
\end{equation}
where $\tau _{\mathrm{b}}$ in \eqref{lemma_expectation_eq4} and $\tau _{\mathrm{u},n}$ in \eqref{lemma_expectation_eq5} are respectively defined as
\begin{equation}\label{tau_b}
  \tau _{\mathrm{b}} =
  \frac{K_{\mathrm{b}}^{2}+1}{\left( K_{\mathrm{b}}+1 \right) ^2}\sigma _{\mathrm{b},0}^{4}-\frac{2K_{\mathrm{b}}}{K_{\mathrm{b}}+1}\sigma _{\mathrm{b},0}^{2}+1,
\end{equation}
\begin{equation}\label{tau_un}
  \tau _{\mathrm{u},n} =
  \frac{K_{\mathrm{u},n}^{2}+1}{\left( K_{\mathrm{u},n}+1 \right) ^2}\sigma _{\mathrm{u},0,n}^{4}-\frac{2K_{\mathrm{u},n}}{K_{\mathrm{u},n}+1}\sigma _{\mathrm{u},0,n}^{2}+1.
\end{equation}

\end{lemma}

Based on Lemma~\ref{lemma_sum} and Lemma~\ref{lemma_expectation},
the detailed steps of the derivations for \eqref{EI_3_gamma_expand} are given as follows:

\subsubsection{\textbf{ The $\mathbf{\Upsilon }_{1112}$ and $\mathbf{\Upsilon }_{2111}$-Related Expectations}} \label{E_1112}

It is noted that
\begin{equation}\label{E_1112_eq1}
  \mathbb{E} \left\{ \mathbf{g}_{\mathrm{u},nt}^{H}\mathbf{\Phi }^H\mathbf{G}_{\mathrm{b}}^{H}\mathbf{\Upsilon }_{2111}\mathbf{G}_{\mathrm{b}}\mathbf{\Phi g}_{\mathrm{u},nt} \right\} =\left( \mathbb{E} \left\{ \mathbf{g}_{\mathrm{u},nt}^{H}\mathbf{\Phi }^H\mathbf{G}_{\mathrm{b}}^{H}\mathbf{\Upsilon }_{1112}\mathbf{G}_{\mathrm{b}}\mathbf{\Phi g}_{\mathrm{u},nt} \right\} \right) ^H .
\end{equation}
Thus, we focus on the expectation $\mathbb{E} \left\{ \mathbf{g}_{\mathrm{u},nt}^{H}\mathbf{\Phi }^H\mathbf{G}_{\mathrm{b}}^{H}\mathbf{\Upsilon }_{1112}\mathbf{G}_{\mathrm{b}}\mathbf{\Phi g}_{\mathrm{u},nt} \right\}$, which can be expanded as
\begin{equation*}
\hspace{-7.5cm}
  \mathbb{E} \left\{ \mathbf{g}_{\mathrm{u},nt}^{H}\mathbf{\Phi }^H\mathbf{G}_{\mathrm{b}}^{H}\mathbf{\Upsilon }_{1112}\mathbf{G}_{\mathrm{b}}\mathbf{\Phi g}_{\mathrm{u},nt} \right\}
\end{equation*}
\begin{equation*}
  = \mathbb{E} \left\{ \left( \bar{\mathbf{g}}_{\mathrm{u},nt}^{H}+\tilde{\mathbf{g}}_{\mathrm{u},nt}^{H} \right) \mathbf{\Phi }^H\left( \bar{\mathbf{G}}_{\mathrm{b}}^{H}+\tilde{\mathbf{G}}_{\mathrm{b}}^{H} \right) \mathbf{\Upsilon }_{1112}\left( \bar{\mathbf{G}}_{\mathrm{b}}+\tilde{\mathbf{G}}_{\mathrm{b}} \right) \mathbf{\Phi }\left( \bar{\mathbf{g}}_{\mathrm{u},nt}+\tilde{\mathbf{g}}_{\mathrm{u},nt} \right) \right\}
\end{equation*}
\begin{equation}\label{E_1112_eq2}
\hspace{-0.6cm}
  \overset{\left( a \right)}{=} \mathbb{E} \left\{ \bar{\mathbf{g}}_{\mathrm{u},nt}^{H}\mathbf{\Phi }^H\bar{\mathbf{G}}_{\mathrm{b}}^{H}\mathbf{\Upsilon }_{1112}\tilde{\mathbf{G}}_{\mathrm{b}}\mathbf{\Phi }\bar{\mathbf{g}}_{\mathrm{u},nt} \right\} +\mathbb{E} \left\{ \tilde{\mathbf{g}}_{\mathrm{u},nt}^{H}\mathbf{\Phi }^H\bar{\mathbf{G}}_{\mathrm{b}}^{H}\mathbf{\Upsilon }_{1112}\tilde{\mathbf{G}}_{\mathrm{b}}\mathbf{\Phi }\tilde{\mathbf{g}}_{\mathrm{u},nt} \right\},
\end{equation}
where step $\left( a \right)$ is obtained by removing the zero-value terms,
which is based on the fact that
for a complex Gaussian random variable $z\sim \mathcal{C} \mathcal{N} \left( 0,\sigma ^2 \right)$,
we have $\mathbb{E} \left\{ z\cdot z \right\} =0$
and $\mathbb{E} \left\{ z\cdot \left| z \right|^2 \right\} =0$.

According to \eqref{gamma_b}, \eqref{gamma_matrix} and \eqref{gamma_matrix_expand}, the $(bN_{\mathrm{c}}+t)$-th element of the diagonal matrix $\mathbf{\Upsilon }_{1112}$ is expressed as
\begin{equation}\label{E_1112_eq3}
  \left[ \mathbf{\Upsilon }_{1112} \right] _{bt,bt}
  =\frac{1}{N_{\mathrm{c}}}\sum_{s=0}^{N_{\mathrm{c}}-1}{\sum_{u=0}^{N_{\mathrm{u}}-1}{p_u\left( \sum_{r_1=0}^{N_{\mathrm{r}}-1}{e^{j\varphi _{r_1}}\bar{g}_{\mathrm{b}}^{\left( b,r_1 \right)}\bar{g}_{\mathrm{u}}^{\left( r_1,u \right)}} \right) \left( \sum_{r_2=0}^{N_{\mathrm{r}}-1}{e^{j\varphi _{r_2}}\tilde{g}_{\mathrm{b},s,s}^{\left( b,r_2 \right)}\bar{g}_{\mathrm{u}}^{\left( r_2,u \right)}} \right) ^{\ast}}}
  \triangleq \gamma _{1112}\left( b \right).
\end{equation}
Thus, the first expectation in \eqref{E_1112_eq2} is calculated as
\begin{equation*}
\hspace{-10.9cm}
  \mathbb{E} \left\{ \bar{\mathbf{g}}_{\mathrm{u},nt}^{H}\mathbf{\Phi }^H\bar{\mathbf{G}}_{\mathrm{b}}^{H}\mathbf{\Upsilon }_{1112}\tilde{\mathbf{G}}_{\mathrm{b}}\mathbf{\Phi }\bar{\mathbf{g}}_{\mathrm{u},nt} \right\}
\end{equation*}
\begin{equation*}
\hspace{-4.8cm}
  =\mathbb{E} \left\{ \sum_{b=0}^{N_{\mathrm{b}}-1}{\gamma _{1112}\left( b \right) \sum_{r_1=0}^{N_{\mathrm{r}}-1}{\sum_{r_2=0}^{N_{\mathrm{r}}-1}{\left( e^{j\varphi _{r_1}}\bar{g}_{\mathrm{u}}^{\left( r_1,n \right)}\bar{g}_{\mathrm{b}}^{\left( b,r_1 \right)} \right) ^{\ast}e^{j\varphi _{r_2}}\bar{g}_{\mathrm{u}}^{\left( r_2,n \right)}\tilde{g}_{\mathrm{b},t,t}^{\left( b,r_2 \right)}}}} \right\}
\end{equation*}
\begin{equation*}
\hspace{-9.7cm}
  \overset{\left( a \right)}{=}\frac{\beta _{\mathrm{u},n}\beta _{\mathrm{b}}\sigma _{\mathrm{u},0,n}^{2}\sigma _{\mathrm{b},0}^{2}K_{\mathrm{u},n}K_{\mathrm{b}}}{N_{\mathrm{c}}\left( K_{\mathrm{u},n}+1 \right) \left( K_{\mathrm{b}}+1 \right)}\left( \varPhi _{N_{\mathrm{r}}}\left( n \right) \right) ^{\ast}
  \times
\end{equation*}
\begin{equation*}
  \sum_{b=0}^{N_{\mathrm{b}}-1}{\sum_{u=0}^{N_{\mathrm{u}}-1}{p_u\sigma _{\mathrm{u},0,u}^{2}\frac{\beta _{\mathrm{u},u}K_{\mathrm{u},u}}{K_{\mathrm{u},u}+1}\varPhi _{N_{\mathrm{r}}}\left( u \right) \sum_{r=0}^{N_{\mathrm{r}}-1}{a_{N_{\mathrm{r}},r}^{\ast}\left( \phi _{\mathrm{r},u}^{aa},\phi _{\mathrm{r},u}^{ea} \right) a_{N_{\mathrm{r}},r}\left( \phi _{\mathrm{r},n}^{aa},\phi _{\mathrm{r},n}^{ea} \right) \mathbb{E} \left\{ \sum_{s=0}^{N_{\mathrm{c}}-1}{\left( \tilde{g}_{\mathrm{b},s,s}^{\left( b,r \right)} \right) ^{\ast}\tilde{g}_{\mathrm{b},t,t}^{\left( b,r \right)}} \right\}}}}
\end{equation*}
\begin{equation*}
\hspace{-1.2cm}
  \overset{\left( b \right)}{=}\frac{\beta _{\mathrm{u},n}\beta _{\mathrm{b}}^{2}\sigma _{\mathrm{u},0,n}^{2}\sigma _{\mathrm{b},0}^{4}K_{\mathrm{u},n}K_{\mathrm{b}}}{\left( K_{\mathrm{u},n}+1 \right) \left( K_{\mathrm{b}}+1 \right) ^2}N_{\mathrm{b}}\left( \varPhi _{N_{\mathrm{r}}}\left( n \right) \right) ^{\ast}\sum_{u=0}^{N_{\mathrm{u}}-1}{p_u\sigma _{\mathrm{u},0,u}^{2}\frac{\beta _{\mathrm{u},u}K_{\mathrm{u},u}}{K_{\mathrm{u},u}+1}\varPhi _{N_{\mathrm{r}}}\left( u \right) \sum_{r=0}^{N_{\mathrm{r}}-1}{\left( \bar{h}_{\mathrm{u}}^{\left( r,u \right)} \right) ^{\ast}\bar{h}_{\mathrm{u}}^{\left( r,n \right)}}}
\end{equation*}
\begin{equation}\label{E_1112_eq4}
\hspace{-0.000001cm}
  \overset{\left( c \right)}{=}\frac{\beta _{\mathrm{u},n}\beta _{\mathrm{b}}^{2}\sigma _{\mathrm{u},0,n}^{2}\sigma _{\mathrm{b},0}^{4}K_{\mathrm{u},n}K_{\mathrm{b}}}{\left( K_{\mathrm{u},n}+1 \right) \left( K_{\mathrm{b}}+1 \right) ^2}N_{\mathrm{b}}\left( \varPhi _{N_{\mathrm{r}}}\left( n \right) \right) ^{\ast}\sum_{u=0}^{N_{\mathrm{u}}-1}{p_u\sigma _{\mathrm{u},0,u}^{2}\frac{\beta _{\mathrm{u},u}K_{\mathrm{u},u}}{K_{\mathrm{u},u}+1}\varPhi _{N_{\mathrm{r}}}\left( u \right) \left( \bar{\mathbf{h}}_{\mathrm{u}}^{\left( \cdot ,u \right)} \right) ^H\bar{\mathbf{h}}_{\mathrm{u}}^{\left( \cdot ,n \right)}}.
\end{equation}
Step $\left( a \right)$ is obtained by substituting \eqref{R-U_FD_channel_ij_tt} and \eqref{B-R_FD_channel_ij_tt} into it and then removing the zero-value terms.
Step $\left( b \right)$ is because of \eqref{lemma_expectation_eq1} in Lemma~\ref{lemma_expectation} and \eqref{lemma_sum_eq1} in Lemma~\ref{lemma_sum}.
And step $\left( c \right)$ is based on the definition in \eqref{h_u_LoS_vector}.
Similarly, the second expectation in \eqref{E_1112_eq2} is obtained as
\begin{equation*}
\hspace{-10.8cm}
  \mathbb{E} \left\{ \tilde{\mathbf{g}}_{\mathrm{u},nt}^{H}\mathbf{\Phi }^H\bar{\mathbf{G}}_{\mathrm{b}}^{H}\mathbf{\Upsilon }_{1112}\tilde{\mathbf{G}}_{\mathrm{b}}\mathbf{\Phi }\tilde{\mathbf{g}}_{\mathrm{u},nt} \right\}
\end{equation*}
\begin{equation*}
\hspace{-4.8cm}
  =\mathbb{E} \left\{ \sum_{b=0}^{N_{\mathrm{b}}-1}{\gamma _{1112}\left( b \right) \sum_{r_1=0}^{N_{\mathrm{r}}-1}{\sum_{r_2=0}^{N_{\mathrm{r}}-1}{\left( e^{j\varphi _{r_1}}\tilde{g}_{\mathrm{u},t,t}^{\left( r_1,n \right)}\bar{g}_{\mathrm{b}}^{\left( b,r_1 \right)} \right) ^{\ast}e^{j\varphi _{r_2}}\tilde{g}_{\mathrm{u},t,t}^{\left( r_2,n \right)}\tilde{g}_{\mathrm{b},t,t}^{\left( b,r_2 \right)}}}} \right\}
\end{equation*}
\begin{equation*}
  \overset{\left( a \right)}{=}\frac{\beta _{\mathrm{u},n}\beta _{\mathrm{b}}\sigma _{\mathrm{b},0}^{2}\varsigma _{\mathrm{u},n}K_{\mathrm{b}}}{N_{\mathrm{c}}\left( K_{\mathrm{b}}+1 \right) \left( K_{\mathrm{u},n}+1 \right)}\sum_{b=0}^{N_{\mathrm{b}}-1}{\sum_{u=0}^{N_{\mathrm{u}}-1}{p_u\frac{\beta _{\mathrm{u},u}K_{\mathrm{u},u}}{K_{\mathrm{u},u}+1}\sigma _{\mathrm{u},0,u}^{2}\varPhi _{N_{\mathrm{r}}}\left( u \right) \sum_{r=0}^{N_{\mathrm{r}}-1}{\left( \varPhi \left( u,r \right) \right) ^{\ast}\mathbb{E} \left\{ \sum_{s=0}^{N_{\mathrm{c}}-1}{\tilde{g}_{\mathrm{b},t,t}^{\left( b,r \right)}\left( \tilde{g}_{\mathrm{b},s,s}^{\left( b,r \right)} \right) ^{\ast}} \right\}}}}
\end{equation*}
\begin{equation}\label{E_1112_eq5}
\hspace{-6.6cm}
  =\frac{\beta _{\mathrm{u},n}\beta _{\mathrm{b}}^{2}\sigma _{\mathrm{b},0}^{4}\varsigma _{\mathrm{u},n}K_{\mathrm{b}}N_{\mathrm{b}}}{\left( K_{\mathrm{u},n}+1 \right) \left( K_{\mathrm{b}}+1 \right) ^2}\sum_{u=0}^{N_{\mathrm{u}}-1}{p_u\frac{\beta _{\mathrm{u},u}K_{\mathrm{u},u}}{K_{\mathrm{u},u}+1}\sigma _{\mathrm{u},0,u}^{2}\left| \varPhi _{N_{\mathrm{r}}}\left( u \right) \right|^2},
\end{equation}
where step $\left( a \right)$ is further based on \eqref{R-U_FD_channel_ij_tt_NLoS}.

Substituting \eqref{E_1112_eq4} and \eqref{E_1112_eq5} into \eqref{E_1112_eq2},
we arrive at
\begin{equation*}
\hspace{-8.9cm}
  \mathbb{E} \left\{ \mathbf{g}_{\mathrm{u},nt}^{H}\mathbf{\Phi }^H\mathbf{G}_{\mathrm{b}}^{H}\mathbf{\Upsilon }_{1112}\mathbf{G}_{\mathrm{b}}\mathbf{\Phi g}_{\mathrm{u},nt} \right\}
\end{equation*}
\begin{equation*}
  =\frac{\beta _{\mathrm{u},n}\beta _{\mathrm{b}}^{2}\sigma _{\mathrm{u},0,n}^{2}\sigma _{\mathrm{b},0}^{4}K_{\mathrm{u},n}K_{\mathrm{b}}}{\left( K_{\mathrm{u},n}+1 \right) \left( K_{\mathrm{b}}+1 \right) ^2}N_{\mathrm{b}}\left( \varPhi _{N_{\mathrm{r}}}\left( n \right) \right) ^{\ast}\sum_{u=0}^{N_{\mathrm{u}}-1}{p_u\sigma _{\mathrm{u},0,u}^{2}\frac{\beta _{\mathrm{u},u}K_{\mathrm{u},u}}{K_{\mathrm{u},u}+1}\varPhi _{N_{\mathrm{r}}}\left( u \right) \left( \bar{\mathbf{h}}_{\mathrm{u}}^{\left( \cdot ,u \right)} \right) ^H\bar{\mathbf{h}}_{\mathrm{u}}^{\left( \cdot ,n \right)}}
\end{equation*}
\begin{equation}\label{E_1112_eq6}
\hspace{-4.7cm}
  +\frac{\beta _{\mathrm{u},n}\beta _{\mathrm{b}}^{2}\sigma _{\mathrm{b},0}^{4}\varsigma _{\mathrm{u},n}K_{\mathrm{b}}N_{\mathrm{b}}}{\left( K_{\mathrm{u},n}+1 \right) \left( K_{\mathrm{b}}+1 \right) ^2}\sum_{u=0}^{N_{\mathrm{u}}-1}{p_u\frac{\beta _{\mathrm{u},u}K_{\mathrm{u},u}}{K_{\mathrm{u},u}+1}\sigma _{\mathrm{u},0,u}^{2}\left| \varPhi _{N_{\mathrm{r}}}\left( u \right) \right|^2} .
\end{equation}

\subsubsection{\textbf{ The $\mathbf{\Upsilon }_{1121}$ and $\mathbf{\Upsilon }_{1211}$-Related Expectations}} \label{E_1121}

Since
\begin{equation}\label{E_1121_eq1}
  \mathbb{E} \left\{ \mathbf{g}_{\mathrm{u},nt}^{H}\mathbf{\Phi }^H\mathbf{G}_{\mathrm{b}}^{H}\mathbf{\Upsilon }_{1211}\mathbf{G}_{\mathrm{b}}\mathbf{\Phi g}_{\mathrm{u},nt} \right\} =\left( \mathbb{E} \left\{ \mathbf{g}_{\mathrm{u},nt}^{H}\mathbf{\Phi }^H\mathbf{G}_{\mathrm{b}}^{H}\mathbf{\Upsilon }_{1121}\mathbf{G}_{\mathrm{b}}\mathbf{\Phi g}_{\mathrm{u},nt} \right\} \right) ^H ,
\end{equation}
we focus on the expectation $\mathbb{E} \left\{ \mathbf{g}_{\mathrm{u},nt}^{H}\mathbf{\Phi }^H\mathbf{G}_{\mathrm{b}}^{H}\mathbf{\Upsilon }_{1121}\mathbf{G}_{\mathrm{b}}\mathbf{\Phi g}_{\mathrm{u},nt} \right\}$.
By removing the zero-value terms, it can be expanded as
\begin{equation*}
\hspace{-7.7cm}
  \mathbb{E} \left\{ \mathbf{g}_{\mathrm{u},nt}^{H}\mathbf{\Phi }^H\mathbf{G}_{\mathrm{b}}^{H}\mathbf{\Upsilon }_{1121}\mathbf{G}_{\mathrm{b}}\mathbf{\Phi g}_{\mathrm{u},nt} \right\}
\end{equation*}
\begin{equation*}
  =\mathbb{E} \left\{ \left( \bar{\mathbf{g}}_{\mathrm{u},nt}^{H}+\tilde{\mathbf{g}}_{\mathrm{u},nt}^{H} \right) \mathbf{\Phi }^H\left( \bar{\mathbf{G}}_{\mathrm{b}}^{H}+\tilde{\mathbf{G}}_{\mathrm{b}}^{H} \right) \mathbf{\Upsilon }_{1121}\left( \bar{\mathbf{G}}_{\mathrm{b}}+\tilde{\mathbf{G}}_{\mathrm{b}} \right) \mathbf{\Phi }\left( \bar{\mathbf{g}}_{\mathrm{u},nt}+\tilde{\mathbf{g}}_{\mathrm{u},nt} \right) \right\}.
\end{equation*}
\begin{equation}\label{E_1121_eq2}
\hspace{-0.9cm}
  =\mathbb{E} \left\{ \bar{\mathbf{g}}_{\mathrm{u},nt}^{H}\mathbf{\Phi }^H\bar{\mathbf{G}}_{\mathrm{b}}^{H}\mathbf{\Upsilon }_{1121}\bar{\mathbf{G}}_{\mathrm{b}}\mathbf{\Phi }\tilde{\mathbf{g}}_{\mathrm{u},nt} \right\} +\mathbb{E} \left\{ \bar{\mathbf{g}}_{\mathrm{u},nt}^{H}\mathbf{\Phi }^H\tilde{\mathbf{G}}_{\mathrm{b}}^{H}\mathbf{\Upsilon }_{1121}\tilde{\mathbf{G}}_{\mathrm{b}}\mathbf{\Phi }\tilde{\mathbf{g}}_{\mathrm{u},nt} \right\}.
\end{equation}
Besides, from \eqref{gamma_b}, \eqref{gamma_matrix} and \eqref{gamma_matrix_expand}, the $(bN_{\mathrm{c}}+t)$-th element of the diagonal matrix $\mathbf{\Upsilon }_{1121}$ is expressed as
\begin{equation}\label{E_1121_eq3}
  \left[ \mathbf{\Upsilon }_{1121} \right] _{bt,bt}=
  \frac{1}{N_{\mathrm{c}}}\sum_{s=0}^{N_{\mathrm{c}}-1}{\sum_{u=0}^{N_{\mathrm{u}}-1}{p_u\left( \sum_{r_1=0}^{N_{\mathrm{r}}-1}{e^{j\varphi _{r_1}}\bar{g}_{\mathrm{b}}^{\left( b,r_1 \right)}\bar{g}_{\mathrm{u}}^{\left( r_1,u \right)}} \right) \left( \sum_{r_2=0}^{N_{\mathrm{r}}-1}{e^{j\varphi _{r_2}}\bar{g}_{\mathrm{b}}^{\left( b,r_2 \right)}\tilde{g}_{\mathrm{u},s,s}^{\left( r_2,u \right)}} \right) ^{\ast}}}
  \triangleq \gamma _{1121}\left( b \right).
\end{equation}

Similar to \eqref{E_1112_eq4} and \eqref{E_1112_eq5},
from \eqref{R-U_FD_channel_ij_tt}, \eqref{B-R_FD_channel_ij_tt}, \eqref{B-R_FD_channel_ij_tt_NLoS}, \eqref{lemma_expectation_eq2} in Lemma~\ref{lemma_expectation}, and \eqref{lemma_sum_eq1} in Lemma~\ref{lemma_sum},
the two expectations in \eqref{E_1121_eq2} can be calculated as
\begin{equation}\label{E_1121_eq4}
  \mathbb{E} \left\{ \bar{\mathbf{g}}_{\mathrm{u},nt}^{H}\mathbf{\Phi }^H\bar{\mathbf{G}}_{\mathrm{b}}^{H}\mathbf{\Upsilon }_{1121}\bar{\mathbf{G}}_{\mathrm{b}}\mathbf{\Phi }\tilde{\mathbf{g}}_{\mathrm{u},nt} \right\}
  = p_n \frac{\beta _{\mathrm{u},n}^{2}\beta _{\mathrm{b}}^{2}\sigma _{\mathrm{u},0,n}^{4}\sigma _{\mathrm{b},0}^{4}K_{\mathrm{u},n}K_{\mathrm{b}}^{2}}{\left( K_{\mathrm{u},n}+1 \right) ^2\left( K_{\mathrm{b}}+1 \right) ^2}N_{\mathrm{b}}N_{\mathrm{r}}\left| \varPhi _{N_{\mathrm{r}}}\left( n \right) \right|^2,
\end{equation}
\begin{equation}\label{E_1121_eq5}
  \mathbb{E} \left\{ \bar{\mathbf{g}}_{\mathrm{u},nt}^{H}\mathbf{\Phi }^H\tilde{\mathbf{G}}_{\mathrm{b}}^{H}\mathbf{\Upsilon }_{1121}\tilde{\mathbf{G}}_{\mathrm{b}}\mathbf{\Phi }\tilde{\mathbf{g}}_{\mathrm{u},nt} \right\}
  = p_n \frac{\beta _{\mathrm{u},n}^{2}\beta _{\mathrm{b}}^{2}\sigma _{\mathrm{u},0,n}^{4}\sigma _{\mathrm{b},0}^{2}\varsigma _{\mathrm{b}}K_{\mathrm{u},n}K_{\mathrm{b}}}{\left( K_{\mathrm{u},n}+1 \right) ^2\left( K_{\mathrm{b}}+1 \right) ^2}N_{\mathrm{b}}\left| \varPhi _{N_{\mathrm{r}}}\left( n \right) \right|^2.
\end{equation}
Therefore, we arrive at
\begin{equation*}
\hspace{-9.9cm}
  \mathbb{E} \left\{ \mathbf{g}_{\mathrm{u},nt}^{H}\mathbf{\Phi }^H\mathbf{G}_{\mathrm{b}}^{H}\mathbf{\Upsilon }_{1121}\mathbf{G}_{\mathrm{b}}\mathbf{\Phi g}_{\mathrm{u},nt} \right\}
\end{equation*}
\begin{equation}\label{E_1121_eq6}
  = p_n \frac{\beta _{\mathrm{u},n}^{2}\beta _{\mathrm{b}}^{2}\sigma _{\mathrm{u},0,n}^{4}\sigma _{\mathrm{b},0}^{4}K_{\mathrm{u},n}K_{\mathrm{b}}^{2}}{\left( K_{\mathrm{u},n}+1 \right) ^2\left( K_{\mathrm{b}}+1 \right) ^2}N_{\mathrm{b}}N_{\mathrm{r}}\left| \varPhi _{N_{\mathrm{r}}}\left( n \right) \right|^2
  + p_n \frac{\beta _{\mathrm{u},n}^{2}\beta _{\mathrm{b}}^{2}\sigma _{\mathrm{u},0,n}^{4}\sigma _{\mathrm{b},0}^{2}\varsigma _{\mathrm{b}}K_{\mathrm{u},n}K_{\mathrm{b}}}{\left( K_{\mathrm{u},n}+1 \right) ^2\left( K_{\mathrm{b}}+1 \right) ^2}N_{\mathrm{b}}\left| \varPhi _{N_{\mathrm{r}}}\left( n \right) \right|^2 .
\end{equation}

\subsubsection{\textbf{ The $\mathbf{\Upsilon }_{1122}$ and $\mathbf{\Upsilon }_{2211}$-Related Expectations}} \label{E_1122}

Because of
\begin{equation}\label{E_1122_eq1}
  \mathbb{E} \left\{ \mathbf{g}_{\mathrm{u},nt}^{H}\mathbf{\Phi }^H\mathbf{G}_{\mathrm{b}}^{H}\mathbf{\Upsilon }_{2211}\mathbf{G}_{\mathrm{b}}\mathbf{\Phi g}_{\mathrm{u},nt} \right\} =\left( \mathbb{E} \left\{ \mathbf{g}_{\mathrm{u},nt}^{H}\mathbf{\Phi }^H\mathbf{G}_{\mathrm{b}}^{H}\mathbf{\Upsilon }_{1122}\mathbf{G}_{\mathrm{b}}\mathbf{\Phi g}_{\mathrm{u},nt} \right\} \right) ^H ,
\end{equation}
we focus on the expectation $\mathbb{E} \left\{ \mathbf{g}_{\mathrm{u},nt}^{H}\mathbf{\Phi }^H\mathbf{G}_{\mathrm{b}}^{H}\mathbf{\Upsilon }_{1122}\mathbf{G}_{\mathrm{b}}\mathbf{\Phi g}_{\mathrm{u},nt} \right\}$.
Similarly, we remove the zero-value terms and transform the expectation into
\begin{equation}\label{E_1122_eq2}
  \mathbb{E} \left\{ \mathbf{g}_{\mathrm{u},nt}^{H}\mathbf{\Phi }^H\mathbf{G}_{\mathrm{b}}^{H}\mathbf{\Upsilon }_{1122}\mathbf{G}_{\mathrm{b}}\mathbf{\Phi g}_{\mathrm{u},nt} \right\}
  =\mathbb{E} \left\{ \bar{\mathbf{g}}_{\mathrm{u},nt}^{H}\mathbf{\Phi }^H\bar{\mathbf{G}}_{\mathrm{b}}^{H}\mathbf{\Upsilon }_{1122}\tilde{\mathbf{G}}_{\mathrm{b}}\mathbf{\Phi }\tilde{\mathbf{g}}_{\mathrm{u},nt} \right\}.
\end{equation}
Besides, the $(bN_{\mathrm{c}}+t)$-th element of the diagonal matrix $\mathbf{\Upsilon }_{1122}$ is expressed as
\begin{equation}\label{E_1122_eq3}
  \left[ \mathbf{\Upsilon }_{1122} \right] _{bt,bt}=
  \frac{1}{N_{\mathrm{c}}}\sum_{s=0}^{N_{\mathrm{c}}-1}{\sum_{u=0}^{N_{\mathrm{u}}-1}{p_u\left( \sum_{r_1=0}^{N_{\mathrm{r}}-1}{e^{j\varphi _{r_1}}\bar{g}_{\mathrm{b}}^{\left( b,r_1 \right)}\bar{g}_{\mathrm{u}}^{\left( r_1,u \right)}} \right) \left( \sum_{r_2=0}^{N_{\mathrm{r}}-1}{e^{j\varphi _{r_2}}\tilde{g}_{\mathrm{b},s,s}^{\left( b,r_2 \right)}\tilde{g}_{\mathrm{u},s,s}^{\left( r_2,u \right)}} \right) ^{\ast}}}
  \triangleq \gamma _{1122}\left( b \right).
\end{equation}

Therefore, the expectation $\mathbb{E} \left\{ \mathbf{g}_{\mathrm{u},nt}^{H}\mathbf{\Phi }^H\mathbf{G}_{\mathrm{b}}^{H}\mathbf{\Upsilon }_{1122}\mathbf{G}_{\mathrm{b}}\mathbf{\Phi g}_{\mathrm{u},nt} \right\}$ can be calculated as
\begin{equation*}
\hspace{-3.3cm}
  \mathbb{E} \left\{ \mathbf{g}_{\mathrm{u},nt}^{H}\mathbf{\Phi }^H\mathbf{G}_{\mathrm{b}}^{H}\mathbf{\Upsilon }_{1122}\mathbf{G}_{\mathrm{b}}\mathbf{\Phi g}_{\mathrm{u},nt} \right\} =\mathbb{E} \left\{ \bar{\mathbf{g}}_{\mathrm{u},nt}^{H}\mathbf{\Phi }^H\bar{\mathbf{G}}_{\mathrm{b}}^{H}\mathbf{\Upsilon }_{1122}\tilde{\mathbf{G}}_{\mathrm{b}}\mathbf{\Phi }\tilde{\mathbf{g}}_{\mathrm{u},nt} \right\}
\end{equation*}
\begin{equation*}
\hspace{-3.3cm}
  =\mathbb{E} \left\{ \sum_{b=0}^{N_{\mathrm{b}}-1}{\gamma _{1122}\left( b \right) \sum_{r_1=0}^{N_{\mathrm{r}}-1}{\sum_{r_2=0}^{N_{\mathrm{r}}-1}{\left( e^{j\varphi _{r_1}}\bar{g}_{\mathrm{u}}^{\left( r_1,n \right)}\bar{g}_{\mathrm{b}}^{\left( b,r_1 \right)} \right) ^{\ast}e^{j\varphi _{r_2}}\tilde{g}_{\mathrm{u},t,t}^{\left( r_2,n \right)}\tilde{g}_{\mathrm{b},t,t}^{\left( b,r_2 \right)}}}} \right\}
\end{equation*}
\begin{equation*}
  \overset{\left( a \right)}{=} p_n \frac{\beta _{\mathrm{u},n}\beta _{\mathrm{b}}\sigma _{\mathrm{u},0,n}^{2}\sigma _{\mathrm{b},0}^{2}K_{\mathrm{u},n}K_{\mathrm{b}}}{N_{\mathrm{c}}\left( K_{\mathrm{u},n}+1 \right) \left( K_{\mathrm{b}}+1 \right)}\left| \varPhi _{N_{\mathrm{r}}}\left( n \right) \right|^2\sum_{b=0}^{N_{\mathrm{b}}-1}{\sum_{r=0}^{N_{\mathrm{r}}-1}{\sum_{s=0}^{N_{\mathrm{c}}-1}{\mathbb{E} \left\{ \left( \tilde{g}_{\mathrm{u},s,s}^{\left( r,n \right)} \right) ^{\ast}\tilde{g}_{\mathrm{u},t,t}^{\left( r,n \right)}\left( \tilde{g}_{\mathrm{b},s,s}^{\left( b,r \right)} \right) ^{\ast}\tilde{g}_{\mathrm{b},t,t}^{\left( b,r \right)} \right\}}}}
\end{equation*}
\begin{equation}\label{E_1122_eq4}
\hspace{-7.1cm}
  \overset{\left( b \right)}{=} p_n \frac{\beta _{\mathrm{u},n}^{2}\beta _{\mathrm{b}}^{2}\sigma _{\mathrm{u},0,n}^{4}\sigma _{\mathrm{b},0}^{4}K_{\mathrm{u},n}K_{\mathrm{b}}}{\left( K_{\mathrm{u},n}+1 \right) ^2\left( K_{\mathrm{b}}+1 \right) ^2}N_{\mathrm{b}}N_{\mathrm{r}}\left| \varPhi _{N_{\mathrm{r}}}\left( n \right) \right|^2,
\end{equation}
where step $\left( a \right)$ is obtained by substituting \eqref{R-U_FD_channel_ij_tt} and \eqref{B-R_FD_channel_ij_tt} into it and then removing the zero-valued terms.
Step $\left( b \right)$ is based on Lemma~\ref{lemma_sum} and Lemma~\ref{lemma_expectation}.

\subsubsection{\textbf{ The $\mathbf{\Upsilon }_{1212}$ and $\mathbf{\Upsilon }_{2121}$-Related Expectations}} \label{E_1212}

Since
\begin{equation}\label{E_1212_eq1}
  \mathbb{E} \left\{ \mathbf{g}_{\mathrm{u},nt}^{H}\mathbf{\Phi }^H\mathbf{G}_{\mathrm{b}}^{H}\mathbf{\Upsilon }_{2121}\mathbf{G}_{\mathrm{b}}\mathbf{\Phi g}_{\mathrm{u},nt} \right\} =\left( \mathbb{E} \left\{ \mathbf{g}_{\mathrm{u},nt}^{H}\mathbf{\Phi }^H\mathbf{G}_{\mathrm{b}}^{H}\mathbf{\Upsilon }_{1212}\mathbf{G}_{\mathrm{b}}\mathbf{\Phi g}_{\mathrm{u},nt} \right\} \right) ^H ,
\end{equation}
we focus on the expectation $\mathbb{E} \left\{ \mathbf{g}_{\mathrm{u},nt}^{H}\mathbf{\Phi }^H\mathbf{G}_{\mathrm{b}}^{H}\mathbf{\Upsilon }_{1212}\mathbf{G}_{\mathrm{b}}\mathbf{\Phi g}_{\mathrm{u},nt} \right\}$.
By removing the zero-value terms, we transform it into
\begin{equation}\label{E_1212_eq2}
  \mathbb{E} \left\{ \mathbf{g}_{\mathrm{u},nt}^{H}\mathbf{\Phi }^H\mathbf{G}_{\mathrm{b}}^{H}\mathbf{\Upsilon }_{1212}\mathbf{G}_{\mathrm{b}}\mathbf{\Phi g}_{\mathrm{u},nt} \right\}
  =\mathbb{E} \left\{ \tilde{\mathbf{g}}_{\mathrm{u},nt}^{H}\mathbf{\Phi }^H\bar{\mathbf{G}}_{\mathrm{b}}^{H}\mathbf{\Upsilon }_{1212}\tilde{\mathbf{G}}_{\mathrm{b}}\mathbf{\Phi }\bar{\mathbf{g}}_{\mathrm{u},nt} \right\}.
\end{equation}
Besides, the $(bN_{\mathrm{c}}+t)$-th element of the diagonal matrix $\mathbf{\Upsilon }_{1212}$ is expressed as
\begin{equation}\label{E_1212_eq3}
  \left[ \mathbf{\Upsilon }_{1212} \right] _{bt,bt}=\frac{1}{N_{\mathrm{c}}}\sum_{s=0}^{N_{\mathrm{c}}-1}{\sum_{u=0}^{N_{\mathrm{u}}-1}{ p_u \left( \sum_{r_1=0}^{N_{\mathrm{r}}-1}{e^{j\varphi _{r_1}}\bar{g}_{\mathrm{b}}^{\left( b,r_1 \right)}\tilde{g}_{\mathrm{u},s,s}^{\left( r_1,u \right)}} \right) \left( \sum_{r_2=0}^{N_{\mathrm{r}}-1}{e^{j\varphi _{r_2}}\tilde{g}_{\mathrm{b},s,s}^{\left( b,r_2 \right)}\bar{g}_{\mathrm{u}}^{\left( r_2,u \right)}} \right) ^{\ast}\triangleq}}\gamma _{1212}\left( b \right).
\end{equation}

Therefore, the expectation $\mathbb{E} \left\{ \mathbf{g}_{\mathrm{u},nt}^{H}\mathbf{\Phi }^H\mathbf{G}_{\mathrm{b}}^{H}\mathbf{\Upsilon }_{1212}\mathbf{G}_{\mathrm{b}}\mathbf{\Phi g}_{\mathrm{u},nt} \right\}$ can be calculated as
\begin{equation*}
\hspace{-4.5cm}
  \mathbb{E} \left\{ \mathbf{g}_{\mathrm{u},nt}^{H}\mathbf{\Phi }^H\mathbf{G}_{\mathrm{b}}^{H}\mathbf{\Upsilon }_{1212}\mathbf{G}_{\mathrm{b}}\mathbf{\Phi g}_{\mathrm{u},nt} \right\} =\mathbb{E} \left\{ \tilde{\mathbf{g}}_{\mathrm{u},nt}^{H}\mathbf{\Phi }^H\bar{\mathbf{G}}_{\mathrm{b}}^{H}\mathbf{\Upsilon }_{1212}\tilde{\mathbf{G}}_{\mathrm{b}}\mathbf{\Phi }\bar{\mathbf{g}}_{\mathrm{u},nt} \right\}
\end{equation*}
\begin{equation*}
\hspace{-4.3cm}
  =\mathbb{E} \left\{ \sum_{b=0}^{N_{\mathrm{b}}-1}{\gamma _{1212}\left( b \right) \sum_{r_1=0}^{N_{\mathrm{r}}-1}{\sum_{r_2=0}^{N_{\mathrm{r}}-1}{\left( e^{j\varphi _{r_1}}\tilde{g}_{\mathrm{u},t,t}^{\left( r_1,n \right)}\bar{g}_{\mathrm{b}}^{\left( b,r_1 \right)} \right) ^{\ast}e^{j\varphi _{r_2}}\bar{g}_{\mathrm{u}}^{\left( r_2,n \right)}\tilde{g}_{\mathrm{b},t,t}^{\left( b,r_2 \right)}}}} \right\}
\end{equation*}
\begin{equation*}
\hspace{-1.3cm}
  \overset{\left( a \right)}{=} p_n \frac{\beta _{\mathrm{u},n}\beta _{\mathrm{b}}\sigma _{\mathrm{u},0,n}^{2}\sigma _{\mathrm{b},0}^{2}K_{\mathrm{u},n}K_{\mathrm{b}}}{N_{\mathrm{c}}\left( K_{\mathrm{u},n}+1 \right) \left( K_{\mathrm{b}}+1 \right)}\sum_{b=0}^{N_{\mathrm{b}}-1}{\sum_{r_1=0}^{N_{\mathrm{r}}-1}{\sum_{r_2=0}^{N_{\mathrm{r}}-1}{\sum_{s=0}^{N_{\mathrm{c}}-1}{\mathbb{E} \left\{ \left( \tilde{g}_{\mathrm{b},s,s}^{\left( b,r_2 \right)} \right) ^{\ast}\tilde{g}_{\mathrm{b},t,t}^{\left( b,r_2 \right)}\tilde{g}_{\mathrm{u},s,s}^{\left( r_1,n \right)}\left( \tilde{g}_{\mathrm{u},t,t}^{\left( r_1,n \right)} \right) ^{\ast} \right\}}}}}
\end{equation*}
\begin{equation*}
\hspace{-9.6cm}
  \overset{\left( b \right)}{=} p_n \frac{\beta _{\mathrm{u},n}\beta _{\mathrm{b}}\sigma _{\mathrm{u},0,n}^{2}\sigma _{\mathrm{b},0}^{2}K_{\mathrm{u},n}K_{\mathrm{b}}}{N_{\mathrm{c}}\left( K_{\mathrm{u},n}+1 \right) \left( K_{\mathrm{b}}+1 \right)}N_{\mathrm{b}}N_{\mathrm{r}}^{2}
  \times
\end{equation*}
\begin{equation*}
\hspace{-2.5cm}
  \sum_{s=0}^{N_{\mathrm{c}}-1}{\left( \frac{\beta _{\mathrm{u},n}\beta _{\mathrm{b}}\sigma _{\mathrm{u},0,n}^{2}\sigma _{\mathrm{b},0}^{2}}{\left( K_{\mathrm{u},n}+1 \right) \left( K_{\mathrm{b}}+1 \right)}+\beta _{\mathrm{u},n}\beta _{\mathrm{b}}\sum_{k_1=1}^{L_{\mathrm{u}}-1}{\sum_{k_2=1}^{L_{\mathrm{b}}-1}{e^{-j\frac{2\pi}{N_{\mathrm{c}}}\left( k_1-k_2 \right) \left( t-s \right)}\sigma _{\mathrm{u},k_1,n}^{2}\sigma _{\mathrm{b},k_2}^{2}}} \right)}
\end{equation*}
\begin{equation}\label{E_1212_eq4}
  \overset{\left( c \right)}{=} p_n \frac{\beta _{\mathrm{u},n}^{2}\beta _{\mathrm{b}}^{2}\sigma _{\mathrm{u},0,n}^{4}\sigma _{\mathrm{b},0}^{4}K_{\mathrm{u},n}K_{\mathrm{b}}}{\left( K_{\mathrm{u},n}+1 \right) ^2\left( K_{\mathrm{b}}+1 \right) ^2}N_{\mathrm{b}}N_{\mathrm{r}}^{2}
  + p_n \frac{\beta _{\mathrm{u},n}^{2}\beta _{\mathrm{b}}^{2}\sigma _{\mathrm{u},0,n}^{2}\sigma _{\mathrm{b},0}^{2}K_{\mathrm{u},n}K_{\mathrm{b}}}{\left( K_{\mathrm{u},n}+1 \right) \left( K_{\mathrm{b}}+1 \right)}N_{\mathrm{b}}N_{\mathrm{r}}^{2}\sum_{k=1}^{L_{1}^{\min}-1}{\sigma _{\mathrm{u},k,n}^{2}\sigma _{\mathrm{b},k}^{2}},
\end{equation}
where $L_{1}^{\min}=\min \left\{ L_{\mathrm{u}},L_{\mathrm{b}} \right\}$.
Step $\left( a \right)$ is obtained by substituting \eqref{R-U_FD_channel_ij_tt} and \eqref{B-R_FD_channel_ij_tt} into it and then removing the zero-valued terms.
Step $\left( b \right)$ is based on Lemma~\ref{lemma_expectation},
and step $\left( c \right)$ is based on Lemma~\ref{lemma_sum}.

\subsubsection{\textbf{ The $\mathbf{\Upsilon }_{1222}$ and $\mathbf{\Upsilon }_{2111}$-Related Expectations}} \label{E_1222}

Since
\begin{equation}\label{E_1222_eq1}
  \mathbb{E} \left\{ \mathbf{g}_{\mathrm{u},nt}^{H}\mathbf{\Phi }^H\mathbf{G}_{\mathrm{b}}^{H}\mathbf{\Upsilon }_{2221}\mathbf{G}_{\mathrm{b}}\mathbf{\Phi g}_{\mathrm{u},nt} \right\} =\left( \mathbb{E} \left\{ \mathbf{g}_{\mathrm{u},nt}^{H}\mathbf{\Phi }^H\mathbf{G}_{\mathrm{b}}^{H}\mathbf{\Upsilon }_{1222}\mathbf{G}_{\mathrm{b}}\mathbf{\Phi g}_{\mathrm{u},nt} \right\} \right) ^H ,
\end{equation}
we focus on the expectation $\mathbb{E} \left\{ \mathbf{g}_{\mathrm{u},nt}^{H}\mathbf{\Phi }^H\mathbf{G}_{\mathrm{b}}^{H}\mathbf{\Upsilon }_{1222}\mathbf{G}_{\mathrm{b}}\mathbf{\Phi g}_{\mathrm{u},nt} \right\}$.
By removing the zero-value terms, we expand it as
\begin{equation*}
\hspace{-7.5cm}
  \mathbb{E} \left\{ \mathbf{g}_{\mathrm{u},nt}^{H}\mathbf{\Phi }^H\mathbf{G}_{\mathrm{b}}^{H}\mathbf{\Upsilon }_{1222}\mathbf{G}_{\mathrm{b}}\mathbf{\Phi g}_{\mathrm{u},nt} \right\}
\end{equation*}
\begin{equation*}
  =\mathbb{E} \left\{ \left( \bar{\mathbf{g}}_{\mathrm{u},nt}^{H}+\tilde{\mathbf{g}}_{\mathrm{u},nt}^{H} \right) \mathbf{\Phi }^H\left( \bar{\mathbf{G}}_{\mathrm{b}}^{H}+\tilde{\mathbf{G}}_{\mathrm{b}}^{H} \right) \mathbf{\Upsilon }_{1222}\left( \bar{\mathbf{G}}_{\mathrm{b}}+\tilde{\mathbf{G}}_{\mathrm{b}} \right) \mathbf{\Phi }\left( \bar{\mathbf{g}}_{\mathrm{u},nt}+\tilde{\mathbf{g}}_{\mathrm{u},nt} \right) \right\}
\end{equation*}
\begin{equation}\label{E_1222_eq2}
\hspace{-0.6cm}
  =\mathbb{E} \left\{ \bar{\mathbf{g}}_{\mathrm{u},nt}^{H}\mathbf{\Phi }^H\bar{\mathbf{G}}_{\mathrm{b}}^{H}\mathbf{\Upsilon }_{1222}\tilde{\mathbf{G}}_{\mathrm{b}}\mathbf{\Phi }\bar{\mathbf{g}}_{\mathrm{u},nt} \right\} +\mathbb{E} \left\{ \tilde{\mathbf{g}}_{\mathrm{u},nt}^{H}\mathbf{\Phi }^H\bar{\mathbf{G}}_{\mathrm{b}}^{H}\mathbf{\Upsilon }_{1222}\tilde{\mathbf{G}}_{\mathrm{b}}\mathbf{\Phi }\tilde{\mathbf{g}}_{\mathrm{u},nt} \right\}.
\end{equation}
Besides, the $(bN_{\mathrm{c}}+t)$-th element of the diagonal matrix $\mathbf{\Upsilon }_{1222}$ is expressed as
\begin{equation}\label{E_1222_eq3}
  \left[ \mathbf{\Upsilon }_{1222} \right] _{bt,bt}=\frac{1}{N_{\mathrm{c}}}\sum_{s=0}^{N_{\mathrm{c}}-1}{\sum_{u=0}^{N_{\mathrm{u}}-1}{ p_u \left( \sum_{r_1=0}^{N_{\mathrm{r}}-1}{e^{j\varphi _{r_1}}\bar{g}_{\mathrm{b}}^{\left( b,r_1 \right)}\tilde{g}_{\mathrm{u},s,s}^{\left( r_1,u \right)}} \right) \left( \sum_{r_2=0}^{N_{\mathrm{r}}-1}{e^{j\varphi _{r_2}}\tilde{g}_{\mathrm{b},s,s}^{\left( b,r_2 \right)}\tilde{g}_{\mathrm{u},s,s}^{\left( r_2,u \right)}} \right) ^{\ast}}}\triangleq \gamma _{1222}\left( b \right) .
\end{equation}

The first expectation in \eqref{E_1222_eq2} is calculated as
\begin{equation*}
\hspace{-10.5cm}
  \mathbb{E} \left\{ \bar{\mathbf{g}}_{\mathrm{u},nt}^{H}\mathbf{\Phi }^H\bar{\mathbf{G}}_{\mathrm{b}}^{H}\mathbf{\Upsilon }_{1222}\tilde{\mathbf{G}}_{\mathrm{b}}\mathbf{\Phi }\bar{\mathbf{g}}_{\mathrm{u},nt} \right\}
\end{equation*}
\begin{equation*}
\hspace{-4.5cm}
  =\mathbb{E} \left\{ \sum_{b=0}^{N_{\mathrm{b}}-1}{\gamma _{1222}\left( b \right) \sum_{r_1=0}^{N_{\mathrm{r}}-1}{\sum_{r_2=0}^{N_{\mathrm{r}}-1}{\left( e^{j\varphi _{r_1}}\bar{g}_{\mathrm{u}}^{\left( r_1,n \right)}\bar{g}_{\mathrm{b}}^{\left( b,r_1 \right)} \right) ^{\ast}e^{j\varphi _{r_2}}\bar{g}_{\mathrm{u}}^{\left( r_2,n \right)}\tilde{g}_{\mathrm{b},t,t}^{\left( b,r_2 \right)}}}} \right\}
\end{equation*}
\begin{equation*}
  \overset{\left( a \right)}{=}c_{1222}^{1}\sum_{b=0}^{N_{\mathrm{b}}-1}{\sum_{u=0}^{N_{\mathrm{u}}-1}{p_u\sum_{r=0}^{N_{\mathrm{r}}-1}{e^{j\varphi _r}a_{N_{\mathrm{r}},r}^{\ast}\left( \phi _{\mathrm{r}}^{ad},\phi _{\mathrm{r}}^{ed} \right) a_{N_{\mathrm{r}},r}\left( \phi _{\mathrm{r},n}^{aa},\phi _{\mathrm{r},n}^{ea} \right) \sum_{s=0}^{N_{\mathrm{c}}-1}{\mathbb{E} \left\{ \left( \tilde{g}_{\mathrm{u},s,s}^{\left( r,u \right)} \right) ^{\ast}\tilde{g}_{\mathrm{u},s,s}^{\left( r,u \right)}\left( \tilde{g}_{\mathrm{b},s,s}^{\left( b,r \right)} \right) ^{\ast}\tilde{g}_{\mathrm{b},t,t}^{\left( b,r \right)} \right\}}}}}
\end{equation*}
\begin{equation*}
  \overset{\left( b \right)}{=}c_{1222}^{1}\sum_{b=0}^{N_{\mathrm{b}}-1}{\sum_{u=0}^{N_{\mathrm{u}}-1}{p_u\sum_{r=0}^{N_{\mathrm{r}}-1}{e^{j\varphi _r}a_{N_{\mathrm{r}},r}^{\ast}\left( \phi _{\mathrm{r}}^{ad},\phi _{\mathrm{r}}^{ed} \right) a_{N_{\mathrm{r}},r}\left( \phi _{\mathrm{r},n}^{aa},\phi _{\mathrm{r},n}^{ea} \right) \sum_{s=0}^{N_{\mathrm{c}}-1}{\frac{\beta _{\mathrm{u},u}\varsigma _{\mathrm{u},u}}{K_{\mathrm{u},u}+1}\mathbb{E} \left\{ \left( \tilde{g}_{\mathrm{b},s,s}^{\left( b,r \right)} \right) ^{\ast}\tilde{g}_{\mathrm{b},t,t}^{\left( b,r \right)} \right\}}}}}
\end{equation*}
\begin{equation*}
\hspace{-2.6cm}
  \overset{\left( c \right)}{=}c_{1222}^{1}\sum_{b=0}^{N_{\mathrm{b}}-1}{\sum_{u=0}^{N_{\mathrm{u}}-1}{p_u\frac{\beta _{\mathrm{u},u}\varsigma _{\mathrm{u},u}}{K_{\mathrm{u},u}+1}\varPhi _{N_{\mathrm{r}}}\left( n \right) \sum_{s=0}^{N_{\mathrm{c}}-1}{\left( \frac{\beta _{\mathrm{b}}\sigma _{\mathrm{b},0}^{2}}{K_{\mathrm{b}}+1}+\beta _{\mathrm{b}}\sum_{k=1}^{L_{\mathrm{b}}-1}{e^{-j\frac{2\pi}{N_{\mathrm{c}}}k\left( t-s \right)}\sigma _{\mathrm{b},k}^{2}} \right)}}}
\end{equation*}
\begin{equation}\label{E_1222_eq4}
\hspace{-6.3cm}
  \overset{\left( d \right)}{=}\frac{\beta _{\mathrm{u},n}\beta _{\mathrm{b}}^{2}\sigma _{\mathrm{u},0,n}^{2}\sigma _{\mathrm{b},0}^{4}K_{\mathrm{u},n}K_{\mathrm{b}}}{\left( K_{\mathrm{u},n}+1 \right) \left( K_{\mathrm{b}}+1 \right) ^2}N_{\mathrm{b}}\left| \varPhi _{N_{\mathrm{r}}}\left( n \right) \right|^2\sum_{u=0}^{N_{\mathrm{u}}-1}{p_u\frac{\beta _{\mathrm{u},u}\varsigma _{\mathrm{u},u}}{K_{\mathrm{u},u}+1}} ,
\end{equation}
where $c_{1222}^{1}=\frac{\beta _{\mathrm{u},n}\beta _{\mathrm{b}}\sigma _{\mathrm{u},0,n}^{2}\sigma _{\mathrm{b},0}^{2}K_{\mathrm{u},n}K_{\mathrm{b}}}{N_{\mathrm{c}}\left( K_{\mathrm{u},n}+1 \right) \left( K_{\mathrm{b}}+1 \right)}\left( \varPhi _{N_{\mathrm{r}}}\left( n \right) \right) ^{\ast}$.
Step $\left( a \right)$ is obtained by substituting \eqref{R-U_FD_channel_ij_tt} and \eqref{B-R_FD_channel_ij_tt} into it and then removing the zero-valued terms.
Step $\left( b \right)$ is due to \eqref{R-U_FD_channel_ij_tt_NLoS}.
Step $\left( c \right)$ is based on Lemma~\ref{lemma_expectation},
and step $\left( d \right)$ is based on Lemma~\ref{lemma_sum}.

The second expectation in \eqref{E_1222_eq2} can be derived as
\begin{equation*}
\hspace{-9.3cm}
  \mathbb{E} \left\{ \tilde{\mathbf{g}}_{\mathrm{u},nt}^{H}\mathbf{\Phi }^H\bar{\mathbf{G}}_{\mathrm{b}}^{H}\mathbf{\Upsilon }_{1222}\tilde{\mathbf{G}}_{\mathrm{b}}\mathbf{\Phi }\tilde{\mathbf{g}}_{\mathrm{u},nt} \right\}
\end{equation*}
\begin{equation*}
\hspace{-3.2cm}
  =\mathbb{E} \left\{ \sum_{b=0}^{N_{\mathrm{b}}-1}{\gamma _{1222}\left( b \right) \sum_{r_1=0}^{N_{\mathrm{r}}-1}{\sum_{r_2=0}^{N_{\mathrm{r}}-1}{\left( e^{j\varphi _{r_1}}\tilde{g}_{\mathrm{u},t,t}^{\left( r_1,n \right)}\bar{g}_{\mathrm{b}}^{\left( b,r_1 \right)} \right) ^{\ast}e^{j\varphi _{r_2}}\tilde{g}_{\mathrm{u},t,t}^{\left( r_2,n \right)}\tilde{g}_{\mathrm{b},t,t}^{\left( b,r_2 \right)}}}} \right\}
\end{equation*}
\begin{equation*}
\hspace{-7cm}
  \overset{\left( a \right)}{=}c_{1222}^{2}\sum_{b=0}^{N_{\mathrm{b}}-1}{\sum_{s=0}^{N_{\mathrm{c}}-1}{\sum_{u=0}^{N_{\mathrm{u}}-1}{p_u\mathbb{E} \left\{ \mathrm{term}_{1222}^{1}\left( b,s,u \right) \right\}}}}
\end{equation*}
\begin{equation*}
\hspace{-2.3cm}
  \overset{\left( b \right)}{=}c_{1222}^{2}\sum_{b=0}^{N_{\mathrm{b}}-1}{\sum_{s=0}^{N_{\mathrm{c}}-1}{\sum_{u=0}^{N_{\mathrm{u}}-1}{p_u\mathbb{E} \left\{ \sum_{r=0}^{N_{\mathrm{r}}-1}{\left( \tilde{g}_{\mathrm{u},t,t}^{\left( r,n \right)} \right) ^{\ast}\tilde{g}_{\mathrm{u},s,s}^{\left( r,u \right)}\left( \tilde{g}_{\mathrm{b},s,s}^{\left( b,r \right)}\tilde{g}_{\mathrm{u},s,s}^{\left( r,u \right)} \right) ^{\ast}\tilde{g}_{\mathrm{u},t,t}^{\left( r,n \right)}\tilde{g}_{\mathrm{b},t,t}^{\left( b,r \right)}} \right\}}}}
\end{equation*}
\begin{equation*}
\hspace{-0.8cm}
  +c_{1222}^{2}\sum_{b=0}^{N_{\mathrm{b}}-1}{\sum_{s=0}^{N_{\mathrm{c}}-1}{\sum_{u=0}^{N_{\mathrm{u}}-1}{p_u\mathbb{E} \left\{ \sum_{r_1=0}^{N_{\mathrm{r}}-1}{\left( \tilde{g}_{\mathrm{u},t,t}^{\left( r_1,n \right)} \right) ^{\ast}\tilde{g}_{\mathrm{u},s,s}^{\left( r_1,u \right)}}\sum_{r_2\ne r_1}^{N_{\mathrm{r}}-1}{\left( \tilde{g}_{\mathrm{b},s,s}^{\left( b,r_2 \right)}\tilde{g}_{\mathrm{u},s,s}^{\left( r_2,u \right)} \right) ^{\ast}\tilde{g}_{\mathrm{u},t,t}^{\left( r_2,n \right)}\tilde{g}_{\mathrm{b},t,t}^{\left( b,r_2 \right)}} \right\}}}}
\end{equation*}
\begin{equation*}
\hspace{-2.4cm}
  \overset{\left( c \right)}{=}c_{1222}^{2}\sum_{b=0}^{N_{\mathrm{b}}-1}{\sum_{s=0}^{N_{\mathrm{c}}-1}{\sum_{u\ne n}^{N_{\mathrm{u}}-1}{p_u\sum_{r=0}^{N_{\mathrm{r}}-1}{\mathbb{E} \left\{ \left( \tilde{g}_{\mathrm{u},t,t}^{\left( r,n \right)} \right) ^{\ast}\tilde{g}_{\mathrm{u},s,s}^{\left( r,u \right)}\left( \tilde{g}_{\mathrm{b},s,s}^{\left( b,r \right)}\tilde{g}_{\mathrm{u},s,s}^{\left( r,u \right)} \right) ^{\ast}\tilde{g}_{\mathrm{u},t,t}^{\left( r,n \right)}\tilde{g}_{\mathrm{b},t,t}^{\left( b,r \right)} \right\}}}}}
\end{equation*}
\begin{equation*}
\hspace{-3.7cm}
  +c_{1222}^{2}p_n\sum_{b=0}^{N_{\mathrm{b}}-1}{\sum_{s=0}^{N_{\mathrm{c}}-1}{\sum_{r=0}^{N_{\mathrm{r}}-1}{\mathbb{E} \left\{ \left( \tilde{g}_{\mathrm{u},t,t}^{\left( r,n \right)} \right) ^{\ast}\tilde{g}_{\mathrm{u},s,s}^{\left( r,n \right)}\left( \tilde{g}_{\mathrm{b},s,s}^{\left( b,r \right)}\tilde{g}_{\mathrm{u},s,s}^{\left( r,n \right)} \right) ^{\ast}\tilde{g}_{\mathrm{u},t,t}^{\left( r,n \right)}\tilde{g}_{\mathrm{b},t,t}^{\left( b,r \right)} \right\}}}}
\end{equation*}
\begin{equation}\label{E_1222_eq5}
  +c_{1222}^{2}p_n\sum_{b=0}^{N_{\mathrm{b}}-1}{\sum_{s=0}^{N_{\mathrm{c}}-1}{\sum_{r_1=0}^{N_{\mathrm{r}}-1}{\mathbb{E} \left\{ \left( \tilde{g}_{\mathrm{u},t,t}^{\left( r_1,n \right)} \right) ^{\ast}\tilde{g}_{\mathrm{u},s,s}^{\left( r_1,n \right)} \right\}}\sum_{r_2\ne r_1}^{N_{\mathrm{r}}-1}{\mathbb{E} \left\{ \left( \tilde{g}_{\mathrm{b},s,s}^{\left( b,r_2 \right)}\tilde{g}_{\mathrm{u},s,s}^{\left( r_2,n \right)} \right) ^{\ast}\tilde{g}_{\mathrm{u},t,t}^{\left( r_2,n \right)}\tilde{g}_{\mathrm{b},t,t}^{\left( b,r_2 \right)} \right\}}}},
\end{equation}
where $c_{1222}^{2}=\frac{\beta _{\mathrm{b}}\sigma _{\mathrm{b},0}^{2}K_{\mathrm{b}}}{N_{\mathrm{c}}\left( K_{\mathrm{b}}+1 \right)}$
and
\begin{equation*}
  \mathrm{term}_{1222}^{1}\left( b,s,u \right) =\sum_{r=0}^{N_{\mathrm{r}}-1}{\left( e^{j\varphi _r}\tilde{g}_{\mathrm{u},t,t}^{\left( r,n \right)}a_{N_{\mathrm{r}},r}^{\ast}\left( \phi _{\mathrm{r}}^{ad},\phi _{\mathrm{r}}^{ed} \right) \right) ^{\ast}}\times \sum_{r_1=0}^{N_{\mathrm{r}}-1}{e^{j\varphi _{r_1}}a_{N_{\mathrm{r}},r_1}^{\ast}\left( \phi _{\mathrm{r}}^{ad},\phi _{\mathrm{r}}^{ed} \right) \tilde{g}_{\mathrm{u},s,s}^{\left( r_1,u \right)}}
\end{equation*}
\begin{equation}
\hspace{-2cm}
  \times \sum_{r_2=0}^{N_{\mathrm{r}}-1}{\left( \tilde{g}_{\mathrm{b},s,s}^{\left( b,r_2 \right)}\tilde{g}_{\mathrm{u},s,s}^{\left( r_2,u \right)} \right) ^{\ast}\tilde{g}_{\mathrm{u},t,t}^{\left( r_2,n \right)}\tilde{g}_{\mathrm{b},t,t}^{\left( b,r_2 \right)}} .
\end{equation}
In \eqref{E_1222_eq5}, step $\left( a \right)$ is obtained by substituting \eqref{R-U_FD_channel_ij_tt} and \eqref{B-R_FD_channel_ij_tt} into it and then removing the zero-valued terms.
Step $\left( b \right)$ extracts non-zero cases from its left hand side:
1) $r=r_1=r_2$ and
2) $r_1=r \neq r_2$.
This is because when $r \neq r_1$, the left hand side of step $\left( b \right)$ becomes zero.
Step $\left( c \right)$ divides the first term of its left hand side into two parts,
depending on whether or not $ u = n $.
For the three terms in \eqref{E_1222_eq5},
using \eqref{R-U_FD_channel_ij_tt_NLoS},
\eqref{B-R_FD_channel_ij_tt_NLoS},
and the conclusions in Lemma~\ref{lemma_sum} and Lemma~\ref{lemma_expectation},
we can calculate them one by one.
Thus, the second expectation in \eqref{E_1222_eq2} is obtained as
\begin{equation*}
\hspace{-4.1cm}
  \mathbb{E} \left\{ \tilde{\mathbf{g}}_{\mathrm{u},nt}^{H}\mathbf{\Phi }^H\bar{\mathbf{G}}_{\mathrm{b}}^{H}\mathbf{\Upsilon }_{1222}\tilde{\mathbf{G}}_{\mathrm{b}}\mathbf{\Phi }\tilde{\mathbf{g}}_{\mathrm{u},nt} \right\} =p_n\frac{\beta _{\mathrm{u},n}^{2}\beta _{\mathrm{b}}^{2}\sigma _{\mathrm{b},0}^{2}K_{\mathrm{b}}}{K_{\mathrm{b}}+1}N_{\mathrm{b}}N_{\mathrm{r}}^{2}\mathrm{term}_{1222}^{2}
\end{equation*}
\begin{equation}\label{E_1222_eq6}
  +\frac{\beta _{\mathrm{b}}^{2}\sigma _{\mathrm{b},0}^{4}K_{\mathrm{b}}}{\left( K_{\mathrm{b}}+1 \right) ^2}\left( \frac{\beta _{\mathrm{u},n}\varsigma _{\mathrm{u},n}}{\left( K_{\mathrm{u},n}+1 \right)}\sum_{u\ne n}^{N_{\mathrm{u}}-1}{p_u\frac{\beta _{\mathrm{u},u}\varsigma _{\mathrm{u},u}}{K_{\mathrm{u},u}+1}}+p_n\frac{\beta _{\mathrm{u},n}^{2}\sigma _{\mathrm{u},0,n}^{4}}{\left( K_{\mathrm{u},n}+1 \right) ^2}\left( N_{\mathrm{r}}-1 \right) +p_n\beta _{\mathrm{u},n}^{2}\tau _{\mathrm{u},n} \right) N_{\mathrm{b}}N_{\mathrm{r}},
\end{equation}
where
\begin{equation}
  \mathrm{term}_{1222}^{2}=\frac{\sigma _{\mathrm{b},0}^{2}}{K_{\mathrm{b}}+1}\sum_{k=1}^{L_{\mathrm{u}}-1}{\sigma _{\mathrm{u},k,n}^{4}}+\frac{\sigma _{\mathrm{u},0,n}^{2}}{K_{\mathrm{u},n}+1}\sum_{k=1}^{L_{1}^{\min}-1}{\sigma _{\mathrm{u},k,n}^{2}\sigma _{\mathrm{b},k}^{2}}+\sum_{k_1=1}^{L_{\mathrm{u}}-1}{\sum_{k_2=k_1+1}^{L_{2}^{\min}-1}{\sigma _{\mathrm{u},k_1,n}^{2}\sigma _{\mathrm{u},k_2,n}^{2}\sigma _{\mathrm{b},k_2-k_1}^{2}}},
\end{equation}
$L_{1}^{\min}=\min \left\{ L_{\mathrm{b}},L_{\mathrm{u}} \right\}$,
$L_{2}^{\min}=\min \left\{ L_{\mathrm{u}},L_{\mathrm{b}}+k_1 \right\}$,
and $\tau _{\mathrm{u},n}$ is defined in \eqref{tau_un}.

Substituting \eqref{E_1222_eq4} and \eqref{E_1222_eq6} into \eqref{E_1222_eq2},
we arrive at
\begin{equation*}
\hspace{-0.3cm}
  \mathbb{E} \left\{ \mathbf{g}_{\mathrm{u},nt}^{H}\mathbf{\Phi }^H\mathbf{G}_{\mathrm{b}}^{H}\mathbf{\Upsilon }_{1222}\mathbf{G}_{\mathrm{b}}\mathbf{\Phi g}_{\mathrm{u},nt} \right\} =\frac{\beta _{\mathrm{u},n}\beta _{\mathrm{b}}^{2}\sigma _{\mathrm{u},0,n}^{2}\sigma _{\mathrm{b},0}^{4}K_{\mathrm{u},n}K_{\mathrm{b}}}{\left( K_{\mathrm{u},n}+1 \right) \left( K_{\mathrm{b}}+1 \right) ^2}N_{\mathrm{b}}\left| \varPhi _{N_{\mathrm{r}}}\left( n \right) \right|^2\sum_{u=0}^{N_{\mathrm{u}}-1}{p_u\frac{\beta _{\mathrm{u},u}\varsigma _{\mathrm{u},u}}{K_{\mathrm{u},u}+1}}
\end{equation*}
\begin{equation*}
  +\frac{\beta _{\mathrm{b}}^{2}\sigma _{\mathrm{b},0}^{4}K_{\mathrm{b}}}{\left( K_{\mathrm{b}}+1 \right) ^2}\left( \frac{\beta _{\mathrm{u},n}\varsigma _{\mathrm{u},n}}{\left( K_{\mathrm{u},n}+1 \right)}\sum_{u\ne n}^{N_{\mathrm{u}}-1}{p_u\frac{\beta _{\mathrm{u},u}\varsigma _{\mathrm{u},u}}{K_{\mathrm{u},u}+1}}+p_n\frac{\beta _{\mathrm{u},n}^{2}\sigma _{\mathrm{u},0,n}^{4}}{\left( K_{\mathrm{u},n}+1 \right) ^2}\left( N_{\mathrm{r}}-1 \right) +p_n\beta _{\mathrm{u},n}^{2}\tau _{\mathrm{u},n} \right) N_{\mathrm{b}}N_{\mathrm{r}}
\end{equation*}
\begin{equation}\label{E_1222_eq7}
\hspace{-9.8cm}
  +p_n\frac{\beta _{\mathrm{u},n}^{2}\beta _{\mathrm{b}}^{2}\sigma _{\mathrm{b},0}^{2}K_{\mathrm{b}}}{K_{\mathrm{b}}+1}N_{\mathrm{b}}N_{\mathrm{r}}^{2}\mathrm{term}_{1222}^{2} .
\end{equation}

\subsubsection{\textbf{ The $\mathbf{\Upsilon }_{2122}$ and $\mathbf{\Upsilon }_{2212}$-Related Expectations}} \label{E_2122}

Because of
\begin{equation}\label{E_2122_eq1}
  \mathbb{E} \left\{ \mathbf{g}_{\mathrm{u},nt}^{H}\mathbf{\Phi }^H\mathbf{G}_{\mathrm{b}}^{H}\mathbf{\Upsilon }_{2212}\mathbf{G}_{\mathrm{b}}\mathbf{\Phi g}_{\mathrm{u},nt} \right\} =\left( \mathbb{E} \left\{ \mathbf{g}_{\mathrm{u},nt}^{H}\mathbf{\Phi }^H\mathbf{G}_{\mathrm{b}}^{H}\mathbf{\Upsilon }_{2122}\mathbf{G}_{\mathrm{b}}\mathbf{\Phi g}_{\mathrm{u},nt} \right\} \right) ^H ,
\end{equation}
we focus on the expectation $\mathbb{E} \left\{ \mathbf{g}_{\mathrm{u},nt}^{H}\mathbf{\Phi }^H\mathbf{G}_{\mathrm{b}}^{H}\mathbf{\Upsilon }_{2122}\mathbf{G}_{\mathrm{b}}\mathbf{\Phi g}_{\mathrm{u},nt} \right\}$, which is expanded as
\begin{equation*}
\hspace{-7.5cm}
  \mathbb{E} \left\{ \mathbf{g}_{\mathrm{u},nt}^{H}\mathbf{\Phi }^H\mathbf{G}_{\mathrm{b}}^{H}\mathbf{\Upsilon }_{2122}\mathbf{G}_{\mathrm{b}}\mathbf{\Phi g}_{\mathrm{u},nt} \right\}
\end{equation*}
\begin{equation*}
  =\mathbb{E} \left\{ \left( \bar{\mathbf{g}}_{\mathrm{u},nt}^{H}+\tilde{\mathbf{g}}_{\mathrm{u},nt}^{H} \right) \mathbf{\Phi }^H\left( \bar{\mathbf{G}}_{\mathrm{b}}^{H}+\tilde{\mathbf{G}}_{\mathrm{b}}^{H} \right) \mathbf{\Upsilon }_{2122}\left( \bar{\mathbf{G}}_{\mathrm{b}}+\tilde{\mathbf{G}}_{\mathrm{b}} \right) \mathbf{\Phi }\left( \bar{\mathbf{g}}_{\mathrm{u},nt}+\tilde{\mathbf{g}}_{\mathrm{u},nt} \right) \right\}
\end{equation*}
\begin{equation}\label{E_2122_eq2}
\hspace{-0.7cm}
  =\mathbb{E} \left\{ \bar{\mathbf{g}}_{\mathrm{u},nt}^{H}\mathbf{\Phi }^H\bar{\mathbf{G}}_{\mathrm{b}}^{H}\mathbf{\Upsilon }_{2122}\bar{\mathbf{G}}_{\mathrm{b}}\mathbf{\Phi }\tilde{\mathbf{g}}_{\mathrm{u},nt} \right\} +\mathbb{E} \left\{ \bar{\mathbf{g}}_{\mathrm{u},nt}^{H}\mathbf{\Phi }^H\tilde{\mathbf{G}}_{\mathrm{b}}^{H}\mathbf{\Upsilon }_{2122}\tilde{\mathbf{G}}_{\mathrm{b}}\mathbf{\Phi }\tilde{\mathbf{g}}_{\mathrm{u},nt} \right\} .
\end{equation}
Besides, the $(bN_{\mathrm{c}}+t)$-th element of the diagonal matrix $\mathbf{\Upsilon }_{2122}$ is expressed as
\begin{equation}\label{E_2122_eq3}
  \left[ \mathbf{\Upsilon }_{1222} \right] _{bt,bt}=\frac{1}{N_{\mathrm{c}}}\sum_{s=0}^{N_{\mathrm{c}}-1}{\sum_{u=0}^{N_{\mathrm{u}}-1}{p_u\left( \sum_{r_1=0}^{N_{\mathrm{r}}-1}{e^{j\varphi _{r_1}}\tilde{g}_{\mathrm{b},s,s}^{\left( b,r_1 \right)}\bar{g}_{\mathrm{u}}^{\left( r_1,u \right)}} \right) \left( \sum_{r_2=0}^{N_{\mathrm{r}}-1}{e^{j\varphi _{r_2}}\tilde{g}_{\mathrm{b},s,s}^{\left( b,r_2 \right)}\tilde{g}_{\mathrm{u},s,s}^{\left( r_2,u \right)}} \right) ^{\ast}}}\triangleq \gamma _{2122}\left( b \right).
\end{equation}

Similar to \eqref{E_1222_eq4},
the first expectation in \eqref{E_2122_eq2} is obtained as
\begin{equation}\label{E_2122_eq4}
  \mathbb{E} \left\{ \bar{\mathbf{g}}_{\mathrm{u},nt}^{H}\mathbf{\Phi }^H\bar{\mathbf{G}}_{\mathrm{b}}^{H}\mathbf{\Upsilon }_{2122}\bar{\mathbf{G}}_{\mathrm{b}}\mathbf{\Phi }\tilde{\mathbf{g}}_{\mathrm{u},nt} \right\}
  = p_n  \frac{\beta _{\mathrm{u},n}^{2}\beta _{\mathrm{b}}^{2}\sigma _{\mathrm{u},0,n}^{4}\sigma _{\mathrm{b},0}^{2}\varsigma _{\mathrm{b}}K_{\mathrm{u},n}K_{\mathrm{b}}}{\left( K_{\mathrm{u},n}+1 \right) ^2\left( K_{\mathrm{b}}+1 \right) ^2}N_{\mathrm{b}}\left| \varPhi _{N_{\mathrm{r}}}\left( n \right) \right|^2 .
\end{equation}
Meanwhile, the second expectation in \eqref{E_2122_eq2} is derived as
\begin{equation*}
\hspace{-8.8cm}
  \mathbb{E} \left\{ \bar{\mathbf{g}}_{\mathrm{u},nt}^{H}\mathbf{\Phi }^H\tilde{\mathbf{G}}_{\mathrm{b}}^{H}\mathbf{\Upsilon }_{2122}\tilde{\mathbf{G}}_{\mathrm{b}}\mathbf{\Phi }\tilde{\mathbf{g}}_{\mathrm{u},nt} \right\}
\end{equation*}
\begin{equation*}
\hspace{-2.9cm}
  =\mathbb{E} \left\{ \sum_{b=0}^{N_{\mathrm{b}}-1}{\gamma _{2122}\left( b \right) \sum_{r_1=0}^{N_{\mathrm{r}}-1}{\sum_{r_2=0}^{N_{\mathrm{r}}-1}{\left( e^{j\varphi _{r_1}}\bar{g}_{\mathrm{u}}^{\left( r_1,n \right)}\tilde{g}_{\mathrm{b},t,t}^{\left( b,r_1 \right)} \right) ^{\ast}e^{j\varphi _{r_2}}\tilde{g}_{\mathrm{u},t,t}^{\left( r_2,n \right)}\tilde{g}_{\mathrm{b},t,t}^{\left( b,r_2 \right)}}}} \right\}
\end{equation*}
\begin{equation*}
\hspace{-1cm}
  =c_{2122}^{1}\sum_{b=0}^{N_{\mathrm{b}}-1}{\sum_{s=0}^{N_{\mathrm{c}}-1}{\mathbb{E} \left\{ \sum_{r_1=0}^{N_{\mathrm{r}}-1}{\tilde{g}_{\mathrm{b},s,s}^{\left( b,r_1 \right)}\left( \tilde{g}_{\mathrm{b},t,t}^{\left( b,r_1 \right)} \right) ^{\ast}}\sum_{r_2=0}^{N_{\mathrm{r}}-1}{\left( \tilde{g}_{\mathrm{u},s,s}^{\left( r_2,n \right)} \right) ^{\ast}\tilde{g}_{\mathrm{u},t,t}^{\left( r_2,n \right)}\left( \tilde{g}_{\mathrm{b},s,s}^{\left( b,r_2 \right)} \right) ^{\ast}\tilde{g}_{\mathrm{b},t,t}^{\left( b,r_2 \right)}} \right\}}}
\end{equation*}
\begin{equation*}
\hspace{-2.8cm}
  \overset{\left( a \right)}{=}c_{2122}^{1}\sum_{b=0}^{N_{\mathrm{b}}-1}{\sum_{s=0}^{N_{\mathrm{c}}-1}{\mathbb{E} \left\{ \sum_{r=0}^{N_{\mathrm{r}}-1}{\left( \tilde{g}_{\mathrm{u},s,s}^{\left( r,n \right)} \right) ^{\ast}\tilde{g}_{\mathrm{u},t,t}^{\left( r,n \right)}\left( \tilde{g}_{\mathrm{b},s,s}^{\left( b,r \right)} \right) ^{\ast}\tilde{g}_{\mathrm{b},t,t}^{\left( b,r \right)}\tilde{g}_{\mathrm{b},s,s}^{\left( b,r \right)}\left( \tilde{g}_{\mathrm{b},t,t}^{\left( b,r \right)} \right) ^{\ast}} \right\}}}
\end{equation*}
\begin{equation}\label{E_2122_eq5}
  +c_{2122}^{1}\sum_{b=0}^{N_{\mathrm{b}}-1}{\sum_{s=0}^{N_{\mathrm{c}}-1}{\mathbb{E} \left\{ \sum_{r_1=0}^{N_{\mathrm{r}}-1}{\tilde{g}_{\mathrm{b},s,s}^{\left( b,r_1 \right)}\left( \tilde{g}_{\mathrm{b},t,t}^{\left( b,r_1 \right)} \right) ^{\ast}\sum_{r_2\ne r_1}^{N_{\mathrm{r}}-1}{\left( \tilde{g}_{\mathrm{u},s,s}^{\left( r_2,n \right)} \right) ^{\ast}\tilde{g}_{\mathrm{u},t,t}^{\left( r_2,n \right)}\left( \tilde{g}_{\mathrm{b},s,s}^{\left( b,r_2 \right)} \right) ^{\ast}\tilde{g}_{\mathrm{b},t,t}^{\left( b,r_2 \right)}}} \right\}}} ,
\end{equation}
where $c_{2122}^{1} = p_n \frac{\beta _{\mathrm{u},n}\sigma _{\mathrm{u},0,n}^{2}K_{\mathrm{u},n}}{N_{\mathrm{c}}\left( K_{\mathrm{u},n}+1 \right)}$.
Step $\left( a \right)$ divides its left hand side into two parts,
depending on whether or not $r_1 = r_2$.
Then, using \eqref{R-U_FD_channel_ij_tt_NLoS},
\eqref{B-R_FD_channel_ij_tt_NLoS},
and the conclusions in Lemma~\ref{lemma_sum} and Lemma~\ref{lemma_expectation},
the second expectation in \eqref{E_2122_eq2} is obtained as
\begin{equation*}
  \mathbb{E} \left\{ \bar{\mathbf{g}}_{\mathrm{u},nt}^{H}\mathbf{\Phi }^H\tilde{\mathbf{G}}_{\mathrm{b}}^{H}\mathbf{\Upsilon }_{2122}\tilde{\mathbf{G}}_{\mathrm{b}}\mathbf{\Phi }\tilde{\mathbf{g}}_{\mathrm{u},nt} \right\}
  = p_n \frac{\beta _{\mathrm{u},n}^{2}\sigma _{\mathrm{u},0,n}^{4}K_{\mathrm{u},n}}{\left( K_{\mathrm{u},n}+1 \right) ^2}\left( \frac{\beta _{\mathrm{b}}^{2}\sigma _{\mathrm{b},0}^{4}}{\left( K_{\mathrm{b}}+1 \right) ^2}\left( N_{\mathrm{r}}-1 \right) +\beta _{\mathrm{b}}^{2}\tau _{\mathrm{b}} \right) N_{\mathrm{b}}N_{\mathrm{r}}
\end{equation*}
\begin{equation}\label{E_2122_eq6}
\hspace{2.4cm}
  + p_n\frac{\beta _{\mathrm{u},n}^{2}\beta _{\mathrm{b}}^{2}\sigma _{\mathrm{u},0,n}^{2}K_{\mathrm{u},n}}{K_{\mathrm{u},n}+1}N_{\mathrm{b}}N_{\mathrm{r}}^{2}\mathrm{term}_{2122}^{1} ,
\end{equation}
where
\begin{equation}
  \mathrm{term}_{2122}^{1}=\frac{\sigma _{\mathrm{u},0,n}^{2}}{K_{\mathrm{u},n}+1}\sum_{k=1}^{L_{\mathrm{b}}-1}{\sigma _{\mathrm{b},k}^{4}}+\frac{\sigma _{\mathrm{b},0}^{2}}{K_{\mathrm{b}}+1}\sum_{k=1}^{L_{1}^{\min}-1}{\sigma _{\mathrm{b},k}^{2}\sigma _{\mathrm{u},k,n}^{2}}+\sum_{k_1=1}^{L_{\mathrm{b}}-1}{\sum_{k_2=k_1+1}^{L_{3}^{\min}-1}{\sigma _{\mathrm{b},k_1}^{2}\sigma _{\mathrm{b},k_2}^{2}\sigma _{\mathrm{u},k_2-k_1,n}^{2}}} ,
\end{equation}
$L_{3}^{\min}=\min \left\{ L_{\mathrm{b}},L_{\mathrm{u}}+k_1 \right\}$,
and $\tau _{\mathrm{b}}$ is defined in \eqref{tau_b}.

Substituting \eqref{E_2122_eq4} and \eqref{E_2122_eq6} into \eqref{E_2122_eq2},
we arrive at
\begin{equation*}
\hspace{-2.8cm}
  \mathbb{E} \left\{ \mathbf{g}_{\mathrm{u},nt}^{H}\mathbf{\Phi }^H\mathbf{G}_{\mathrm{b}}^{H}\mathbf{\Upsilon }_{2122}\mathbf{G}_{\mathrm{b}}\mathbf{\Phi g}_{\mathrm{u},nt} \right\}
  = p_n \frac{\beta _{\mathrm{u},n}^{2}\beta _{\mathrm{b}}^{2}\sigma _{\mathrm{u},0,n}^{4}\sigma _{\mathrm{b},0}^{2}\varsigma _{\mathrm{b}}K_{\mathrm{u},n}K_{\mathrm{b}}}{\left( K_{\mathrm{u},n}+1 \right) ^2\left( K_{\mathrm{b}}+1 \right) ^2}N_{\mathrm{b}}\left| \varPhi _{N_{\mathrm{r}}}\left( n \right) \right|^2
\end{equation*}
\begin{equation}\label{E_2122_eq7}
  +p_n\frac{\beta _{\mathrm{u},n}^{2}\sigma _{\mathrm{u},0,n}^{4}K_{\mathrm{u},n}}{\left( K_{\mathrm{u},n}+1 \right) ^2}\left( \frac{\beta _{\mathrm{b}}^{2}\sigma _{\mathrm{b},0}^{4}}{\left( K_{\mathrm{b}}+1 \right) ^2}\left( N_{\mathrm{r}}-1 \right) +\beta _{\mathrm{b}}^{2}\tau _{\mathrm{b}} \right) N_{\mathrm{b}}N_{\mathrm{r}}+p_n\frac{\beta _{\mathrm{u},n}^{2}\beta _{\mathrm{b}}^{2}\sigma _{\mathrm{u},0,n}^{2}K_{\mathrm{u},n}}{K_{\mathrm{u},n}+1}N_{\mathrm{b}}N_{\mathrm{r}}^{2}\mathrm{term}_{2122}^{1} .
\end{equation}

\subsubsection{\textbf{ The $\mathbf{\Upsilon }_{1111}$-Related Expectation}} \label{E_1111}

The expectation $\mathbb{E} \left\{ \mathbf{g}_{\mathrm{u},nt}^{H}\mathbf{\Phi }^H\mathbf{G}_{\mathrm{b}}^{H}\mathbf{\Upsilon }_{1111}\mathbf{G}_{\mathrm{b}}\mathbf{\Phi g}_{\mathrm{u},nt} \right\}$ can be expanded as
\begin{equation*}
\hspace{-7.5cm}
  \mathbb{E} \left\{ \mathbf{g}_{\mathrm{u},nt}^{H}\mathbf{\Phi }^H\mathbf{G}_{\mathrm{b}}^{H}\mathbf{\Upsilon }_{1111}\mathbf{G}_{\mathrm{b}}\mathbf{\Phi g}_{\mathrm{u},nt} \right\}
\end{equation*}
\begin{equation*}
  =\mathbb{E} \left\{ \left( \bar{\mathbf{g}}_{\mathrm{u},nt}^{H}+\tilde{\mathbf{g}}_{\mathrm{u},nt}^{H} \right) \mathbf{\Phi }^H\left( \bar{\mathbf{G}}_{\mathrm{b}}^{H}+\tilde{\mathbf{G}}_{\mathrm{b}}^{H} \right) \mathbf{\Upsilon }_{1111}\left( \bar{\mathbf{G}}_{\mathrm{b}}+\tilde{\mathbf{G}}_{\mathrm{b}} \right) \mathbf{\Phi }\left( \bar{\mathbf{g}}_{\mathrm{u},nt}+\tilde{\mathbf{g}}_{\mathrm{u},nt} \right) \right\}
\end{equation*}
\begin{equation*}
\hspace{-0.8cm}
  \overset{\left( a \right)}{=}\mathbb{E} \left\{ \bar{\mathbf{g}}_{\mathrm{u},nt}^{H}\mathbf{\Phi }^H\bar{\mathbf{G}}_{\mathrm{b}}^{H}\mathbf{\Upsilon }_{1111}\bar{\mathbf{G}}_{\mathrm{b}}\mathbf{\Phi }\bar{\mathbf{g}}_{\mathrm{u},nt} \right\}
  +\mathbb{E} \left\{ \bar{\mathbf{g}}_{\mathrm{u},nt}^{H}\mathbf{\Phi }^H\tilde{\mathbf{G}}_{\mathrm{b}}^{H}\mathbf{\Upsilon }_{1111}\tilde{\mathbf{G}}_{\mathrm{b}}\mathbf{\Phi }\bar{\mathbf{g}}_{\mathrm{u},nt} \right\}
\end{equation*}
\begin{equation}\label{E_1111_eq1}
\hspace{-0.7cm}
  +\mathbb{E} \left\{ \tilde{\mathbf{g}}_{\mathrm{u},nt}^{H}\mathbf{\Phi }^H\bar{\mathbf{G}}_{\mathrm{b}}^{H}\mathbf{\Upsilon }_{1111}\bar{\mathbf{G}}_{\mathrm{b}}\mathbf{\Phi }\tilde{\mathbf{g}}_{\mathrm{u},nt} \right\}
  +\mathbb{E} \left\{ \tilde{\mathbf{g}}_{\mathrm{u},nt}^{H}\mathbf{\Phi }^H\tilde{\mathbf{G}}_{\mathrm{b}}^{H}\mathbf{\Upsilon }_{1111}\tilde{\mathbf{G}}_{\mathrm{b}}\mathbf{\Phi }\tilde{\mathbf{g}}_{\mathrm{u},nt} \right\} ,
\end{equation}
where step $\left( a \right)$ is obtained by removing the zero-value terms.
Meanwhile, the $(bN_{\mathrm{c}}+t)$-th element of the diagonal matrix $\mathbf{\Upsilon }_{1111}$ is expressed as
\begin{equation}\label{E_1111_eq2}
  \left[ \mathbf{\Upsilon }_{1111} \right] _{bt,bt}=\frac{1}{N_{\mathrm{c}}}\sum_{s=0}^{N_{\mathrm{c}}-1}{\sum_{u=0}^{N_{\mathrm{u}}-1}{p_u\left( \sum_{r_1=0}^{N_{\mathrm{r}}-1}{e^{j\varphi _{r_1}}\bar{g}_{\mathrm{b}}^{\left( b,r_1 \right)}\bar{g}_{\mathrm{u}}^{\left( r_1,u \right)}} \right) \left( \sum_{r_2=0}^{N_{\mathrm{r}}-1}{e^{j\varphi _{r_2}}\bar{g}_{\mathrm{b}}^{\left( b,r_2 \right)}\bar{g}_{\mathrm{u}}^{\left( r_2,u \right)}} \right) ^{\ast}}}\triangleq \gamma _{1111}\left( b \right) .
\end{equation}

Similarly, we use \eqref{R-U_FD_channel_ij_tt}-\eqref{B-R_FD_channel_ij_tt_NLoS}, Lemma~\ref{lemma_sum} and Lemma~\ref{lemma_expectation} to derive the four expectations in \eqref{E_1111_eq1},
which can be obtained as
\begin{equation*}
\hspace{-7cm}
  \mathbb{E} \left\{ \bar{\mathbf{g}}_{\mathrm{u},nt}^{H}\mathbf{\Phi }^H\bar{\mathbf{G}}_{\mathrm{b}}^{H}\mathbf{\Upsilon }_{1111}\bar{\mathbf{G}}_{\mathrm{b}}\mathbf{\Phi }\bar{\mathbf{g}}_{\mathrm{u},nt} \right\}
\end{equation*}
\begin{equation*}
\hspace{-1.3cm}
  =\sum_{b=0}^{N_{\mathrm{b}}-1}{\gamma _{1111}\left( b \right) \sum_{r_1=0}^{N_{\mathrm{r}}-1}{\sum_{r_2=0}^{N_{\mathrm{r}}-1}{\left( \left( e^{j\varphi _{r_1}}\bar{g}_{\mathrm{u}}^{\left( r_1,n \right)}\bar{g}_{\mathrm{b}}^{\left( b,r_1 \right)} \right) ^{\ast}e^{j\varphi _{r_2}}\bar{g}_{\mathrm{u}}^{\left( r_2,n \right)}\bar{g}_{\mathrm{b}}^{\left( b,r_2 \right)} \right)}}}
\end{equation*}
\begin{equation}\label{E_1111_eq3}
  =\frac{\beta _{\mathrm{u},n}\beta _{\mathrm{b}}^{2}\sigma _{\mathrm{u},0,n}^{2}\sigma _{\mathrm{b},0}^{4}K_{\mathrm{u},n}K_{\mathrm{b}}^{2}}{\left( K_{\mathrm{u},n}+1 \right) \left( K_{\mathrm{b}}+1 \right) ^2}N_{\mathrm{b}}\left| \varPhi _{N_{\mathrm{r}}}\left( n \right) \right|^2\sum_{u=0}^{N_{\mathrm{u}}-1}{p_u\frac{\beta _{\mathrm{u},u}K_{\mathrm{u},u}}{K_{\mathrm{u},u}+1}\sigma _{\mathrm{u},0,u}^{2}\left| \varPhi _{N_{\mathrm{r}}}\left( u \right) \right|^2},
\end{equation}
\begin{equation*}
\hspace{-7cm}
  \mathbb{E} \left\{ \bar{\mathbf{g}}_{\mathrm{u},nt}^{H}\mathbf{\Phi }^H\tilde{\mathbf{G}}_{\mathrm{b}}^{H}\mathbf{\Upsilon }_{1111}\tilde{\mathbf{G}}_{\mathrm{b}}\mathbf{\Phi }\bar{\mathbf{g}}_{\mathrm{u},nt} \right\}
\end{equation*}
\begin{equation*}
\hspace{-0.4cm}
  =\mathbb{E} \left\{ \sum_{b=0}^{N_{\mathrm{b}}-1}{\gamma _{1111}\left( b \right) \sum_{r_1=0}^{N_{\mathrm{r}}-1}{\sum_{r_2=0}^{N_{\mathrm{r}}-1}{\left( \left( e^{j\varphi _{r_1}}\bar{g}_{\mathrm{u}}^{\left( r_1,n \right)}\tilde{g}_{\mathrm{b},t,t}^{\left( b,r_1 \right)} \right) ^{\ast}e^{j\varphi _{r_2}}\bar{g}_{\mathrm{u}}^{\left( r_2,n \right)}\tilde{g}_{\mathrm{b},t,t}^{\left( b,r_2 \right)} \right)}}} \right\}
\end{equation*}
\begin{equation}\label{E_1111_eq4}
\hspace{-1.1cm}
  =\frac{\beta _{\mathrm{u},n}\beta _{\mathrm{b}}^{2}\sigma _{\mathrm{u},0,n}^{2}\sigma _{\mathrm{b},0}^{2}\varsigma _{\mathrm{b}}K_{\mathrm{u},n}K_{\mathrm{b}}}{\left( K_{\mathrm{u},n}+1 \right) \left( K_{\mathrm{b}}+1 \right) ^2}N_{\mathrm{b}}N_{\mathrm{r}}\sum_{u=0}^{N_{\mathrm{u}}-1}{p_u\frac{\beta _{\mathrm{u},u}K_{\mathrm{u},u}}{K_{\mathrm{u},u}+1}\sigma _{\mathrm{u},0,u}^{2}\left| \varPhi _{N_{\mathrm{r}}}\left( u \right) \right|^2} ,
\end{equation}
\begin{equation*}
\hspace{-7.1cm}
  \mathbb{E} \left\{ \tilde{\mathbf{g}}_{\mathrm{u},nt}^{H}\mathbf{\Phi }^H\bar{\mathbf{G}}_{\mathrm{b}}^{H}\mathbf{\Upsilon }_{1111}\bar{\mathbf{G}}_{\mathrm{b}}\mathbf{\Phi }\tilde{\mathbf{g}}_{\mathrm{u},nt} \right\}
\end{equation*}
\begin{equation*}
\hspace{-0.5cm}
  =\mathbb{E} \left\{ \sum_{b=0}^{N_{\mathrm{b}}-1}{\gamma _{1111}\left( b \right)}\sum_{r_1=0}^{N_{\mathrm{r}}-1}{\sum_{r_2=0}^{N_{\mathrm{r}}-1}{\left( \left( e^{j\varphi _{r_1}}\tilde{g}_{\mathrm{u},t,t}^{\left( r_1,n \right)}\bar{g}_{\mathrm{b}}^{\left( b,r_1 \right)} \right) ^{\ast}e^{j\varphi _{r_2}}\tilde{g}_{\mathrm{u},t,t}^{\left( r_2,n \right)}\bar{g}_{\mathrm{b}}^{\left( b,r_2 \right)} \right)}} \right\}
\end{equation*}
\begin{equation}\label{E_1111_eq5}
\hspace{-1.9cm}
  =\frac{\beta _{\mathrm{u},n}\beta _{\mathrm{b}}^{2}\sigma _{\mathrm{b},0}^{4}\varsigma _{\mathrm{u},n}K_{\mathrm{b}}^{2}}{\left( K_{\mathrm{u},n}+1 \right) \left( K_{\mathrm{b}}+1 \right) ^2}N_{\mathrm{b}}N_{\mathrm{r}}\sum_{u=0}^{N_{\mathrm{u}}-1}{p_u\frac{\beta _{\mathrm{u},u}K_{\mathrm{u},u}}{K_{\mathrm{u},u}+1}\sigma _{\mathrm{u},0,u}^{2}\left| \varPhi _{N_{\mathrm{r}}}\left( u \right) \right|^2} ,
\end{equation}
\begin{equation*}
\hspace{-7cm}
  \mathbb{E} \left\{ \tilde{\mathbf{g}}_{\mathrm{u},nt}^{H}\mathbf{\Phi }^H\tilde{\mathbf{G}}_{\mathrm{b}}^{H}\mathbf{\Upsilon }_{1111}\tilde{\mathbf{G}}_{\mathrm{b}}\mathbf{\Phi }\tilde{\mathbf{g}}_{\mathrm{u},nt} \right\}
\end{equation*}
\begin{equation*}
\hspace{-0.6cm}
  =\mathbb{E} \left\{ \sum_{b=0}^{N_{\mathrm{b}}-1}{\gamma _{1111}\left( b \right) \sum_{r_1=0}^{N_{\mathrm{r}}-1}{\sum_{r_2=0}^{N_{\mathrm{r}}-1}{\left( \left( e^{j\varphi _{r_1}}\tilde{g}_{\mathrm{u},t,t}^{\left( r_1,n \right)}\tilde{g}_{\mathrm{b},t,t}^{\left( b,r_1 \right)} \right) ^{\ast}e^{j\varphi _{r_2}}\tilde{g}_{\mathrm{u},t,t}^{\left( r_2,n \right)}\tilde{g}_{\mathrm{b},t,t}^{\left( b,r_2 \right)} \right)}}} \right\}
\end{equation*}
\begin{equation}\label{E_1111_eq6}
\hspace{-2.1cm}
  =\frac{\beta _{\mathrm{u},n}\beta _{\mathrm{b}}^{2}\sigma _{\mathrm{b},0}^{2}\varsigma _{\mathrm{u},n}\varsigma _{\mathrm{b}}K_{\mathrm{b}}}{\left( K_{\mathrm{u},n}+1 \right) \left( K_{\mathrm{b}}+1 \right) ^2}N_{\mathrm{b}}N_{\mathrm{r}}\sum_{u=0}^{N_{\mathrm{u}}-1}{p_u\frac{\beta _{\mathrm{u},u}K_{\mathrm{u},u}}{K_{\mathrm{u},u}+1}\sigma _{\mathrm{u},0,u}^{2}\left| \varPhi _{N_{\mathrm{r}}}\left( u \right) \right|^2}  .
\end{equation}

Substituting \eqref{E_1111_eq3}-\eqref{E_1111_eq6} into \eqref{E_1111_eq1}, we arrive at
\begin{equation*}
\hspace{-0.9cm}
  \mathbb{E} \left\{ \mathbf{g}_{\mathrm{u},nt}^{H}\mathbf{\Phi }^H\mathbf{G}_{\mathrm{b}}^{H}\mathbf{\Upsilon }_{1111}\mathbf{G}_{\mathrm{b}}\mathbf{\Phi g}_{\mathrm{u},nt} \right\}
  =\frac{\beta _{\mathrm{u},n}\beta _{\mathrm{b}}^{2}K_{\mathrm{b}}N_{\mathrm{b}}}{\left( K_{\mathrm{u},n}+1 \right) \left( K_{\mathrm{b}}+1 \right) ^2}
  \bigg(
  \sigma _{\mathrm{u},0,n}^{2}\sigma _{\mathrm{b},0}^{4}K_{\mathrm{u},n}K_{\mathrm{b}}\left| \varPhi _{N_{\mathrm{r}}}\left( n \right) \right|^2
\end{equation*}
\begin{equation}\label{E_gamma_1111_value}
  +\sigma _{\mathrm{u},0,n}^{2}\sigma _{\mathrm{b},0}^{2}\varsigma _{\mathrm{b}}K_{\mathrm{u},n}N_{\mathrm{r}}+\sigma _{\mathrm{b},0}^{4}\varsigma _{\mathrm{u},n}K_{\mathrm{b}}N_{\mathrm{r}}+\sigma _{\mathrm{b},0}^{2}\varsigma _{\mathrm{u},n}\varsigma _{\mathrm{b}}N_{\mathrm{r}}
  \bigg)
  \sum_{u=0}^{N_{\mathrm{u}}-1}{ p_u \frac{\beta _{\mathrm{u},u}K_{\mathrm{u},u}}{K_{\mathrm{u},u}+1}\sigma _{\mathrm{u},0,u}^{2}\left| \varPhi _{N_{\mathrm{r}}}\left( u \right) \right|^2} .
\end{equation}

\subsubsection{\textbf{ The $\mathbf{\Upsilon }_{1221}$-Related Expectation}} \label{E_1221}
The expectation $\mathbb{E} \left\{ \mathbf{g}_{\mathrm{u},nt}^{H}\mathbf{\Phi }^H\mathbf{G}_{\mathrm{b}}^{H}\mathbf{\Upsilon }_{1221}\mathbf{G}_{\mathrm{b}}\mathbf{\Phi g}_{\mathrm{u},nt} \right\}$ is expanded as
\begin{equation*}
\hspace{-7.5cm}
  \mathbb{E} \left\{ \mathbf{g}_{\mathrm{u},nt}^{H}\mathbf{\Phi }^H\mathbf{G}_{\mathrm{b}}^{H}\mathbf{\Upsilon }_{1221}\mathbf{G}_{\mathrm{b}}\mathbf{\Phi g}_{\mathrm{u},nt} \right\}
\end{equation*}
\begin{equation*}
  =\mathbb{E} \left\{ \left( \bar{\mathbf{g}}_{\mathrm{u},nt}^{H}+\tilde{\mathbf{g}}_{\mathrm{u},nt}^{H} \right) \mathbf{\Phi }^H\left( \bar{\mathbf{G}}_{\mathrm{b}}^{H}+\tilde{\mathbf{G}}_{\mathrm{b}}^{H} \right) \mathbf{\Upsilon }_{1221}\left( \bar{\mathbf{G}}_{\mathrm{b}}+\tilde{\mathbf{G}}_{\mathrm{b}} \right) \mathbf{\Phi }\left( \bar{\mathbf{g}}_{\mathrm{u},nt}+\tilde{\mathbf{g}}_{\mathrm{u},nt} \right) \right\}
\end{equation*}
\begin{equation*}
\hspace{-0.9cm}
  =\mathbb{E} \left\{ \bar{\mathbf{g}}_{\mathrm{u},nt}^{H}\mathbf{\Phi }^H\bar{\mathbf{G}}_{\mathrm{b}}^{H}\mathbf{\Upsilon }_{1221}\bar{\mathbf{G}}_{\mathrm{b}}\mathbf{\Phi }\bar{\mathbf{g}}_{\mathrm{u},nt} \right\}
  +\mathbb{E} \left\{ \bar{\mathbf{g}}_{\mathrm{u},nt}^{H}\mathbf{\Phi }^H\tilde{\mathbf{G}}_{\mathrm{b}}^{H}\mathbf{\Upsilon }_{1221}\tilde{\mathbf{G}}_{\mathrm{b}}\mathbf{\Phi }\bar{\mathbf{g}}_{\mathrm{u},nt} \right\}
\end{equation*}
\begin{equation}\label{E_1221_eq1}
\hspace{-0.8cm}
  +\mathbb{E} \left\{ \tilde{\mathbf{g}}_{\mathrm{u},nt}^{H}\mathbf{\Phi }^H\bar{\mathbf{G}}_{\mathrm{b}}^{H}\mathbf{\Upsilon }_{1221}\bar{\mathbf{G}}_{\mathrm{b}}\mathbf{\Phi }\tilde{\mathbf{g}}_{\mathrm{u},nt} \right\}
  +\mathbb{E} \left\{ \tilde{\mathbf{g}}_{\mathrm{u},nt}^{H}\mathbf{\Phi }^H\tilde{\mathbf{G}}_{\mathrm{b}}^{H}\mathbf{\Upsilon }_{1221}\tilde{\mathbf{G}}_{\mathrm{b}}\mathbf{\Phi }\tilde{\mathbf{g}}_{\mathrm{u},nt} \right\}.
\end{equation}
Besides, the $(bN_{\mathrm{c}}+t)$-th element of the diagonal matrix $\mathbf{\Upsilon }_{1221}$ is expressed as
\begin{equation}\label{E_1221_eq2}
  \left[ \mathbf{\Upsilon }_{1221} \right] _{bt,bt}=\frac{1}{N_{\mathrm{c}}}\sum_{s=0}^{N_{\mathrm{c}}-1}{\sum_{u=0}^{N_{\mathrm{u}}-1}{p_u\left( \sum_{r_1=0}^{N_{\mathrm{r}}-1}{e^{j\varphi _{r_1}}\bar{g}_{\mathrm{b}}^{\left( b,r_1 \right)}\tilde{g}_{\mathrm{u},s,s}^{\left( r_1,u \right)}} \right) \left( \sum_{r_2=0}^{N_{\mathrm{r}}-1}{e^{j\varphi _{r_2}}\bar{g}_{\mathrm{b}}^{\left( b,r_2 \right)}\tilde{g}_{\mathrm{u},s,s}^{\left( r_2,u \right)}} \right) ^{\ast}}}\triangleq \gamma _{1221}\left( b \right).
\end{equation}

Based on \eqref{R-U_FD_channel_ij_tt}-\eqref{B-R_FD_channel_ij_tt_NLoS}, Lemma~\ref{lemma_sum} and Lemma~\ref{lemma_expectation},
the first and second expectations in \eqref{E_1221_eq1} can be obtained as
\begin{equation}\label{E_1221_eq3}
  \mathbb{E} \left\{ \bar{\mathbf{g}}_{\mathrm{u},nt}^{H}\mathbf{\Phi }^H\bar{\mathbf{G}}_{\mathrm{b}}^{H}\mathbf{\Upsilon }_{1221}\bar{\mathbf{G}}_{\mathrm{b}}\mathbf{\Phi }\bar{\mathbf{g}}_{\mathrm{u},nt} \right\} =\frac{\beta _{\mathrm{u},n}\beta _{\mathrm{b}}^{2}\sigma _{\mathrm{u},0,n}^{2}\sigma _{\mathrm{b},0}^{4}K_{\mathrm{u},n}K_{\mathrm{b}}^{2}}{\left( K_{\mathrm{u},n}+1 \right) \left( K_{\mathrm{b}}+1 \right) ^2}N_{\mathrm{b}}N_{\mathrm{r}}\left| \varPhi _{N_{\mathrm{r}}}\left( n \right) \right|^2\sum_{u=0}^{N_{\mathrm{u}}-1}{ p_u \frac{\beta _{\mathrm{u},u}\varsigma _{\mathrm{u},u}}{K_{\mathrm{u},u}+1}},
\end{equation}
\begin{equation}\label{E_1221_eq4}
  \mathbb{E} \left\{ \bar{\mathbf{g}}_{\mathrm{u},nt}^{H}\mathbf{\Phi }^H\tilde{\mathbf{G}}_{\mathrm{b}}^{H}\mathbf{\Upsilon }_{1221}\tilde{\mathbf{G}}_{\mathrm{b}}\mathbf{\Phi }\bar{\mathbf{g}}_{\mathrm{u},nt} \right\} =\frac{\beta _{\mathrm{u},n}\beta _{\mathrm{b}}^{2}\sigma _{\mathrm{u},0,n}^{2}\sigma _{\mathrm{b},0}^{2}\varsigma _{\mathrm{b}}K_{\mathrm{u},n}K_{\mathrm{b}}}{\left( K_{\mathrm{u},n}+1 \right) \left( K_{\mathrm{b}}+1 \right) ^2}N_{\mathrm{b}}N_{\mathrm{r}}^{2}\sum_{u=0}^{N_{\mathrm{u}}-1}{ p_u  \frac{\beta _{\mathrm{u},u}\varsigma _{\mathrm{u},u}}{K_{\mathrm{u},u}+1}}.
\end{equation}

The third expectation in \eqref{E_1221_eq1} is further expanded as
\begin{equation*}
\hspace{-5.8cm}
  \mathbb{E} \left\{ \tilde{\mathbf{g}}_{\mathrm{u},nt}^{H}\mathbf{\Phi }^H\bar{\mathbf{G}}_{\mathrm{b}}^{H}\mathbf{\Upsilon }_{1221}\bar{\mathbf{G}}_{\mathrm{b}}\mathbf{\Phi }\tilde{\mathbf{g}}_{\mathrm{u},nt} \right\}
\end{equation*}
\begin{equation*}
  =\mathbb{E} \left\{ \sum_{b=0}^{N_{\mathrm{b}}-1}{\gamma _{1221}\left( b \right) \sum_{r_1=0}^{N_{\mathrm{r}}-1}{\sum_{r_2=0}^{N_{\mathrm{r}}-1}{\left( e^{j\varphi _{r_1}}\tilde{g}_{\mathrm{u},t,t}^{\left( r_1,n \right)}\bar{g}_{\mathrm{b}}^{\left( b,r_1 \right)} \right) ^{\ast}e^{j\varphi _{r_2}}\tilde{g}_{\mathrm{u},t,t}^{\left( r_2,n \right)}\bar{g}_{\mathrm{b}}^{\left( b,r_2 \right)}}}} \right\}
\end{equation*}
\begin{equation*}
\hspace{-2.2cm}
  =\frac{\beta _{\mathrm{b}}^{2}\sigma _{\mathrm{b},0}^{4}K_{\mathrm{b}}^{2}}{N_{\mathrm{c}}\left( K_{\mathrm{b}}+1 \right) ^2}\mathbb{E} \left\{ \sum_{b=0}^{N_{\mathrm{b}}-1}{\sum_{s=0}^{N_{\mathrm{c}}-1}{\sum_{u=0}^{N_{\mathrm{u}}-1}{p_u\mathrm{term}_{1221}^{1}\left( b,s,u \right)}}} \right\}
\end{equation*}
\begin{equation*}
\hspace{-3.5cm}
  \overset{\left( a \right)}{=}p_n\frac{\beta _{\mathrm{b}}^{2}\sigma _{\mathrm{b},0}^{4}K_{\mathrm{b}}^{2}}{N_{\mathrm{c}}\left( K_{\mathrm{b}}+1 \right) ^2}\mathbb{E} \left\{ \sum_{b=0}^{N_{\mathrm{b}}-1}{\sum_{s=0}^{N_{\mathrm{c}}-1}{\mathrm{term}_{1221}^{2}\left( b,s \right)}} \right\}
\end{equation*}
\begin{equation}\label{E_1221_eq5}
  +\frac{\beta _{\mathrm{b}}^{2}\sigma _{\mathrm{b},0}^{4}K_{\mathrm{b}}^{2}}{N_{\mathrm{c}}\left( K_{\mathrm{b}}+1 \right) ^2}\mathbb{E} \left\{ \sum_{b=0}^{N_{\mathrm{b}}-1}{\sum_{s=0}^{N_{\mathrm{c}}-1}{\sum_{u\ne n}^{N_{\mathrm{u}}-1}{p_u\sum_{r_1=0}^{N_{\mathrm{r}}-1}{\left| \tilde{g}_{\mathrm{u},s,s}^{\left( r_1,u \right)} \right|^2}}}\sum_{r_2=0}^{N_{\mathrm{r}}-1}{\left| \tilde{g}_{\mathrm{u},t,t}^{\left( r_2,n \right)} \right|^2}} \right\},
\end{equation}
where
\begin{equation*}
  \mathrm{term}_{1221}^{1}\left( b,s,u \right) =\sum_{r_1=0}^{N_{\mathrm{r}}-1}{e^{j\varphi _{r_1}}a_{N_{\mathrm{r}},r_1}^{\ast}\left( \phi _{\mathrm{r}}^{ad},\phi _{\mathrm{r}}^{ed} \right) \tilde{g}_{\mathrm{u},s,s}^{\left( r_1,u \right)}}\sum_{r_2=0}^{N_{\mathrm{r}}-1}{\left( e^{j\varphi _{r_2}}a_{N_{\mathrm{r}},r_2}^{\ast}\left( \phi _{\mathrm{r}}^{ad},\phi _{\mathrm{r}}^{ed} \right) \tilde{g}_{\mathrm{u},s,s}^{\left( r_2,u \right)} \right) ^{\ast}}
\end{equation*}
\begin{equation}
  \times \sum_{r_{11}=0}^{N_{\mathrm{r}}-1}{\left( e^{j\varphi _{r_{11}}}a_{N_{\mathrm{r}},r_{11}}^{\ast}\left( \phi _{\mathrm{r}}^{ad},\phi _{\mathrm{r}}^{ed} \right) \tilde{g}_{\mathrm{u},t,t}^{\left( r_{11},n \right)} \right) ^{\ast}}\sum_{r_{22}=0}^{N_{\mathrm{r}}-1}{e^{j\varphi _{r_{22}}}a_{N_{\mathrm{r}},r_{22}}^{\ast}\left( \phi _{\mathrm{r}}^{ad},\phi _{\mathrm{r}}^{ed} \right) \tilde{g}_{\mathrm{u},t,t}^{\left( r_{22},n \right)}} ,
\end{equation}
\begin{equation*}
  \mathrm{term}_{1221}^{2}\left( b,s \right) =\sum_{r_1=0}^{N_{\mathrm{r}}-1}{e^{j\varphi _{r_1}}a_{N_{\mathrm{r}},r_1}^{\ast}\left( \phi _{\mathrm{r}}^{ad},\phi _{\mathrm{r}}^{ed} \right) \tilde{g}_{\mathrm{u},s,s}^{\left( r_1,n \right)}}\sum_{r_2=0}^{N_{\mathrm{r}}-1}{\left( e^{j\varphi _{r_2}}a_{N_{\mathrm{r}},r_2}^{\ast}\left( \phi _{\mathrm{r}}^{ad},\phi _{\mathrm{r}}^{ed} \right) \tilde{g}_{\mathrm{u},s,s}^{\left( r_2,n \right)} \right) ^{\ast}}
\end{equation*}
\begin{equation}
  \times \sum_{r_{11}=0}^{N_{\mathrm{r}}-1}{\left( e^{j\varphi _{r_{11}}}a_{N_{\mathrm{r}},r_{11}}^{\ast}\left( \phi _{\mathrm{r}}^{ad},\phi _{\mathrm{r}}^{ed} \right) \tilde{g}_{\mathrm{u},t,t}^{\left( r_{11},n \right)} \right) ^{\ast}}\sum_{r_{22}=0}^{N_{\mathrm{r}}-1}{e^{j\varphi _{r_{22}}}a_{N_{\mathrm{r}},r_{22}}^{\ast}\left( \phi _{\mathrm{r}}^{ad},\phi _{\mathrm{r}}^{ed} \right) \tilde{g}_{\mathrm{u},t,t}^{\left( r_{22},n \right)}} .
\end{equation}
Step $\left( a \right)$ divides the equation \eqref{E_1221_eq5} into two parts, which depends on whether the parameter $u$ is equal to $n$ or not.
When $u \neq n$, from \eqref{R-U_FD_channel_ij_tt_NLoS}, the second expectation in \eqref{E_1221_eq5} can be easily calculated as
\begin{equation}\label{E_1221_eq6}
  \mathbb{E} \left\{ \sum_{b=0}^{N_{\mathrm{b}}-1}{\sum_{s=0}^{N_{\mathrm{c}}-1}{\sum_{u\ne n}^{N_{\mathrm{u}}-1}{\sum_{r_1=0}^{N_{\mathrm{r}}-1}{\left| \tilde{g}_{\mathrm{u},s,s}^{\left( r_1,u \right)} \right|^2}}}\sum_{r_2=0}^{N_{\mathrm{r}}-1}{\left| \tilde{g}_{\mathrm{u},t,t}^{\left( r_2,n \right)} \right|^2}} \right\} =\frac{\beta _{\mathrm{u},n}\varsigma _{\mathrm{u},n}}{K_{\mathrm{u},n}+1}N_{\mathrm{c}}N_{\mathrm{b}}N_{\mathrm{r}}^{2}\sum_{u\ne n}^{N_{\mathrm{u}}-1}{p_u\frac{\beta _{\mathrm{u},u}\varsigma _{\mathrm{u},u}}{K_{\mathrm{u},u}+1}}  .
\end{equation}
When $u = n$, the first expectation in \eqref{E_1221_eq5} is further expanded as
\begin{equation*}
  \mathbb{E} \left\{ \sum_{b=0}^{N_{\mathrm{b}}-1}{\sum_{s=0}^{N_{\mathrm{c}}-1}{\mathrm{term}_{1221}^{2}\left( b,s \right)}} \right\}
  \overset{\left( a \right)}{=}\mathbb{E} \left\{ \sum_{b=0}^{N_{\mathrm{b}}-1}{\sum_{s=0}^{N_{\mathrm{c}}-1}{\sum_{r=0}^{N_{\mathrm{r}}-1}{\left| \tilde{g}_{\mathrm{u},s,s}^{\left( r,n \right)} \right|^2\left| \tilde{g}_{\mathrm{u},t,t}^{\left( r,n \right)} \right|^2}}} \right\}
\end{equation*}
\begin{equation*}
\hspace{-4.2cm}
  +\mathbb{E} \left\{ \sum_{b=0}^{N_{\mathrm{b}}-1}{\sum_{s=0}^{N_{\mathrm{c}}-1}{\sum_{r_1=0}^{N_{\mathrm{r}}-1}{\left| \tilde{g}_{\mathrm{u},s,s}^{\left( r_1,n \right)} \right|^2}\sum_{r_2\ne r_1}^{N_{\mathrm{r}}-1}{\left| \tilde{g}_{\mathrm{u},t,t}^{\left( r_2,n \right)} \right|^2}}} \right\}
\end{equation*}
\begin{equation}\label{E_1221_eq7}
\hspace{-1.8cm}
  +\mathbb{E} \left\{ \sum_{b=0}^{N_{\mathrm{b}}-1}{\sum_{s=0}^{N_{\mathrm{c}}-1}{\sum_{r_1=0}^{N_{\mathrm{r}}-1}{\tilde{g}_{\mathrm{u},s,s}^{\left( r_1,n \right)}\left( \tilde{g}_{\mathrm{u},t,t}^{\left( r_1,n \right)} \right) ^{\ast}}\sum_{r_2\ne r_1}^{N_{\mathrm{r}}-1}{\left( \tilde{g}_{\mathrm{u},s,s}^{\left( r_2,n \right)} \right) ^{\ast}\tilde{g}_{\mathrm{u},t,t}^{\left( r_2,n \right)}}}} \right\} ,
\end{equation}
where step $\left( a \right)$ divides the equation into three parts:
1) $ r_1 = r_2 = r_{11} = r_{22} $,
2) $ r_1 = r_2 \neq r_{11} = r_{22} $,
and 3) $ r_1 = r_{11} \neq r_2 = r_{22} $.
Based on \eqref{R-U_FD_channel_ij_tt}-\eqref{B-R_FD_channel_ij_tt_NLoS}, Lemma~\ref{lemma_sum} and Lemma~\ref{lemma_expectation},
the three terms in \eqref{E_1221_eq7} is calculated respectively as
\begin{equation}\label{E_1221_eq8}
\hspace{-1.4cm}
  \mathbb{E} \left\{ \sum_{b=0}^{N_{\mathrm{b}}-1}{\sum_{s=0}^{N_{\mathrm{c}}-1}{\sum_{r=0}^{N_{\mathrm{r}}-1}{\left| \tilde{g}_{\mathrm{u},s,s}^{\left( r,n \right)} \right|^2\left| \tilde{g}_{\mathrm{u},t,t}^{\left( r,n \right)} \right|^2}}} \right\} =\beta _{\mathrm{u},n}^{2}N_{\mathrm{c}}N_{\mathrm{b}}N_{\mathrm{r}}\left( \tau _{\mathrm{u},n}+\sum_{k=1}^{L_{\mathrm{u}}-1}{\sigma _{\mathrm{u},k,n}^{4}} \right) ,
\end{equation}
\begin{equation}\label{E_1221_eq9}
  \mathbb{E} \left\{ \sum_{b=0}^{N_{\mathrm{b}}-1}{\sum_{s=0}^{N_{\mathrm{c}}-1}{\sum_{r_1=0}^{N_{\mathrm{r}}-1}{\left| \tilde{g}_{\mathrm{u},s,s}^{\left( r_1,n \right)} \right|^2}\sum_{r_2\ne r_1}^{N_{\mathrm{r}}-1}{\left| \tilde{g}_{\mathrm{u},t,t}^{\left( r_2,n \right)} \right|^2}}} \right\} =\frac{\beta _{\mathrm{u},n}^{2}\varsigma _{\mathrm{u},n}^{2}}{\left( K_{\mathrm{u},n}+1 \right) ^2}N_{\mathrm{c}}N_{\mathrm{b}}N_{\mathrm{r}}\left( N_{\mathrm{r}}-1 \right) ,
\end{equation}
\begin{equation*}
\hspace{-4.3cm}
  \mathbb{E} \left\{ \sum_{b=0}^{N_{\mathrm{b}}-1}{\sum_{s=0}^{N_{\mathrm{c}}-1}{\sum_{r_1=0}^{N_{\mathrm{r}}-1}{\tilde{g}_{\mathrm{u},s,s}^{\left( r_1,n \right)}\left( \tilde{g}_{\mathrm{u},t,t}^{\left( r_1,n \right)} \right) ^{\ast}}\sum_{r_2\ne r_1}^{N_{\mathrm{r}}-1}{\left( \tilde{g}_{\mathrm{u},s,s}^{\left( r_2,n \right)} \right) ^{\ast}\tilde{g}_{\mathrm{u},t,t}^{\left( r_2,n \right)}}}} \right\}
\end{equation*}
\begin{equation}\label{E_1221_eq10}
\hspace{-3.5cm}
  =\beta _{\mathrm{u},n}^{2}N_{\mathrm{c}}N_{\mathrm{b}}N_{\mathrm{r}}\left( N_{\mathrm{r}}-1 \right) \left( \frac{\sigma _{\mathrm{u},0,n}^{4}}{\left( K_{\mathrm{u},n}+1 \right) ^2}+\sum_{k=1}^{L_{\mathrm{u}}-1}{\sigma _{\mathrm{u},k,n}^{4}} \right).
\end{equation}
Substituting \eqref{E_1221_eq8}-\eqref{E_1221_eq10} into \eqref{E_1221_eq7},
and then substituting \eqref{E_1221_eq6} and \eqref{E_1221_eq7} into \eqref{E_1221_eq5},
we have the third expectation in \eqref{E_1221_eq1} calculated as
\begin{equation*}
  \mathbb{E} \left\{ \tilde{\mathbf{g}}_{\mathrm{u},nt}^{H}\mathbf{\Phi }^H\bar{\mathbf{G}}_{\mathrm{b}}^{H}\mathbf{\Upsilon }_{1221}\bar{\mathbf{G}}_{\mathrm{b}}\mathbf{\Phi }\tilde{\mathbf{g}}_{\mathrm{u},nt} \right\}
  = p_n\frac{\beta _{\mathrm{u},n}^{2}\beta _{\mathrm{b}}^{2}\sigma _{\mathrm{b},0}^{4}K_{\mathrm{b}}^{2}}{\left( K_{\mathrm{b}}+1 \right) ^2}N_{\mathrm{b}}N_{\mathrm{r}}\left( \tau _{\mathrm{u},n}+N_{\mathrm{r}}\sum_{k=1}^{L_{\mathrm{u}}-1}{\sigma _{\mathrm{u},k,n}^{4}}+\frac{\sigma _{\mathrm{u},0,n}^{4}\left( N_{\mathrm{r}}-1 \right)}{\left( K_{\mathrm{u},n}+1 \right) ^2} \right)
\end{equation*}
\begin{equation}\label{E_1221_eq11}
\hspace{-3.1cm}
  +\frac{\beta _{\mathrm{u},n}\beta _{\mathrm{b}}^{2}\sigma _{\mathrm{b},0}^{4}\varsigma _{\mathrm{u},n}K_{\mathrm{b}}^{2}}{\left( K_{\mathrm{u},n}+1 \right) \left( K_{\mathrm{b}}+1 \right) ^2}N_{\mathrm{b}}N_{\mathrm{r}}\left( N_{\mathrm{r}}\sum_{u\ne n}^{N_{\mathrm{u}}-1}{p_u\frac{\beta _{\mathrm{u},u}\varsigma _{\mathrm{u},u}}{K_{\mathrm{u},u}+1}}+p_n\frac{\beta _{\mathrm{u},n}\varsigma _{\mathrm{u},n}}{K_{\mathrm{u},n}+1}\left( N_{\mathrm{r}}-1 \right) \right) .
\end{equation}

Furthermore, the fourth expectation in \eqref{E_1221_eq1} can be derived as
\begin{equation*}
\hspace{-6cm}
  \mathbb{E} \left\{ \tilde{\mathbf{g}}_{\mathrm{u},nt}^{H}\mathbf{\Phi }^H\tilde{\mathbf{G}}_{\mathrm{b}}^{H}\mathbf{\Upsilon }_{1221}\tilde{\mathbf{G}}_{\mathrm{b}}\mathbf{\Phi }\tilde{\mathbf{g}}_{\mathrm{u},nt} \right\}
\end{equation*}
\begin{equation*}
  =\mathbb{E} \left\{ \sum_{b=0}^{N_{\mathrm{b}}-1}{\gamma _{1221}\left( b \right) \sum_{r_1=0}^{N_{\mathrm{r}}-1}{\sum_{r_2=0}^{N_{\mathrm{r}}-1}{\left( e^{j\varphi _{r_1}}\tilde{g}_{\mathrm{u},t,t}^{\left( r_1,n \right)}\tilde{g}_{\mathrm{b},t,t}^{\left( b,r_1 \right)} \right) ^{\ast}e^{j\varphi _{r_2}}\tilde{g}_{\mathrm{u},t,t}^{\left( r_2,n \right)}\tilde{g}_{\mathrm{b},t,t}^{\left( b,r_2 \right)}}}} \right\}
\end{equation*}
\begin{equation*}
\hspace{-0.5cm}
  =\frac{\beta _{\mathrm{b}}^{2}\sigma _{\mathrm{b},0}^{2}\varsigma _{\mathrm{b}}K_{\mathrm{b}}}{N_{\mathrm{c}}\left( K_{\mathrm{b}}+1 \right) ^2}\mathbb{E} \left\{ \sum_{b=0}^{N_{\mathrm{b}}-1}{\sum_{s=0}^{N_{\mathrm{c}}-1}{\sum_{u=0}^{N_{\mathrm{u}}-1}{p_u\sum_{r_1=0}^{N_{\mathrm{r}}-1}{\left| \tilde{g}_{\mathrm{u},s,s}^{\left( r_1,u \right)} \right|^2}\sum_{r_2=0}^{N_{\mathrm{r}}-1}{\left| \tilde{g}_{\mathrm{u},t,t}^{\left( r_2,n \right)} \right|^2}}}} \right\}
\end{equation*}
\begin{equation*}
\hspace{-1.4cm}
  \overset{\left( a \right)}{=} p_n \frac{\beta _{\mathrm{b}}^{2}\sigma _{\mathrm{b},0}^{2}\varsigma _{\mathrm{b}}K_{\mathrm{b}}}{N_{\mathrm{c}}\left( K_{\mathrm{b}}+1 \right) ^2}\mathbb{E} \left\{ \sum_{b=0}^{N_{\mathrm{b}}-1}{\sum_{s=0}^{N_{\mathrm{c}}-1}{\sum_{r_1=0}^{N_{\mathrm{r}}-1}{\left| \tilde{g}_{\mathrm{u},s,s}^{\left( r_1,n \right)} \right|^2\sum_{r_2=0}^{N_{\mathrm{r}}-1}{\left| \tilde{g}_{\mathrm{u},t,t}^{\left( r_2,n \right)} \right|^2}}}} \right\}
\end{equation*}
\begin{equation}\label{E_1221_eq12}
\hspace{-0.5cm}
  +\frac{\beta _{\mathrm{b}}^{2}\sigma _{\mathrm{b},0}^{2}\varsigma _{\mathrm{b}}K_{\mathrm{b}}}{N_{\mathrm{c}}\left( K_{\mathrm{b}}+1 \right) ^2}\mathbb{E} \left\{ \sum_{b=0}^{N_{\mathrm{b}}-1}{\sum_{s=0}^{N_{\mathrm{c}}-1}{\sum_{u\ne n}^{N_{\mathrm{u}}-1}{p_u\sum_{r_1=0}^{N_{\mathrm{r}}-1}{\left| \tilde{g}_{\mathrm{u},s,s}^{\left( r_1,u \right)} \right|^2}\sum_{r_2=0}^{N_{\mathrm{r}}-1}{\left| \tilde{g}_{\mathrm{u},t,t}^{\left( r_2,n \right)} \right|^2}}}} \right\} .
\end{equation}
Step $\left( a \right)$ divides \eqref{E_1221_eq12} into two parts,
based on whether or not the parameter $u$ is equal to $n$.
According to \eqref{E_1221_eq8} and \eqref{E_1221_eq9},
the first expectation in \eqref{E_1221_eq12} is obtained as
\begin{equation}\label{E_1221_eq13}
  \mathbb{E} \left\{ \sum_{b=0}^{N_{\mathrm{b}}-1}{\sum_{s=0}^{N_{\mathrm{c}}-1}{\sum_{r_1=0}^{N_{\mathrm{r}}-1}{\left| \tilde{g}_{\mathrm{u},s,s}^{\left( r_1,n \right)} \right|^2}\sum_{r_2=0}^{N_{\mathrm{r}}-1}{\left| \tilde{g}_{\mathrm{u},t,t}^{\left( r_2,n \right)} \right|^2}}} \right\} =\beta _{\mathrm{u},n}^{2}N_{\mathrm{c}}N_{\mathrm{b}}N_{\mathrm{r}}\left( \tau _{\mathrm{u},n}+\sum_{k=1}^{L_{\mathrm{u}}-1}{\sigma _{\mathrm{u},k,n}^{4}}+\frac{\varsigma _{\mathrm{u},n}^{2}\left( N_{\mathrm{r}}-1 \right)}{\left( K_{\mathrm{u},n}+1 \right) ^2} \right) .
\end{equation}
From \eqref{R-U_FD_channel_ij_tt_NLoS},
the second expectation in \eqref{E_1221_eq12} is calculated as
\begin{equation}\label{E_1221_eq14}
  \mathbb{E} \left\{ \sum_{b=0}^{N_{\mathrm{b}}-1}{\sum_{s=0}^{N_{\mathrm{c}}-1}{\sum_{u\ne n}^{N_{\mathrm{u}}-1}{p_u\sum_{r_1=0}^{N_{\mathrm{r}}-1}{\left| \tilde{g}_{\mathrm{u},s,s}^{\left( r_1,u \right)} \right|^2}\sum_{r_2=0}^{N_{\mathrm{r}}-1}{\left| \tilde{g}_{\mathrm{u},t,t}^{\left( r_2,n \right)} \right|^2}}}} \right\} =\frac{\beta _{\mathrm{u},n}\varsigma _{\mathrm{u},n}}{K_{\mathrm{u},n}+1}N_{\mathrm{c}}N_{\mathrm{b}}N_{\mathrm{r}}^{2}\sum_{u\ne n}^{N_{\mathrm{u}}-1}{p_u\frac{\beta _{\mathrm{u},u}\varsigma _{\mathrm{u},u}}{K_{\mathrm{u},u}+1}}  .
\end{equation}
Thus, substituting \eqref{E_1221_eq13} and \eqref{E_1221_eq14} into \eqref{E_1221_eq12},
we have
\begin{equation*}
\hspace{-0.3cm}
  \mathbb{E} \left\{ \tilde{\mathbf{g}}_{\mathrm{u},nt}^{H}\mathbf{\Phi }^H\tilde{\mathbf{G}}_{\mathrm{b}}^{H}\mathbf{\Upsilon }_{1221}\tilde{\mathbf{G}}_{\mathrm{b}}\mathbf{\Phi }\tilde{\mathbf{g}}_{\mathrm{u},nt} \right\}
  =  p_n  \frac{\beta _{\mathrm{u},n}^{2}\beta _{\mathrm{b}}^{2}\sigma _{\mathrm{b},0}^{2}\varsigma _{\mathrm{b}}K_{\mathrm{b}}}{\left( K_{\mathrm{b}}+1 \right) ^2}N_{\mathrm{b}}N_{\mathrm{r}}\left( \tau _{\mathrm{u},n}+\sum_{k=1}^{L_{\mathrm{u}}-1}{\sigma _{\mathrm{u},k,n}^{4}} \right)
\end{equation*}
\begin{equation}\label{E_1221_eq15}
  +\frac{\beta _{\mathrm{u},n}\beta _{\mathrm{b}}^{2}\sigma _{\mathrm{b},0}^{2}\varsigma _{\mathrm{u},n}\varsigma _{\mathrm{b}}K_{\mathrm{b}}}{\left( K_{\mathrm{u},n}+1 \right) \left( K_{\mathrm{b}}+1 \right) ^2}N_{\mathrm{b}}N_{\mathrm{r}}\left( N_{\mathrm{r}}\sum_{u\ne n}^{N_{\mathrm{u}}-1}{p_u\frac{\beta _{\mathrm{u},u}\varsigma _{\mathrm{u},u}}{K_{\mathrm{u},u}+1}}+p_n\frac{\beta _{\mathrm{u},n}\varsigma _{\mathrm{u},n}}{K_{\mathrm{u},n}+1}\left( N_{\mathrm{r}}-1 \right) \right)
 .
\end{equation}

Substituting \eqref{E_1221_eq3}, \eqref{E_1221_eq4}, \eqref{E_1221_eq11} and \eqref{E_1221_eq15} into \eqref{E_1221_eq1},
we arrive at
\begin{equation*}
\hspace{-7.5cm}
  \mathbb{E} \left\{ \mathbf{g}_{\mathrm{u},nt}^{H}\mathbf{\Phi }^H\mathbf{G}_{\mathrm{b}}^{H}\mathbf{\Upsilon }_{1221}\mathbf{G}_{\mathrm{b}}\mathbf{\Phi g}_{\mathrm{u},nt} \right\}
\end{equation*}
\begin{equation*}
  =\frac{\beta _{\mathrm{u},n}\beta _{\mathrm{b}}^{2}\sigma _{\mathrm{u},0,n}^{2}\sigma _{\mathrm{b},0}^{2}K_{\mathrm{u},n}K_{\mathrm{b}}}{\left( K_{\mathrm{u},n}+1 \right) \left( K_{\mathrm{b}}+1 \right) ^2}N_{\mathrm{b}}N_{\mathrm{r}}\left( \sigma _{\mathrm{b},0}^{2}K_{\mathrm{b}}\left| \varPhi _{N_{\mathrm{r}}}\left( n \right) \right|^2+\varsigma _{\mathrm{b}}N_{\mathrm{r}} \right) \sum_{u=0}^{N_{\mathrm{u}}-1}{ p_u  \frac{\beta _{\mathrm{u},u}\varsigma _{\mathrm{u},u}}{K_{\mathrm{u},u}+1}}
\end{equation*}
\begin{equation*}
  +  p_n \frac{\beta _{\mathrm{u},n}^{2}\beta _{\mathrm{b}}^{2}\sigma _{\mathrm{b},0}^{2}K_{\mathrm{b}}}{\left( K_{\mathrm{b}}+1 \right) ^2}N_{\mathrm{b}}N_{\mathrm{r}}\left( \left( \sigma _{\mathrm{b},0}^{2}K_{\mathrm{b}}+\varsigma _{\mathrm{b}} \right) \tau _{\mathrm{u},n}+\left( \sigma _{\mathrm{b},0}^{2}K_{\mathrm{b}}N_{\mathrm{r}}+\varsigma _{\mathrm{b}} \right) \sum_{k=1}^{L_{\mathrm{u}}-1}{\sigma _{\mathrm{u},k,n}^{4}} \right)
\end{equation*}
\begin{equation*}
  +  p_n  \frac{\beta _{\mathrm{u},n}^{2}\beta _{\mathrm{b}}^{2}\sigma _{\mathrm{b},0}^{2}K_{\mathrm{b}}}{\left( K_{\mathrm{u},n}+1 \right) ^2\left( K_{\mathrm{b}}+1 \right) ^2}\left( \sigma _{\mathrm{u},0,n}^{4}\sigma _{\mathrm{b},0}^{2}K_{\mathrm{b}}+\sigma _{\mathrm{b},0}^{2}\varsigma _{\mathrm{u},n}^{2}K_{\mathrm{b}}+\varsigma _{\mathrm{u},n}^{2}\varsigma _{\mathrm{b}} \right) N_{\mathrm{b}}N_{\mathrm{r}}\left( N_{\mathrm{r}}-1 \right)
\end{equation*}
\begin{equation}\label{E_1221_eq16}
\hspace{-3cm}
  +\frac{\beta _{\mathrm{u},n}\beta _{\mathrm{b}}^{2}\sigma _{\mathrm{b},0}^{2}\varsigma _{\mathrm{u},n}K_{\mathrm{b}}}{\left( K_{\mathrm{u},n}+1 \right) \left( K_{\mathrm{b}}+1 \right) ^2}N_{\mathrm{b}}N_{\mathrm{r}}^{2}\left( \sigma _{\mathrm{b},0}^{2}K_{\mathrm{b}}+\varsigma _{\mathrm{b}} \right) \sum_{u\ne n}^{N_{\mathrm{u}}-1}{  p_u  \frac{\beta _{\mathrm{u},u}\varsigma _{\mathrm{u},u}}{K_{\mathrm{u},u}+1}} .
\end{equation}

\subsubsection{\textbf{ The $\mathbf{\Upsilon }_{2112}$-Related Expectation}} \label{E_2112}

The expectation $\mathbb{E} \left\{ \mathbf{g}_{\mathrm{u},nt}^{H}\mathbf{\Phi }^H\mathbf{G}_{\mathrm{b}}^{H}\mathbf{\Upsilon }_{2112}\mathbf{G}_{\mathrm{b}}\mathbf{\Phi g}_{\mathrm{u},nt} \right\}$ can be expanded as
\begin{equation*}
\hspace{-7.5cm}
  \mathbb{E} \left\{ \mathbf{g}_{\mathrm{u},nt}^{H}\mathbf{\Phi }^H\mathbf{G}_{\mathrm{b}}^{H}\mathbf{\Upsilon }_{2112}\mathbf{G}_{\mathrm{b}}\mathbf{\Phi g}_{\mathrm{u},nt} \right\}
\end{equation*}
\begin{equation*}
  =\mathbb{E} \left\{ \left( \bar{\mathbf{g}}_{\mathrm{u},nt}^{H}+\tilde{\mathbf{g}}_{\mathrm{u},nt}^{H} \right) \mathbf{\Phi }^H\left( \bar{\mathbf{G}}_{\mathrm{b}}^{H}+\tilde{\mathbf{G}}_{\mathrm{b}}^{H} \right) \mathbf{\Upsilon }_{2112}\left( \bar{\mathbf{G}}_{\mathrm{b}}+\tilde{\mathbf{G}}_{\mathrm{b}} \right) \mathbf{\Phi }\left( \bar{\mathbf{g}}_{\mathrm{u},nt}+\tilde{\mathbf{g}}_{\mathrm{u},nt} \right) \right\}
\end{equation*}
\begin{equation*}
\hspace{-0.9cm}
  =\mathbb{E} \left\{ \bar{\mathbf{g}}_{\mathrm{u},nt}^{H}\mathbf{\Phi }^H\bar{\mathbf{G}}_{\mathrm{b}}^{H}\mathbf{\Upsilon }_{2112}\bar{\mathbf{G}}_{\mathrm{b}}\mathbf{\Phi }\bar{\mathbf{g}}_{\mathrm{u},nt} \right\}
  +\mathbb{E} \left\{ \bar{\mathbf{g}}_{\mathrm{u},nt}^{H}\mathbf{\Phi }^H\tilde{\mathbf{G}}_{\mathrm{b}}^{H}\mathbf{\Upsilon }_{2112}\tilde{\mathbf{G}}_{\mathrm{b}}\mathbf{\Phi }\bar{\mathbf{g}}_{\mathrm{u},nt} \right\}
\end{equation*}
\begin{equation}\label{E_2112_eq1}
\hspace{-0.8cm}
  +\mathbb{E} \left\{ \tilde{\mathbf{g}}_{\mathrm{u},nt}^{H}\mathbf{\Phi }^H\bar{\mathbf{G}}_{\mathrm{b}}^{H}\mathbf{\Upsilon }_{2112}\bar{\mathbf{G}}_{\mathrm{b}}\mathbf{\Phi }\tilde{\mathbf{g}}_{\mathrm{u},nt} \right\}
  +\mathbb{E} \left\{ \tilde{\mathbf{g}}_{\mathrm{u},nt}^{H}\mathbf{\Phi }^H\tilde{\mathbf{G}}_{\mathrm{b}}^{H}\mathbf{\Upsilon }_{2112}\tilde{\mathbf{G}}_{\mathrm{b}}\mathbf{\Phi }\tilde{\mathbf{g}}_{\mathrm{u},nt} \right\}.
\end{equation}
Meanwhile, the $(bN_{\mathrm{c}}+t)$-th element of the diagonal matrix $\mathbf{\Upsilon }_{2112}$ is expressed as
\begin{equation}\label{E_2112_eq2}
  \left[ \mathbf{\Upsilon }_{2112} \right] _{bt,bt}=\frac{1}{N_{\mathrm{c}}}\sum_{s=0}^{N_{\mathrm{c}}-1}{\sum_{u=0}^{N_{\mathrm{u}}-1}{p_u\left( \sum_{r_1=0}^{N_{\mathrm{r}}-1}{e^{j\varphi _{r_1}}\tilde{g}_{\mathrm{b},s,s}^{\left( b,r_1 \right)}\bar{g}_{\mathrm{u}}^{\left( r_1,u \right)}} \right) \left( \sum_{r_2=0}^{N_{\mathrm{r}}-1}{e^{j\varphi _{r_2}}\tilde{g}_{\mathrm{b},s,s}^{\left( b,r_2 \right)}\bar{g}_{\mathrm{u}}^{\left( r_2,u \right)}} \right) ^{\ast}}}\triangleq \gamma _{2112}\left( b \right).
\end{equation}

From \eqref{R-U_FD_channel_ij_tt}-\eqref{B-R_FD_channel_ij_tt_NLoS}, Lemma~\ref{lemma_sum} and Lemma~\ref{lemma_expectation},
the first and third expectations in \eqref{E_2112_eq1} can be easily obtained as
\begin{equation*}
\hspace{-6cm}
  \mathbb{E} \left\{ \bar{\mathbf{g}}_{\mathrm{u},nt}^{H}\mathbf{\Phi }^H\bar{\mathbf{G}}_{\mathrm{b}}^{H}\mathbf{\Upsilon }_{2112}\bar{\mathbf{G}}_{\mathrm{b}}\mathbf{\Phi }\bar{\mathbf{g}}_{\mathrm{u},nt} \right\}
\end{equation*}
\begin{equation}\label{E_2112_eq3}
  =\frac{\beta _{\mathrm{u},n}\beta _{\mathrm{b}}^{2}\sigma _{\mathrm{u},0,n}^{2}\sigma _{\mathrm{b},0}^{2}\varsigma _{\mathrm{b}}K_{\mathrm{u},n}K_{\mathrm{b}}}{\left( K_{\mathrm{u},n}+1 \right) \left( K_{\mathrm{b}}+1 \right) ^2}\left| \varPhi _{N_{\mathrm{r}}}\left( n \right) \right|^2N_{\mathrm{b}}N_{\mathrm{r}}\sum_{u=0}^{N_{\mathrm{u}}-1}{  p_u \frac{\beta _{\mathrm{u},u}K_{\mathrm{u},u}}{K_{\mathrm{u},u}+1}\sigma _{\mathrm{u},0,u}^{2}},
\end{equation}
\begin{equation}\label{E_2112_eq4}
  \mathbb{E} \left\{ \tilde{\mathbf{g}}_{\mathrm{u},nt}^{H}\mathbf{\Phi }^H\bar{\mathbf{G}}_{\mathrm{b}}^{H}\mathbf{\Upsilon }_{2112}\bar{\mathbf{G}}_{\mathrm{b}}\mathbf{\Phi }\tilde{\mathbf{g}}_{\mathrm{u},nt} \right\} =\frac{\beta _{\mathrm{u},n}\beta _{\mathrm{b}}^{2}\sigma _{\mathrm{b},0}^{2}\varsigma _{\mathrm{u},n}\varsigma _{\mathrm{b}}K_{\mathrm{b}}}{\left( K_{\mathrm{u},n}+1 \right) \left( K_{\mathrm{b}}+1 \right) ^2}N_{\mathrm{b}}N_{\mathrm{r}}^{2}\sum_{u=0}^{N_{\mathrm{u}}-1}{  p_u \frac{\beta _{\mathrm{u},u}K_{\mathrm{u},u}}{K_{\mathrm{u},u}+1}\sigma _{\mathrm{u},0,u}^{2}} .
\end{equation}

The second expectation in \eqref{E_2112_eq1} can be further expanded as
\begin{equation*}
  \mathbb{E} \left\{ \bar{\mathbf{g}}_{\mathrm{u},nt}^{H}\mathbf{\Phi }^H\tilde{\mathbf{G}}_{\mathrm{b}}^{H}\mathbf{\Upsilon }_{2112}\tilde{\mathbf{G}}_{\mathrm{b}}\mathbf{\Phi }\bar{\mathbf{g}}_{\mathrm{u},nt} \right\}
  =c_{2112}^{2}\mathbb{E} \left\{ \sum_{b=0}^{N_{\mathrm{b}}-1}{\sum_{s=0}^{N_{\mathrm{c}}-1}{\sum_{u=0}^{N_{\mathrm{u}}-1}{p_u\frac{\beta _{\mathrm{u},u}\sigma _{\mathrm{u},0,u}^{2}K_{\mathrm{u},u}}{K_{\mathrm{u},u}+1}\mathrm{term}_{2112}^{1}\left( b,s,u \right)}}} \right\}
\end{equation*}
\begin{equation*}
\hspace{-5cm}
  \overset{\left( a \right)}{=}c_{2112}^{2}\mathbb{E} \left\{ \sum_{b=0}^{N_{\mathrm{b}}-1}{\sum_{s=0}^{N_{\mathrm{c}}-1}{\sum_{r=0}^{N_{\mathrm{r}}-1}{\left| \tilde{g}_{\mathrm{b},s,s}^{\left( b,r \right)} \right|^2\left| \tilde{g}_{\mathrm{b},t,t}^{\left( b,r \right)} \right|^2}}} \right\} \sum_{u=0}^{N_{\mathrm{u}}-1}{p_u\frac{\beta _{\mathrm{u},u}\sigma _{\mathrm{u},0,u}^{2}K_{\mathrm{u},u}}{K_{\mathrm{u},u}+1}}
\end{equation*}
\begin{equation*}
\hspace{-4cm}
  +c_{2112}^{2}\mathbb{E} \left\{ \sum_{b=0}^{N_{\mathrm{b}}-1}{\sum_{s=0}^{N_{\mathrm{c}}-1}{\sum_{r_1=0}^{N_{\mathrm{r}}-1}{\left| \tilde{g}_{\mathrm{b},s,s}^{\left( b,r_1 \right)} \right|^2}}\sum_{r_2\ne r_1}^{N_{\mathrm{r}}-1}{\left| \tilde{g}_{\mathrm{b},t,t}^{\left( b,r_2 \right)} \right|^2}} \right\} \sum_{u=0}^{N_{\mathrm{u}}-1}{p_u\frac{\beta _{\mathrm{u},u}\sigma _{\mathrm{u},0,u}^{2}K_{\mathrm{u},u}}{K_{\mathrm{u},u}+1}}
\end{equation*}
\begin{equation}\label{E_2112_eq5}
\hspace{-5.6cm}
  +c_{2112}^{2}\mathbb{E} \left\{ \sum_{b=0}^{N_{\mathrm{b}}-1}{\sum_{s=0}^{N_{\mathrm{c}}-1}{\sum_{u=0}^{N_{\mathrm{u}}-1}{p_u\frac{\beta _{\mathrm{u},u}\sigma _{\mathrm{u},0,u}^{2}K_{\mathrm{u},u}}{K_{\mathrm{u},u}+1}\mathrm{term}_{2112}^{2}\left( b,s,u \right)}}} \right\} ,
\end{equation}
where $c_{2112}^{2}=\frac{\beta _{\mathrm{u},n}\sigma _{\mathrm{u},0,n}^{2}K_{\mathrm{u},n}}{N_{\mathrm{c}}\left( K_{\mathrm{u},n}+1 \right)}$,
\begin{equation*}
\hspace{-4.3cm}
  \mathrm{term}_{2112}^{1}\left( b,s,u \right) =\sum_{r_1=0}^{N_{\mathrm{r}}-1}{e^{j\varphi _{r_1}}\bar{h}_{\mathrm{u}}^{\left( r_1,u \right)}\tilde{g}_{\mathrm{b},s,s}^{\left( b,r_1 \right)}}\sum_{r_2=0}^{N_{\mathrm{r}}-1}{\left( e^{j\varphi _{r_2}}\bar{h}_{\mathrm{u}}^{\left( r_2,u \right)}\tilde{g}_{\mathrm{b},s,s}^{\left( b,r_2 \right)} \right) ^{\ast}}
\end{equation*}
\begin{equation}
\hspace{-0.33cm}
  \times
  \sum_{r_{11}=0}^{N_{\mathrm{r}}-1}{\left( e^{j\varphi _{r_{11}}}\bar{h}_{\mathrm{u}}^{\left( r_{11},n \right)}\tilde{g}_{\mathrm{b},t,t}^{\left( b,r_{11} \right)} \right) ^{\ast}}
  \sum_{r_{22}=0}^{N_{\mathrm{r}}-1}{e^{j\varphi _{r_{22}}}\bar{h}_{\mathrm{u}}^{\left( r_{22},n \right)}\tilde{g}_{\mathrm{b},t,t}^{\left( b,r_{22} \right)}} ,
\end{equation}
\begin{equation}
  \mathrm{term}_{2112}^{2}\left( b,s,u \right) =\sum_{r_1=0}^{N_{\mathrm{r}}-1}{\bar{h}_{\mathrm{u}}^{\left( r_1,u \right)}\tilde{g}_{\mathrm{b},s,s}^{\left( b,r_1 \right)}\left( \bar{h}_{\mathrm{u}}^{\left( r_1,n \right)}\tilde{g}_{\mathrm{b},t,t}^{\left( b,r_1 \right)} \right) ^{\ast}}\sum_{r_2\ne r_1}^{N_{\mathrm{r}}-1}{\left( \bar{h}_{\mathrm{u}}^{\left( r_2,u \right)}\tilde{g}_{\mathrm{b},s,s}^{\left( b,r_2 \right)} \right) ^{\ast}\bar{h}_{\mathrm{u}}^{\left( r_2,n \right)}\tilde{g}_{\mathrm{b},t,t}^{\left( b,r_2 \right)}}  .
\end{equation}
Similar to \eqref{E_1221_eq7},
step $\left( a \right)$ in \eqref{E_2112_eq5} divides the equation into three parts:
1) $ r_1 = r_2 = r_{11} = r_{22} $,
2) $ r_1 = r_2 \neq r_{11} = r_{22} $,
and
3) $ r_1 = r_{11} \neq r_2 = r_{22} $.
Also, from \eqref{R-U_FD_channel_ij_tt}-\eqref{B-R_FD_channel_ij_tt_NLoS}, Lemma~\ref{lemma_sum}, and Lemma~\ref{lemma_expectation},
the three expectations in \eqref{E_2112_eq5} are calculated respectively as
\begin{equation}\label{E_2112_eq6}
\hspace{-4.1cm}
  \mathbb{E} \left\{ \sum_{b=0}^{N_{\mathrm{b}}-1}{\sum_{s=0}^{N_{\mathrm{c}}-1}{\sum_{r=0}^{N_{\mathrm{r}}-1}{\left| \tilde{g}_{\mathrm{b},s,s}^{\left( b,r \right)} \right|^2\left| \tilde{g}_{\mathrm{b},t,t}^{\left( b,r \right)} \right|^2}}} \right\} =\beta _{\mathrm{b}}^{2}\left( \tau _{\mathrm{b}}+\sum_{k=1}^{L_{\mathrm{b}}-1}{\sigma _{\mathrm{b},k}^{4}} \right) N_{\mathrm{c}}N_{\mathrm{b}}N_{\mathrm{r}} ,
\end{equation}
\begin{equation}\label{E_2112_eq7}
\hspace{-3.1cm}
  \mathbb{E} \left\{ \sum_{b=0}^{N_{\mathrm{b}}-1}{\sum_{s=0}^{N_{\mathrm{c}}-1}{\sum_{r_1=0}^{N_{\mathrm{r}}-1}{\left| \tilde{g}_{\mathrm{b},s,s}^{\left( b,r_1 \right)} \right|^2}}\sum_{r_2\ne r_1}^{N_{\mathrm{r}}-1}{\left| \tilde{g}_{\mathrm{b},t,t}^{\left( b,r_2 \right)} \right|^2}} \right\} =\frac{\beta _{\mathrm{b}}^{2}\varsigma _{\mathrm{b}}^{2}}{\left( K_{\mathrm{b}}+1 \right) ^2}N_{\mathrm{c}}N_{\mathrm{b}}N_{\mathrm{r}}\left( N_{\mathrm{r}}-1 \right) ,
\end{equation}
\begin{equation*}
\hspace{-6.8cm}
  \mathbb{E} \left\{ \sum_{b=0}^{N_{\mathrm{b}}-1}{\sum_{s=0}^{N_{\mathrm{c}}-1}{\sum_{u=0}^{N_{\mathrm{u}}-1}{p_u\frac{\beta _{\mathrm{u},u}\sigma _{\mathrm{u},0,u}^{2}K_{\mathrm{u},u}}{K_{\mathrm{u},u}+1}\mathrm{term}_{2112}^{2}\left( b,s,u \right)}}} \right\}
\end{equation*}
\begin{equation}\label{E_2112_eq8}
  =\beta _{\mathrm{b}}^{2}\left( \frac{\sigma _{\mathrm{b},0}^{4}}{\left( K_{\mathrm{b}}+1 \right) ^2}+\sum_{k=1}^{L_{\mathrm{b}}-1}{\sigma _{\mathrm{b},k}^{4}} \right) N_{\mathrm{c}}N_{\mathrm{b}}\sum_{u=0}^{N_{\mathrm{u}}-1}{p_u\frac{\beta _{\mathrm{u},u}K_{\mathrm{u},u}}{K_{\mathrm{u},u}+1}\sigma _{\mathrm{u},0,u}^{2}\left( \left| \left( \bar{\mathbf{h}}_{\mathrm{u}}^{\left( \cdot ,n \right)} \right) ^H\bar{\mathbf{h}}_{\mathrm{u}}^{\left( \cdot ,u \right)} \right|^2-N_{\mathrm{r}} \right)} .
\end{equation}
Thus, substituting \eqref{E_2112_eq6}-\eqref{E_2112_eq8} into \eqref{E_2112_eq5}, we have \vspace{0.3cm}
\begin{equation*}
\hspace{-10.2cm}
  \mathbb{E} \left\{ \bar{\mathbf{g}}_{\mathrm{u},nt}^{H}\mathbf{\Phi }^H\tilde{\mathbf{G}}_{\mathrm{b}}^{H}\mathbf{\Upsilon }_{2112}\tilde{\mathbf{G}}_{\mathrm{b}}\mathbf{\Phi }\bar{\mathbf{g}}_{\mathrm{u},nt} \right\}
\end{equation*}
\begin{equation*}
\hspace{-1cm}
  =\frac{\beta _{\mathrm{u},n}\beta _{\mathrm{b}}^{2}\sigma _{\mathrm{u},0,n}^{2}K_{\mathrm{u},n}}{\left( K_{\mathrm{u},n}+1 \right) \left( K_{\mathrm{b}}+1 \right) ^2}N_{\mathrm{b}}N_{\mathrm{r}}\left( \tau _{\mathrm{b}}\left( K_{\mathrm{b}}+1 \right) ^2+\varsigma _{\mathrm{b}}^{2}\left( N_{\mathrm{r}}-1 \right) -\sigma _{\mathrm{b},0}^{4} \right) \sum_{u=0}^{N_{\mathrm{u}}-1}{\frac{\beta _{\mathrm{u},u}K_{\mathrm{u},u}}{K_{\mathrm{u},u}+1}\sigma _{\mathrm{u},0,u}^{2}}
\end{equation*}
\begin{equation}\label{E_2112_eq9}
  +\frac{\beta _{\mathrm{u},n}\beta _{\mathrm{b}}^{2}\sigma _{\mathrm{u},0,n}^{2}K_{\mathrm{u},n}}{K_{\mathrm{u},n}+1}\left( \frac{\sigma _{\mathrm{b},0}^{4}}{\left( K_{\mathrm{b}}+1 \right) ^2}+\sum_{k=1}^{L_{\mathrm{b}}-1}{\sigma _{\mathrm{b},k}^{4}} \right) N_{\mathrm{b}}\sum_{u=0}^{N_{\mathrm{u}}-1}{\frac{\beta _{\mathrm{u},u}\sigma _{\mathrm{u},0,u}^{2}K_{\mathrm{u},u}}{K_{\mathrm{u},u}+1}\left| \left( \bar{\mathbf{h}}_{\mathrm{u}}^{\left( \cdot ,n \right)} \right) ^H\bar{\mathbf{h}}_{\mathrm{u}}^{\left( \cdot ,u \right)} \right|^2} .
\end{equation}

From \eqref{R-U_FD_channel_ij_tt}-\eqref{B-R_FD_channel_ij_tt_NLoS},
the fourth expectation in \eqref{E_2112_eq1} is calculated as
\begin{equation*}
\hspace{-8.3cm}
  \mathbb{E} \left\{ \tilde{\mathbf{g}}_{\mathrm{u},nt}^{H}\mathbf{\Phi }^H\tilde{\mathbf{G}}_{\mathrm{b}}^{H}\mathbf{\Upsilon }_{2112}\tilde{\mathbf{G}}_{\mathrm{b}}\mathbf{\Phi }\tilde{\mathbf{g}}_{\mathrm{u},nt} \right\}
\end{equation*}
\begin{equation*}
\hspace{-2.2cm}
  =\mathbb{E} \left\{ \sum_{b=0}^{N_{\mathrm{b}}-1}{\gamma _{2112}\left( b \right) \sum_{r_1=0}^{N_{\mathrm{r}}-1}{\sum_{r_2=0}^{N_{\mathrm{r}}-1}{\left( e^{j\varphi _{r_1}}\tilde{g}_{\mathrm{u},t,t}^{\left( r_1,n \right)}\tilde{g}_{\mathrm{b},t,t}^{\left( b,r_1 \right)} \right) ^{\ast}e^{j\varphi _{r_2}}\tilde{g}_{\mathrm{u},t,t}^{\left( r_2,n \right)}\tilde{g}_{\mathrm{b},t,t}^{\left( b,r_2 \right)}}}} \right\}
\end{equation*}
\begin{equation*}
  =\frac{\beta _{\mathrm{u},n}\varsigma _{\mathrm{u},n}}{N_{\mathrm{c}}\left( K_{\mathrm{u},n}+1 \right)}\mathbb{E} \left\{ \sum_{b=0}^{N_{\mathrm{b}}-1}{\sum_{s=0}^{N_{\mathrm{c}}-1}{\sum_{r_1=0}^{N_{\mathrm{r}}-1}{\left| \tilde{g}_{\mathrm{b},s,s}^{\left( b,r_1 \right)} \right|^2}\sum_{r_2=0}^{N_{\mathrm{r}}-1}{\left| \tilde{g}_{\mathrm{b},t,t}^{\left( b,r_2 \right)} \right|^2}}} \right\} \sum_{u=0}^{N_{\mathrm{u}}-1}{  p_u  \frac{\beta _{\mathrm{u},u}K_{\mathrm{u},u}}{K_{\mathrm{u},u}+1}\sigma _{\mathrm{u},0,u}^{2}}
\end{equation*}
\begin{equation}\label{E_2112_eq10}
\hspace{-1.2cm}
  \overset{\left( a \right)}{=}\frac{\beta _{\mathrm{u},n}\beta _{\mathrm{b}}^{2}\varsigma _{\mathrm{u},n}}{K_{\mathrm{u},n}+1}\left( \tau _{\mathrm{b}}+\sum_{k=1}^{L_{\mathrm{b}}-1}{\sigma _{\mathrm{b},k}^{4}}+\frac{\varsigma _{\mathrm{b}}^{2}\left( N_{\mathrm{r}}-1 \right)}{\left( K_{\mathrm{b}}+1 \right) ^2} \right) N_{\mathrm{b}}N_{\mathrm{r}}\sum_{u=0}^{N_{\mathrm{u}}-1}{ p_u    \frac{\beta _{\mathrm{u},u}K_{\mathrm{u},u}}{K_{\mathrm{u},u}+1}\sigma _{\mathrm{u},0,u}^{2}} .
\end{equation}
Step $\left( a \right)$ in \eqref{E_2112_eq10} is based on
\begin{equation}\label{E_2112_eq11}
  \mathbb{E} \left\{ \sum_{b=0}^{N_{\mathrm{b}}-1}{\sum_{s=0}^{N_{\mathrm{c}}-1}{\sum_{r_1=0}^{N_{\mathrm{r}}-1}{\left| \tilde{g}_{\mathrm{b},s,s}^{\left( b,r_1 \right)} \right|^2}\sum_{r_2=0}^{N_{\mathrm{r}}-1}{\left| \tilde{g}_{\mathrm{b},t,t}^{\left( b,r_2 \right)} \right|^2}}} \right\} =\beta _{\mathrm{b}}^{2}N_{\mathrm{c}}N_{\mathrm{b}}N_{\mathrm{r}}\left( \tau _{\mathrm{b}}+\sum_{k=1}^{L_{\mathrm{b}}-1}{\sigma _{\mathrm{b},k}^{4}}+\frac{\varsigma _{\mathrm{b}}^{2}\left( N_{\mathrm{r}}-1 \right)}{\left( K_{\mathrm{b}}+1 \right) ^2} \right) ,
\end{equation}
which can be obtained by using \eqref{E_2112_eq6} and \eqref{E_2112_eq7}.

Substituting \eqref{E_2112_eq3}, \eqref{E_2112_eq4}, \eqref{E_2112_eq9} and \eqref{E_2112_eq10} into \eqref{E_2112_eq1}, we arrive at
\begin{equation*}
  \mathbb{E} \left\{ \mathbf{g}_{\mathrm{u},nt}^{H}\mathbf{\Phi }^H\mathbf{G}_{\mathrm{b}}^{H}\mathbf{\Upsilon }_{2112}\mathbf{G}_{\mathrm{b}}\mathbf{\Phi g}_{\mathrm{u},nt} \right\} =\frac{\beta _{\mathrm{u},n}\beta _{\mathrm{b}}^{2}N_{\mathrm{b}}N_{\mathrm{r}}}{\left( K_{\mathrm{u},n}+1 \right) \left( K_{\mathrm{b}}+1 \right) ^2}\mathrm{term}_{2112}^{3}\sum_{u=0}^{N_{\mathrm{u}}-1}{p_u\frac{\beta _{\mathrm{u},u}K_{\mathrm{u},u}}{K_{\mathrm{u},u}+1}\sigma _{\mathrm{u},0,u}^{2}}
\end{equation*}
\begin{equation}\label{E_2112_eq12}
  +\frac{\beta _{\mathrm{u},n}\beta _{\mathrm{b}}^{2}\sigma _{\mathrm{u},0,n}^{2}K_{\mathrm{u},n}}{K_{\mathrm{u},n}+1}\left( \frac{\sigma _{\mathrm{b},0}^{4}}{\left( K_{\mathrm{b}}+1 \right) ^2}+\sum_{k=1}^{L_{\mathrm{b}}-1}{\sigma _{\mathrm{b},k}^{4}} \right) N_{\mathrm{b}}\sum_{u=0}^{N_{\mathrm{u}}-1}{  p_u   \frac{\beta _{\mathrm{u},u}K_{\mathrm{u},u}}{K_{\mathrm{u},u}+1}\sigma _{\mathrm{u},0,u}^{2}\left| \left( \bar{\mathbf{h}}_{\mathrm{u}}^{\left( \cdot ,n \right)} \right) ^H\bar{\mathbf{h}}_{\mathrm{u}}^{\left( \cdot ,u \right)} \right|^2} ,
\end{equation}
where \vspace{0.3cm}
\begin{equation*}
\hspace{-1.8cm}
  \mathrm{term}_{2112}^{3}=\sigma _{\mathrm{u},0,n}^{2}\sigma _{\mathrm{b},0}^{2}\varsigma _{\mathrm{b}}K_{\mathrm{u},n}K_{\mathrm{b}}\left| \varPhi _{N_{\mathrm{r}}}\left( n \right) \right|^2+\sigma _{\mathrm{u},0,n}^{2}\varsigma _{\mathrm{b}}^{2}K_{\mathrm{u},n}\left( N_{\mathrm{r}}-1 \right) +\sigma _{\mathrm{b},0}^{2}\varsigma _{\mathrm{u},n}\varsigma _{\mathrm{b}}K_{\mathrm{b}}N_{\mathrm{r}}
\end{equation*}
\begin{equation}
  +\varsigma _{\mathrm{u},n}\varsigma _{\mathrm{b}}^{2}\left( N_{\mathrm{r}}-1 \right) -\sigma _{\mathrm{u},0,n}^{2}\sigma _{\mathrm{b},0}^{4}K_{\mathrm{u},n}+\left( K_{\mathrm{b}}+1 \right) ^2\left( \sigma _{\mathrm{u},0,n}^{2}\tau _{\mathrm{b}}K_{\mathrm{u},n}+\varsigma _{\mathrm{u},n}\tau _{\mathrm{b}}+\varsigma _{\mathrm{u},n}\sum_{k=1}^{L_{\mathrm{b}}-1}{\sigma _{\mathrm{b},k}^{4}} \right).
\end{equation}

\subsubsection{\textbf{ The $\mathbf{\Upsilon }_{2222}$-Related Expectation}} \label{E_2222}

With the zero-value terms removed,
the expectation $\mathbb{E} \left\{ \mathbf{g}_{\mathrm{u},nt}^{H}\mathbf{\Phi }^H\mathbf{G}_{\mathrm{b}}^{H}\mathbf{\Upsilon }_{2222}\mathbf{G}_{\mathrm{b}}\mathbf{\Phi g}_{\mathrm{u},nt} \right\}$ is expressed as
\begin{equation*}
\hspace{-7.2cm}
  \mathbb{E} \left\{ \mathbf{g}_{\mathrm{u},nt}^{H}\mathbf{\Phi }^H\mathbf{G}_{\mathrm{b}}^{H}\mathbf{\Upsilon }_{2222}\mathbf{G}_{\mathrm{b}}\mathbf{\Phi g}_{\mathrm{u},nt} \right\}
\end{equation*}
\begin{equation*}
  =\mathbb{E} \left\{ \left( \bar{\mathbf{g}}_{\mathrm{u},nt}^{H}+\tilde{\mathbf{g}}_{\mathrm{u},nt}^{H} \right) \mathbf{\Phi }^H\left( \bar{\mathbf{G}}_{\mathrm{b}}^{H}+\tilde{\mathbf{G}}_{\mathrm{b}}^{H} \right) \mathbf{\Upsilon }_{2222}\left( \bar{\mathbf{G}}_{\mathrm{b}}+\tilde{\mathbf{G}}_{\mathrm{b}} \right) \mathbf{\Phi }\left( \bar{\mathbf{g}}_{\mathrm{u},nt}+\tilde{\mathbf{g}}_{\mathrm{u},nt} \right) \right\}
\end{equation*}
\begin{equation*}
\hspace{-0.8cm}
  =\mathbb{E} \left\{ \bar{\mathbf{g}}_{\mathrm{u},nt}^{H}\mathbf{\Phi }^H\bar{\mathbf{G}}_{\mathrm{b}}^{H}\mathbf{\Upsilon }_{2222}\bar{\mathbf{G}}_{\mathrm{b}}\mathbf{\Phi }\bar{\mathbf{g}}_{\mathrm{u},nt} \right\}
  +\mathbb{E} \left\{ \bar{\mathbf{g}}_{\mathrm{u},nt}^{H}\mathbf{\Phi }^H\tilde{\mathbf{G}}_{\mathrm{b}}^{H}\mathbf{\Upsilon }_{2222}\tilde{\mathbf{G}}_{\mathrm{b}}\mathbf{\Phi }\bar{\mathbf{g}}_{\mathrm{u},nt} \right\}
\end{equation*}
\begin{equation}\label{E_2222_eq1}
\hspace{-0.7cm}
  +\mathbb{E} \left\{ \tilde{\mathbf{g}}_{\mathrm{u},nt}^{H}\mathbf{\Phi }^H\bar{\mathbf{G}}_{\mathrm{b}}^{H}\mathbf{\Upsilon }_{2222}\bar{\mathbf{G}}_{\mathrm{b}}\mathbf{\Phi }\tilde{\mathbf{g}}_{\mathrm{u},nt} \right\}
  +\mathbb{E} \left\{ \tilde{\mathbf{g}}_{\mathrm{u},nt}^{H}\mathbf{\Phi }^H\tilde{\mathbf{G}}_{\mathrm{b}}^{H}\mathbf{\Upsilon }_{2222}\tilde{\mathbf{G}}_{\mathrm{b}}\mathbf{\Phi }\tilde{\mathbf{g}}_{\mathrm{u},nt} \right\}.
\end{equation}
Besides, the $(bN_{\mathrm{c}}+t)$-th element of the diagonal matrix $\mathbf{\Upsilon }_{2222}$ is expressed as
\begin{equation}\label{E_2222_eq2}
  \left[ \mathbf{\Upsilon }_{2222} \right] _{bt,bt}=\frac{1}{N_{\mathrm{c}}}\sum_{s=0}^{N_{\mathrm{c}}-1}{\sum_{u=0}^{N_{\mathrm{u}}-1}{p_u\left( \sum_{r_1=0}^{N_{\mathrm{r}}-1}{e^{j\varphi _{r_1}}\tilde{g}_{\mathrm{b},s,s}^{\left( b,r_1 \right)}\tilde{g}_{\mathrm{u},s,s}^{\left( r_1,u \right)}} \right) \left( \sum_{r_2=0}^{N_{\mathrm{r}}-1}{e^{j\varphi _{r_2}}\tilde{g}_{\mathrm{b},s,s}^{\left( b,r_2 \right)}\tilde{g}_{\mathrm{u},s,s}^{\left( r_2,u \right)}} \right) ^{\ast}}}\triangleq \gamma _{2222}\left( b \right).
\end{equation}

From \eqref{R-U_FD_channel_ij_tt}-\eqref{B-R_FD_channel_ij_tt_NLoS},
the first expectation in \eqref{E_2222_eq1} can be easily obtained as
\begin{equation}\label{E_2222_eq3}
  \mathbb{E} \left\{ \bar{\mathbf{g}}_{\mathrm{u},nt}^{H}\mathbf{\Phi }^H\bar{\mathbf{G}}_{\mathrm{b}}^{H}\mathbf{\Upsilon }_{2222}\bar{\mathbf{G}}_{\mathrm{b}}\mathbf{\Phi }\bar{\mathbf{g}}_{\mathrm{u},nt} \right\}
  =\frac{\beta _{\mathrm{u},n}\beta _{\mathrm{b}}^{2}\sigma _{\mathrm{u},0,n}^{2}\sigma _{\mathrm{b},0}^{2}\varsigma _{\mathrm{b}}K_{\mathrm{u},n}K_{\mathrm{b}}}{\left( K_{\mathrm{u},n}+1 \right) \left( K_{\mathrm{b}}+1 \right) ^2}N_{\mathrm{b}}N_{\mathrm{r}}\left| \varPhi _{N_{\mathrm{r}}}\left( n \right) \right|^2\sum_{u=0}^{N_{\mathrm{u}}-1}{ p_u \frac{\beta _{\mathrm{u},u}\varsigma _{\mathrm{u},u}}{K_{\mathrm{u},u}+1}}.
\end{equation}
Similar to \eqref{E_2112_eq10},
the second expectation in \eqref{E_2222_eq1} is calculated as
\begin{equation*}
\hspace{-8.4cm}
  \mathbb{E} \left\{ \bar{\mathbf{g}}_{\mathrm{u},nt}^{H}\mathbf{\Phi }^H\tilde{\mathbf{G}}_{\mathrm{b}}^{H}\mathbf{\Upsilon }_{2222}\tilde{\mathbf{G}}_{\mathrm{b}}\mathbf{\Phi }\bar{\mathbf{g}}_{\mathrm{u},nt} \right\}
\end{equation*}
\begin{equation*}
\hspace{-0.9cm}
  =\frac{\beta _{\mathrm{u},n}\sigma _{\mathrm{u},0,n}^{2}K_{\mathrm{u},n}}{N_{\mathrm{c}}\left( K_{\mathrm{u},n}+1 \right)}\mathbb{E} \left\{ \sum_{b=0}^{N_{\mathrm{b}}-1}{\sum_{s=0}^{N_{\mathrm{c}}-1}{\sum_{r_1=0}^{N_{\mathrm{r}}-1}{\left| \tilde{g}_{\mathrm{b},s,s}^{\left( b,r_1 \right)} \right|^2}}\sum_{r_2=0}^{N_{\mathrm{r}}-1}{\left| \tilde{g}_{\mathrm{b},t,t}^{\left( b,r_2 \right)} \right|^2}} \right\} \sum_{u=0}^{N_{\mathrm{u}}-1}{ p_u \frac{\beta _{\mathrm{u},u}\varsigma _{\mathrm{u},u}}{K_{\mathrm{u},u}+1}}
\end{equation*}
\begin{equation}\label{E_2222_eq4}
  = \frac{\beta _{\mathrm{u},n}\beta _{\mathrm{b}}^{2}\sigma _{\mathrm{u},0,n}^{2}K_{\mathrm{u},n}}{K_{\mathrm{u},n}+1}\left( \tau _{\mathrm{b}}+\sum_{k=1}^{L_{\mathrm{b}}-1}{\sigma _{\mathrm{b},k}^{4}}+\frac{\varsigma _{\mathrm{b}}^{2}\left( N_{\mathrm{r}}-1 \right)}{\left( K_{\mathrm{b}}+1 \right) ^2} \right) N_{\mathrm{b}}N_{\mathrm{r}}\sum_{u=0}^{N_{\mathrm{u}}-1}{ p_u  \frac{\beta _{\mathrm{u},u}\varsigma _{\mathrm{u},u}}{K_{\mathrm{u},u}+1}} .
\end{equation}
Furthermore, similar to \eqref{E_1221_eq12},
the third expectation in \eqref{E_2222_eq1} can be derived as
\begin{equation*}
\hspace{-5.9cm}
  \mathbb{E} \left\{ \tilde{\mathbf{g}}_{\mathrm{u},nt}^{H}\mathbf{\Phi }^H\bar{\mathbf{G}}_{\mathrm{b}}^{H}\mathbf{\Upsilon }_{2222}\bar{\mathbf{G}}_{\mathrm{b}}\mathbf{\Phi }\tilde{\mathbf{g}}_{\mathrm{u},nt} \right\}
\end{equation*}
\begin{equation*}
  =\frac{\beta _{\mathrm{b}}^{2}\sigma _{\mathrm{b},0}^{2}\varsigma _{\mathrm{b}}K_{\mathrm{b}}}{N_{\mathrm{c}}\left( K_{\mathrm{b}}+1 \right) ^2}\mathbb{E} \left\{ \sum_{b=0}^{N_{\mathrm{b}}-1}{\sum_{s=0}^{N_{\mathrm{c}}-1}{\sum_{u=0}^{N_{\mathrm{u}}-1}{p_u\sum_{r_1=0}^{N_{\mathrm{r}}-1}{\left| \tilde{g}_{\mathrm{u},s,s}^{\left( r_1,u \right)} \right|^2}}}\sum_{r_2=0}^{N_{\mathrm{r}}-1}{\left| \tilde{g}_{\mathrm{u},t,t}^{\left( r_2,n \right)} \right|^2}} \right\}
\end{equation*}
\begin{equation*}
\hspace{-0.9cm}
  =  p_n \frac{\beta _{\mathrm{b}}^{2}\sigma _{\mathrm{b},0}^{2}\varsigma _{\mathrm{b}}K_{\mathrm{b}}}{N_{\mathrm{c}}\left( K_{\mathrm{b}}+1 \right) ^2}\mathbb{E} \left\{ \sum_{b=0}^{N_{\mathrm{b}}-1}{\sum_{s=0}^{N_{\mathrm{c}}-1}{\sum_{r_1=0}^{N_{\mathrm{r}}-1}{\left| \tilde{g}_{\mathrm{u},s,s}^{\left( r_1,n \right)} \right|^2}\sum_{r_2=0}^{N_{\mathrm{r}}-1}{\left| \tilde{g}_{\mathrm{u},t,t}^{\left( r_2,n \right)} \right|^2}}} \right\}
\end{equation*}
\begin{equation*}
  +\frac{\beta _{\mathrm{b}}^{2}\sigma _{\mathrm{b},0}^{2}\varsigma _{\mathrm{b}}K_{\mathrm{b}}}{N_{\mathrm{c}}\left( K_{\mathrm{b}}+1 \right) ^2}\mathbb{E} \left\{ \sum_{b=0}^{N_{\mathrm{b}}-1}{\sum_{s=0}^{N_{\mathrm{c}}-1}{\sum_{u\ne n}^{N_{\mathrm{u}}-1}{p_u\sum_{r_1=0}^{N_{\mathrm{r}}-1}{\left| \tilde{g}_{\mathrm{u},s,s}^{\left( r_1,u \right)} \right|^2}}}\sum_{r_2=0}^{N_{\mathrm{r}}-1}{\left| \tilde{g}_{\mathrm{u},t,t}^{\left( r_2,n \right)} \right|^2}} \right\}
\end{equation*}
\begin{equation*}
\hspace{-0.4cm}
  = p_n  \frac{\beta _{\mathrm{u},n}^{2}\beta _{\mathrm{b}}^{2}\sigma _{\mathrm{b},0}^{2}\varsigma _{\mathrm{b}}K_{\mathrm{b}}}{\left( K_{\mathrm{b}}+1 \right) ^2}\left( \tau _{\mathrm{u},n}+\sum_{k=1}^{L_{\mathrm{u}}-1}{\sigma _{\mathrm{u},k,n}^{4}}+\frac{\varsigma _{\mathrm{u},n}^{2}\left( N_{\mathrm{r}}-1 \right)}{\left( K_{\mathrm{u},n}+1 \right) ^2} \right) N_{\mathrm{b}}N_{\mathrm{r}}
\end{equation*}
\begin{equation}\label{E_2222_eq5}
\hspace{-1.7cm}
  +\frac{\beta _{\mathrm{u},n}\beta _{\mathrm{b}}^{2}\sigma _{\mathrm{b},0}^{2}\varsigma _{\mathrm{u},n}\varsigma _{\mathrm{b}}K_{\mathrm{b}}}{\left( K_{\mathrm{u},n}+1 \right) \left( K_{\mathrm{b}}+1 \right) ^2}N_{\mathrm{b}}N_{\mathrm{r}}\left( N_{\mathrm{r}}-1 \right) \sum_{u\ne n}^{N_{\mathrm{u}}-1}{ p_u   \frac{\beta _{\mathrm{u},u}\varsigma _{\mathrm{u},u}}{K_{\mathrm{u},u}+1}} .
\end{equation}

The fourth expectation in \eqref{E_2222_eq1} can be expanded as \vspace{0.3cm}
\begin{equation*}
\hspace{-10.4cm}
  \mathbb{E} \left\{ \tilde{\mathbf{g}}_{\mathrm{u},nt}^{H}\mathbf{\Phi }^H\tilde{\mathbf{G}}_{\mathrm{b}}^{H}\mathbf{\Upsilon }_{2222}\tilde{\mathbf{G}}_{\mathrm{b}}\mathbf{\Phi }\tilde{\mathbf{g}}_{\mathrm{u},nt} \right\}
\end{equation*}
\begin{equation*}
\hspace{-4.3cm}
  =\mathbb{E} \left\{ \sum_{b=0}^{N_{\mathrm{b}}-1}{\gamma _{2222}\left( b \right) \sum_{r_1=0}^{N_{\mathrm{r}}-1}{\sum_{r_2=0}^{N_{\mathrm{r}}-1}{\left( e^{j\varphi _{r_1}}\tilde{g}_{\mathrm{u},t,t}^{\left( r_1,n \right)}\tilde{g}_{\mathrm{b},t,t}^{\left( b,r_1 \right)} \right) ^{\ast}e^{j\varphi _{r_2}}\tilde{g}_{\mathrm{u},t,t}^{\left( r_2,n \right)}\tilde{g}_{\mathrm{b},t,t}^{\left( b,r_2 \right)}}}} \right\}
\end{equation*}
\begin{equation*}
\hspace{-8.3cm}
  =\frac{1}{N_{\mathrm{c}}}\mathbb{E} \left\{ \sum_{b=0}^{N_{\mathrm{b}}-1}{\sum_{s=0}^{N_{\mathrm{c}}-1}{\sum_{u=0}^{N_{\mathrm{u}}-1}{p_u\mathrm{term}_{2222}^{1}\left( b,s,u \right)}}} \right\}
\end{equation*}
\begin{equation*}
\hspace{-4.9cm}
  \overset{\left( a \right)}{=}\frac{1}{N_{\mathrm{c}}}\mathbb{E} \left\{ \sum_{b=0}^{N_{\mathrm{b}}-1}{\sum_{s=0}^{N_{\mathrm{c}}-1}{\sum_{u=0}^{N_{\mathrm{u}}-1}{p_u\sum_{r=0}^{N_{\mathrm{r}}-1}{\left| \tilde{g}_{\mathrm{b},s,s}^{\left( b,r \right)} \right|^2\left| \tilde{g}_{\mathrm{b},t,t}^{\left( b,r \right)} \right|^2\left| \tilde{g}_{\mathrm{u},s,s}^{\left( r,u \right)} \right|^2\left| \tilde{g}_{\mathrm{u},t,t}^{\left( r,n \right)} \right|^2}}}} \right\}
\end{equation*}
\begin{equation*}
\hspace{-3.6cm}
  +\frac{1}{N_{\mathrm{c}}}\mathbb{E} \left\{ \sum_{b=0}^{N_{\mathrm{b}}-1}{\sum_{s=0}^{N_{\mathrm{c}}-1}{\sum_{u=0}^{N_{\mathrm{u}}-1}{p_u\sum_{r_1=0}^{N_{\mathrm{r}}-1}{\left| \tilde{g}_{\mathrm{b},s,s}^{\left( b,r_1 \right)} \right|^2\left| \tilde{g}_{\mathrm{u},s,s}^{\left( r_1,u \right)} \right|^2}\sum_{r_2\ne r_1}^{N_{\mathrm{r}}-1}{\left| \tilde{g}_{\mathrm{u},t,t}^{\left( r_2,n \right)} \right|^2\left| \tilde{g}_{\mathrm{b},t,t}^{\left( b,r_2 \right)} \right|^2}}}} \right\}
\end{equation*}
\begin{equation}\label{E_2222_eq6}
  +\frac{1}{N_{\mathrm{c}}}\mathbb{E} \left\{ \sum_{b=0}^{N_{\mathrm{b}}-1}{\sum_{s=0}^{N_{\mathrm{c}}-1}{\sum_{u=0}^{N_{\mathrm{u}}-1}{p_u\sum_{r_1=0}^{N_{\mathrm{r}}-1}{\tilde{g}_{\mathrm{b},s,s}^{\left( b,r_1 \right)}\tilde{g}_{\mathrm{u},s,s}^{\left( r_1,u \right)}\left( \tilde{g}_{\mathrm{b},t,t}^{\left( b,r_1 \right)}\tilde{g}_{\mathrm{u},t,t}^{\left( r_1,n \right)} \right) ^{\ast}}\sum_{r_2\ne r_1}^{N_{\mathrm{r}}-1}{\left( \tilde{g}_{\mathrm{b},s,s}^{\left( b,r_2 \right)}\tilde{g}_{\mathrm{u},s,s}^{\left( r_2,u \right)} \right) ^{\ast}\tilde{g}_{\mathrm{b},t,t}^{\left( b,r_2 \right)}\tilde{g}_{\mathrm{u},t,t}^{\left( r_2,n \right)}}}}} \right\}  ,
\end{equation}
where
\begin{equation*}
  \mathrm{term}_{2222}^{1}\left( b,s,u \right) =\sum_{r_1=0}^{N_{\mathrm{r}}-1}{e^{j\varphi _{r_1}}\tilde{g}_{\mathrm{b},s,s}^{\left( b,r_1 \right)}\tilde{g}_{\mathrm{u},s,s}^{\left( r_1,u \right)}}\sum_{r_2=0}^{N_{\mathrm{r}}-1}{\left( e^{j\varphi _{r_2}}\tilde{g}_{\mathrm{b},s,s}^{\left( b,r_2 \right)}\tilde{g}_{\mathrm{u},s,s}^{\left( r_2,u \right)} \right) ^{\ast}}
\end{equation*}
\begin{equation}
\hspace{5.5cm}
  \times \sum_{r_{11}=0}^{N_{\mathrm{r}}-1}{\left( e^{j\varphi _{r_{11}}}\tilde{g}_{\mathrm{b},t,t}^{\left( b,r_{11} \right)}\tilde{g}_{\mathrm{u},t,t}^{\left( r_{11},n \right)} \right) ^{\ast}}\sum_{r_{22}=0}^{N_{\mathrm{r}}-1}{e^{j\varphi _{r_{22}}}\tilde{g}_{\mathrm{b},t,t}^{\left( b,r_{22} \right)}\tilde{g}_{\mathrm{u},t,t}^{\left( r_{22},n \right)}} .
\end{equation}
Step $\left( a \right)$ divides \eqref{E_2222_eq6} into three parts:
1) $ r_1 = r_2 = r_{11} = r_{22} $,
2) $ r_1 = r_2 \neq r_{11} = r_{22} $,
and
3) $ r_1 = r_{11} \neq r_2 = r_{22} $.
The first term in \eqref{E_2222_eq6} is calculated as
\begin{equation*}
\hspace{-4.5cm}
  \frac{1}{N_{\mathrm{c}}}\mathbb{E} \left\{ \sum_{b=0}^{N_{\mathrm{b}}-1}{\sum_{s=0}^{N_{\mathrm{c}}-1}{\sum_{u=0}^{N_{\mathrm{u}}-1}{p_u\sum_{r=0}^{N_{\mathrm{r}}-1}{\left| \tilde{g}_{\mathrm{b},s,s}^{\left( b,r \right)} \right|^2\left| \tilde{g}_{\mathrm{b},t,t}^{\left( b,r \right)} \right|^2\left| \tilde{g}_{\mathrm{u},s,s}^{\left( r,u \right)} \right|^2\left| \tilde{g}_{\mathrm{u},t,t}^{\left( r,n \right)} \right|^2}}}} \right\}
\end{equation*}
\begin{equation*}
\hspace{-3.8cm}
  \overset{\left( a \right)}{=}\frac{1}{N_{\mathrm{c}}}p_n\sum_{b=0}^{N_{\mathrm{b}}-1}{\sum_{r=0}^{N_{\mathrm{r}}-1}{\sum_{s=0}^{N_{\mathrm{c}}-1}{\mathbb{E} \left\{ \left| \tilde{g}_{\mathrm{b},s,s}^{\left( b,r \right)} \right|^2\left| \tilde{g}_{\mathrm{b},t,t}^{\left( b,r \right)} \right|^2 \right\} \mathbb{E} \left\{ \left| \tilde{g}_{\mathrm{u},s,s}^{\left( r,n \right)} \right|^2\left| \tilde{g}_{\mathrm{u},t,t}^{\left( r,n \right)} \right|^2 \right\}}}}
\end{equation*}
\begin{equation*}
\hspace{-2.1cm}
  +\frac{1}{N_{\mathrm{c}}}\sum_{b=0}^{N_{\mathrm{b}}-1}{\sum_{r=0}^{N_{\mathrm{r}}-1}{\sum_{u\ne n}^{N_{\mathrm{u}}-1}{p_u\sum_{s=0}^{N_{\mathrm{c}}-1}{\mathbb{E} \left\{ \left| \tilde{g}_{\mathrm{b},s,s}^{\left( b,r \right)} \right|^2\left| \tilde{g}_{\mathrm{b},t,t}^{\left( b,r \right)} \right|^2 \right\} \mathbb{E} \left\{ \left| \tilde{g}_{\mathrm{u},s,s}^{\left( r,u \right)} \right|^2 \right\} \mathbb{E} \left\{ \left| \tilde{g}_{\mathrm{u},t,t}^{\left( r,n \right)} \right|^2 \right\}}}}}
\end{equation*}
\begin{equation*}
\hspace{-5.4cm}
  \overset{\left( b \right)}{=} p_n  \beta _{\mathrm{b}}^{2}\beta _{\mathrm{u},n}^{2}\left( \tau _{\mathrm{b}}+\sum_{k=1}^{L_{\mathrm{b}}-1}{\sigma _{\mathrm{b},k}^{4}} \right) \left( \tau _{\mathrm{u},n}+\sum_{k=1}^{L_{\mathrm{u}}-1}{\sigma _{\mathrm{u},k,n}^{4}} \right) N_{\mathrm{b}}N_{\mathrm{r}}
\end{equation*}
\begin{equation}\label{E_2222_eq7}
  +\beta _{\mathrm{b}}^{2}\frac{\beta _{\mathrm{u},n}\varsigma _{\mathrm{u},n}}{K_{\mathrm{u},n}+1}\left( \tau _{\mathrm{b}}+\sum_{k=1}^{L_{\mathrm{b}}-1}{\sigma _{\mathrm{b},k}^{4}} \right) N_{\mathrm{b}}N_{\mathrm{r}}\sum_{u\ne n}^{N_{\mathrm{u}}-1}{p_u\frac{\beta _{\mathrm{u},u}\varsigma _{\mathrm{u},u}}{K_{\mathrm{u},u}+1}}+2p_n\beta _{\mathrm{b}}^{2}\beta _{\mathrm{u},n}^{2}N_{\mathrm{b}}N_{\mathrm{r}}\mathrm{term}_{2222}^{2},
\end{equation}
where
\begin{equation*}
\hspace{-1.2cm}
  \mathrm{term}_{2222}^{2}=\frac{\sigma _{\mathrm{b},0}^{2}}{K_{\mathrm{b}}+1}\frac{\sigma _{\mathrm{u},0,n}^{2}}{K_{\mathrm{u},n}+1}\sum_{k=1}^{L_{1}^{\min}-1}{\sigma _{\mathrm{u},k,n}^{2}\sigma _{\mathrm{b},k}^{2}}+\frac{\sigma _{\mathrm{b},0}^{2}}{K_{\mathrm{b}}+1}\sum_{k_1=1}^{L_{\mathrm{u}}-1}{\sum_{k_2=k_1+1}^{L_{2}^{\min}-1}{\sigma _{\mathrm{u},k_1,n}^{2}\sigma _{\mathrm{u},k_2,n}^{2}\sigma _{\mathrm{b},k_2-k_1}^{2}}}
\end{equation*}
\begin{equation}
  +\frac{\sigma _{\mathrm{u},0,n}^{2}}{K_{\mathrm{u},n}+1}\sum_{k_1=1}^{L_{\mathrm{b}}-1}{\sum_{k_2=k_1+1}^{L_{3}^{\min}-1}{\sigma _{\mathrm{b},k_1}^{2}\sigma _{\mathrm{b},k_2}^{2}\sigma _{\mathrm{u},k_2-k_1,n}^{2}}}+\sum_{k_1=1}^{L_{\mathrm{b}}-1}{\sum_{k_2=k_1+1}^{L_{4}^{\min}-1}{\sum_{k_3=k_2-k_1+1}^{L_{\mathrm{u}}-1}{\sigma _{\mathrm{b},k_1}^{2}\sigma _{\mathrm{b},k_2}^{2}\sigma _{\mathrm{u},k_3,n}^{2}\sigma _{\mathrm{u},k_1-k_2+k_3,n}^{2}}}}
\end{equation}
and $L_{4}^{\min}=\min \left\{ L_{\mathrm{b}},L_{\mathrm{u}}+k_1-1 \right\}$.
Step $\left( a \right)$ in \eqref{E_2222_eq7} divides the equation into two parts,
based on whether or not $u = n$.
Step $\left( b \right)$ is according to \eqref{R-U_FD_channel_ij_tt}-\eqref{B-R_FD_channel_ij_tt_NLoS}, Lemma~\ref{lemma_sum} and Lemma~\ref{lemma_expectation}.
The second term in \eqref{E_2222_eq6} can be easily obtained as
\begin{equation*}
  \frac{1}{N_{\mathrm{c}}}\mathbb{E} \left\{ \sum_{b=0}^{N_{\mathrm{b}}-1}{\sum_{s=0}^{N_{\mathrm{c}}-1}{\sum_{u=0}^{N_{\mathrm{u}}-1}{p_u\sum_{r_1=0}^{N_{\mathrm{r}}-1}{\left| \tilde{g}_{\mathrm{b},s,s}^{\left( b,r_1 \right)} \right|^2\left| \tilde{g}_{\mathrm{u},s,s}^{\left( r_1,u \right)} \right|^2}\sum_{r_2\ne r_1}^{N_{\mathrm{r}}-1}{\left| \tilde{g}_{\mathrm{u},t,t}^{\left( r_2,n \right)} \right|^2\left| \tilde{g}_{\mathrm{b},t,t}^{\left( b,r_2 \right)} \right|^2}}}} \right\}
\end{equation*}
\begin{equation}\label{E_2222_eq8}
  =\frac{\beta _{\mathrm{u},n}\beta _{\mathrm{b}}^{2}\varsigma _{\mathrm{u},n}\varsigma _{\mathrm{b}}^{2}}{\left( K_{\mathrm{u},n}+1 \right) \left( K_{\mathrm{b}}+1 \right) ^2}N_{\mathrm{b}}N_{\mathrm{r}}\left( N_{\mathrm{r}}-1 \right) \sum_{u=0}^{N_{\mathrm{u}}-1}{ p_u  \frac{\beta _{\mathrm{u},u}\varsigma _{\mathrm{u},u}}{K_{\mathrm{u},u}+1}}.
\end{equation}
Moreover, from Lemma~\ref{lemma_sum} and Lemma~\ref{lemma_expectation},
the third term in \eqref{E_2222_eq6} is calculated as
\begin{equation*}
  \frac{1}{N_{\mathrm{c}}}\mathbb{E} \left\{ \sum_{b=0}^{N_{\mathrm{b}}-1}{\sum_{s=0}^{N_{\mathrm{c}}-1}{\sum_{u=0}^{N_{\mathrm{u}}-1}{p_u\sum_{r_1=0}^{N_{\mathrm{r}}-1}{\tilde{g}_{\mathrm{b},s,s}^{\left( b,r_1 \right)}\tilde{g}_{\mathrm{u},s,s}^{\left( r_1,u \right)}\left( \tilde{g}_{\mathrm{b},t,t}^{\left( b,r_1 \right)}\tilde{g}_{\mathrm{u},t,t}^{\left( r_1,n \right)} \right) ^{\ast}}\sum_{r_2\ne r_1}^{N_{\mathrm{r}}-1}{\left( \tilde{g}_{\mathrm{b},s,s}^{\left( b,r_2 \right)}\tilde{g}_{\mathrm{u},s,s}^{\left( r_2,u \right)} \right) ^{\ast}\tilde{g}_{\mathrm{b},t,t}^{\left( b,r_2 \right)}\tilde{g}_{\mathrm{u},t,t}^{\left( r_2,n \right)}}}}} \right\}
\end{equation*}
\begin{equation}\label{E_2222_eq9}
  =p_n\frac{\beta _{\mathrm{u},n}^{2}\beta _{\mathrm{b}}^{2}\sigma _{\mathrm{u},0,n}^{4}\sigma _{\mathrm{b},0}^{4}}{\left( K_{\mathrm{u},n}+1 \right) ^2\left( K_{\mathrm{b}}+1 \right) ^2}N_{\mathrm{b}}N_{\mathrm{r}}\left( N_{\mathrm{r}}-1 \right) +p_n\beta _{\mathrm{u},n}^{2}\beta _{\mathrm{b}}^{2}N_{\mathrm{b}}N_{\mathrm{r}}\left( N_{\mathrm{r}}-1 \right) \mathrm{term}_{2222}^{3} ,
\end{equation}
where  \vspace{0.3cm}
\begin{equation*}
\hspace{-2.6cm}
  \mathrm{term}_{2222}^{3}=\frac{\sigma _{\mathrm{u},0,n}^{4}}{\left( K_{\mathrm{u},n}+1 \right) ^2}\sum_{k=1}^{L_{\mathrm{b}}-1}{\sigma _{\mathrm{b},k}^{4}}+\frac{\sigma _{\mathrm{b},0}^{4}}{\left( K_{\mathrm{b}}+1 \right) ^2}\sum_{k=1}^{L_{\mathrm{u}}-1}{\sigma _{\mathrm{u},k,n}^{4}}+\sum_{k_1=1}^{L_{\mathrm{b}}-1}{\sum_{k_2=1}^{L_{\mathrm{u}}-1}{\sigma _{\mathrm{b},k_1}^{4}\sigma _{\mathrm{u},k_2,n}^{4}}}
\end{equation*}
\begin{equation*}
\hspace{-2.4cm}
  +\frac{2\sigma _{\mathrm{u},0,n}^{2}\sigma _{\mathrm{b},0}^{2}}{\left( K_{\mathrm{u},n}+1 \right) \left( K_{\mathrm{b}}+1 \right)}\sum_{k=1}^{L_{1}^{\min}-1}{\sigma _{\mathrm{u},k,n}^{2}\sigma _{\mathrm{b},k}^{2}}+\frac{2\sigma _{\mathrm{u},0,n}^{2}}{K_{\mathrm{u},n}+1}\sum_{k_1=1}^{L_{\mathrm{b}}-1}{\sum_{k_2=k_1+1}^{L_{3}^{\min}-1}{\sigma _{\mathrm{b},k_1}^{2}\sigma _{\mathrm{b},k_2}^{2}\sigma _{\mathrm{u},k_2-k_1,n}^{2}}}
\end{equation*}
\begin{equation}
  +\frac{2\sigma _{\mathrm{b},0}^{2}}{K_{\mathrm{b}}+1}\sum_{k_1=1}^{L_{\mathrm{u}}-1}{\sum_{k_2=k_1+1}^{L_{2}^{\min}-1}{\sigma _{\mathrm{u},k_1,n}^{2}\sigma _{\mathrm{u},k_2,n}^{2}\sigma _{\mathrm{b},k_2-k_1}^{2}}}+2\sum_{k_1=1}^{L_{\mathrm{b}}-1}{\sum_{k_2=k_1+1}^{L_{4}^{\min}-1}{\sum_{k_3=k_2-k_1+1}^{L_{\mathrm{u}}-1}{\sigma _{\mathrm{b},k_1}^{2}\sigma _{\mathrm{b},k_2}^{2}\sigma _{\mathrm{u},k_3,n}^{2}\sigma _{\mathrm{u},k_1-k_2+k_3,n}^{2}}}}.
\end{equation}
Therefore, substituting \eqref{E_2222_eq7}-\eqref{E_2222_eq9} into \eqref{E_2222_eq6},
we have the fourth expectation in \eqref{E_2222_eq1} as
\begin{equation*}
\hspace{-6cm}
  \mathbb{E} \left\{ \tilde{\mathbf{g}}_{\mathrm{u},nt}^{H}\mathbf{\Phi }^H\tilde{\mathbf{G}}_{\mathrm{b}}^{H}\mathbf{\Upsilon }_{2222}\tilde{\mathbf{G}}_{\mathrm{b}}\mathbf{\Phi }\tilde{\mathbf{g}}_{\mathrm{u},nt} \right\} =2p_n\beta _{\mathrm{u},n}^{2}\beta _{\mathrm{b}}^{2}N_{\mathrm{b}}N_{\mathrm{r}}^{2}\mathrm{term}_{2222}^{4}
\end{equation*}
\begin{equation*}
\hspace{-1.9cm}
  +  p_n \beta _{\mathrm{u},n}^{2}\beta _{\mathrm{b}}^{2}N_{\mathrm{b}}N_{\mathrm{r}}\left( \left( \tau _{\mathrm{b}}+\sum_{k=1}^{L_{\mathrm{b}}-1}{\sigma _{\mathrm{b},k}^{4}} \right) \left( \tau _{\mathrm{u},n}+\sum_{k=1}^{L_{\mathrm{u}}-1}{\sigma _{\mathrm{u},k,n}^{4}} \right) +\frac{\sigma _{\mathrm{u},0,n}^{4}\sigma _{\mathrm{b},0}^{4}\left( N_{\mathrm{r}}-1 \right)}{\left( K_{\mathrm{u},n}+1 \right) ^2\left( K_{\mathrm{b}}+1 \right) ^2} \right)
\end{equation*}
\begin{equation*}
\hspace{-1.1cm}
  +\frac{\beta _{\mathrm{u},n}\varsigma _{\mathrm{u},n}}{K_{\mathrm{u},n}+1}\beta _{\mathrm{b}}^{2}N_{\mathrm{b}}N_{\mathrm{r}}\left( \left( \tau _{\mathrm{b}}+\sum_{k=1}^{L_{\mathrm{b}}-1}{\sigma _{\mathrm{b},k}^{4}} \right) \sum_{u\ne n}^{N_{\mathrm{u}}-1}{p_u\frac{\beta _{\mathrm{u},u}\varsigma _{\mathrm{u},u}}{K_{\mathrm{u},u}+1}}+\frac{\varsigma _{\mathrm{b}}^{2}\left( N_{\mathrm{r}}-1 \right)}{\left( K_{\mathrm{b}}+1 \right) ^2}\sum_{u=0}^{N_{\mathrm{u}}-1}{p_u\frac{\beta _{\mathrm{u},u}\varsigma _{\mathrm{u},u}}{K_{\mathrm{u},u}+1}} \right)
\end{equation*}
\begin{equation}\label{E_2222_eq10}
  + p_n\beta _{\mathrm{u},n}^{2}\beta _{\mathrm{b}}^{2}N_{\mathrm{b}}N_{\mathrm{r}} \! \left( N_{\mathrm{r}} \! - \! 1 \right) \! \left( \frac{\sigma _{\mathrm{u},0,n}^{4}}{\left( K_{\mathrm{u},n}+1 \right) ^2} \!\! \sum_{k=1}^{L_{\mathrm{b}}-1}{\sigma _{\mathrm{b},k}^{4}}+\frac{\sigma _{\mathrm{b},0}^{4}}{\left( K_{\mathrm{b}}+1 \right) ^2} \!\! \sum_{k=1}^{L_{\mathrm{u}}-1}{\sigma _{\mathrm{u},k,n}^{4}}+\sum_{k_1=1}^{L_{\mathrm{b}}-1}{ \!  \sum_{k_2=1}^{L_{\mathrm{u}}-1}{\sigma _{\mathrm{b},k_1}^{4}\sigma _{\mathrm{u},k_2,n}^{4}}} \right) ,
\end{equation}
where
\begin{equation*}
\hspace{-0.6cm}
  \mathrm{term}_{2222}^{4}=\frac{\sigma _{\mathrm{u},0,n}^{2}\sigma _{\mathrm{b},0}^{2}}{\left( K_{\mathrm{u},n}+1 \right) \left( K_{\mathrm{b}}+1 \right)}\sum_{k=1}^{L_{1}^{\min}-1}{\sigma _{\mathrm{u},k,n}^{2}\sigma _{\mathrm{b},k}^{2}}+\frac{\sigma _{\mathrm{u},0,n}^{2}}{K_{\mathrm{u},n}+1}\sum_{k_1=1}^{L_{\mathrm{b}}-1}{\sum_{k_2=k_1+1}^{L_{3}^{\min}-1}{\sigma _{\mathrm{b},k_1}^{2}\sigma _{\mathrm{b},k_2}^{2}\sigma _{\mathrm{u},k_2-k_1,n}^{2}}}
\end{equation*}
\begin{equation}
  +\frac{\sigma _{\mathrm{b},0}^{2}}{K_{\mathrm{b}}+1}\sum_{k_1=1}^{L_{\mathrm{u}}-1}{\sum_{k_2=k_1+1}^{L_{2}^{\min}-1}{\sigma _{\mathrm{u},k_1,n}^{2}\sigma _{\mathrm{u},k_2,n}^{2}\sigma _{\mathrm{b},k_2-k_1}^{2}}}+\sum_{k_1=1}^{L_{\mathrm{b}}-1}{\sum_{k_2=k_1+1}^{L_{4}^{\min}-1}{\sum_{k_3=k_2-k_1+1}^{L_{\mathrm{u}}-1}{\sigma _{\mathrm{b},k_1}^{2}\sigma _{\mathrm{b},k_2}^{2}\sigma _{\mathrm{u},k_3,n}^{2}\sigma _{\mathrm{u},k_1-k_2+k_3,n}^{2}}}} ,
\end{equation}
with $L_{1}^{\min}=\min \left\{ L_{\mathrm{b}},L_{\mathrm{u}} \right\}$,
$L_{2}^{\min}=\min \left\{ L_{\mathrm{u}},L_{\mathrm{b}}+k_1 \right\}$,
$L_{3}^{\min}=\min \left\{ L_{\mathrm{b}},L_{\mathrm{u}}+k_1 \right\}$,
$L_{4}^{\min}=\min \left\{ L_{\mathrm{b}},L_{\mathrm{u}}+k_1-1 \right\}$.

Substituting \eqref{E_2222_eq3}, \eqref{E_2222_eq4}, \eqref{E_2222_eq5} and \eqref{E_2222_eq10} into \eqref{E_2222_eq1},
we arrive at
\begin{equation*}
\hspace{-1.6cm}
  \mathbb{E} \left\{ \mathbf{g}_{\mathrm{u},nt}^{H}\mathbf{\Phi }^H\mathbf{G}_{\mathrm{b}}^{H}\mathbf{\Upsilon }_{2222}\mathbf{G}_{\mathrm{b}}\mathbf{\Phi g}_{\mathrm{u},nt} \right\} =2p_n\beta _{\mathrm{u},n}^{2}\beta _{\mathrm{b}}^{2}N_{\mathrm{b}}N_{\mathrm{r}}^{2}\mathrm{term}_{2222}^{4}+p_n\beta _{\mathrm{u},n}^{2}\beta _{\mathrm{b}}^{2}N_{\mathrm{b}}N_{\mathrm{r}}\mathrm{term}_{2222}^{5}
\end{equation*}
\begin{equation*}
\hspace{-8.1cm}
  +\frac{\beta _{\mathrm{u},n}\beta _{\mathrm{b}}^{2}N_{\mathrm{b}}N_{\mathrm{r}}}{\left( K_{\mathrm{u},n}+1 \right) \left( K_{\mathrm{b}}+1 \right) ^2}\mathrm{term}_{2222}^{6}\sum_{u=0}^{N_{\mathrm{u}}-1}{p_u\frac{\beta _{\mathrm{u},u}\varsigma _{\mathrm{u},u}}{K_{\mathrm{u},u}+1}}
\end{equation*}
\begin{equation*}
\hspace{-3.5cm}
  +\frac{\beta _{\mathrm{u},n}\beta _{\mathrm{b}}^{2}\varsigma _{\mathrm{u},n}}{K_{\mathrm{u},n}+1}N_{\mathrm{b}}N_{\mathrm{r}}\left( \tau _{\mathrm{b}}+\sum_{k=1}^{L_{\mathrm{b}}-1}{\sigma _{\mathrm{b},k}^{4}}+\frac{\sigma _{\mathrm{b},0}^{2}\varsigma _{\mathrm{b}}K_{\mathrm{b}}}{\left( K_{\mathrm{b}}+1 \right) ^2}\left( N_{\mathrm{r}}-1 \right) \right) \sum_{u\ne n}^{N_{\mathrm{u}}-1}{ p_u  \frac{\beta _{\mathrm{u},u}\varsigma _{\mathrm{u},u}}{K_{\mathrm{u},u}+1}}
\end{equation*}
\begin{equation}\label{E_2222_eq11}
  + p_n\beta _{\mathrm{u},n}^{2}\beta _{\mathrm{b}}^{2}N_{\mathrm{b}}N_{\mathrm{r}} \! \left( N_{\mathrm{r}} \! - \! 1 \right) \! \left( \frac{\sigma _{\mathrm{u},0,n}^{4}}{\left( K_{\mathrm{u},n}+1 \right) ^2} \!\! \sum_{k=1}^{L_{\mathrm{b}}-1}{\sigma _{\mathrm{b},k}^{4}}+\frac{\sigma _{\mathrm{b},0}^{4}}{\left( K_{\mathrm{b}}+1 \right) ^2} \!\! \sum_{k=1}^{L_{\mathrm{u}}-1}{\sigma _{\mathrm{u},k,n}^{4}}+\sum_{k_1=1}^{L_{\mathrm{b}}-1}{ \!  \sum_{k_2=1}^{L_{\mathrm{u}}-1}{\sigma _{\mathrm{b},k_1}^{4}\sigma _{\mathrm{u},k_2,n}^{4}}} \right) ,
\end{equation}
where \vspace{0.3cm}
\begin{equation*}
\hspace{-1.1cm}
  \mathrm{term}_{2222}^{5}=\left( \tau _{\mathrm{b}}+\sum_{k=1}^{L_{\mathrm{b}}-1}{\sigma _{\mathrm{b},k}^{4}} \right) \left( \tau _{\mathrm{u},n}+\sum_{k=1}^{L_{\mathrm{u}}-1}{\sigma _{\mathrm{u},k,n}^{4}} \right) +\frac{\sigma _{\mathrm{u},0,n}^{4}\sigma _{\mathrm{b},0}^{4}\left( N_{\mathrm{r}}-1 \right)}{\left( K_{\mathrm{u},n}+1 \right) ^2\left( K_{\mathrm{b}}+1 \right) ^2}
\end{equation*}
\begin{equation}
\hspace{-1.1cm}
  +\frac{\sigma _{\mathrm{b},0}^{2}\varsigma _{\mathrm{b}}K_{\mathrm{b}}}{\left( K_{\mathrm{b}}+1 \right) ^2}\left( \tau _{\mathrm{u},n}+\sum_{k=1}^{L_{\mathrm{u}}-1}{\sigma _{\mathrm{u},k,n}^{4}}+\frac{\varsigma _{\mathrm{u},n}^{2}\left( N_{\mathrm{r}}-1 \right)}{\left( K_{\mathrm{u},n}+1 \right) ^2} \right),
\end{equation}
\begin{equation*}
  \mathrm{term}_{2222}^{6}=\sigma _{\mathrm{u},0,n}^{2}\sigma _{\mathrm{b},0}^{2}\varsigma _{\mathrm{b}}K_{\mathrm{u},n}K_{\mathrm{b}}\left| \varPhi _{N_{\mathrm{r}}}\left( n \right) \right|^2+\sigma _{\mathrm{u},0,n}^{2}K_{\mathrm{u},n}\left( K_{\mathrm{b}}+1 \right) ^2\left( \tau _{\mathrm{b}}+\sum_{k=1}^{L_{\mathrm{b}}-1}{\sigma _{\mathrm{b},k}^{4}} \right)
\end{equation*}
\begin{equation}
\hspace{-4.1cm}
  +\left( \sigma _{\mathrm{u},0,n}^{2}K_{\mathrm{u},n}+\varsigma _{\mathrm{u},n} \right) \varsigma _{\mathrm{b}}^{2}\left( N_{\mathrm{r}}-1 \right).
\end{equation}

Based on \eqref{EI_3_gamma_expand},
we can obtain the expectation $\mathbb{E} \left\{ \mathbf{g}_{\mathrm{u},nt}^{H}\mathbf{\Phi }^H\mathbf{G}_{\mathrm{b}}^{H}\mathbf{\Upsilon G}_{\mathrm{b}}\mathbf{\Phi g}_{\mathrm{u},nt} \right\}$,
which is expressed as
\begin{equation*}
\hspace{-1.4cm}
  \mathbb{E} \left\{ \mathbf{g}_{\mathrm{u},nt}^{H}\mathbf{\Phi }^H\mathbf{G}_{\mathrm{b}}^{H}\mathbf{\Upsilon G}_{\mathrm{b}}\mathbf{\Phi g}_{\mathrm{u},nt} \right\} =\mathrm{term}_{\xi}^{1}K_{\mathrm{b}}c_{\xi}^{1}\sum_{u=0}^{N_{\mathrm{u}}-1}{\frac{p_u\beta _{\mathrm{u},u}\sigma _{\mathrm{u},0,u}^{2}K_{\mathrm{u},u}}{K_{\mathrm{u},u}+1}\left| \varPhi _{N_{\mathrm{r}}}\left( u \right) \right|^2}
\end{equation*}
\begin{equation*}
  +2\sigma _{\mathrm{u},0,n}^{2}\sigma _{\mathrm{b},0}^{4}K_{\mathrm{u},n}K_{\mathrm{b}}c_{\xi}^{1}\mathrm{Re}\left( \left( \varPhi _{N_{\mathrm{r}}}\left( n \right) \right) ^{\ast}\sum_{u=0}^{N_{\mathrm{u}}-1}{\frac{p_u\beta _{\mathrm{u},u}\sigma _{\mathrm{u},0,u}^{2}K_{\mathrm{u},u}}{K_{\mathrm{u},u}+1}\varPhi _{N_{\mathrm{r}}}\left( u \right) \left( \bar{\mathbf{h}}_{\mathrm{u}}^{\left( \cdot ,u \right)} \right) ^H\bar{\mathbf{h}}_{\mathrm{u}}^{\left( \cdot ,n \right)}} \right)
\end{equation*}
\begin{equation*}
\hspace{-3.9cm}
  +\mathrm{term}_{\xi}^{2}N_{\mathrm{r}}c_{\xi}^{1}\sum_{u=0}^{N_{\mathrm{u}}-1}{\frac{p_u\beta _{\mathrm{u},u}\sigma _{\mathrm{u},0,u}^{2}K_{\mathrm{u},u}}{K_{\mathrm{u},u}+1}}+\mathrm{term}_{\xi}^{3}c_{\xi}^{1}\sum_{u=0}^{N_{\mathrm{u}}-1}{\frac{p_u\beta _{\mathrm{u},u}\varsigma _{\mathrm{u},u}}{K_{\mathrm{u},u}+1}}
\end{equation*}
\begin{equation*}
\hspace{-0.6cm}
  +\left( \frac{\sigma _{\mathrm{b},0}^{4}}{\left( K_{\mathrm{b}}+1 \right) ^2}+\sum_{k=1}^{L_{\mathrm{b}}-1}{\sigma _{\mathrm{b},k}^{4}} \right) \sigma _{\mathrm{u},0,n}^{2}K_{\mathrm{u},n}c_{\xi}^{2}\sum_{u=0}^{N_{\mathrm{u}}-1}{\frac{p_u\beta _{\mathrm{u},u}\sigma _{\mathrm{u},0,u}^{2}K_{\mathrm{u},u}}{K_{\mathrm{u},u}+1}\left| \left( \bar{\mathbf{h}}_{\mathrm{u}}^{\left( \cdot ,n \right)} \right) ^H\bar{\mathbf{h}}_{\mathrm{u}}^{\left( \cdot ,u \right)} \right|^2}
\end{equation*}
\begin{equation*}
\hspace{-4.6cm}
  +\mathrm{term}_{\xi}^{4}\varsigma _{\mathrm{u},n}N_{\mathrm{r}}c_{\xi}^{2}\sum_{u\ne n}^{N_{\mathrm{u}}-1}{\frac{p_u\beta _{\mathrm{u},u}\varsigma _{\mathrm{u},u}}{K_{\mathrm{u},u}+1}}+\mathrm{term}_{\xi}^{5}N_{\mathrm{r}}c_{\xi}^{3}+\mathrm{term}_{\xi}^{6}c_{\xi}^{4}\,\,
\end{equation*}
\begin{equation}\label{xi_n}
\hspace{-3.6cm}
  +\left( \sigma _{\mathrm{b},0}^{2}K_{\mathrm{b}}N_{\mathrm{r}}+\sigma _{\mathrm{b},0}^{2}N_{\mathrm{r}}+2\varsigma _{\mathrm{b}} \right) 2\sigma _{\mathrm{u},0,n}^{4}K_{\mathrm{u},n}K_{\mathrm{b}}c_{\xi}^{3}\left| \varPhi _{N_{\mathrm{r}}}\left( n \right) \right|^2\triangleq \xi _n ,
\end{equation}
where $c_{\xi}^{1}=\frac{\beta _{\mathrm{u},n}\beta _{\mathrm{b}}^{2}N_{\mathrm{b}}}{\left( K_{\mathrm{u},n}+1 \right) \left( K_{\mathrm{b}}+1 \right) ^2}$,
$c_{\xi}^{2}=\frac{\beta _{\mathrm{u},n}\beta _{\mathrm{b}}^{2}N_{\mathrm{b}}}{K_{\mathrm{u},n}+1}$,
$c_{\xi}^{3}=\frac{p_n\beta _{\mathrm{u},n}^{2}\beta _{\mathrm{b}}^{2}\sigma _{\mathrm{b},0}^{2}N_{\mathrm{b}}}{\left( K_{\mathrm{u},n}+1 \right) ^2\left( K_{\mathrm{b}}+1 \right) ^2}$,
$c_{\xi}^{4}=p_n\beta _{\mathrm{u},n}^{2}\beta _{\mathrm{b}}^{2}N_{\mathrm{b}}N_{\mathrm{r}}$,
\begin{equation}
  \mathrm{term}_{\xi}^{1}=\sigma _{\mathrm{u},0,n}^{2}\sigma _{\mathrm{b},0}^{4}K_{\mathrm{u},n}K_{\mathrm{b}}\left| \varPhi _{N_{\mathrm{r}}}\left( n \right) \right|^2+\sigma _{\mathrm{u},0,n}^{2}\sigma _{\mathrm{b},0}^{2}\varsigma _{\mathrm{b}}K_{\mathrm{u},n}N_{\mathrm{r}}+\sigma _{\mathrm{b},0}^{4}\varsigma _{\mathrm{u},n}K_{\mathrm{b}}N_{\mathrm{r}}
  +\sigma _{\mathrm{b},0}^{2}\varsigma _{\mathrm{u},n}\varsigma _{\mathrm{b}}N_{\mathrm{r}}+2\sigma _{\mathrm{b},0}^{4}\varsigma _{\mathrm{u},n},
\end{equation}
\begin{equation*}
  \mathrm{term}_{\xi}^{2}=\sigma _{\mathrm{u},0,n}^{2}\sigma _{\mathrm{b},0}^{2}\varsigma _{\mathrm{b}}K_{\mathrm{u},n}K_{\mathrm{b}}\left| \varPhi _{N_{\mathrm{r}}}\left( n \right) \right|^2+\sigma _{\mathrm{u},0,n}^{2}\varsigma _{\mathrm{b}}^{2}K_{\mathrm{u},n}\left( N_{\mathrm{r}}-1 \right) +\sigma _{\mathrm{b},0}^{2}\varsigma _{\mathrm{u},n}\varsigma _{\mathrm{b}}K_{\mathrm{b}}N_{\mathrm{r}}
  +\varsigma _{\mathrm{u},n}\varsigma _{\mathrm{b}}^{2}\left( N_{\mathrm{r}}-1 \right)
\end{equation*}
\begin{equation}
\hspace{-2cm}
  -\sigma _{\mathrm{u},0,n}^{2}\sigma _{\mathrm{b},0}^{4}K_{\mathrm{u},n}+\left( K_{\mathrm{b}}+1 \right) ^2\left( \sigma _{\mathrm{u},0,n}^{2}\tau _{\mathrm{b}}K_{\mathrm{u},n}+\varsigma _{\mathrm{u},n}\tau _{\mathrm{b}}+\varsigma _{\mathrm{u},n}\sum_{k=1}^{L_{\mathrm{b}}-1}{\sigma _{\mathrm{b},k}^{4}} \right),
\end{equation}
\begin{equation*}
\hspace{-1.2cm}
  \mathrm{term}_{\xi}^{3}=\left( \varsigma _{\mathrm{b}}N_{\mathrm{r}}+2\sigma _{\mathrm{b},0}^{2}+\sigma _{\mathrm{b},0}^{2}K_{\mathrm{b}}N_{\mathrm{r}} \right) \sigma _{\mathrm{u},0,n}^{2}\sigma _{\mathrm{b},0}^{2}K_{\mathrm{u},n}K_{\mathrm{b}}\left| \varPhi _{N_{\mathrm{r}}}\left( n \right) \right|^2+\sigma _{\mathrm{u},0,n}^{2}\sigma _{\mathrm{b},0}^{2}\varsigma _{\mathrm{b}}K_{\mathrm{u},n}K_{\mathrm{b}}N_{\mathrm{r}}^{2}
\end{equation*}
\begin{equation}
\hspace{1.2cm}
  +\sigma _{\mathrm{u},0,n}^{2}K_{\mathrm{u},n}\left( K_{\mathrm{b}}+1 \right) ^2\left( \tau _{\mathrm{b}}+\sum_{k=1}^{L_{\mathrm{b}}-1}{\sigma _{\mathrm{b},k}^{4}} \right) N_{\mathrm{r}}+\left( \sigma _{\mathrm{u},0,n}^{2}K_{\mathrm{u},n}+\varsigma _{\mathrm{u},n} \right) \varsigma _{\mathrm{b}}^{2}N_{\mathrm{r}}\left( N_{\mathrm{r}}-1 \right),
\end{equation}
\begin{equation}
\hspace{-0.01cm}
  \mathrm{term}_{\xi}^{4}=\tau _{\mathrm{b}}+\sum_{k=1}^{L_{\mathrm{b}}-1}{\sigma _{\mathrm{b},k}^{4}}+\frac{\sigma _{\mathrm{b},0}^{2}\varsigma _{\mathrm{b}}K_{\mathrm{b}}}{\left( K_{\mathrm{b}}+1 \right) ^2}\left( N_{\mathrm{r}}-1 \right) +\frac{2\sigma _{\mathrm{b},0}^{4}K_{\mathrm{b}}}{\left( K_{\mathrm{b}}+1 \right) ^2}+\frac{\sigma _{\mathrm{b},0}^{2}K_{\mathrm{b}}}{\left( K_{\mathrm{b}}+1 \right) ^2}N_{\mathrm{r}}\left( \sigma _{\mathrm{b},0}^{2}K_{\mathrm{b}}+\varsigma _{\mathrm{b}} \right),
\end{equation}
\begin{equation*}
\hspace{-1.6cm}
  \mathrm{term}_{\xi}^{5}=\sigma _{\mathrm{u},0,n}^{4}\sigma _{\mathrm{b},0}^{2}\left( N_{\mathrm{r}}-1 \right) +2\sigma _{\mathrm{u},0,n}^{4}\sigma _{\mathrm{b},0}^{2}K_{\mathrm{u},n}K_{\mathrm{b}}N_{\mathrm{r}}+2\sigma _{\mathrm{u},0,n}^{4}\sigma _{\mathrm{b},0}^{2}\left( K_{\mathrm{b}}+K_{\mathrm{u},n} \right) \left( N_{\mathrm{r}}-1 \right)
\end{equation*}
\begin{equation}
\hspace{-4.9cm}
  +K_{\mathrm{b}}\left( \sigma _{\mathrm{u},0,n}^{4}\sigma _{\mathrm{b},0}^{2}K_{\mathrm{b}}+\sigma _{\mathrm{b},0}^{2}\varsigma _{\mathrm{u},n}^{2}K_{\mathrm{b}}+2\varsigma _{\mathrm{u},n}^{2}\varsigma _{\mathrm{b}} \right) \left( N_{\mathrm{r}}-1 \right),
\end{equation}
\begin{equation*}
\hspace{-3.9cm}
  \mathrm{term}_{\xi}^{6}=\tau _{\mathrm{b}}\tau _{\mathrm{u},n}+\frac{2\sigma _{\mathrm{u},0,n}^{4}K_{\mathrm{u},n}}{\left( K_{\mathrm{u},n}+1 \right) ^2}\tau _{\mathrm{b}}+\frac{\sigma _{\mathrm{b},0}^{2}K_{\mathrm{b}}}{\left( K_{\mathrm{b}}+1 \right) ^2}\left( \sigma _{\mathrm{b},0}^{2}K_{\mathrm{b}}+2\sigma _{\mathrm{b},0}^{2}+2\varsigma _{\mathrm{b}} \right) \tau _{\mathrm{u},n}
\end{equation*}
\begin{equation*}
\hspace{-0.8cm}
  +\left( \tau _{\mathrm{b}}+\frac{\sigma _{\mathrm{b},0}^{2}}{\left( K_{\mathrm{b}}+1 \right) ^2}\left( \sigma _{\mathrm{b},0}^{2}\left( K_{\mathrm{b}}^{2}N_{\mathrm{r}}+2K_{\mathrm{b}}N_{\mathrm{r}}+N_{\mathrm{r}}-1 \right) +2\varsigma _{\mathrm{b}}K_{\mathrm{b}} \right) \right) \sum_{k=1}^{L_{\mathrm{u}}-1}{\sigma _{\mathrm{u},k,n}^{4}}
\end{equation*}
\begin{equation*}
  +\left( \tau _{\mathrm{u},n}+\frac{\sigma _{\mathrm{u},0,n}^{4}}{\left( K_{\mathrm{u},n}+1 \right) ^2}\left( 2K_{\mathrm{u},n}N_{\mathrm{r}}+N_{\mathrm{r}}-1 \right) \right) \sum_{k=1}^{L_{\mathrm{b}}-1}{\sigma _{\mathrm{b},k}^{4}}+N_{\mathrm{r}}\sum_{k_1=1}^{L_{\mathrm{b}}-1}{\sum_{k_2=1}^{L_{\mathrm{u}}-1}{\sigma _{\mathrm{b},k_1}^{4}\sigma _{\mathrm{u},k_2,n}^{4}}}
\end{equation*}
\begin{equation}
\hspace{-4.4cm}
  +2N_{\mathrm{r}}\sum_{k_1=0}^{L_{\mathrm{b}}-1}{\sum_{k_2=k_1+1}^{L_{3}^{\min}-1}{\sum_{k_3=k_2-k_1}^{L_{\mathrm{u}}-1}{\sigma _{\mathrm{b},k_1}^{2}\sigma _{\mathrm{b},k_2}^{2}\sigma _{\mathrm{u},k_3,n}^{2}\sigma _{\mathrm{u},k_1-k_2+k_3,n}^{2}}}}  .
\end{equation}
Thus, from \eqref{EI_3_ex}, we have
\begin{equation}\label{EI_3_value}
  \mathbb{E} \left\{ \mathbf{g}_{\mathrm{u},nt}^{H}\mathbf{\Phi }^H\mathbf{G}_{\mathrm{b}}^{H}\left( \mathbf{I}_{N_{\mathrm{b}}}\otimes \mathbf{F} \right) \mathbf{R}_{\mathbf{z}_{\mathrm{q}}}\left( \mathbf{I}_{N_{\mathrm{b}}}\otimes \mathbf{F}^H \right) \mathbf{G}_{\mathrm{b}}\mathbf{\Phi g}_{\mathrm{u},nt} \right\}
  = \alpha \left( 1-\alpha \right) \xi _n+\sigma _{\mathrm{noise}}^{2}\alpha \left( 1-\alpha \right) \epsilon _n .
\end{equation}

By substituting \eqref{omega_bar_n} into \eqref{R_nt_S_E}, and substituting \eqref{eta_n_u}, \eqref{epsilon_n} and \eqref{xi_n} into \eqref{R_nt_I_E},
we arrive at
\begin{equation}\label{last_1}
  \tilde{R}_{n,t}=\log _2\left( 1+\frac{\alpha ^2p_n\varpi _n}{\alpha ^2\sum_{u\ne n}^{N_{\mathrm{u}}-1}{p_u\eta _{n,u}}+\alpha _n\left( 1-\alpha \right) \xi _n+\sigma _{\mathrm{noise}}^{2}\alpha \epsilon _n} \right).
\end{equation}
It can be observed from \eqref{last_1} that $\tilde{R}_{n,t}$ is independent of the sub-carrier number $t$.
Thus we have
\begin{equation}\label{last_2}
  R_n\approx \frac{1}{N_{\mathrm{cp}}+N_{\mathrm{c}}}\sum_{t=0}^{N_{\mathrm{c}}-1}{\tilde{R}_{n,t}}=\frac{N_{\mathrm{c}}}{N_{\mathrm{cp}}+N_{\mathrm{c}}}\tilde{R}_{n,t},
\end{equation}
and complete the proof.

\end{appendices}

\vspace{-0.2cm}

\bibliographystyle{IEEEtran}
\bibliography{IEEEabrv,Reference}

\end{document}